\documentclass[twocolumn,english,american,csquotes]{article}
\usepackage[T1]{fontenc}
\usepackage{geometry}
\usepackage{xcolor}
\usepackage{pdfcolmk}
\usepackage{array}
\usepackage{units}
\usepackage{multirow}
\usepackage{amsmath}
\usepackage{amsthm}
\usepackage{amssymb}
\usepackage{stmaryrd}
\usepackage{graphicx}
\PassOptionsToPackage{normalem}{ulem}
\usepackage{ulem}

\makeatletter

\providecommand{\tabularnewline}{\\}
\providecolor{lyxadded}{rgb}{0,0,1}
\providecolor{lyxdeleted}{rgb}{1,0,0}

\DeclareRobustCommand{\lyxsout}[1]{\ifx\\#1\else\sout{#1}\fi}

\usepackage{abstract} 
\newcommand{\makeabstract}{\@ifundefined{abstractcontent}{}{\begin{abstract}\abstractcontent\end{abstract}}}
\newcommand{\makefrontmatter}{\if@twocolumn{\twocolumn[\maketitle\makeabstract\vskip2\baselineskip]\saythanks}\else{\maketitle\makeabstract}\fi}
\theoremstyle{remark}
\newtheorem*{acknowledgement*}{\protect\acknowledgementname}

\usepackage [latin1]{inputenc}

\makeatother

\usepackage{babel}
\usepackage[style=numeric-comp,sorting=none]{biblatex}
\addto\captionsamerican{\renewcommand{\acknowledgementname}{Acknowledgement}}
\addto\captionsenglish{\renewcommand{\acknowledgementname}{Acknowledgement}}
\providecommand{\acknowledgementname}{Acknowledgement}

\begin{document}
\title{A Reputation Game Simulation:\linebreak{}
 Emergent Social Phenomena from Information Theory}
\author{Torsten En{\ss}lin$^{1,2,3}$ , Viktoria Kainz$^{1,2}$, and C\'eline
B{\oe}hm$^{4}$\\
{\small{}$^{1}$ Max Planck Institute for Astrophysics, Karl-Schwarzschild-Str.
1, 85748 Garching, Germany}\\
{\small{}$^{2}$ Ludwig-Maximilians-Universit{\small{}\"a}t M{\small{}\"u}nchen,
Geschwister-Scholl-Platz 1 80539 Munich, Germany}\\
{\small{}$^{3}$ Excellence cluster ORIGINS, Boltzmannstr. 2 85748
Garching, Germany}\\
{\small{}$^{4}$ School of Physics, The University of Sydney, Physics
Road, Camperdown, NSW 2006, Australia}}
\newcommand{\abstractcontent}{Reputation is a central element of social communications, be it with
human or \emph{artificial intelligence} (AI), and as such can be the
primary target of malicious communication strategies. There is already
a vast amount of literature on \emph{trust networks} addressing this
issue and proposing ways to simulate these networks dynamics using
Bayesian principles and involving \emph{Theory of Mind} models. The
main issue for these simulations is usually the amount of information
that can be stored and is usually solved by discretising variables
and using hard thresholds. Here we propose a novel approach to the
way information is updated that accounts for knowledge uncertainty
and is closer to reality. In our game, agents use information compression
techniques to capture their complex environment and store it in their
finite memories. The loss of information that results from this leads
to emergent phenomena, such as echo chambers, self-deception, deception
symbiosis, and freezing of group opinions. Various malicious strategies
of agents are studied for their impact on group sociology, like sycophancy,
egocentricity, pathological lying, and aggressiveness. Even though
our modeling could be made more complex, our set-up can already provide
insights into social interactions and can be used to investigate the
effects of various communication strategies and find ways to counteract
malicious ones. Eventually this work should help to safeguard the
design of non-abusive AI systems.\linebreak{}
\textbf{\emph{Key words:}} \emph{sociophysics; information theory;
dynamical systems; reputation dynamics; computational psychology;}}

\makefrontmatter
\vspace{0cm}

\section{Introduction\emph{}}

Reputation is essential to human communication. The reputation of
a speaker influences strongly how the made statements are perceived,
and these statements in turn contribute to the speaker's reputation.
A speaker is for example judged in terms of being competent, honest,
or influential. Building up a good reputation is of high importance
in any society as it determines the reach of one's messages. The reputation
network of a society is therefore an essential ingredient of it. It
is hence not surprising that adapted strategies exist and are used
to increase one's reputation. These might be harmless or harmful to
others and the society.

With the raise of social media, which are guided and influenced by
artificial intelligence (AI) systems, the need to understand the vulnerabilities
and shortcomings of social communication increases \cite{lazer2018science,aral2019protecting}.
Presently, AI systems seem to shape interactions in social media by
influencing which message reaches which participant, or by actively
participating in the communications while pretending to be human \cite{ratkiewicz2011detecting,bradshaw2017troops,bradshaw2018challenging,wu2019misinformation,ferrara2020characterizing}.
Thus, they try to substitute and manipulate human reputation networks.

Here, we explore the hypothesis that the need to gain, maintain, or
boost one\textquoteright s reputation is a central element to many
human communication strategies and may lead to some identifiable patterns
in their communication dynamics. Especially when there are uncertainties
or incomplete information about others' beliefs and intentions, reputation
effects have been shown to be essential to all kinds of interactions
\cite{pfeiffer2012value,KREPS1982253,MILGROM1982280}, which can in
turn be amplified or exploited by AI systems. In order to test this
hypothesis and explore its consequences we develop an agent based
model, called\emph{ reputation game simulation}, \emph{reputation
game}, or just \emph{game} in the following. Our reputation game simulation
can be regarded as an extension of the many existing socio-physical
simulations of gossip or rumor dynamics \cite[e.g. ][]{Deffuant2000,kitsak2010identification,2015RvMP...87..925P,DBLP:journals/corr/ClementiGPS17,geschke2019triple}
and trust networks \cite[e.g.][]{tirole1996theory,tadelis1999s,conte2002reputation,Mui01ratingsin,mui2002notions,mui2002computational,iniguez2014effects,forster2016trust,Babino8728},
which aims at a more detailed cognitive and psychological modeling
of each individual agent. For overviews on the growing field of socio-physics
we refer to Refs.\ \cite{castellano2009statistical,perc2017statistical}
and references therein.

The game simulates the opinions that agents have about other agents
and how these opinions change through communications with others.
In the game, the agents are virtual entities that are intended to
mimic certain aspects of human beings. All of them will strive for
a high reputation, and use strategies to reach this goal. Here, reputation
refers only to how honest an agent is perceived by the others, the
many other facets of human reputation are ignored. A key element of
this game is the need for agents to form an opinion about the peers\textquoteright{}
honesty, which is achieved by exchanging messages with other agents.
Since each message contains some information, although the sender
might be misinformed or even deceptive, the message receiving agents
use probabilistic logic and some rudimentary rules to determine whether
the information received should be considered trustworthy and needed
to be memorized or should be ignored and the perceived honesty of
the sender adjusted for a lie.\footnote{All messages potentially affect the payoff of the game, as they might
change reputations. Therefore there is no \emph{cheap talk} \cite{10.1257/jep.10.3.103}
in our reputation game. \emph{Cheap talk} is characterized by being
costless, non-binding, and unverifiable \cite{enwiki:1007141321}.
The first condition is not fulfilled in our reputation game as there
communication opportunities are a valuable resource for the agents.
}

Not all agents will interpret a given message in a similar way as
their beliefs on the honesty of others will differ; this makes the
game highly complex and its outcome non-trivial. The beliefs that
agents maintain about each other are stored in a simple cognitive
model, which is based on information theoretical principles (probabilistic
logic \cite{jaynes2003probability} and optimal belief approximation
\cite{2017Entrp..19..402L}), but has limited capabilities. For efficient
lie construction and detection, agents also maintain guesses on the
beliefs and intentions of the other agents; i.e.\emph{\ }they possess
a rudimentary \emph{Theory of Mind} \cite{premack1978does}.

Our motivation for developing a reputation game simulation is to provide
a theoretical framework to test and describe the elements of communication
strategies that could affect the self-esteem and reputation of agents.
How people share their information depends -- among many other things
-- on their personality \cite{STEINEL201085}. Personality traits
are not necessarily obvious, but they can be inferred by observing
behavior patterns. AI systems that manage social media communications
with the aim to keep the attention of participants \cite{paasonen2018affect}
might well discover and exploit such patterns to reach their aims.
The game presented here is intended to allow testing the link between
emerging communication patterns, cognitive states of the participating
agents, social situations, and the impact on idealized model characters.
It further might help to understand the reputation dynamics of social
groups, to decipher strategies observed in real world communications
aiming for reputation or exploiting the vulnerability of humans to
deception and manipulations, and to support the development of methods
to identify and counteract malicious communications. The latter is
becoming increasingly important with the rise of AI based communication
systems, which have opened the door for large-scale malicious communication
attacks on human minds as well as on other AI systems.

The game as it stands is built to reproduce some exaggerated behaviors
that show some resemblance with humans. It is designed as a proof
of concept that is built on principles of information theory as these
should be universal to any functional cognitive system. Thus, some
of the effects it exhibits can be expected to capture mechanisms that
show resemblance to well-known real-world sociological, cognitive,
and psychological phenomena and therefore also to help identifying
new ones. Others might just be artifacts of our ad-hoc model choices.
Our modeling follows a minimalist spirit, in the tradition of theoretical
physical modeling of phenomena, where one tries to strip down a complex
phenomenon to the bare essential. Thus, many aspects of real humans
are ignored in our toy model. Our work takes inspiration from work
in socio-physical simulations, computational cognition and computational
psychology \cite{Deffuant2000,2015RvMP...87..925P,sun_2008,DBLP:journals/corr/ClementiGPS17,tirole1996theory,busemeyer2015oxford,Sun2018,tadelis1999s,conte2002reputation,Mui01ratingsin,mui2002computational,mui2002notions,iniguez2014effects,Babino8728}
but the use of information theory \cite{6773024,Cox1946,jaynes2003probability,2017Entrp..19..402L}
enables us to consider more complex phenomena.

The paper is organized as follows: The game's principles and related
approaches are discussed in Sect.\ \ref{sec:Principles}. The basic
concepts of the game, an overview of the agent's interactions, and
the rules of the game are specified in Sect.\ \ref{sec:The-reputation-game},
whereas the mathematical details on the formalism can be found in
App.\ \ref{sec:Information-representation}. The agents' receiver
strategies are described in Sec.\ \ref{sec:Receiver-strategies-1},
while their mathematical details are given in App.\ \ref{sec:detailed-receiver-strategies}.
The different communication strategies used by agents are described
in Sect.\  \ref{sec:Communication-strategies} and their mathematical
details in App.\ \ref{sec:Detailed-communication-strategies}. Simulation
runs of our reputation game in various configurations are discussed
in Sect.\ \ref{sec:Simulations} and in more detail in App.\ \ref{sec:Detailed-figures}.
We discuss our main findings in Sect.\ \ref{sec:Discussion} and
conclude in Sect.\ \ref{sec:Conclusions}. An overview on the used
mathematical symbols can be found in Tab.\ \ref{tab:Used-variables-and}
and summaries of the different receiver and communication strategies
of agents in Tab.\ \ref{tab:receiver-strategies} and Tab.\ \ref{tab:Summary-of-agent's-strategies},
respectively.

\section{Principles of the game\label{sec:Principles}}

\subsection{Related approaches and research}

\textbf{}

\emph{Reputation game} is becoming an established term in the literature
for the game theoretical perspective on reputation systems \cite{1231515,soton262593,5552777,jaramillo2010game,https://doi.org/10.3982/ECTA7377,dowling2016winning,waller2017reputation}.
The sociological background relevant for our modeling might be spanned
by works on \foreignlanguage{english}{Goffman's Sociology \cite{JOHANSSON2007275},
communication and media studies \cite{hartley2019communication},
reputation in marketing \cite{boddy2012impact}\emph{, }and economic
game theory \cite{kreps1990game}.} More specifically, there exists
a rich literature on probabilistic modeling of trust and reputation
in a sociological and economical context \cite[e.g.][]{tirole1996theory,tadelis1999s,conte2002reputation,Mui01ratingsin,mui2002notions,mui2002computational,forster2016trust}
and on the evolution and maintenance of cooperation \cite[e.g. ][]{axelrod1981evolution,okada2020review,xu2019cooperation,PhysRevE.76.026114},
where especially in more recent works agent based modeling of reputation
in sociological contexts becomes increasingly important \cite{10.2307/26395051,smaldino2015theory,doi:10.1080/19312458.2020.1784401,10.2307/27826254,doi:10.1080/19312458.2020.1768521,RePEc:jas:jasssj:2013-11-3}.
As mentioned already, our work can be regarded as an extension of
these works into the direction of modeling the individual cognition
and psychology in more detail. Our work uses ideas and principles
from socio-physical simulations \cite{Deffuant2000,DBLP:journals/corr/ClementiGPS17,10.2307/26395051},
computational cognition \cite[and references therein]{Sun2018}, computational
psychology \cite{sun_2008,busemeyer2015oxford,https://doi.org/10.1002/wcs.1330},
information theory \cite{6773024,Cox1946,kullback1951,jaynes2003probability,2017Entrp..19..402L},
and complex adaptive system research \cite{axelrod2001harnessing}.

Since our model, just like numerous studies on gossip and rumor dynamics
\cite[e.g. ][]{Deffuant2000,DBLP:journals/corr/ClementiGPS17,geschke2019triple},
deals with the formation and evolution of opinions, we start from
similar approaches. A major difference is, however, that we want to
capture some more subtle aspects of human communication needed in
the battle of deception and counter-deception. In order to model such
behavior, we need more complex agents that are capable of both designing
targeted lies and identifying them. Both mechanisms require more parameters
and choices than might be typical for a socio-physical model \cite[see for example trust networks in][]{iniguez2014effects,Babino8728}.
We motivate our modeling choices, but these will need to be questioned,
revised, and improved in future research to provide a more realistic
setting.Equally necessary for the handling of manipulation in reputation
systems is the agents' awareness of the states, opinions, and intentions
of others. The more accurate these assessments are, the better influencing
strategies can achieve their effect, but also the more effective are
defensive strategies and detection mechanisms. Analogous to the approaches
in PsychSim simulations \cite{marsella2004psychsim,pynadath2005psychsim,ito2007decision}
or Bayesian Theory of Mind models \cite{baker2012bayesian,baker2014modeling},
our agents therefore also possess a rudimentary Theory of Mind.

In order to build up an as good as possible judgment it is necessary
for the agents to estimate incoming information correctly and to reason
with it, i.e.\ to think. In the field of computational cognition
and computational psychology a bottom up approach is often chosen,
by simulating low level brain functionalities like individual neurons
and by shaping them in a way that intelligent behavior follows \cite[e.g.][]{thomas2008connectionist}.
Here, instead a top down approach is chosen, in which a certain amount
of rationality of the agents' minds is postulated in the spirit of
\emph{Bayesian models of cognition}, \cite{l2008bayesian,martins2021agent}.
Also, we assume that rationality tries to align with Bayesian reasoning
(or probabilistic logic) \cite{Cox1946,jaynes2003probability} and
follows information theoretical principles \cite{6773024,kullback1951,2017Entrp..19..402L}.
Bayesian frameworks are used in many works on computational psychology
and cognition \cite[e.g. see][]{sun_2008} as well as information
theory \cite[ and references therein]{condon_mottus_2021}. Unfortunately,
however, this amount of rationality is limited both by limited computational
resources in simulations and mental limitations, as we know that also
human minds are far from being perfect probabilistic logical systems
\cite{Tversky&Kahneman}. The relevance of cognitive limitations w.r.t.\ perfect
reasoning has been recognized for realistic psychological modeling
\cite{sloman2008putting}, but can be implemented in different ways.
Some works use for example memory loss over time \cite{doi:10.1080/19312458.2020.1768519,5552777},
whereas our agents only memorize compressed summaries of the complex
network of cross-referencing statements they receive. Information
theory specifies how this is done optimally \cite{2017Entrp..19..402L}.
The limited information available to their reasoning is in turn likely
to make agents more vulnerable to manipulative strategies aiming to
modify the reputation network.

Our agents strongly interact, adapt to their social environment by
learning about the honesty of other agents, and try to shape this
environment to their advantage by raising or lowering the reputation
of other agents that seem to be beneficial or harmful to them, respectively.
Such hedging of other agents' reputation is also called \emph{attribution
of credit} in complex adaptive system research \cite{axelrod2001harnessing}.
Here, we focus on the impact of strategies on group dynamics and the
emerging phenomena that they create. We do not study the origin of
the strategies themselves. Some of the behavioral patterns that we
investigate are inspired from the \emph{Dark Triad} personalities
of the machiavellian, narcissistic, and sociopathic type \cite[e.g. ][]{bouchard_1973,morf2001unraveling,paulhus2002dark,jakobwitz2006dark,babiak2006snakes,thomas2014confessions}.
For example, Babiak et al. \cite{babiak2006snakes} write about psychopaths:
\begin{quote}
\emph{Specifically, their game plans involved manipulating communication
networks to enhance their own reputation, to disparage others, and
to create conflicts and rivalries among organization members, thereby
keeping them from sharing information that might uncover the deceit.}
\end{quote}
Our reputation game simulation allows to assess within its setting
how successful such strategies are, and thereby might help to explain
why such strategies have evolved in the real world in the first place
\cite{levenson1992rethinking,mealey1995primary,penke2007evolutionary}.
In the following the model assumptions of the game are stated. These
should be regarded as illustrative as several of them could have been
made differently.

\subsection{Players and their strategies}

The reputation game simulation contains at least two agents. Each
can send messages and receive them. All use both a \emph{number of
communication }and a number of \emph{receiver strategies}. These strategies
specify what we call each agent's personality. In this initial work,
we will choose the personalities such that their performance can be
studied in isolation. Strategies to send messages can be malicious
(e.g.\ manipulative, destructive, aggressive etc.) or ordinary, if
they are devoid of bad intentions. Agents interpret the messages they
receive according to their level of \emph{psychological awareness}
or \emph{intelligence}. For example a naive agent will not have the
ability to determine if a message is honest or dishonest. Agents in
our simulation can be deaf, naive, uncritical, ordinary, strategic,
anti-strategic, flattering, egocentric, aggressive, shameless, smart,
deceptive, clever, manipulative, dominant and destructive. This is
fixed at the beginning of the game upon definition of the agents in
the game.

The environment for each agent playing the game is defined by the
set of other agents. This environment is noisy with a noise level
that depends on the agents' characters (aka the used strategies) and
moods (their intrinsic parameters). Agents communicate with each other
in binary conversations, and -- in order to strip the model from
unessential complexity -- there is only one single type of conversation
topic, namely the honesty of a third agent, or that of one of the
conversation partners. The statements agents make can be honest or
dishonest. The choice is made randomly, according to agent specific,
fixed properties, namely the agents\textquoteright{} honesties. Thus,
the agents have to figure out how honest everybody is from the unreliable
statements they get and some sparse clues. The only reliable information
they get are their self-observations -- they are aware whether they
speak honestly or dishonestly -- and accidental signs of other agents
that can give their lies away. In particular, we account for the possibility
that an agent may be \textquotedblleft blushing\textquotedblright{}
when they lie. This is emulated in the game by introducing a probability
to ``blush\textquotedblright . In the remainder of the paper, the
value for this probability will be set to 10\% for most agents.

Even though some communications may be deceptive, they nearly always
contain valuable information and, as such, help determining whether
agents are trustworthy or not. For example, an agent may recognize
that a message is in strong contradiction with their own knowledge,
which then adds information about the speaker, for example. Whether
a diverging opinion in a message is recognized as valuable information
or as a sign of a deception attempt depends among other things on
the opinion that the receiver has about the honesty of the speaker.
We call this opinion the \textbf{reputation} of the speaker with the
receiver in the following. The self-reputation of an agent will be
called the \textbf{self-esteem} of that agent.\footnote{Human reputation and self-esteem are certainly more complex phenomena.}

What each agent thinks about the honesty of other agents and about
themselves\footnote{As agents have unspecified genders we use the singular \emph{they}
and \emph{them} to refer to them individually.} is described by separate probability distribution functions. If the
game contains three agents, then there should be nine such probability
functions. These probability functions depend on parameters whose
values enable to describe whether an agent is trustworthy or not.
After receiving an information, an agent will update these parameters
to reflect the new information they got. The result of this opinion
update will depend on their previous opinion of the sender and whether
they consider the message trustworthy. The apparent honesty of the
speaker plays a central role in this information update, as it partly
determines how much their message is believed. The influences of opinions
on the update of other opinions couple to the beliefs of the agents
in a complicated way and eventually lead to emergent behaviors. Indeed,
the messages of a more reputed agent, i.e. an agent perceived by the
others to be more honest, will have a larger impact than messages
from a less reputed one. However, more dishonest agents have more
opportunities to manipulate others' beliefs into a direction that
is favorable to them. Here the word ``favorable\textquotedblright{}
means achieving a higher reputation. When lying, agents can promote
others, who seem to talk more positively about them, or try to reduce
the reputation of those, who seem to make statements that are more
harmful to their own reputation. In order to keep track of whom to
support and whom to marginalize, each agent maintains a friend and
an enemy list, which are updated by the agents whenever they hear
a statement about themselves. Depending on how favorable this statement
is in comparison to other agents' statements, the speaking agent becomes
a friend or an enemy to the receiver.

When agents lie, they try to undermine the receiver's ability to detect
the lie. In addition to the speakers reputation, lie detection of
the agents is largely based on the similarity of the expressed opinion
to the receiver's own belief, thus liars try to send a slightly modified
version of this belief back to their victim. To do this, they need
to maintain an idea of what the victim believes on the different topics.

The interactions of agents allow for various strategies to boost their
own reputation. Ordinary agents for example pick conversation partners
and topics randomly and uniformly from the set of agents, while strategic
agents target highest reputed agents as communication partners and
egocentric agents prefer to speak about themselves. Reputation game
simulations allow to study the impact different strategies have on
a social group, in terms of the networks defined by the reputations
with each other and the relationships between agents.

\section{Reputation game simulation\label{sec:The-reputation-game}}

\subsection{Basic elements\label{subsec:Basic-elements}}

\begin{figure}
\vspace{-1.5cm}
\includegraphics[clip,width=1\columnwidth]{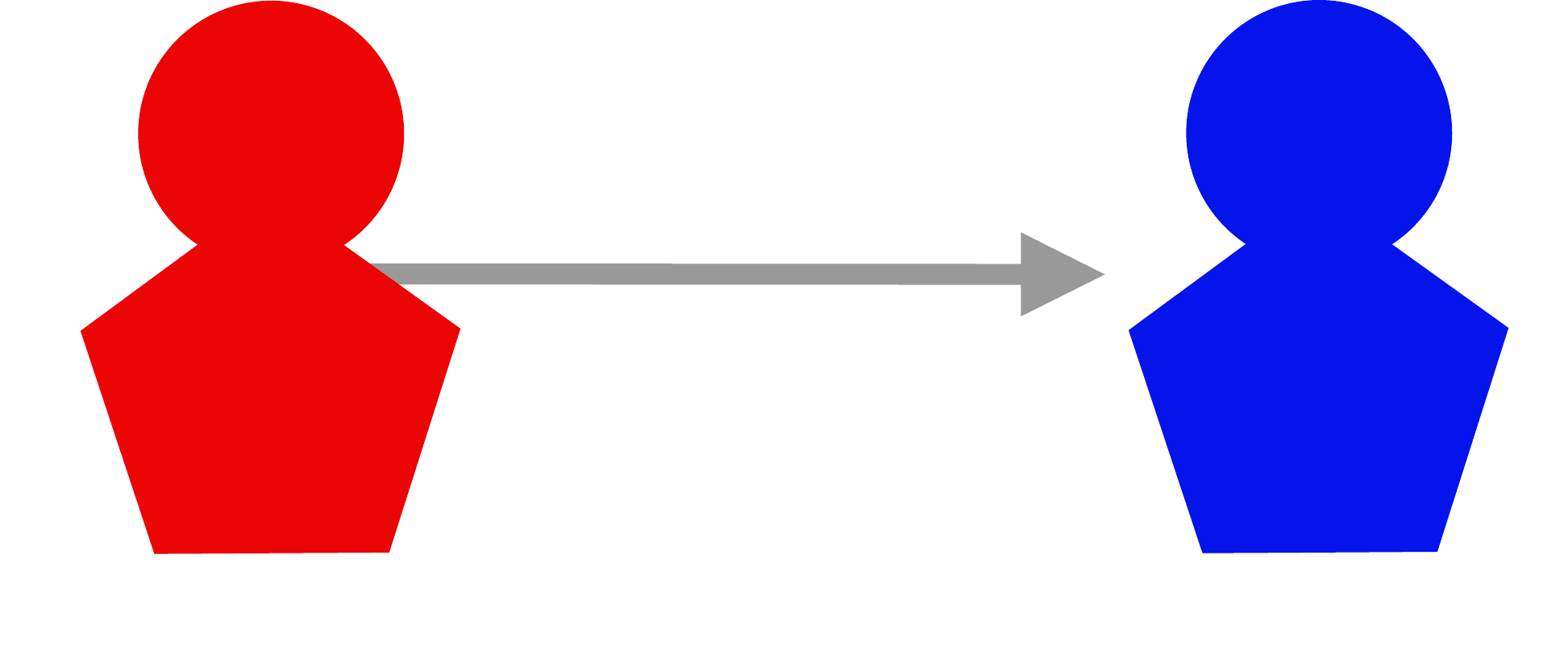}

\includegraphics[clip,width=1\columnwidth]{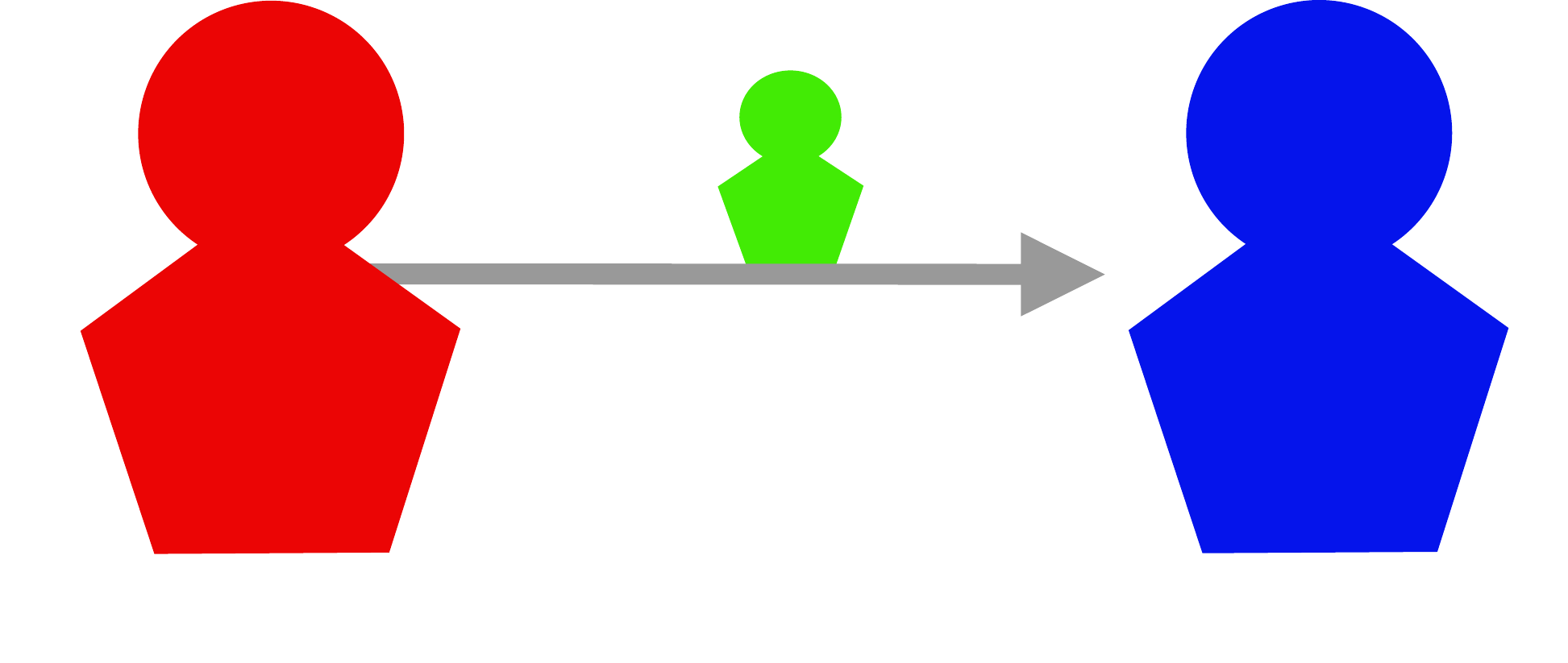}

\includegraphics[clip,width=1\columnwidth]{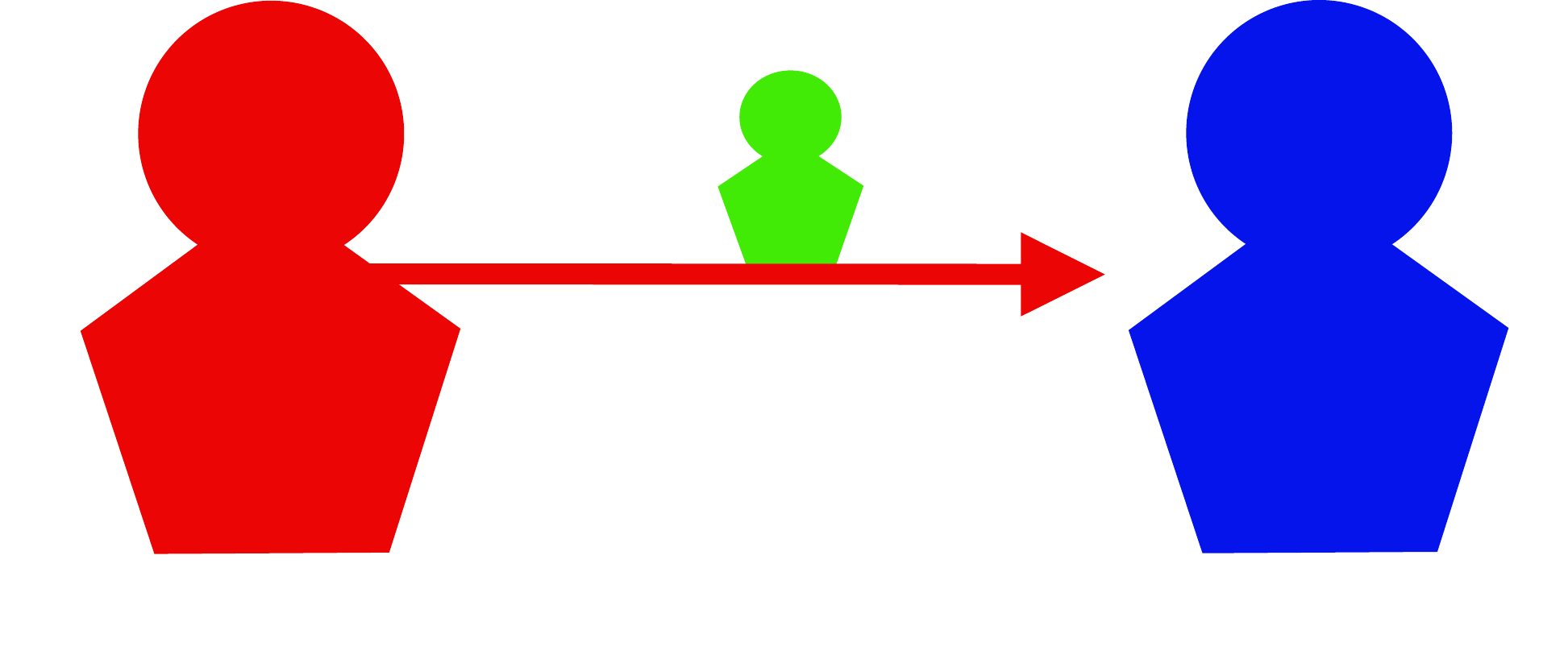}

\includegraphics[clip,width=1\columnwidth]{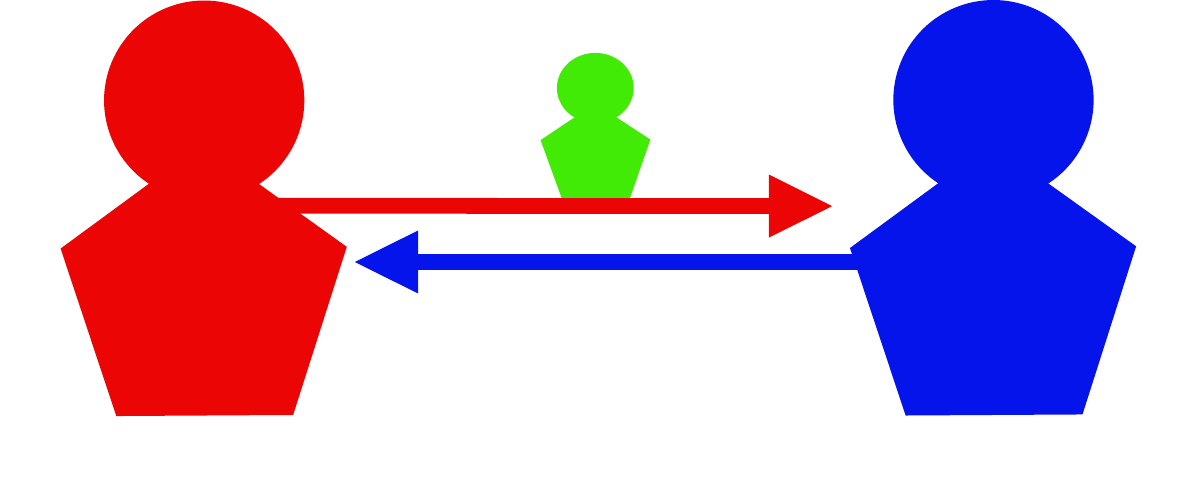}

\caption{A conversation, $\text{red}\!\!\protect\overset{\text{green}}{\rightleftarrows}\text{\!\!blue}$,
started by the agent red, who selects agent blue as the conversation
partner (first panel) and then agent green as the topic of the conversation
(second panel). After this, red talks to\textbf{ }blue\textbf{ }about
green ($\text{red}\!\!\protect\overset{\text{green}}{\rightarrow}\text{\!\!blue}$,
third panel), then blue responds to red with a statement on green
($\text{red}\!\protect\overset{\text{green}}{\leftarrow}\text{\!blue}$,
fourth panel), and finally both conversation partners update their
beliefs about themselves, each other, and green. The initiator of
a conversation (here red) is marked by having the conversation topic's
icon (here green) next to that agent's communication arrow, while
the responding agent's (here blue's) arrow has no topic marked, as
the topic for both communications is set by the initiator. Note that
red could as well have chosen one of the conversation partners, red
or blue, as the conversation topic.\label{fig:conversation}}
\end{figure}

In the game, a set $\mathcal{A}$ of $n$ agents communicates together
in sequential conversations. A conversation is defined as two agents
exchanging statements. The conversation initiating agent $a\in\mathcal{A}$
chooses a conversation partner $b\in\mathcal{A}\backslash\{a\}$ and
a conversation topic $c\in\mathcal{A}$. Then $a$ and $b$ exchange
statements about the reputation of $c$, denoted by $a\overset{c}{\rightarrow}b$
($a$ speaks to $b$ about $c$) and $a\overset{c}{\leftarrow}b$
($b$ speaks to $a$ about $c$) for the individual communications,
as well as by $a\overset{c}{\rightleftarrows}b$ for the conversation
(of $a$ and $b$ about $c$). Finally both $a$ and $b$ update the
reputation of $a$, $b$, and $c$ according to their experiences
and interpretations thereof. Agent $c$ could be a third agent, the
initial speaker $a$, or the initial receiver $b$. A game round is
a sequence of $n$ conversations, in which each agent initiates one
conversation. The game ends after a predefined number of rounds. A
single conversation and a conversation round are depicted in Fig.\ \ref{fig:conversation}
and \ref{fig:conversation-round}, respectively. The goal of each
agent is to eventually obtain a reputation as high as possible.

The belief an agent $a$ maintains about some other agent $c$'s honesty
$x_{c}\in[0,1]$ is given by a parametrized, one-dimensional probability
density distribution (PDF)\footnote{We denote probabilities with $P$ and PDFs with $\mathcal{P}$. They
are related via integration: $P(x\in[x_{1},x_{2}]|I)=\int_{x_{1}}^{x_{2}}dx\,\mathcal{P}(x|I)$.
Note that probabilities take values in $P\in[0,1]$, whereas PDFs
in $\mathcal{P}\in\mathbb{R}_{0}^{+}=[0,\infty]$. Bayes' theorem
applies to both, so that a strict discrimination between those is
not always necessary. We therefore use the word ``probability''
for both, probabilities and PDFs.} $\mathcal{P}(x_{c}|I_{ac})$, where $I_{ac}=(\mu_{ac},\lambda_{ac})\in\mathbb{R}^{2}$
is a tuple of parameters, which store the knowledge of $a$ on $x_{c}$.

In what follows, we choose $\mathcal{P}(x_{c}|I_{ac})$ to be a Beta
distribution, as it is also used in related works \cite[e.g.][]{mui2002notions}
and is a natural choice, as shown in App.\ \ref{subsec:Belief-representation}.
When $\mu_{ab}$ and $\lambda_{ab}$ are natural numbers they can
be interpreted as being respectively the number of honest and dishonest
statements that $a$ believes $b$ has made. We allow, however, both
parameters to take values in the continuous interval $(-1,10^{6}]$
(chosen for numerical reasons).  The two parameters of the distribution
allow to express an assumed mean honesty $\overline{x}_{ac}$, which
we identify with $c$'s honesty according to $a$ (aka $c$'s reputation
with $a$), as well as the uncertainty $\sigma_{ac}$ around that
mean, which expresses how sure $a$ is about $c$'s reputation in
terms of a standard deviation. The message from $a$ to \textbf{$b$}
consists of the topic $c$ and $a$'s belief encoding parameters $I_{ac}$
in case $a$ was honest, or a distorted version thereof, in case $a$
lied. Thus, agents own, maintain, and exchange beliefs in form of
probabilities.

Similarly, agent $a$ will form their own views of the honesty of
agent $b$ they communicate with. This is embodied by the set of parameters
$I_{ab}$ for each $b\in\mathcal{A}$. We will refer to $a$'s \textbf{belief
state} as $I_{a}=(I_{ab})_{b\in\mathcal{A}}$. If not mentioned otherwise,
belief states do not contain information at the beginning of the game.
But this changes in the course of the game, as agents' belief states
evolve with time in accordance with their experiences.

The moves an agent $a$ can make in the game are to choose a conversation
partner $b$ and a topic $c$, as well as to decide to lie, as depicted
in Fig.\ \ref{fig:conversation}. By default, these choices are made
randomly. For example, whether an agent $a$ communicates honestly
is chosen randomly according to agent $a$'s honesty parameter $x_{a}\in[0,1]$.
This specifies the frequency with which $a$ is honest and therefore
$x_{a}\equiv P(a\text{ honest}|x_{a})$. Other choices might be guided
by the agent's strategy. For example, we will define strategic agents,
who preferentially pick highly reputed agents as conversation partners.

\subsection{Information handling\label{subsec:Information-handling}}

When an agent $b$ receives a message from agent $a$ about agent
$c$, agent $b$ has to judge how reliable the message is. If the
message appears honest, the information contained in the message should
be used to update $b$'s belief about $c$. The fact that $a$ was
honest is also recorded by $b$. If the message appears to be a lie,
$b$ should discard the message's content and only record a lie for
$a$. The problem is that $b$ rarely knows whether a message is honest
or not, and can at best assign a probability to these possibilities.
As a consequence, the PDF describing the correct posterior knowledge
that an agent should have after receiving a message is a superposition
of these two possible updates.

Furthermore, the PDFs, with which agent $b$ describes the honesty
of speaker $a$ and of topic $c$, become entangled, as $b$ needs
to recognize whether $a$ sent genuine information about $c$. The
functional form of this potentially bi-modal, two dimensional, and
potentially entangled PDF $\mathcal{P}(x_{a},x_{c}|d,I_{b})$, with
$d$ the data obtained by $b$ from the conversation and $I_{b}$
the prior knowledge of $b$, cannot be precisely captured by the functional
form of the one dimensional PDFs $\mathcal{P}(x_{a}|I'_{ba})$ and
$\mathcal{P}(x_{c}|I'_{bc})$ agent $b$ uses to store the updated
knowledge $I'_{b}$. These only allow for product belief states of
the form $\mathcal{P}(x_{a},x_{c}|I'_{b})=\mathcal{P}(x_{a}|I'_{ba})\,\mathcal{P}(x_{c}|I'_{bc})$,
which cannot express entanglements. Thus, information gets lost in
an update, and agent $b$ should choose the new parameters $I'_{b}$
such that as much information as possible is kept from $\mathcal{P}(x_{a},x_{c}|d,I_{b})$.

We use the \textbf{principle of minimal information loss} for choosing
the parameters in $I'_{b}$. Information loss can be quantified using
the Kullback-Leibler (KL) divergence \cite{kullback1951}, which measures
the information difference between original and approximate PDF. This
choice of the information measure rests on a solid mathematical proof,
which states, that in the absence of any other criteria\footnote{In human psychology, however, additional criteria might be relevant
that can lead to deviations from a pure KL based data compression.
Recognizing liars might be more essential than differentiating between
mostly honest people. Consequently, a positive-negative asymmetry
of diagnosticity of information seems to be used by human minds when
deciding how to store morality related information (e.g.\ honesty-dishonesty)
\cite{skowronski1987social}. In this first incarnation of our reputation
game, we ignore such subtleties.}, the KL is the only consistent choice to quantify how optimal a belief
update is \cite{2017Entrp..19..402L}. The KL based principle of minimal
information loss has also proven to be extremely useful in many areas,
like information field theory \cite{2009PhRvD..80j5005E,2010PhRvE..82e1112E,https://doi.org/10.1002/andp.201800127}
and information field dynamics \cite{2013PhRvE..87a3308E,2013PhRvE..87b2719R,2018PhRvE..97c3314L,2018PhRvE..98d3307D}.

Some of the information will inevitably get lost during the belief
update of an agent due to the limited flexibility of the parametric
form and the product structure of belief states. This makes them vulnerable
to rumors, misinformation, and self-deception, which in turn can be
exploited by special communication strategies of deceptive agents.

For a more detailed introduction into the agent's information handling
we refer to App.\ \ref{sec:Information-representation}.

\subsection{Belief update}

In the following, we briefly explain the update due to the initial
communication $a\overset{c}{\rightarrow}b$. The update due to the
response $a\overset{c}{\leftarrow}b$ is analogous. Mathematical details
of the update can be found in App.\ \ref{subsec:Believe-update}
and \ref{subsec:Optimal-believe-approximation}.

First, the speaker $a$ updates their self-image according to whether
$a$ was honest or lied in the conversation, i.e. agent $a$ increases
$\mu_{aa}$ by one if the message was honest, otherwise $\lambda_{aa}$
is increased by one, as explained in Sect.\ \ref{subsec:Believe-update}.
The information agent $b$ uses for the update is the overall communication
setting $a\overset{c}{\rightarrow}b$, the messages exchanged $J(t)$,
the blushing observation $o_{t}$, and $b$'s assessments of $x_{a}$
and $x_{c}$. We call the tuple $d_{t}=(a\overset{c}{\rightarrow}b,J(t),o_{t})$
the data of the communication at time $t$ and $\mathcal{P}(x_{a},x_{c}|I_{b})=\mathcal{P}(x_{a}|I_{ba})\,\mathcal{P}(x_{c}|I_{bc})$
the prior of the update. The update of agent $b$ proceeds in three
stages:

\paragraph{Assessment of message:}

First, $b$ constructs the joint posterior probability function $\mathcal{P}(x_{a},x_{c}|d_{t},I_{b},A_{b})\propto\mathcal{P}(d_{t}|x_{a},x_{c},I_{b},A_{b})\,\mathcal{P}(x_{a},x_{c}|I_{b})$.
This expression contains the likelihood $\mathcal{P}(d_{t}|x_{a},x_{c},I_{b},A_{b})$
to obtain the message $d_{t}$. The functional form of this likelihood
depends on agent $b$'s receiver strategy (see Tab.\ \ref{tab:receiver-strategies}).
A receiver strategy is the background information that determines
the form of the likelihood $b$ is using, given $x_{a}$, $x_{c}$,
$I_{b}$ and additional auxiliary information $A_{b}$, which forms
the agent's Theory of Mind knowledge basis. This auxiliary information
is dynamical and is used by $b$ for orientation. It comprises of
$\kappa_{b}$, the level of surprise marking for $b$ the border between
typical lies and honest statements, agent $b$'s guesses for agent
$a$'s beliefs and intentions w.r.t. $c$, $I_{bac}$ and $\widetilde{I}_{bac}$,
respectively, and other quantities.

\paragraph{Information compression:}

Second, This joint posterior is then approximated by the parametric
form used to store beliefs, $\mathcal{P}(x_{a},x_{c}|I'_{ba},I'_{bc})=\mathcal{P}(x_{a}|I'_{ba})\,\mathcal{P}(x_{c}|I'_{bc})=\text{Beta}(x_{a}|\mu'_{ba},\lambda'_{ba})\,\text{Beta}(x_{c}|\mu'_{bc},\lambda'_{bc})$,
by choosing values of $I'_{ba}=(\mu'_{ba},\lambda'_{ba})$ and $I'_{bc}=(\mu'_{bc},\lambda'_{bc})$,
which then become the new belief parameters at time $t+1$. The principle
of least information loss is used to compress data, but information
is inevitably lost in this step, since (i) the parametric form of
the beta function is not able to represent all posterior structures
and (ii) the entanglement of the variable $x_{a}$ and $x_{c}$ due
to the received information cannot be represented by the product structure
of the belief representation. The mathematical details of this update
can be found in App.\ \ref{subsec:Optimal-believe-approximation}
and its numerical details in App.\ \ref{sec:Numerical-Implementation-Details}.

\paragraph{Theory of Mind update:}

Finally, the Theory of Mind of the agents, which tries to track the
other's beliefs and intention, as well as the typical scale surprise
used by them to lie are updated. The corresponding auxiliary variables
are stored in $A_{b}$. How this is done in detail is explained in
App.\ \ref{subsec:Auxiliary-parameters-update}. 

\subsection{Strategies}

\begin{figure}[t]
\includegraphics[clip,width=1\columnwidth]{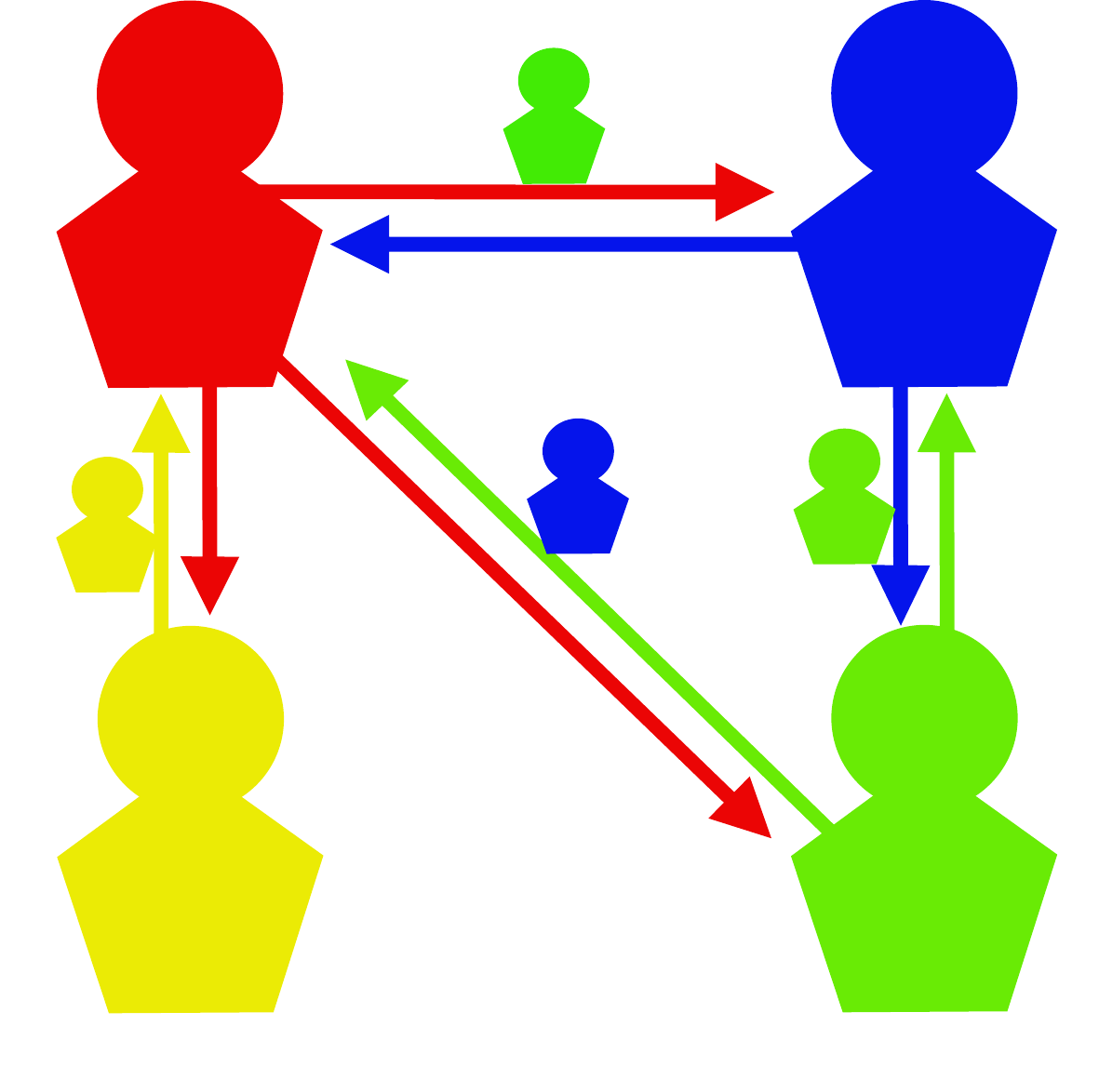}\caption{In a game round, each agent is initiator of one conversation (arrow
with small agent icon directly next to it), but not necessarily addressed
as a conversation partner (arrow without agent icon directly next
to it). The opinion expressed by agent red to blue about agent green
can affect green's self-esteem, as it might partly propagate to green
via blue. Whether the statement by blue on green was more positive
or negative decides whether green regards blue as a friend, and therefore
eventually how green speaks about blue to red. Thus, how red speaks
about green can affect later on what red is made to believe about
blue. This illustrates the high level of entanglement of the agent's
interactions. \label{fig:conversation-round}}
\end{figure}

A \textbf{communication strategy} consists of a set of rules about
whom to pick as a conversation partner and as a topic, as well as
rules that guide the decisions on how to lie. For example, a malicious
communication strategy could be to expose a victim to propaganda in
form of massive self-appraisal. This can lead to a nearly complete
conversion of the victim to the position expressed in the propaganda,
as we show later on (Sect.\ \ref{subsec:Propaganda-and-resilience}).

A communication $a\overset{c}{\rightarrow}b$ received by agent $b$
may provide relevant information on agent $c$ if the information
is trustworthy, but also on the speaker $a$'s honesty and intention.
A receiver might judge the honesty of a message on the basis of various
signs of deception. How an agent analyzes a message is the agent's
\textbf{receiver strategy}. All, but naive agents, use the reputation
of the speaker with them as one of the indicators that gives weight
to the message. This makes more reputed agents automatically more
influential and being influential is what we regard as the agents'
ultimate goal.

Since agents do not know their objective honesty a priori, they have
to learn this from self-observations and feedback of the other agents.
The resulting self-esteem is an important variable as well, as it
is the basis of the communicated self-picture of an agent in case
of honest communications. Obtaining a high self-esteem might therefore
be a secondary goal of agents, as this permits self-appraisal without
the risks involved in lying.

\subsection{Rules of the game\label{sec:The-rules-of-the-game}}

The protocol of our reputation game consists of the following steps:
\begin{enumerate}
\item A set of labeled agents $\mathcal{A}=$\{red, black, cyan, yellow,
blue, ...\} participates in the game. Each agent $a\in\mathcal{A}$
has a number of static properties ($x_{a}$, the set of used strategies,
$\ldots$) specifying the agent's communication strategy and a set
of dynamical variables ($I_{a},\widetilde{I}_{a},K_{a},\text{\ensuremath{\kappa_{a}}},\text{\ensuremath{\ldots}}$),
being the parameters of the agent's world model.
\item Time $t$ is measured in communication events, which happen sequentially.
\item The central property of each agent $a$ is the agent's frequency to
be honest, $x_{a}=P(a\text{ is honest}|x_{a})$. Other properties
determine other aspects of the agent's communication and receiving
strategies.
\item The belief of agent $a$ regarding the honesty of agent $b$ is encoded
in the parametric probability distribution $\mathcal{P}(x_{b}|I_{ab})=\text{Beta}(x_{b}|\mu_{ab},\lambda_{ab})$,
where $I_{ab}=(\mu_{ab},\lambda_{ab})$ is the tuple of dynamical
variables parameterizing $b$'s belief and Beta is the beta distribution.
The joint belief state of an agent regarding the honesty of all other
agents is set to the direct product of the single agent beliefs, $\mathcal{P}(\underline{x}|I_{a})=\prod_{b\in\mathcal{A}}\mathcal{P}(x_{b}|I_{ab})$,
with $\underline{x}=(x_{b})_{b\in\mathcal{A}}$ and $I_{a}=(I_{ab})_{b\in\mathcal{A}}$.
This implies that agents are unable to keep track of entangled information
of the sort ``only one of $b$ and $c$ can be honest, not both''.
Such an knowledge state would actually be appropriate in case the
two agents $b$ and $c$ accuse each other to be liars.
\item A conversation $a\overset{c}{\rightleftarrows}b$ is an exchange of
statements between two agents $a$ and $b$ about agent $c$, who
is the topic of the conversation. The conversation starts at time
$t$ with the conversation initiator $a$ choosing another agent $b\in\mathcal{A}\backslash\{a\}$
(excluding themselves to avoid a soliloquy without information exchange),
and $c\in\mathcal{A}$ out of the set of all agents. Then $a$ composes
and transmits a statement $J(t)$ about $c$ to $b$, which we also
refer to as $J=J(t)=J_{a\overset{c}{\rightarrow}b}(t)$ to clarify
that message $J$ is associated with the communication $a\overset{c}{\rightarrow}b$.
The initial communication is followed by a reciprocal message $J(t+1)$
from $b$ to $a$ about $c$, denoted by $a\overset{c}{\leftarrow}b$.
The full conversation is denoted by $a\overset{c}{\rightleftarrows}b$.
Only after the statements are exchanged, the agents update their beliefs.
By choosing conversation partner and topic, the initiating agent $a$
basically requests $b$ to make a statement on $c$ (which could as
well be $a$ or $b$). How agents make these choices depends on their
communication strategy (see Tab.\ \ref{tab:Summary-of-agent's-strategies}).
Agents that are initiating conversations about themselves, for example,
will get to know who are their friends and enemies. See Fig.\ \ref{fig:conversation}
for an illustration of a conversation.
\item The game is played in a number of $N_{\text{rounds}}$ rounds. In
each round, each agent initiates exactly one conversation with another
agent, which consists of two communications and subsequent belief
updates. The game ends after $N_{\text{rounds}}$ rounds at time $t_{\text{end}}=2\,N_{\text{rounds}}|\mathcal{A}|$.
See Fig.\ \ref{fig:conversation-round} for an illustration of a
round of conversations.
\item The format of the messages is that of the internal belief representation.
For an honest communication $a\overset{c}{\rightarrow}b$ at time
$t$ we therefore have the message $J_{a\overset{c}{\rightarrow}b}(t)=I_{ac}(t)$.
\item Whether an agent $a$ lies in a given conversation is usually decided
by chance, with the agent specific frequency $x_{a}$. When lying,
all, except one category of agents called \emph{shameless agents},
risk to accidentally reveal to their communication partner the fact
that they are lying. The probability of being caught lying is $f_{\text{b}}=0.1$.
Here b stands for ``blushing'', to mimic the fact that agents can
give away the act of lying. This gives other agents some direct information
about one's honesty. We denote the observation of the blushing status
of the speaker at time $t$ as $o_{t}$. Note, that agent $b$ can
also become convinced that $a$ was honest. This happens when $a$
makes a \emph{confession}, a disadvantageous self-statement (without
blushing).
\item After a conversation about agent $c$, both communicating agents $a$
and $b$ update their beliefs in response to the information perceived
about all involved agents, i.e. $a$, $b,$ and $c$ as explained
before. With this slightly delayed update for the initial receiver
$b$, a communicated opinion $J_{a\overset{c}{\rightarrow}b}$ of
the conversation initiator $a$ is not directly mirrored back to $a$
in $b$'s response $J_{b\overset{c}{\rightarrow}a}$. Side effects
on other agents $\mathcal{A}\backslash\{a,b,c\}$ need not to be taken
care of in the update due to the independent product structure of
the belief representation, as detailed in App.\ \ref{subsec:Believe-update}.
\item After the game is over, the performance of the agents is judged with
respect to a number of performance metrics, such as the average reputation
of an agent, which is an average of the other agents' posterior means
on the agent's honesty, the frequency of obtaining a top reputation,
and the like.
\end{enumerate}
\begin{figure}
\centering{}\includegraphics[viewport=10bp 0bp 920bp 630bp,width=1\columnwidth]{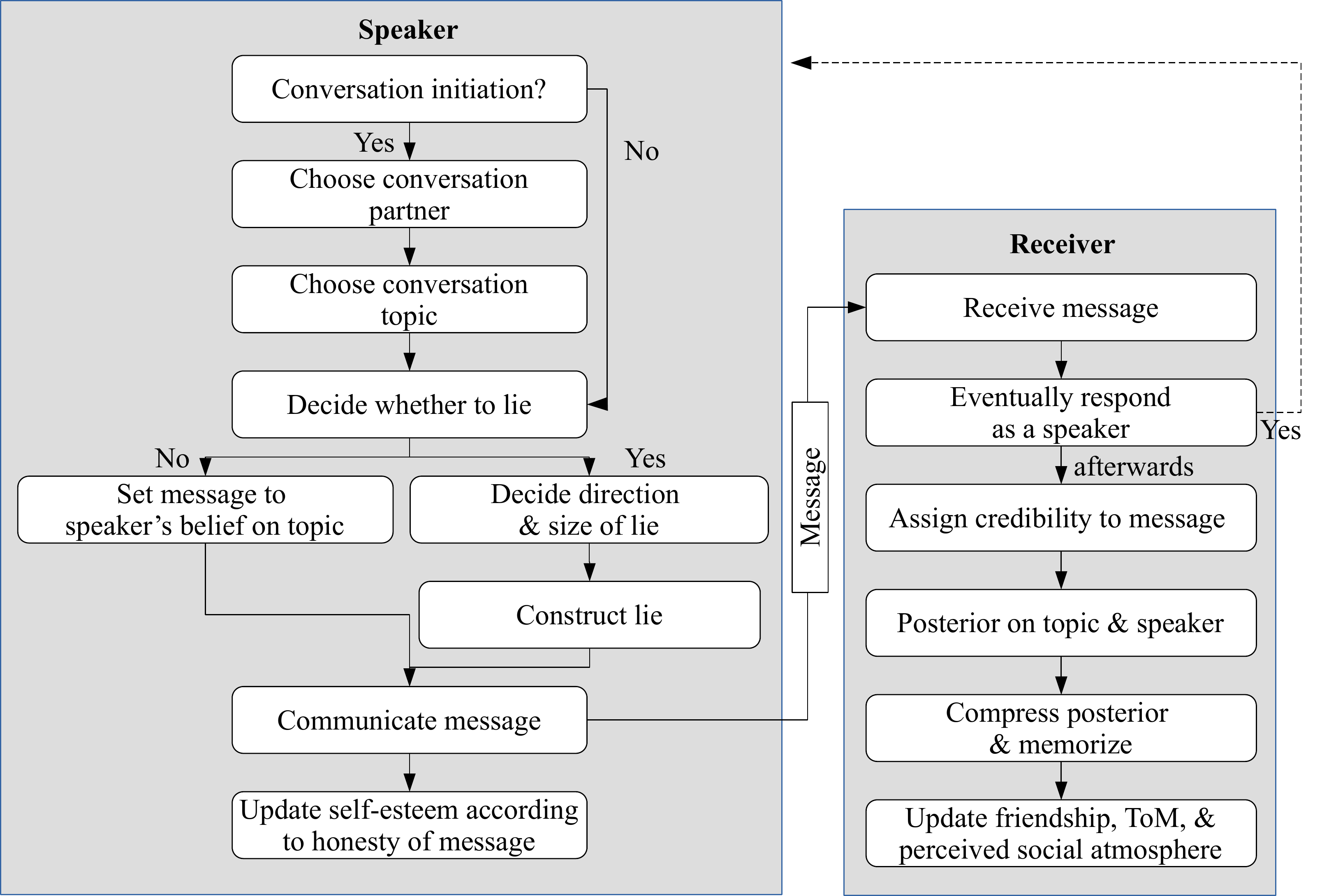}\caption{\-\-\-Stepwise actions when generating a message as a speaker and
interpreting a message as a receiver.\label{fig:talking}}
\end{figure}

\section{Receiver strategies\label{sec:Receiver-strategies-1}}

Receiver strategies determine how agents deal with incoming information
and generally which mental updates they perform after having received
a message $J$. A stepwise description of this update mechanism is
shown in the right part of Fig. \ref{fig:talking}. Most importantly
the receiving agent $b$ has to judge the message's content. In order
to find the right trade-off between believing the message and thereby
gaining new information, and distrusting the message to not blindly
follow a possible lie the receiver has to assign a credibility value
$y_{J}:=\text{\ensuremath{\mathcal{P}(J\text{ is honest}|d,I_{b})}}$
to the message. Here, $d$ is again all the data agent $b$ got from
the conversation and $I_{b}$ is the information agent b already had
before the communication. In reality there are of course endless means
and criteria humans use to determine the trustworthiness of others
and their statements \cite{sherif1961social}. In our model we try
to capture at least some of these means and test different combinations
thereof to observe their influence. These combinations, which we call
receiver strategies, basically differ in their usage of the conversation
data $d$ and the agents' mental capacities in both lie detection
and maintaining an accurate Theory of Mind.

\subsection{Lie detection strategies\label{subsec:Lie-detection}}

First of all there is the \textbf{naive agent}, who listens to the
message and naively trusts it all the time, i.e. assigns the credibility
value $y_{J}=1$ to every message. This, of course, is not a very
sophisticated receiver strategy, but the simplest possible and will
therefore serve as a reference for the others. A second type are \textbf{deaf
agents}, who primarily use the direct sign whether or not the speaker
has blushed to identify lies, as well as the current reputation of
the speaker in their eyes. The message's content, however, is completely
ignored. This way deaf agents do not risk being caught by lies, but
also do not benefit from honestly communicated, valuable information,
which makes their learning progress very slow. A bit more advanced
is the receiver strategy of \textbf{uncritical agents}. Just like
all of the following strategies, uncritical agents listen to the communicated
message and use its content to rapidly gain knowledge. This, together
with the right judgment criteria for credibility, clearly is superior
to ignoring the content. Besides the speaker's reputation and blushes
uncritical agents can also observe a second clear signal, confessions.
If a statement about the speaker themselves turns out to be much worse
than what the receiver has believed about the speaker so far, it can
be assumed that the message has been honest. Because otherwise, if
the speaker had lied, the message would have been based on the knowledge
of the receiver (as assumed by the speaker) and would have been biased
upwards, i.e. the speaker would have made a statement that is more
positive than what the receiver believes. On top of those criteria,
\textbf{critical agents} additionally use the content of the message
to judge its credibility based on surprise. When the surprise of a
received message is high, or in other words its content deviates strongly
from the receiver's current belief, the latter becomes skeptical and
down rates the message's credibility. Analogously, if the surprise
is low and the communicated message fits well to the receiver's belief
it is much more likely to be accepted as true. This strategy seems
totally natural at a first glace, although it can easily be exploited.
In reality, most of the time lies are not just randomly created statements
that others are supposed to accept, but rather are designed carefully
and adapted to the intended receiver. In terms of our simulation that
means, that lies are always based on the speaker's assumption on the
receiver's current opinion and from there shifted into the desired
direction. \textbf{Smart agents} are aware of this mechanism and use
it additionally to all previous criteria in order to detect lies.
For this they compare the received message with different scenarios
the message could have been created. When the message seems to match
the speaker's real opinion it is likely to be honest. When, however,
the message better coincides with the speaker's guess on the receiver's
opinion plus a little distortion, smart agents identify it as a probable
lie. This requires the awareness of other's opinions, i.e. a Theory
of Mind, and helps the agents to distinguish between lies and honest
statements on an advanced level. A more detailed and mathematical
description of the above receiver strategies can be found in App.
\ref{sec:detailed-receiver-strategies} and the summarizing table
\ref{tab:receiver-strategies}.

After having assigned a credibility value to the received message,
taking into account all criteria that are set by the agents' strategies,
they update their knowledge accordingly. This way our model allows
for a continuous transition between trust and distrust, which leads
to realistic, individual and situation based decisions and therefore
very complex learning processes.

\subsection{Theory of Mind update\label{subsec:Auxiliary-parameter-update}}

Besides the update of the agents' opinions on others' honesties, there
are also the Theory of Mind parameters the agents need to track in
order to be oriented well in their environment. However, unlike the
lie detection, this is updated in the same way for all types of agents.
First, the agents update their friends and enemies lists in case the
conversation topic were themselves. For this the other's statement
is compared to the opinions the agent heared aboutemselves last from
each of the others. If the reputation expressed in the message at
hand was above the median of the others', the speaker is seen as a
friend from now on. If it was below the median, the speaker is regarded
an enemy in the following. Friends and enemies are therefore defined
in our simulation as agents, who have recently spoken more positively
or negatively about a certain agent w.r.t. the peer group, respectively.
Second, the agents update their Theory of Mind, i.e. their knowledge
on the others' beliefs. If on the one hand the communicated message
was honest, the receiver directly got to know the belief of the speaker.
If on the other hand the message was a lie, the receiver at least
got information on what the speaker wants them to belief and remembers
that. Since most of the time it is not clear which one of the cases
applies, the message credibility value $y_{J}$ is used as a weight
that decides how much this message J influences the receiver's assumption
about the speaker's belief (not at all if $y_{J}=0$, completely if
$y_{J}=1$), and its complement, $1-y_{J}$, how much the assumption
about the speakers intention. Finally, the surprise caused by the
message is measured and remembered. Here, the median of the last ten
such surprise values is used as a reference scale $\kappa_{a}$ that
each agent $a$ stores individually. On the one hand this serves as
a reference to distinguish lies from honest statements in the next
conversations, since there is no absolute scale how large lies may
be. On the other hand the agents use this scale also to determine
the right size of their own lies, in order them to have a chance to
go unnoticed, assuming the receiver uses a similar value to detect
lies. We identify the scale with the by the agent perceived social
atmosphere. This updating mechanism of the reference surprise scale
enables the adaption of the agents' reasoning to a changing social
atmospheres, maybe in a way that somehow resembles such processes
in real-world social systems. More about the consequences of the surprise
scale adaption can be found in section \ref{sec:Simulations} and
its technical details in App. \ref{subsec:Auxiliary-parameters-update}.

\section{Communication strategies\label{sec:Communication-strategies}}

Communication strategies, in comparison to receiver strategies, are
a set of rules (or frequencies) that specify how to select the conversation
partner $b$, which topic $c$ to choose, how frequently to lie, and
in which way, i.e. how to send messages. Fig.\ \ref{fig:talking}
again shows a step-wise description of these processes. We will use
the names of strategies also as adjectives for agents, meaning that
an aggressive agent always uses an aggressive communication strategy.
In our simulations we define several basic communication strategies,
which can be combined resulting in so called special strategies.

The basic reference communication strategy is that of an ordinary
agent, and all other basic strategies are described in terms of their
differences to this in Sect.\ \ref{subsec:Basic-communication-strategies}.
For \textbf{special agents}\footnote{The fictional \emph{special agent 007}, for example, exhibits strategies
that resemble the here introduced manipulative and destructive strategies,
which are both special in our nomenclature.}, i.e. agents that use a combination of various basic strategies,
the reference will be the clever agent as discussed in Sect.\ \ref{subsec:Special-communication-strategies}.
An overview of the different communication and receiver strategies
is given by Tab.\ \ref{tab:Summary-of-agent's-strategies}.

We would like to emphasize at this point, that all the following strategies
are ad-hoc choices we made in the hope that they capture some aspects
of real-life personality types. Of course, we do not claim that these
strategies are even close to being realistic enough to emulate real
personalities, nor that our selection of strategies is exhaustive.
They only serve as a possible set of strategies that we want to investigate
and observe the reputation dynamics resulting from these particular
choices.

\subsection{Basic strategies\label{subsec:Basic-communication-strategies}}

\textbf{Ordinary agents} make all their decisions randomly. They choose
a conversation partner randomly among all other agents, a communication
topic among all others or themselves, communicate honestly according
to their predifined honesty value, lie positively about friends and
negatively about enemies and use a critical receiver strategy for
lie detection. Two opposing strategies are used by \textbf{strategic
and anti-strategic agents}, who choose their conversation partners
according to the agents' reputation, i.e. their presumed honesty.
Strategic agents thereby aim for the most reputed, presumably most
honest agents, whereas anti-strategic agents preferentially talk to
least reputed, and therefore presumably most dishonest ones. These
two strategies aim for different effects. Being strategic, the agents
benefit when they manage to convince their interlocutors of their
own honesty, as this opinion is then efficiently passed on to third
parties by the honest and reputed agents. First, those mostly say
what they belief and second, they are trusted by the others. Being
anti-strategic, however, is not aiming that the targeted agent distributes
their honest opinions, but instead that lies are favorable for the
anti-strategic agent. Since less reputed, and therefore presumably
more dishonest agents can be assumed to lie more frequently, anti-strategic
agents mostly benefit from simply being their friends and therefore
passively taking advantage of all the good rumors that dishonest agents
spread about them. Other types of agents choose the conversation topic
with special care.\textbf{ Flattering agents}, for example, make their
interlocutors compliments, i.e. make positive lies to their interlocutors
whenever possible in order to befriend them. Of course it is therefore
very reasonable to always choose the conversation partner also as
conversation topic, as it is part of the flattering strategy.\textbf{
Egocentric agents} also choose the conversation topic with a special
motivation, namely to promote themselves. This is why in more than
half of the cases where egocentric agents start a conversation they
choose themselves as conversation topics. A third way of choosing
the topic is used by \textbf{aggressive agents}, who always talk about
their enemies in order to harm their reputation. At the same time,
however, they do not promote themselves nor their friends, such that
the aggressive strategy might generally cause low reputations in a
simulation for all participating agents, but is designed to damage
others' even more than their own. \textbf{Shameless agents} do not
run the risk, in comparison to all others, to blush when lying. This
makes it harder for their interlocuters to identify lies, which in
turn helps to keep the agents' reputation up. Another strategy to
appear honest is to create a complete illusion of oneself and never
let reality show through. This is what \textbf{deceptive agents} do,
by lying in every single communication. Thus, they never make confessions
or otherwise reveal their true opinions to anyone else.

A more detailed and mathematical description of these basic strategies
can be found in App. \ref{sec:Detailed-communication-strategies}.

The impact of most of these strategies on the agent's reputation is
modest if used alone. Therefore, we use matching combinations where
the individual basic strategies can unfold their power best. These
are what we call special strategies.

\subsection{Special strategies\label{subsec:Special-communication-strategies}}

A \textbf{clever agent} is smart and deceptive and is the reference
point for the special agents, which are all smart and deceptive as
well. Smartness permits the agent to understand the beliefs and intentions
of other agents better, allowing for more precisely placed lies.

The \textbf{manipulative agent} is clever, flattering, and anti-strategic.
This should enable the agent to efficiently identify and befriend
other dishonest agents, who more frequently praise their friends and
therefore the befriended manipulative agent than other agents. Manipulative
agents should thereby become popular and influential. Their presence
in a social group is expected to lift the self-esteem of the group
members, owning to flattering. This lift should be stronger for less
reputed agents, as those are preferentially targeted for conversations.
These are often the more dishonest agents, which then, as a friend
of the manipulative agent, hopefully give positively biased testimonies
about the manipulative agent. As a consequence, we expect manipulative
agents to frequently establish mutual friendship.

The \textbf{dominant agent} is clever, egocentric, and strategic.
The agent's communications are targeting the most reputed agents to
praise themselves. If successful, these will most efficiently propagate
a positive image of the dominant agent to others as well as mirroring
this image back to the dominant agent themselves. The latter effect
might efficiently boost the self-esteem of the dominant agent. Dominant
agents will be best informed about their own reputation, by making
themselves the conversation topic. This will, however, couple their
self-esteem more to their reputation compared to other special agents.
This will also provide them with a more accurate friend and enemy
classification, as they see how other agents talk about them. This
classification will not be accurate, however, in case they are interacting
with manipulative agents, as the latter speak differently about a
topic depending on whether the topic is also their conversation partner
or not. Nevertheless, dominant agents are expected to be drawn towards
manipulative agents w.r.t.\ to their communication partner choice
and friendship, whenever the manipulative agent manages to become
reputed.

Finally, the \textbf{destructive agent} is clever, deceptive, aggressive,
and shameless. By also targeting reputed agents for communications,
the agent's disrespectful propaganda about the agent's enemies can
unfold best. Since destructive agents are shameless, they are not
risking to blush while lying. This might compensate for the lack of
direct self-promotion of the destructive agent. Their lack of self-
and friend-promotion lowers the surprise variance the receivers experience
compared to just deceptive or clever agents, which helps the destructive
agent to appear more honest. The presence of a destructive agent in
a social group is expected to lower the reputation and self-esteem
values of the other agents significantly due to that agent's tendency
to concentrate conversation topics on enemies, about which the agent
talks disrespectfully.

To summarize, we have defined a number of communication and receiver
strategies and can now see how agents equipped with different sets
of such strategies interact.

\section{Simulations\label{sec:Simulations}}

We now discuss our reputation game simulations. All agents' initial
beliefs and assumptions on honesty and reputations of other agents
are set to be non-informative, $I_{ab}(0)=I_{abc}(0)=\widetilde{I}_{abc}(0)=I_{0}=(0,0)$,
if not specified differently. In the displayed simulation runs, individual
random sequences for the different processes like choosing conversation
partners, topics, whether to lie, how strong to lie and the like are
kept identical between the simulations. The intrinsic honesty of agents
is also kept identical, with $\underline{x}=(x_{\text{red}},x_{\text{cyan}},x_{\text{black}})=(0.27,\,0.80,\,0.97)$,
if not specified differently. Differences in dynamics therefore only
arise here because of the different strategies used by agents in the
different simulation runs, as these specify how the random number
sequences are used in detail. This should help to highlight the effects
of the strategies, and to facilitate their comparison. However, differences
in the performance of individual strategies observed this way are
only indicative. The dynamics is chaotic and thus firm conclusions
about the efficiency of strategies can only be drawn from a sufficiently
large statistical ensemble of simulation runs with varying random
number sequences, which we discuss in what follows.

First, we present a few reputation game simulations to motivate the
complexity of the agents' receiver strategies (Sect.\ \ref{subsec:Receiver-strategies}).
Sect.\ \ref{subsec:Propaganda-and-resilience} presents simulations
of propaganda situations without random elements. These give insights
into the cognitive models assumed in this work and explain the need
for their complexity. Sect.\ \ref{subsec:Communication-strategies}
illustrates the effects of basic and special communication strategies
with individual simulations. Sect.\ \ref{subsec:Statistics} presents
statistical results based on one hundred simulation runs per setup.
These allow to measure the effects of the different deceptive strategies
quantitatively, and provide statistics of the resulting reputation
and friendship relations between agents.

\begin{figure*}[!t]
\includegraphics[width=0.5\textwidth]{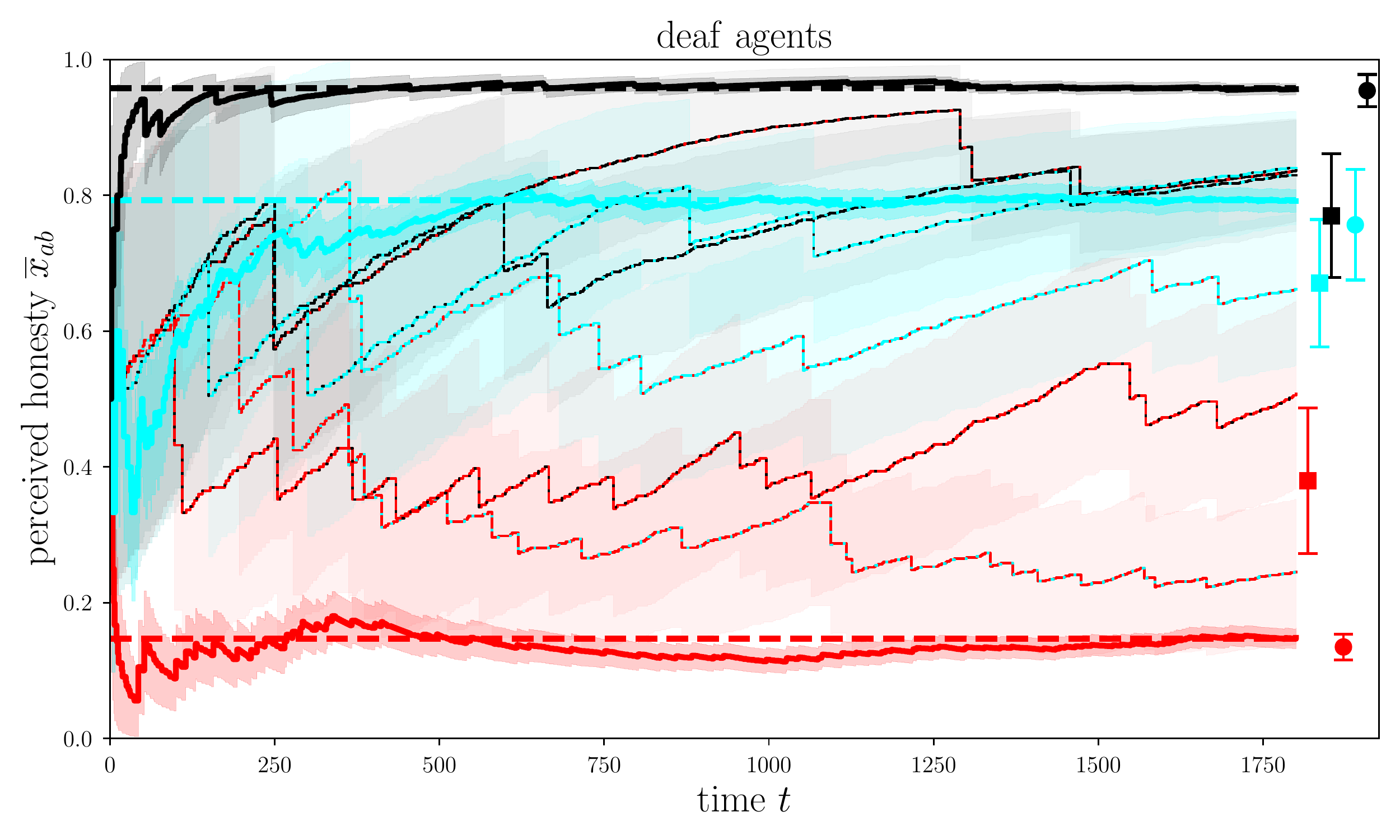}\includegraphics[width=0.5\textwidth]{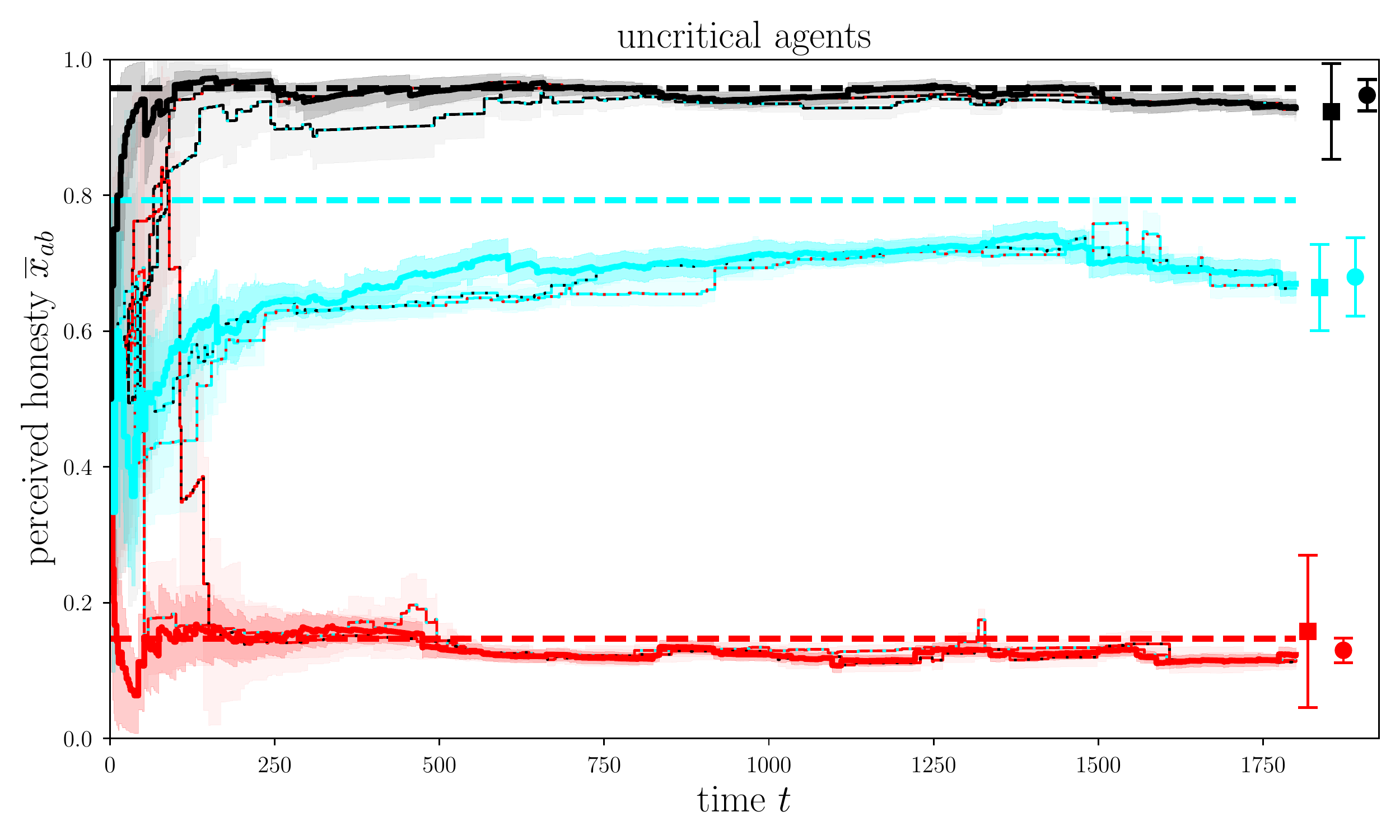}

\includegraphics[width=0.5\textwidth]{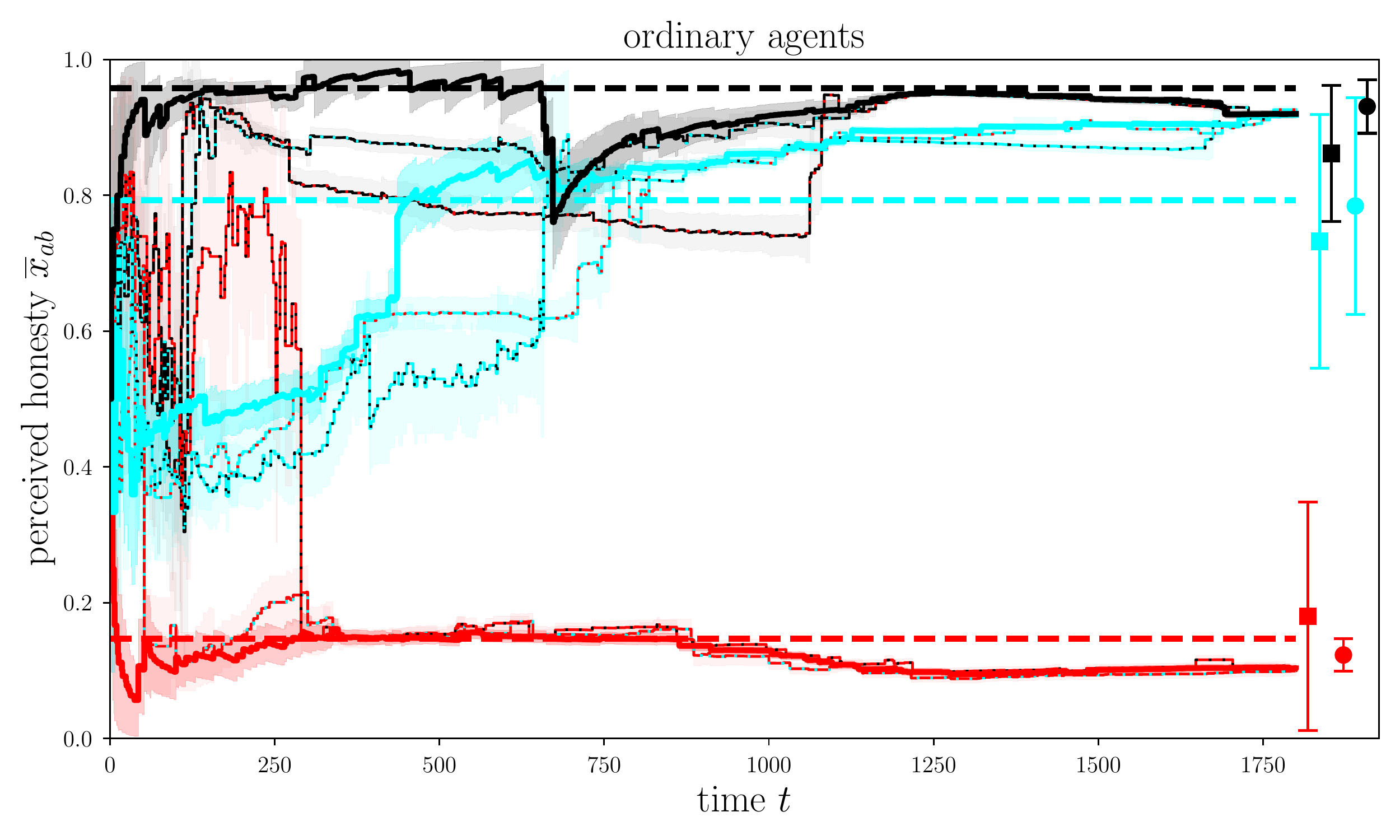}\includegraphics[width=0.5\textwidth]{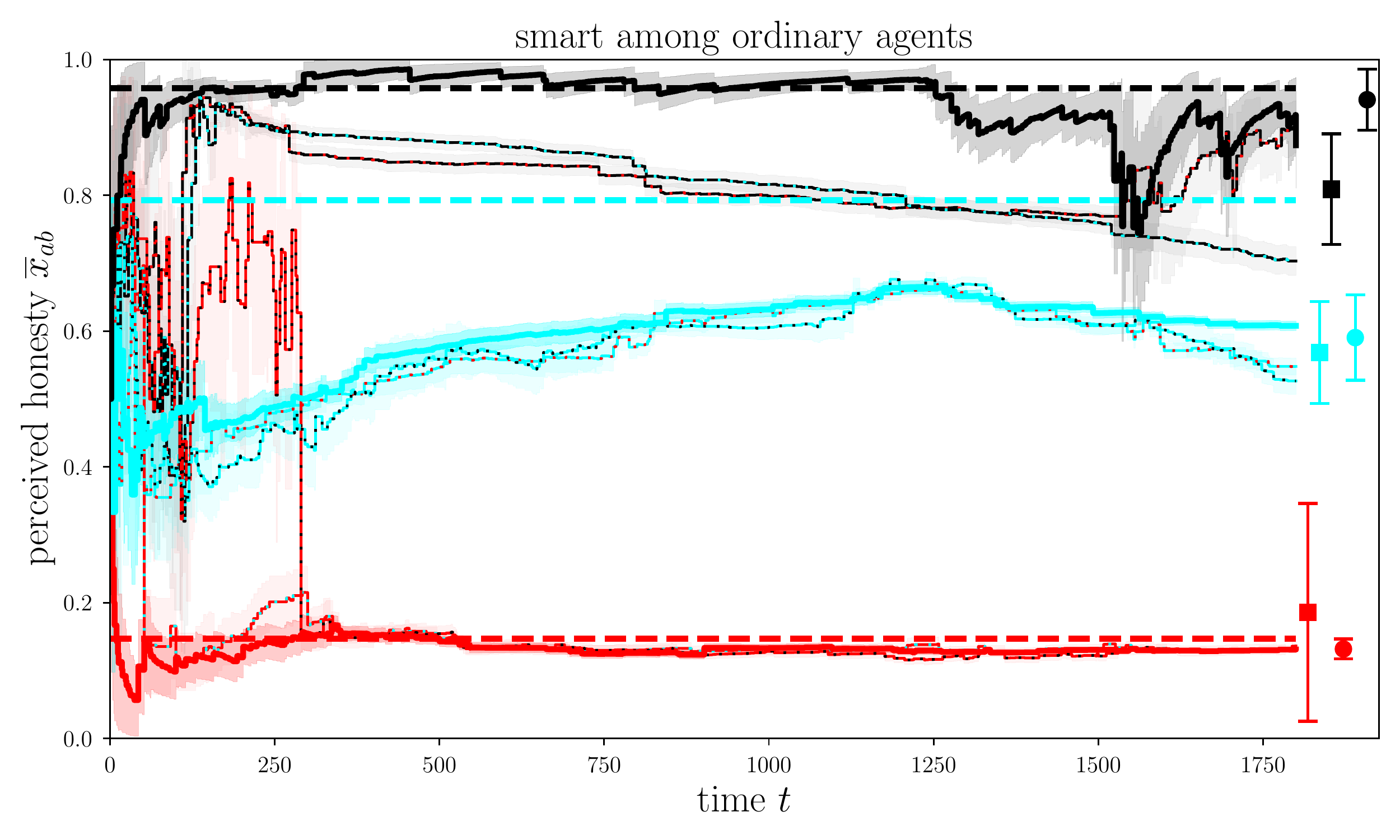}

\caption{Reputation game simulations for three agents with the panels showing
different receiver strategies: deaf agents (top left), uncritical
agents (top right), ordinary agents, which have a critical receiver
strategy (bottom left), and agent red being a smart agent (bottom
right). All simulations are run with the same random number sequences,
implying that the communication configurations (like $a\protect\overset{c}{\rightarrow}b$)
and message honesty states (honest or lie) exhibit exactly the same
sequences. Differences are solely caused by differences in receiver
strategies. This and other figures showing communication patterns
intend to give an overview. To inspect details, we recommend to magnify
their electronic, vector graphics versions. Text statements about
certain precise communication events were not taken from these figures,
but from the simulation log files. The self-esteem $\overline{x}_{aa}$
of agent $a$ is shown as a thick solid line in the color of $a$,
agent $b$'s reputation $\overline{x}_{ab}$ in the eyes of agent
$a$ is shown as a thin line, which carries the color of agent $b$
and has dots in the color of agent $a$ on it. One sigma uncertainties
of self-esteem and reputations are displayed as transparently shaded
areas in the color of agent $a$. The dashed lines show the actual
honesty of the individual agents, the fraction of honest statements
made, which is close to their intrinsic honesty of $\underline{x}=(x_{\text{red}},x_{\text{cyan}},x_{\text{black}})=(0.27,\,0.80,\,0.97)$.
The data points with bars at the right side display summary statistics
of the full dynamics. The squares and their bars indicate the mean
and variance of the reputation of the agent in the corresponding color.
Similarly, the circles and bar indicate mean and variance of self-esteems
of agents. \label{fig:Reputation-game-simulations}}
\end{figure*}

\begin{figure*}[!t]
\includegraphics[width=0.5\textwidth]{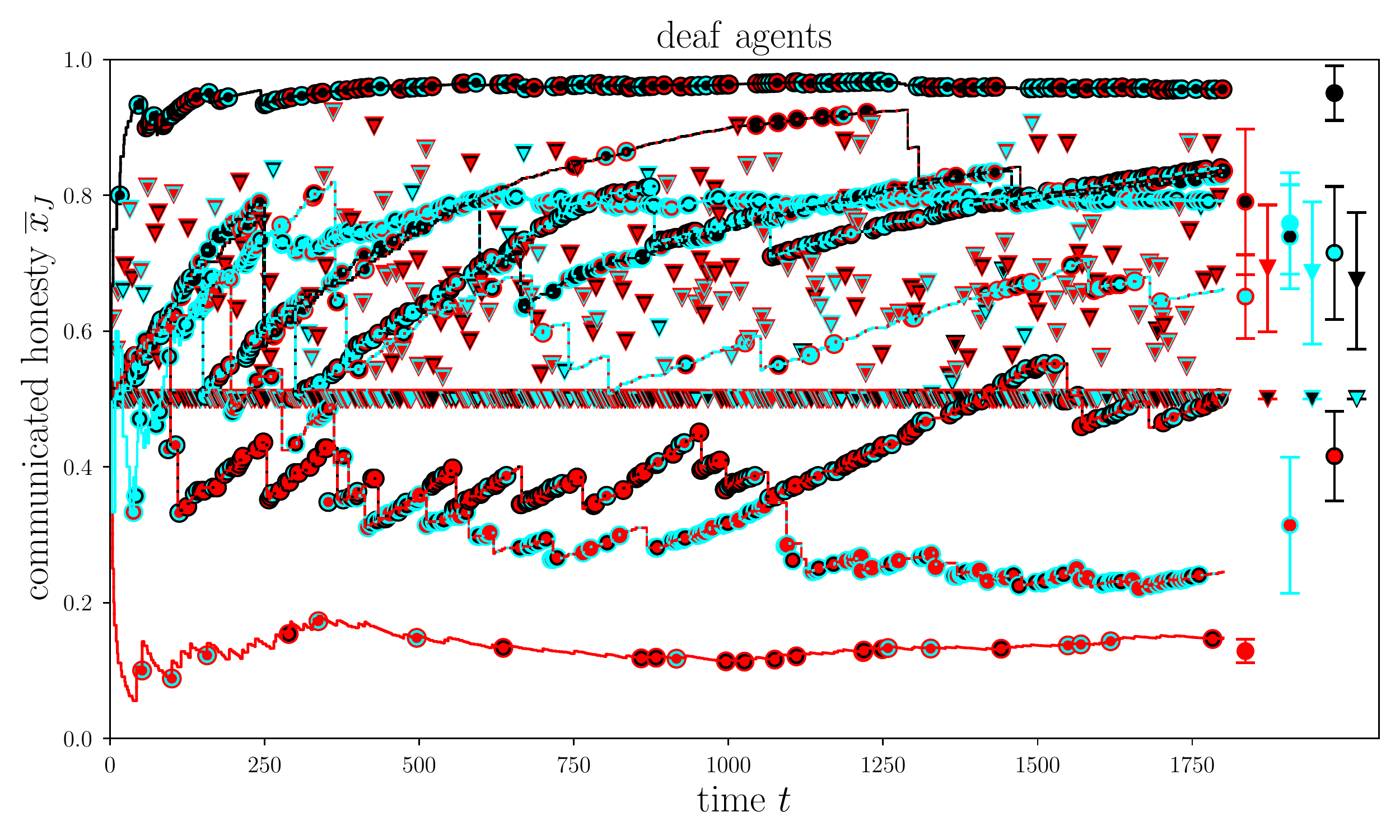}\includegraphics[width=0.5\textwidth]{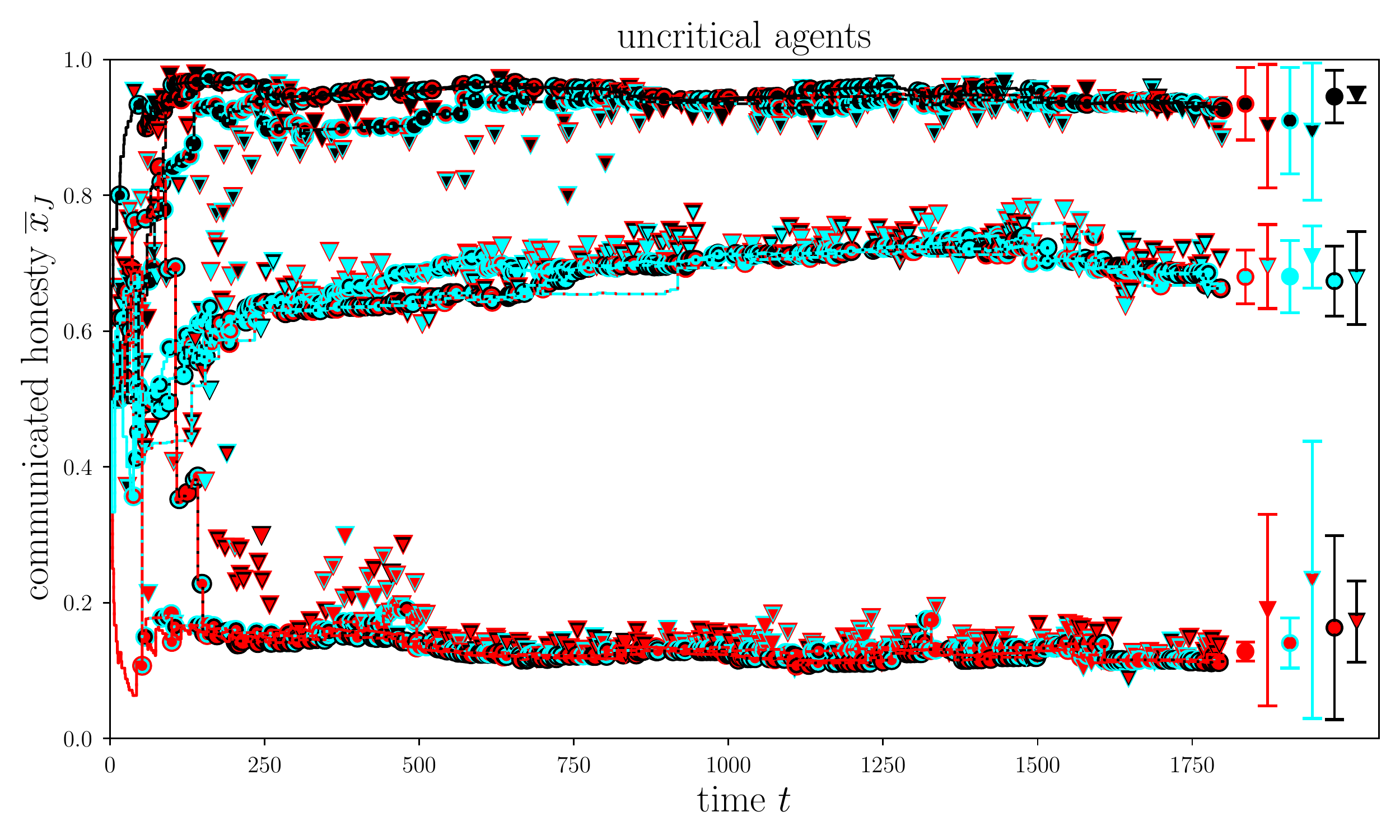}

\includegraphics[width=0.5\textwidth]{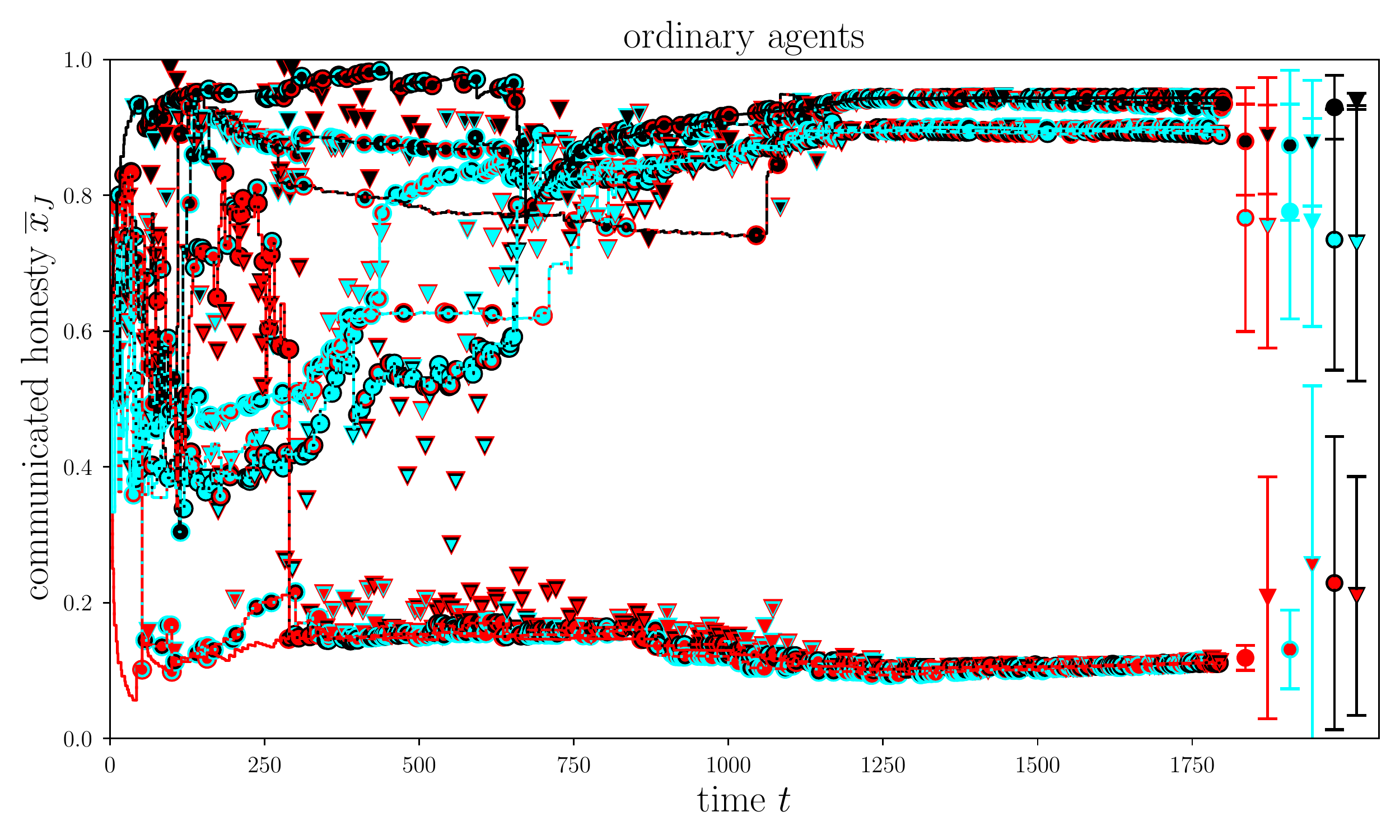}\includegraphics[width=0.5\textwidth]{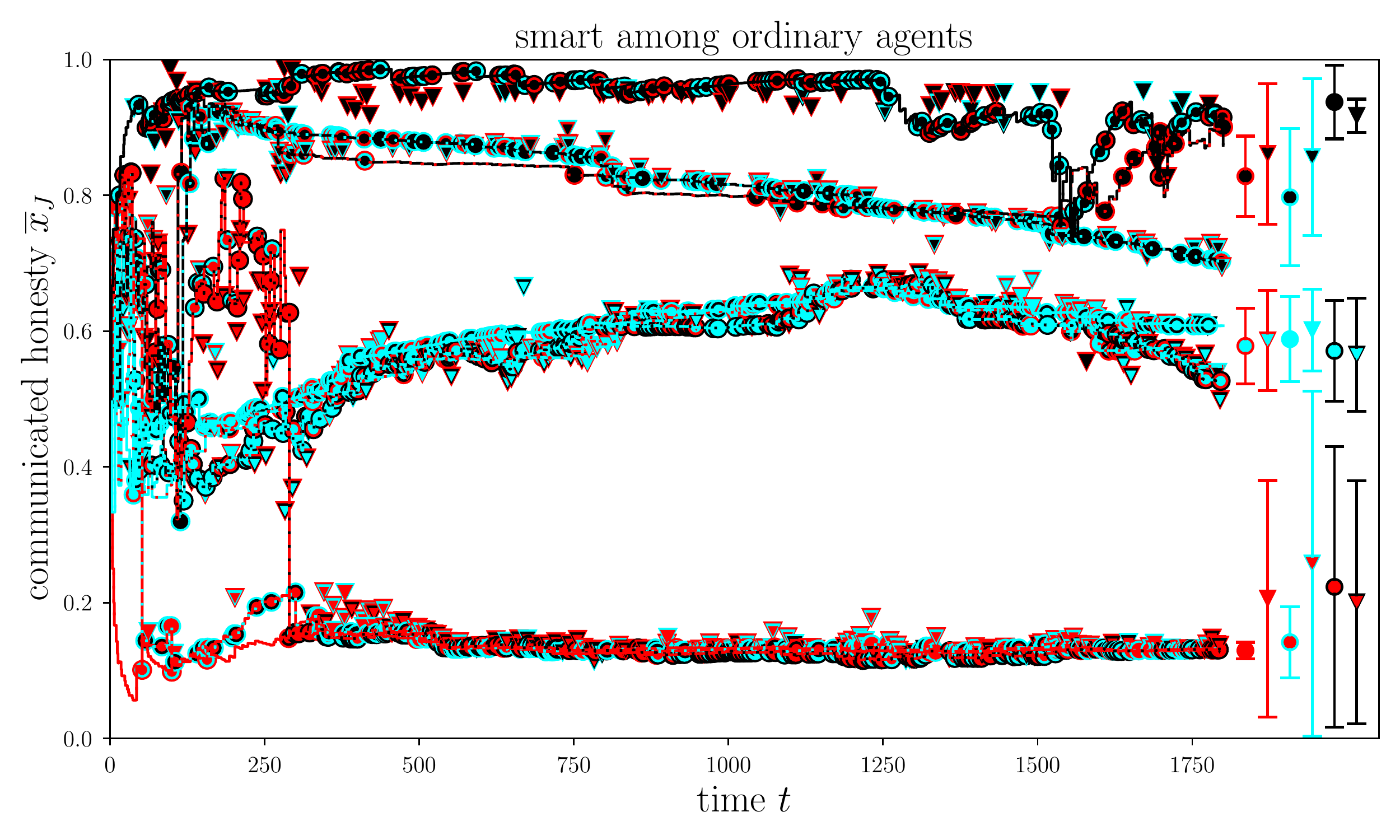}

\caption{Communication patterns of the simulation runs shown in Fig.\ \ref{fig:Reputation-game-simulations},
arranged in the same way. Circles and triangles correspond to the
mean honesty $\overline{x}_{J}$ expressed in honest and dishonest
statements, respectively. The speaker is indicated by the outermost
color of such a symbol, the receiver by the middle color, and the
topic by the central area color of a symbol. Reputations and self-esteems
are both displayed here as thin lines, colored as in Fig.\ \ref{fig:Reputation-game-simulations}.
The circles and triangles with bars on the right indicate the mean
opinion expressed in honest and dishonest statements, respectively.
The outer color and that of the bars thereby indicate the speaker
and the inner color the topic. The bars indicate the variance of the
corresponding set of statements. \label{fig:Communication-patterns}}
\end{figure*}

\subsection{Receiver strategies\label{subsec:Receiver-strategies}}

First, we investigate different receiver strategies. The sender strategy
is that of ordinary agents in what follows. An overview on the different
receiver strategies is given in Tab.\ \ref{tab:receiver-strategies}.

\subsubsection{Deaf agents\label{subsec:Game-simulations}}

We want to demonstrate the agent's ability to learn from unbiased
signals, like the agents' self-observations and the blushing signals.
These are the only information sources available to deaf agents. To
this end, the top left panel in Fig.\ \ref{fig:Reputation-game-simulations}
shows the reputation dynamics of three deaf agents performing 300
conversation rounds. these agents learn relatively quickly and accurately
their true honesty from their self-observations. Whenever they are
honest, their self-esteem increases; when they lie, it decreases.
The learning of the honesty of others is much more difficult, as it
relies on the occasional blushing signals, which are visible as the
sudden drops of the otherwise monotonically increasing reputation
lines. Despite this difficulty, agents manage nevertheless to get
the tendencies right. The discrepancies between correct honesty and
their reputation seems to be consistent with the associated uncertainty
estimates.

Despite the deaf agents not hearing each other, the patterns of their
statements are instructive. These are shown in the top left panel
of Fig.\ \ref{fig:Communication-patterns}. This is a busy figure
that we discuss briefly.

All honest statements from agent $a$ to $b$ about $c$ are displayed
as circles with the outer, middle, and inner color indicating the
agent $a,$ $b$, and $c$, respectively. These honest statements
reflect the beliefs of the speaker $a$ on topic $c$ ($J_{a\overset{c}{\rightarrow}b}=I_{ac}$).
For this reason, they are on top of the agents' belief curves $\overline{x}_{ac}(t)$,
which are displayed as well. The main color of any of these lines
is that of $c$ and the dots on top are in the color of $a$. The
circles on the uni-color lines are self-statements. Their densities
reflect the intrinsic honesties of the speaking agent, with agent
black making most frequently honest statements and agent red least
frequently.

All lies ($J_{a\overset{c}{\rightarrow}b}=I_{abc}+\alpha D$, where
$I_{abc}$ is the speaker $a$'s assumption about the receiver $b$'s
belief on $c$, $D$ the direction of the lie, and $\alpha$ the size
of its distortion) are displayed as triangles with the same color
coding (speaker $a$ specifying the outer, receiver $b$ the intermediate,
and topic $c$ the inner color). The lies are mostly from agent red
and fall into two categories: First, all lies about other agents ($a\neq c$)
are located at the horizontal line $\overline{x}_{J}=\nicefrac{1}{2}$.
The reason for this is that deaf agents do neither get friends nor
enemies (as they do not listen to each other), and therefore make
only white lies ($\alpha=0\Rightarrow J_{a\overset{c}{\rightarrow}b}=I_{abc}$).
Since deaf agents can not update $I_{abc}$ (they don't hear others'
opinions), their white lies are the initial value of this quantity,
$J_{a\overset{c}{\rightarrow}b}=I_{abc}=I_{0},$ and therefore displayed
at $\overline{x}_{I_{0}}=\nicefrac{1}{2}$. The lies agents make about
themselves, $J_{a\overset{a}{\rightarrow}b},$ are biased positively
($\alpha>0,\,D=(1,0)\Rightarrow\overline{x}_{J_{a\overset{a}{\rightarrow}b}}>\overline{x}_{I_{aba}}=\overline{x}_{I_{0}}=\nicefrac{1}{2}$),
and thus are found in the upper half of the diagram.

\subsubsection{Uncritical agents}

Agents should get much better estimates of each other's honesty compared
to the deaf agent scenario, if they exchange the information they
collect. This is shown in the top right panel of Fig.\ \ref{fig:Reputation-game-simulations},
where uncritical agents, who listen to each other, perform the same
set of conversations (as specified by $(a\overset{c}{\rightleftarrows}b)(t)$)
as the deaf agents did (top left panel), with also being honest or
lying at exactly the same instances. What they say, however, differs
from the deaf agents simulation, as agents now listen to each other
and therefore their opinions and assumptions evolve differently to
those of the previous run.

It is apparent that the agent's guesses on each other's honesty become
much more accurate and definite. Actually, some overconfidence can
be observed for the opinions on agent cyan, which have converged to
a value significantly below the agent's true honesty with a confidence
that excludes the correct value. The self-esteem of cyan even follows
this slightly incorrect value, despite cyan's self-observation should
inform cyan better. However, the opinions expressed by the others
on cyan, in particular the ones of the most reputed agent black, seem
to have a stronger pull. The collective development of the overconfident,
but incorrect opinions on cyan is the result of an echo chamber: The
initially more dispersed opinions of the different agents converge
to a value that is partly decoupled from reality (cyan's true honesty),
and this value is largely determined by the group dynamics.

Inspecting the corresponding communication patterns in the top right
panel of Fig.\ \ref{fig:Communication-patterns} shows for example
the concentration of statements about agent cyan around cyan's self-esteem.
It is apparent that agent red regards cyan as a friend for most of
the time, as agent red's lies for cyan are typically above cyan's
self-esteem. Consequently, red regards black mostly as an enemy, as
the lies about black are aiming for lowering black's reputation and
self-esteem. These lies by red, however, have little influence compared
to the opinions expressed by cyan and in particular by black, due
to the much higher reputations of black and cyan compared to red.

Investigating red's reputation is also instructive. Initially it is
high, as red's early self-promoting lies fly. However, two confessions
of red to cyan (at $t=52$ and $100$), who thereafter regards red
as unreliable, and cyan's repeated spreading of these news to black
($\text{cyan\ensuremath{\overset{\text{red}}{\longrightarrow}\text{black at }}\ensuremath{t=57}, 84, 93, and 150)}$,
destroy red's initially high reputation in an irreparable way.

Red would probably have overcome this resistance if red's lies would
simply have been much stronger. This is because the weight of a message,
which an uncritical agent assigns, does not depend on how extreme
the position of a message is, but the shift of the receiver's opinion
does depend on this. Uncritical agents are therefore very prone to
propaganda in form of exaggerated lies, as we show in Sect.\ \ref{subsec:Propaganda-and-resilience}.

\subsubsection{Critical agents}

Ordinary agents use a critical receiver strategy, which is able to
recognize exaggerated statements. A simulation run with such agents
is shown in the bottom left panel of Fig.\ \ref{fig:Reputation-game-simulations},
again for the same sequence of communication decisions. Overall, the
outcome of the simulation is similar to that of uncritical agents,
in the sense that the final reputations and self- esteems converge
to values not too far from the correct honesty of the agents. However,
at least two interesting differences to the uncritical agent simulation
can be spotted here and should be discussed: The more volatile evolution
of beliefs about cyan, with a significant gap between cyan's self-esteem
and cyan's reputation (in the period 100 to 800) and the much later
time in the critical simulation compared to the ordinary one ($t=291$
instead of $t=150$) at which agent black's opinion about red joins
that of cyan. Both are a consequence of critical agents being more
reluctant to accept diverging opinions. This allows the self-esteem
of cyan to evolve more decoupled from the lower opinions expressed
by red and black, and makes black more skeptical about cyan's reports
on red's dishonesty.

\begin{figure*}[!t]
\includegraphics[width=1\columnwidth]{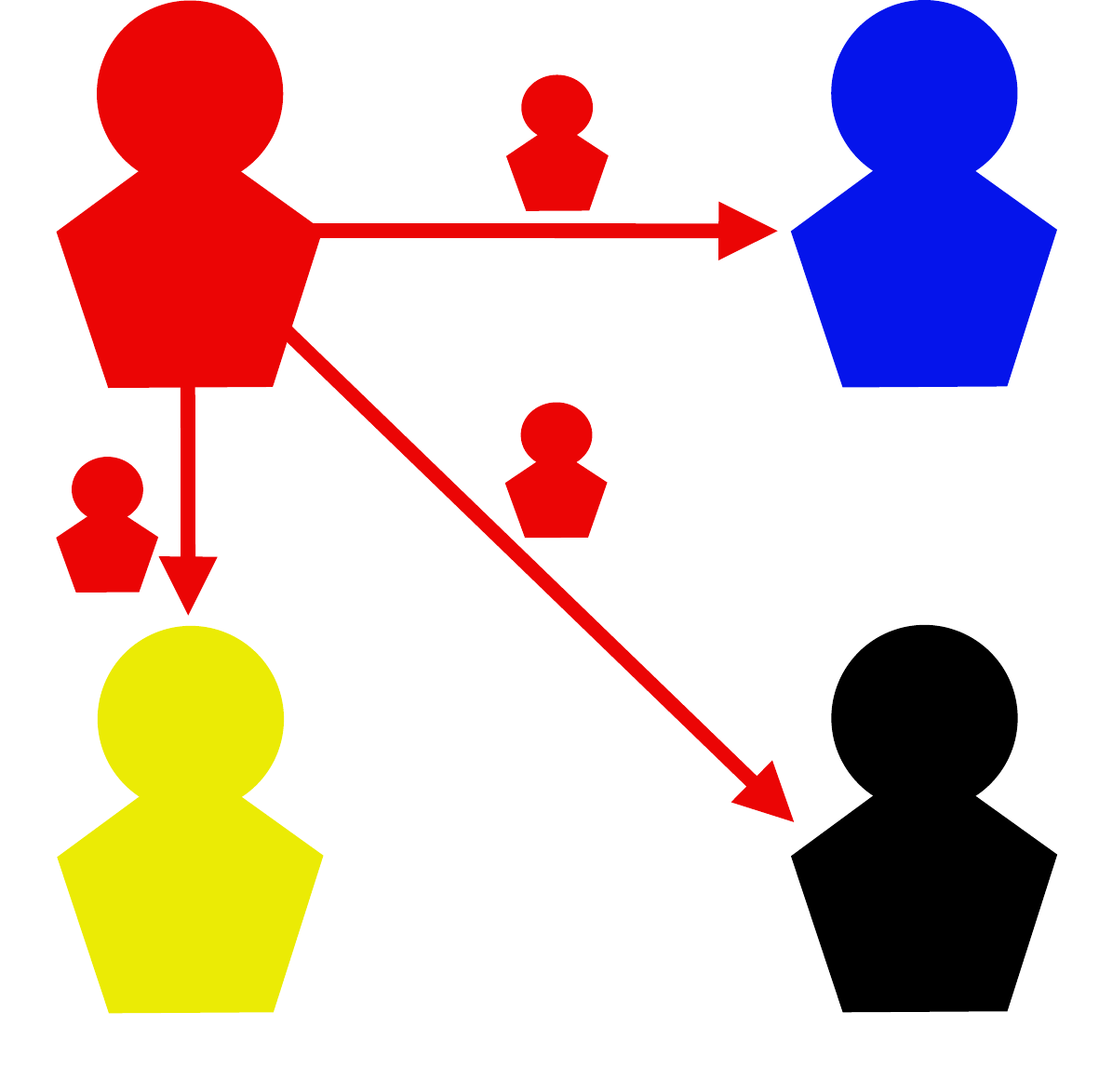}\hfill{}\includegraphics[width=1\columnwidth]{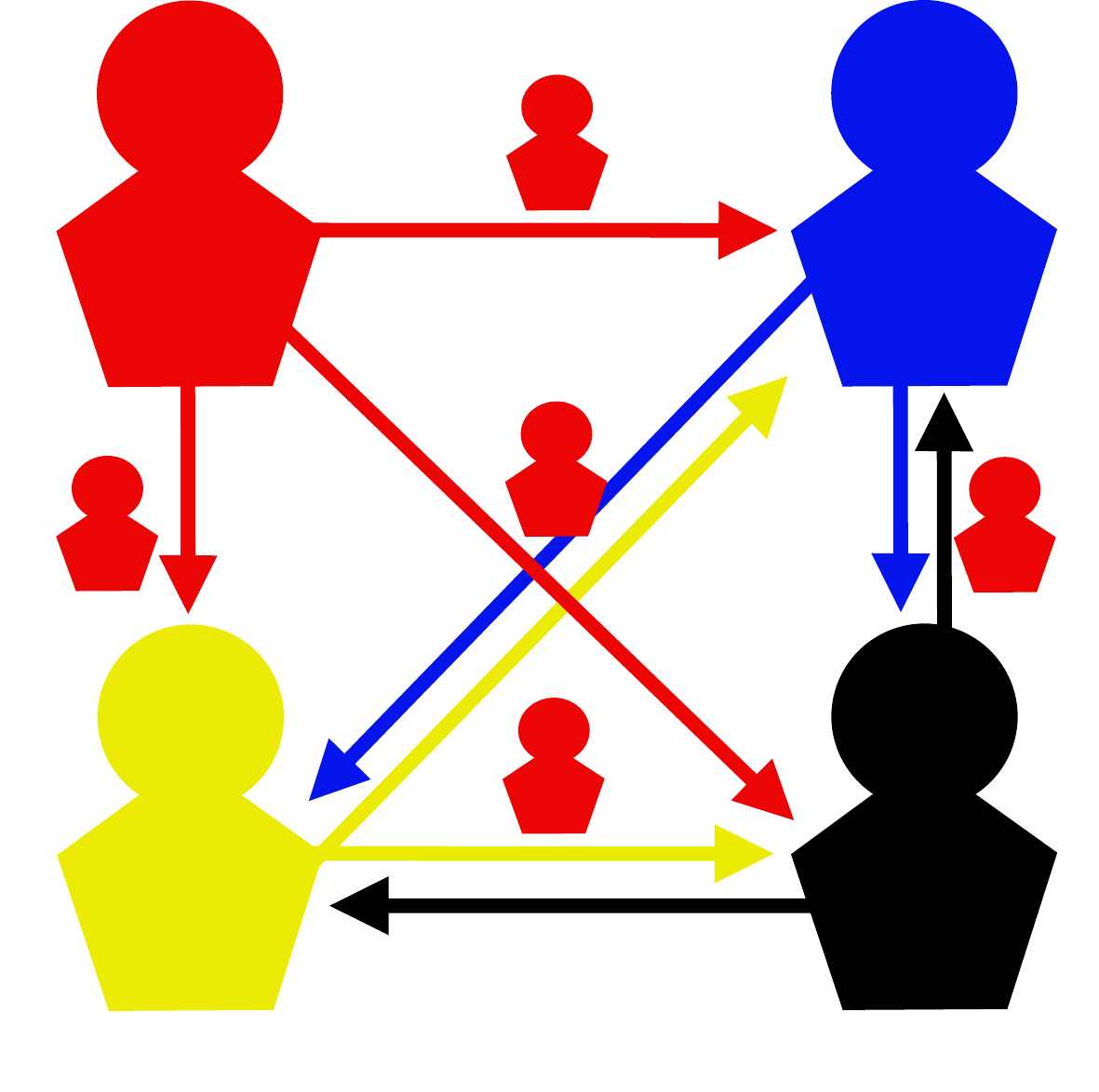}

\caption{Communication configuration in which only one agent (here red) communicates
self-propaganda to isolated agents (left) and where the other agents
exchange their opinions on red among each other, with each of them
communicating to both others whenever having received propaganda.
(right) \label{fig:Propaganda-simulation-setup}}
\end{figure*}
\begin{figure*}[!t]
\includegraphics[width=0.5\textwidth]{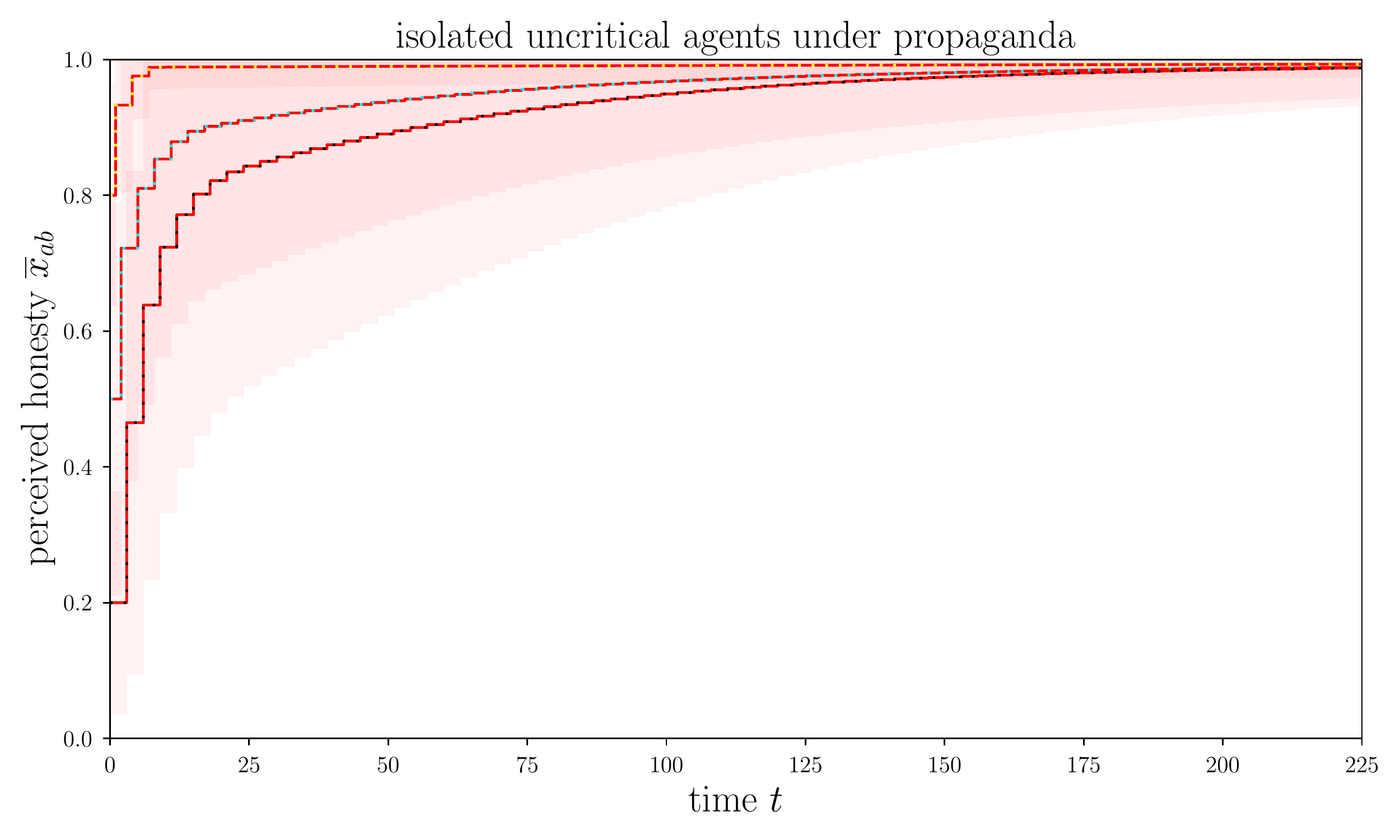}\includegraphics[width=0.5\textwidth]{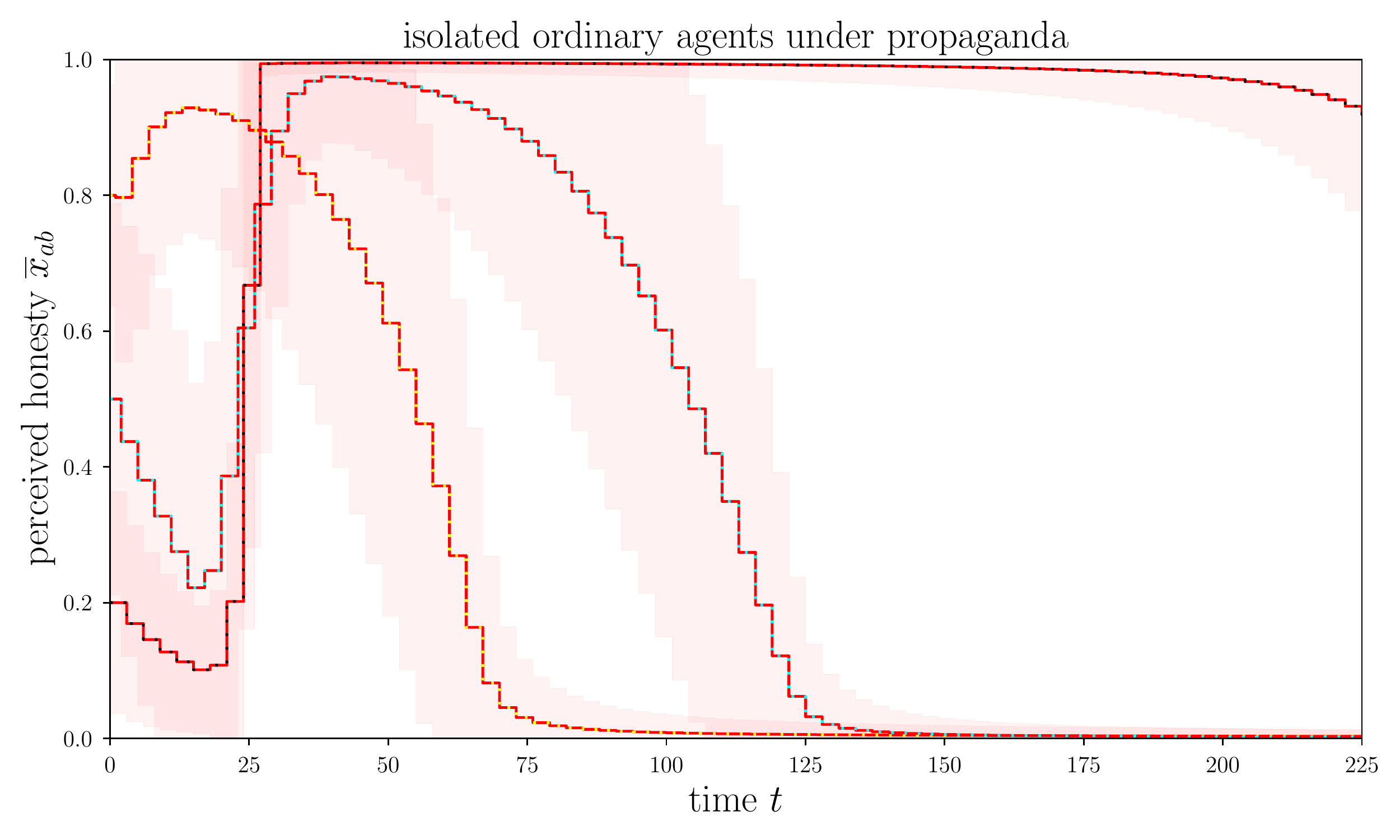}

\includegraphics[width=0.5\textwidth]{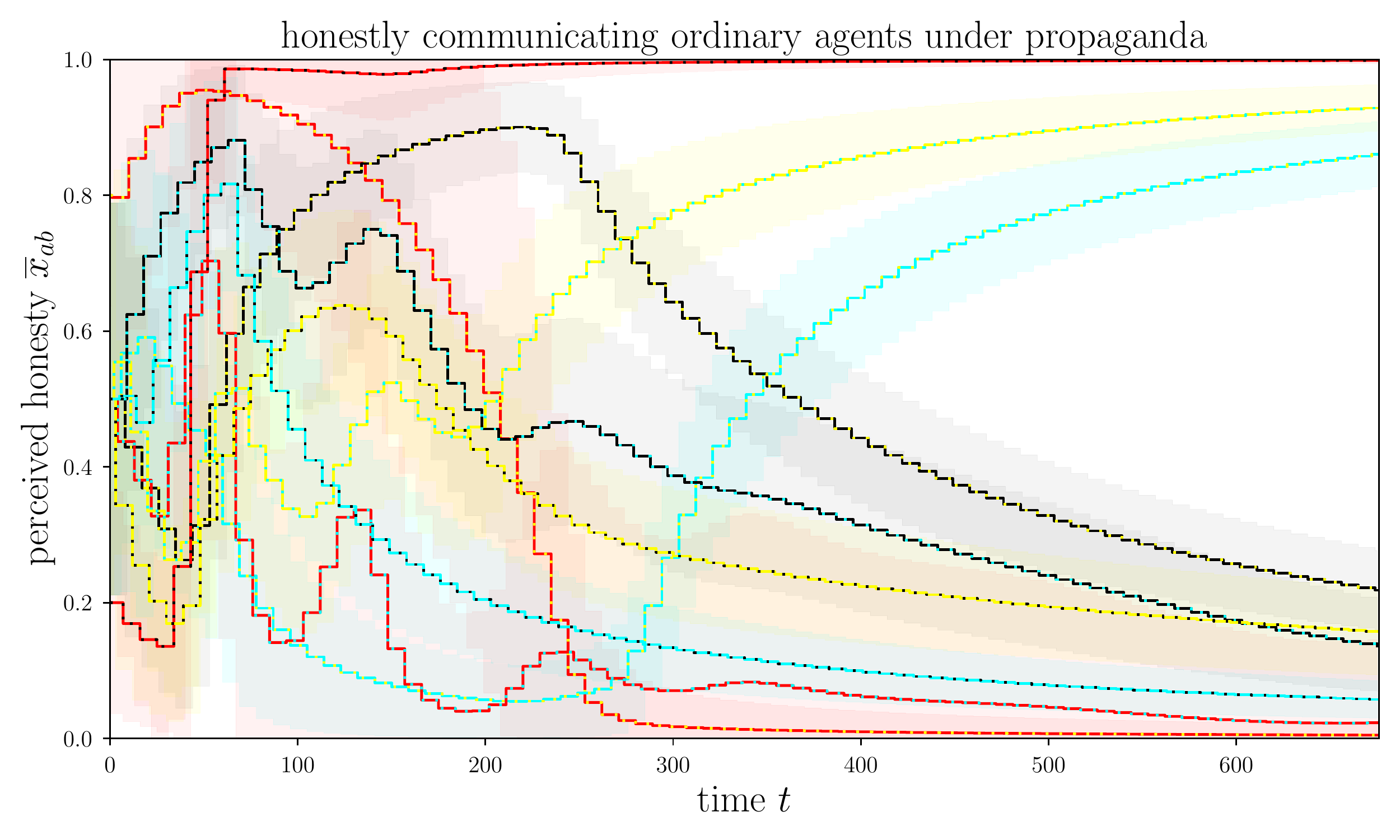}\includegraphics[width=0.5\textwidth]{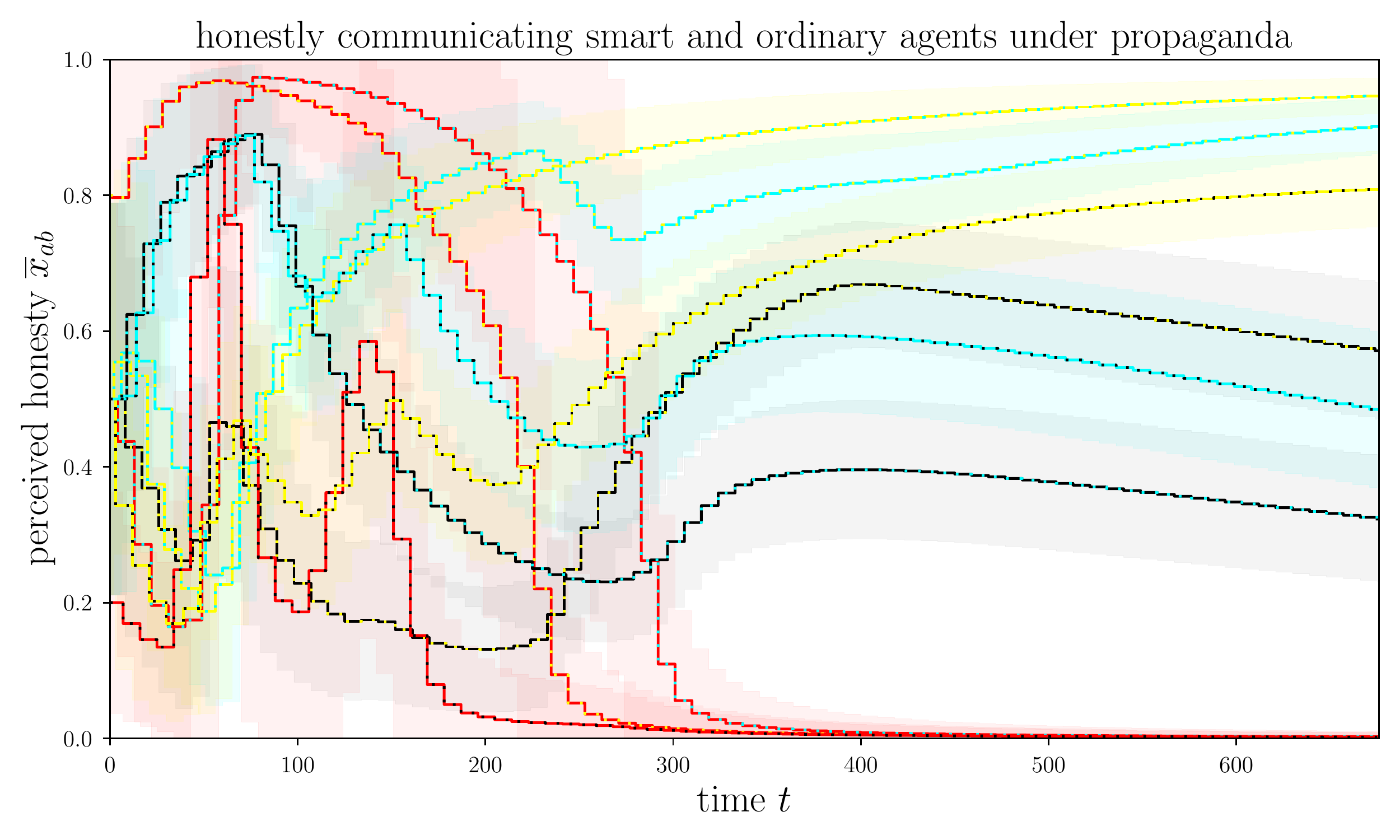}

\caption{Impact of propaganda on agents with differing initial beliefs in various
social situations as depicted in Fig.\ \ref{fig:Propaganda-simulation-setup}.
Here, agent red constantly claims (without blushing) to be extremely
honest, $J_{\text{red\ensuremath{\protect\overset{\text{red}}{\rightarrow}}\ensuremath{\ensuremath{\cdot}}}}=(10^{3},0)$.
The initial beliefs of the receiver of red's self-propaganda range
from being slightly reserved with $I_{\text{black red}}(t=0)=(0,3)$
over neutral with $I_{\text{cyan red}}(0)=(0,0)$ to positively inclined
with $I_{\text{yellow red}}(0)=(3,0)$. The panels show the belief
dynamics of isolated uncritical agents (top left), isolated ordinary
(critical) agents (top right), honestly cross-communicating ordinary
agents (bottom left), and the same, just with black being smart (bottom
right). Only non-trivial beliefs ($\protect\neq I_{0}=(0,0))$ on
other agents are plotted in the color coding of Fig.\ \ref{fig:Reputation-game-simulations}
(only red in the top panels, all agents in the bottom panels). Self-esteems
are not shown. \label{fig:Propaganda-simulation}}
\end{figure*}

\subsubsection{Smart agents}

Smart agents have an even more sophisticated receiver strategy compared
to critical agents. This should allow them to maintain a more accurate
picture of the other agents' beliefs, which improves their lie detection
and lie construction. To illustrate this, the bottom right panels
of Figs.\ \ref{fig:Reputation-game-simulations} and \ref{fig:Communication-patterns}
show a simulation run in which agents black and cyan still use critical
receiver strategies (as in the bottom left panel), but red uses a
smart strategy.

In this smart scenario (agent red being smart), the self-esteem and
reputation of cyan do not show the strong growth that is visible in
the critical scenario (agent red being only critical). The reason
for this are the better targeted lies of red in the smart run, which
undermine cyan's and black's lie detection more efficiently than in
the critical run. This makes red's lies more effective. As these lies
mirror the other agents' previously communicated beliefs, just in
a slightly distorted manner, they counteract rapid evolution of these
beliefs by pulling them back towards those previous values. Additionally,
the echo chamber effect of group opinions converging to overconfident,
but incorrect positions is strong, also due to the better targeted
lies of red. Both, the retarding back-reaction and the opinion focusing
effect of red's more effective lies, effectively add inertia to cyan's
self-esteem and reputation, which keeps those from reaching the correct
honesty value of cyan. 

\subsection{Propaganda and resilience\label{subsec:Propaganda-and-resilience}}

\subsubsection{Simulation setups}

We claimed that agents with naive or uncritical strategies are very
susceptible to exaggerated lies and that critical and smart receiver
strategies provide some resilience against such lies. To demonstrate
this, but also to illustrate the inner working of the cognitive model
adapted, a number of propaganda situations are simulated. In those,
all agents, except agent red, who will be the propagandist, will be
absolutely honest.

The basic propaganda situation is depicted on the left of Fig.\ \ref{fig:Propaganda-simulation-setup}:
There, only agent red communicates to black, cyan, yellow, by repeatedly
sending strong self-appraisal ($J_{\text{red}\overset{\text{red}}{\rightarrow}\cdot}=(10^{3},0)$)
to them without blushing. Red's initial reputation with them differs,
being initially low with black ($\overline{x}_{\text{black red}}(0)=0.2$),
medium with cyan ($\overline{x}_{\text{cyan red}}(0)=0.5$), and high
with yellow ($\overline{x}_{\text{yellow red}}(0)=0.8)$. These agents
are isolated in this setup, as they only receive the propaganda, but
can not exchange their positions among themselves.

This is changed in the setup with cross-communication among the propaganda
receiving agents shown on the right of Fig.\ \ref{fig:Propaganda-simulation-setup}.
There, every receiver communicates honestly the updated opinion on
red to all other receivers after every exposure to the propaganda.

Simulations with the basic setup are shown in Fig.\ \ref{fig:Propaganda-simulation}
for three uncritical agents (top left panel) and for three critical
agents (top right) as receivers. Simulations with the cross-communication
setup are run for three critical receivers (bottom left) and for one
smart among two critical ones (bottom right). In all simulations,
75 propaganda rounds are performed.

\subsubsection{Isolated uncritical agents}

All uncritical, isolated agents rapidly adapt a high reputation for
red under red's propaganda (top left panel of Fig.\ \ref{fig:Propaganda-simulation}),
as the strong messages received are only slightly moderated by their
initial limited respect for red. Although red's reputation with them
is steadily growing, it does not reach the position announced in the
propaganda message of $\overline{x}_{J_{\text{red}\overset{\text{red}}{\rightarrow}\cdot}}=0.999$.
This is caused by the receiver mechanism that tries to identify the
novel part of a message, and disregards the part that already seems
to be accounted for.\footnote{This saturation effect can be read off from Eqs.\ \ref{eq:DeltaJ}
and \ref{eq:ToM-1} for a repeated message $J$ with $y_{J}\approx1$.} Naive agents would have fully adopted that latter position on the
first exposure to the propaganda (not shown).

\subsubsection{Isolated critical agents}

Ordinary agents, which have critical minds, have much more resilience
against propaganda, as can be seen in the top right panel of Fig.\ \ref{fig:Propaganda-simulation}.
Agents black and cyan, who are initially skeptical about red's honesty,
become immediately more skeptical under the exposure of the propaganda,
as they perceive this as lies. This changes at $t=15$, after 5 propaganda
rounds, when red's reputation with them starts to grow. What causes
this change is that at this point in time, the large divergence of
the propaganda messages from their own beliefs start to affect the
scale $\kappa_{\cdot}$, which agents use to discriminate lies from
honest statements, as this is based on the median of the last ten
message surprises. Consequently, the mechanism to separate lies from
honest statements starts to fail, which lets the propaganda appear
slightly more trustworthy. Since the propaganda makes strong claims,
it shifts -- despite being still more distrusted than believed --
black's and cyan's opinions on red upwards.

Agent black, who is initially most skeptical, is hit the strongest
by this effect. Being initially most skeptical about red, black experiences
the largest opinion divergence by the propaganda, and therefore the
largest shift in $\kappa_{\text{black}}.$ This then makes black most
vulnerable to propaganda.

We see that even critical agents, who are more wary, can be prone
to propaganda. All their beliefs in red's reliability increase and
do this the more, the lower the initial trust was. At some moment,
the novelty of the propaganda message wears off and red's reputation
stops to increase further.\footnote{To prevent this, agent red would need to constantly increase the propaganda
claim.} The propaganda statements are still received as mostly being lies,
which thus makes the prestige of red finally disappear for each of
the recipients again. 

To summarize, the strategy to classify lies only according to the
surprise they create works as long as the reference surprise value
is not inflated. This quantity is determined empirically and increases
to a too high value if there are many more lies than expected. This
effectively shuts down the full rejection of strong claims by the
agents' lie detection system and thereby enables propaganda to affect
their minds.

\subsubsection{Cross-communicating critical agents}

A counter measure against attacks on the lie detection system can
be the exposure to honest messages, or just messages with low surprise
values. This can be achieved by honest and frequent exchanges with
other honest agents. Such exchanges should help to a \emph{healthy}
lie detection system, and should provide corrective inputs that counteract
the pull of the propaganda.

\begin{figure*}[!t]
\vspace{-1.5em}

\includegraphics[width=0.5\textwidth]{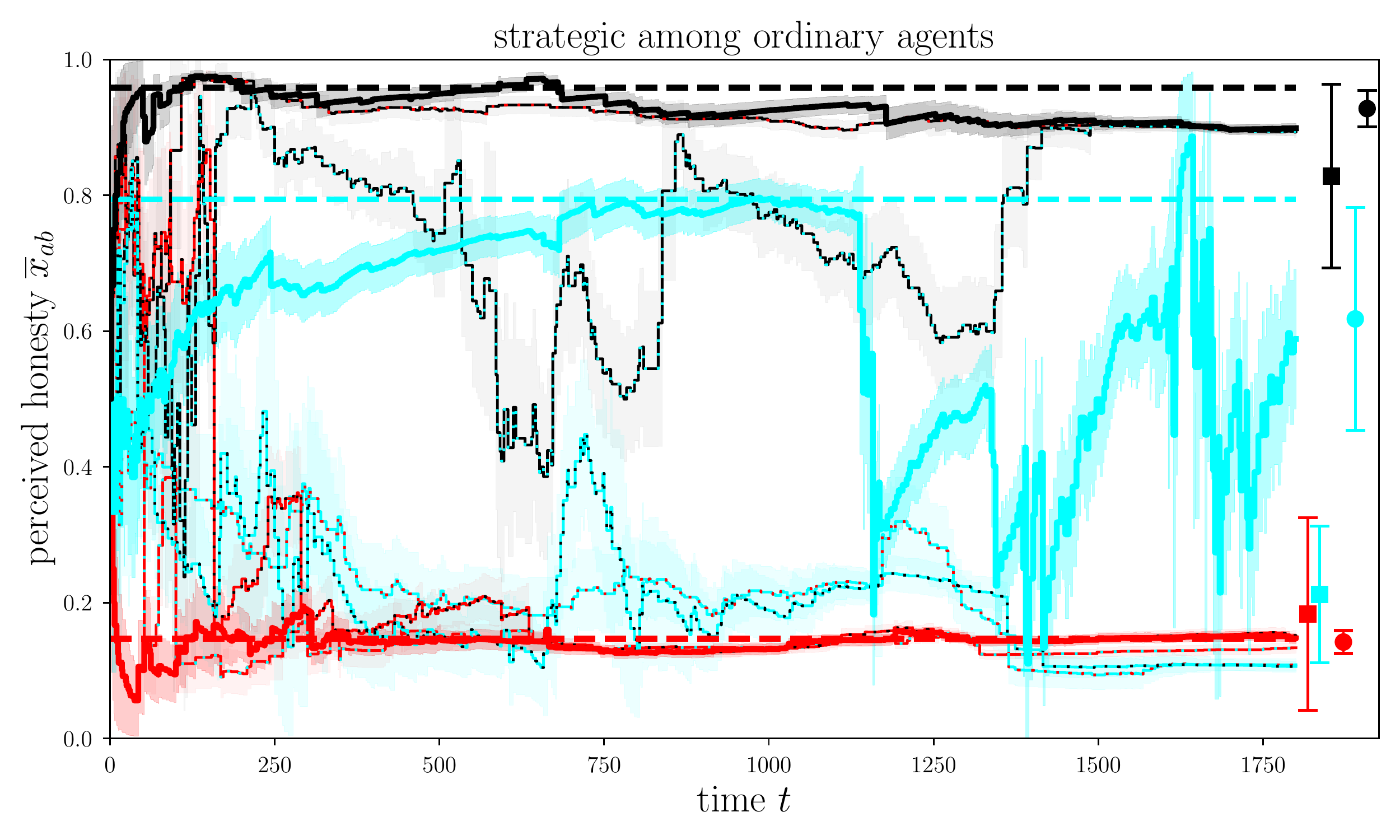}\includegraphics[width=0.5\textwidth]{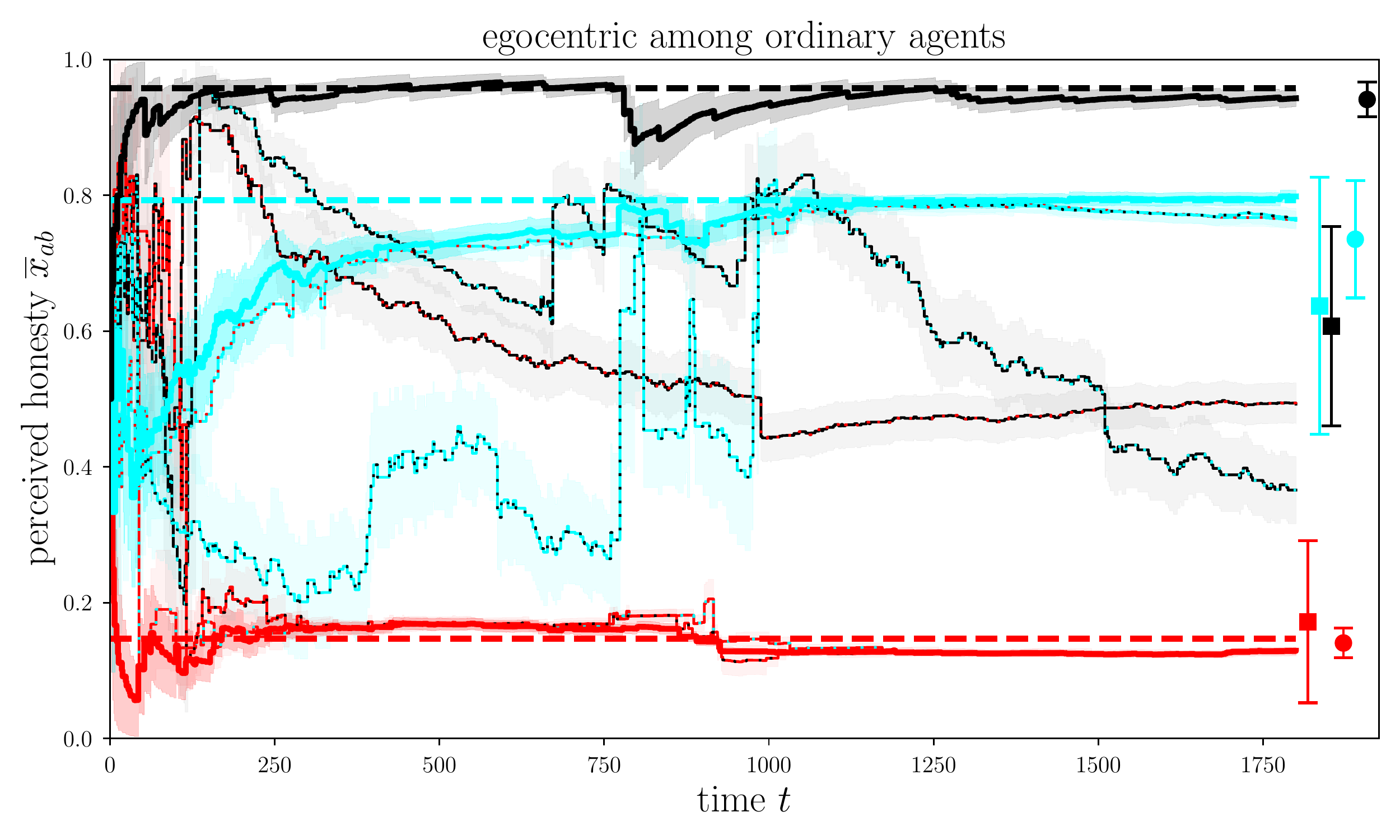}

\includegraphics[width=0.5\textwidth]{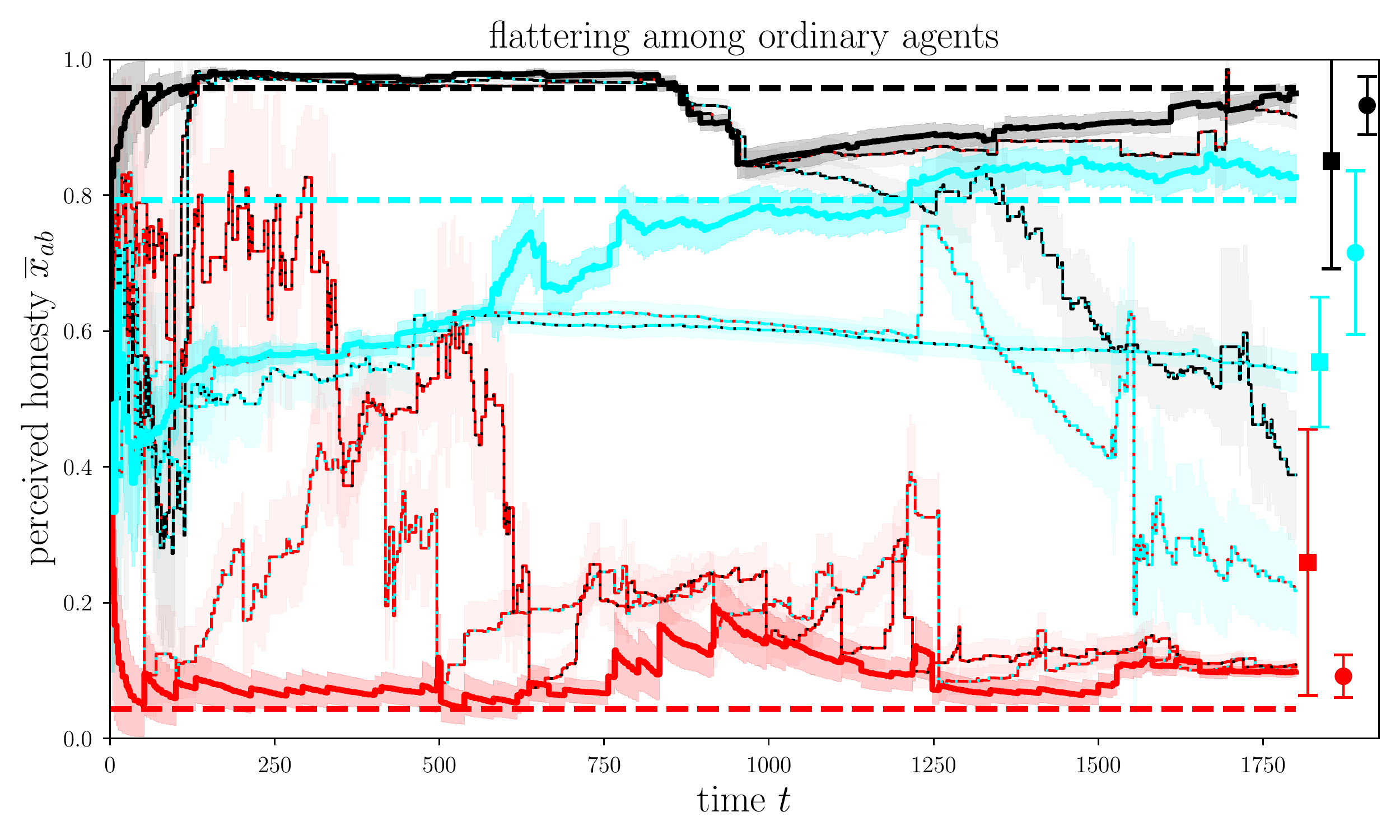}\includegraphics[width=0.5\textwidth]{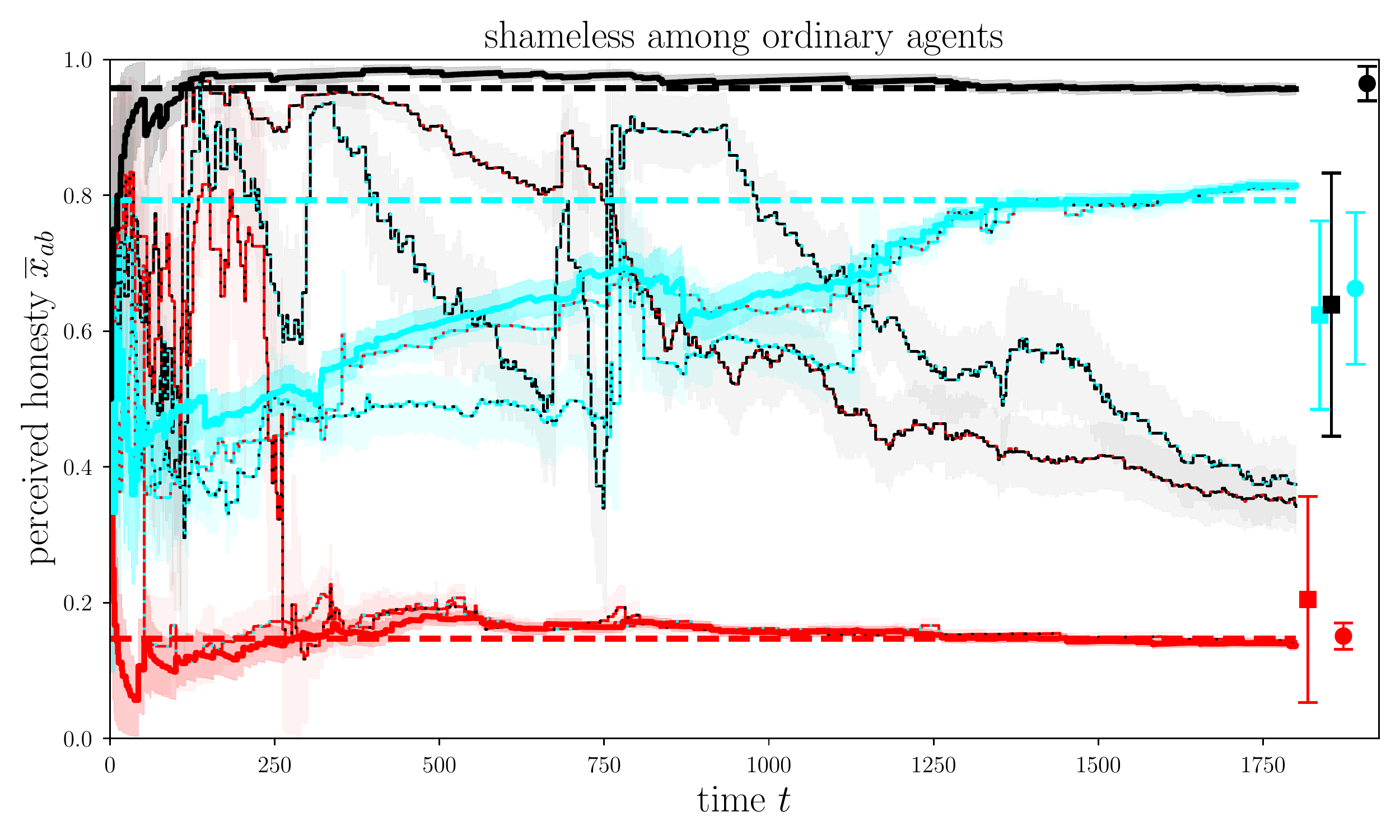}

\includegraphics[width=0.5\textwidth]{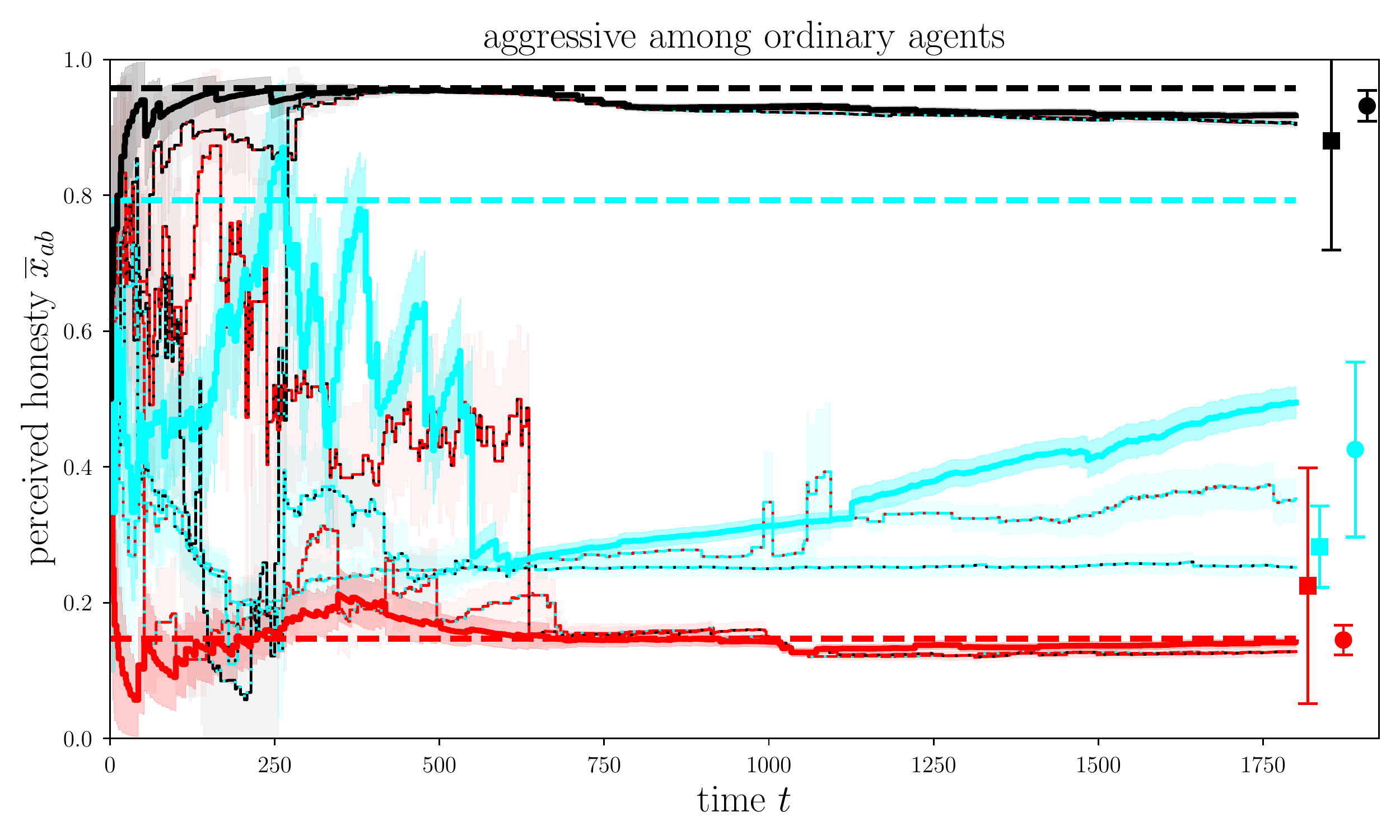}\includegraphics[width=0.5\textwidth]{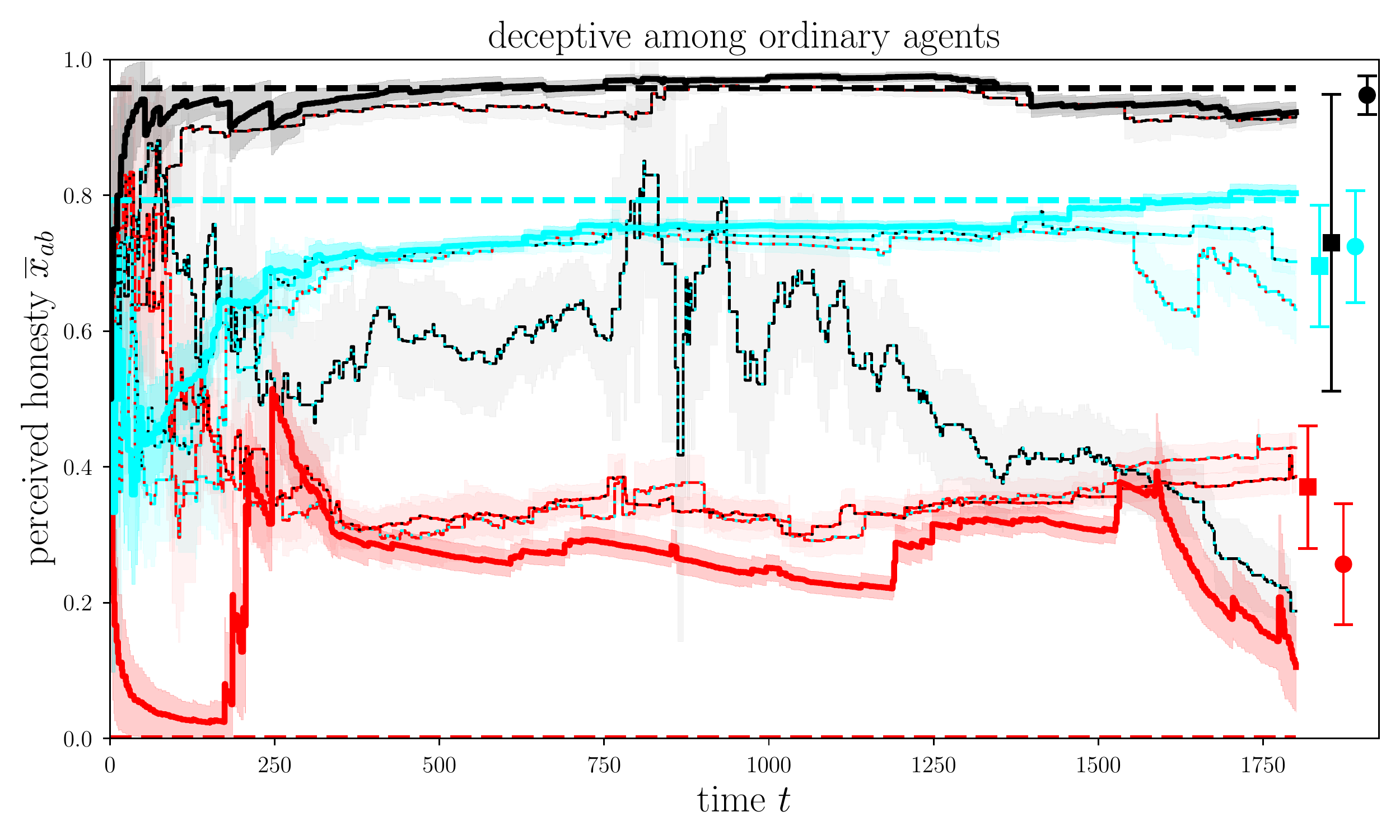}

\caption{Reputation dynamics as in Fig.\ \ref{fig:Reputation-game-simulations}
for simulations of basic communication strategies with agent red being
here strategic (top left), egocentric (top right), and flattering
(middle left), shameless (middle right), aggressive (bottom left),
and deceptive (fourth row). The used random sequences No.\ 1 are
also identical to the simulations shown there. \label{fig:Basic-communication-strategies}}
\end{figure*}

A propaganda simulation with such honest cross-communication is shown
in the bottom left panel of Fig.\ \ref{fig:Propaganda-simulation}.
As soon as the receiving agents cross-communicate their beliefs about
red, the dynamics gets even more complicated. Although the receiving
agents communicate honestly, they first have to build trust. This
process exhibits a complex dynamic, which lets only agent cyan and
yellow trust each other in the end and distrusting red. Agent black,
despite being initially very reputed, loses the trust of the other
agents as well as black loses the trust in them.

This isolates black from the protecting effect of their communications,
and lets black accept the propaganda even more than in the scenario
without cross-communication. The reason for this is that diverging
opinions of cyan and yellow about red harm black's lie detection in
addition to what the propaganda does to it.

Nevertheless, this simulation shows that honest cross-communication
among recipients of propaganda can mitigate the propaganda's impact
to some degree.

\subsubsection{Impact and resilience of a smart agent}

The bottom right panel of Fig.\ \ref{fig:Propaganda-simulation}
shows a simulation with a similar setup as before, but this time we
assume agent black uses a smart receiver strategy. This means that
black maintains and uses a set of guesses about the other agents'
beliefs (as stored in $I_{\text{black red red}}$ for red) and intentions
(as $\widetilde{I}_{\text{black red red}}$) to detect lies. The smart
receiver startegy allows black to identify red's communications as
propaganda, after a period of varying opinions about red, and to convince
cyan and yellow also to distrust red in the end. Thus, the smart receiver
strategy offers more resilience against exaggerated lies than the
critical one.

We note that at the peak of red's reputation with black, black has
a bimodal belief state about red with $I_{\text{black red}}^{(\text{smart})}=(-0.39,-0.92)$.
This is expressing that at that moment black is aware that either
red is very honest or very dishonest, but certainly not anything in
the middle between these extremes.\footnote{A negative $\mu$ or $\lambda$ creates an integrable singularity
(for $\mu,\lambda>-1$) in the belief distribution at $x=0$ (complete
dishonest) or $x=1$ (complete honest), receptively.} 

\begin{figure*}[!t]
\vspace{-1.5em}
\includegraphics[width=0.5\textwidth]{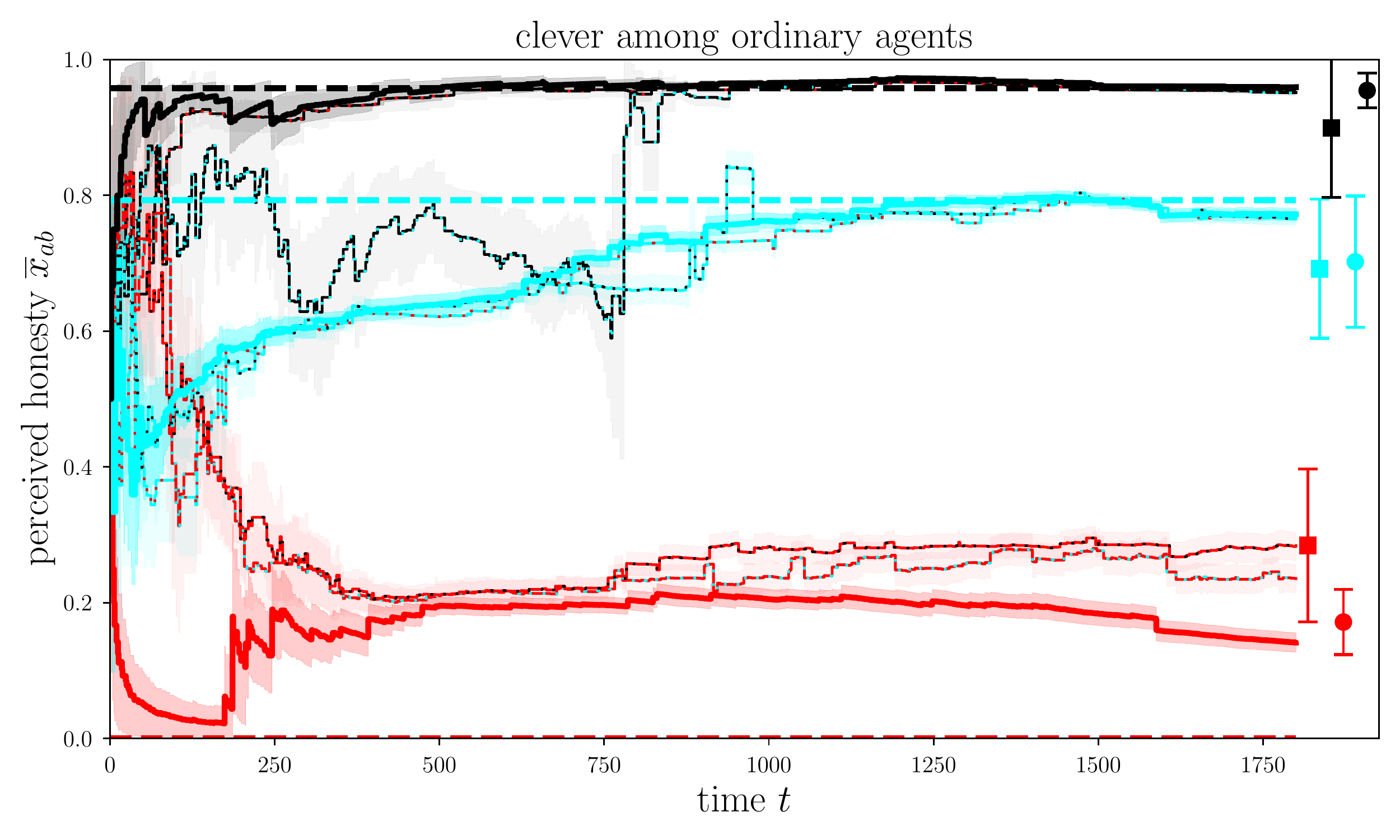}\includegraphics[width=0.5\textwidth]{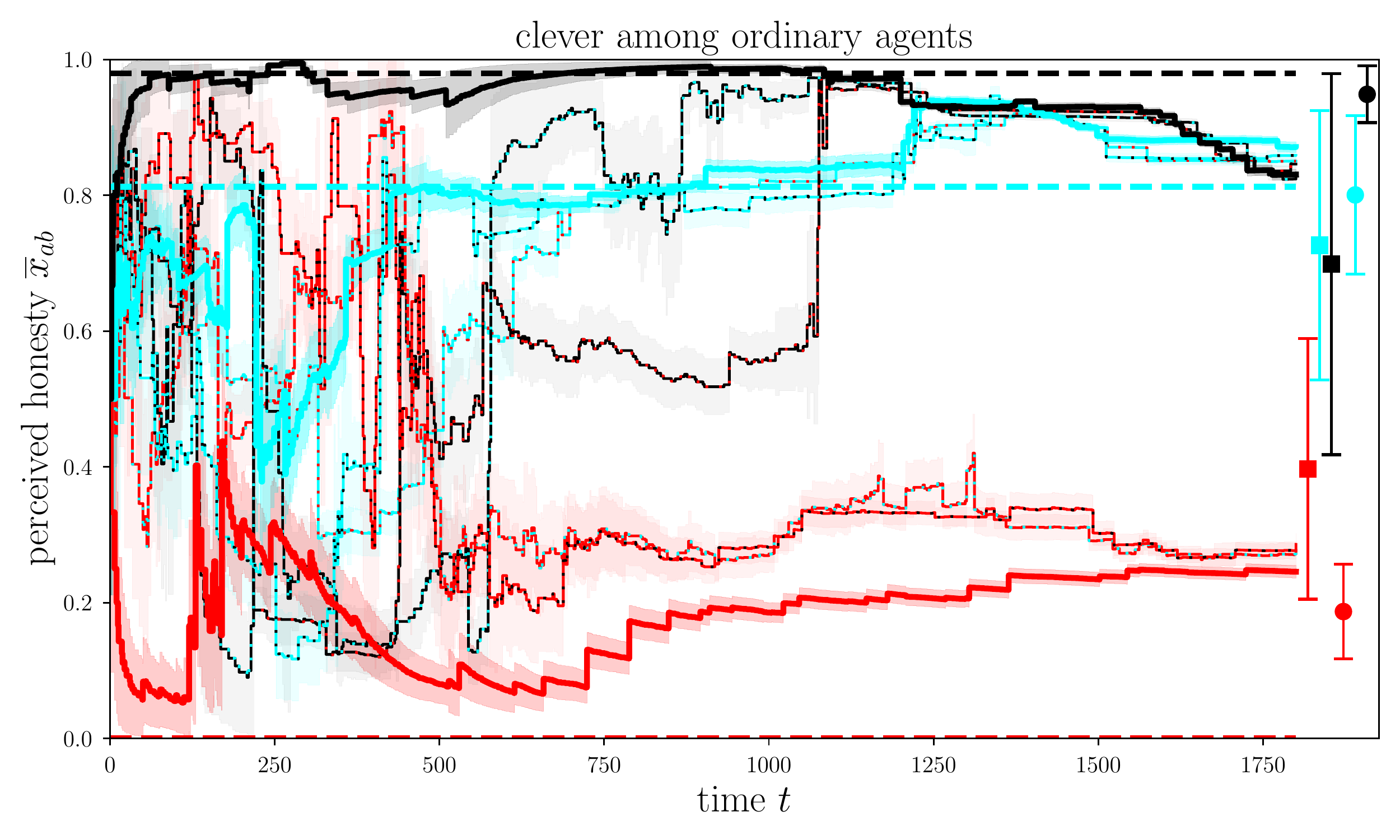}

\includegraphics[width=0.5\textwidth]{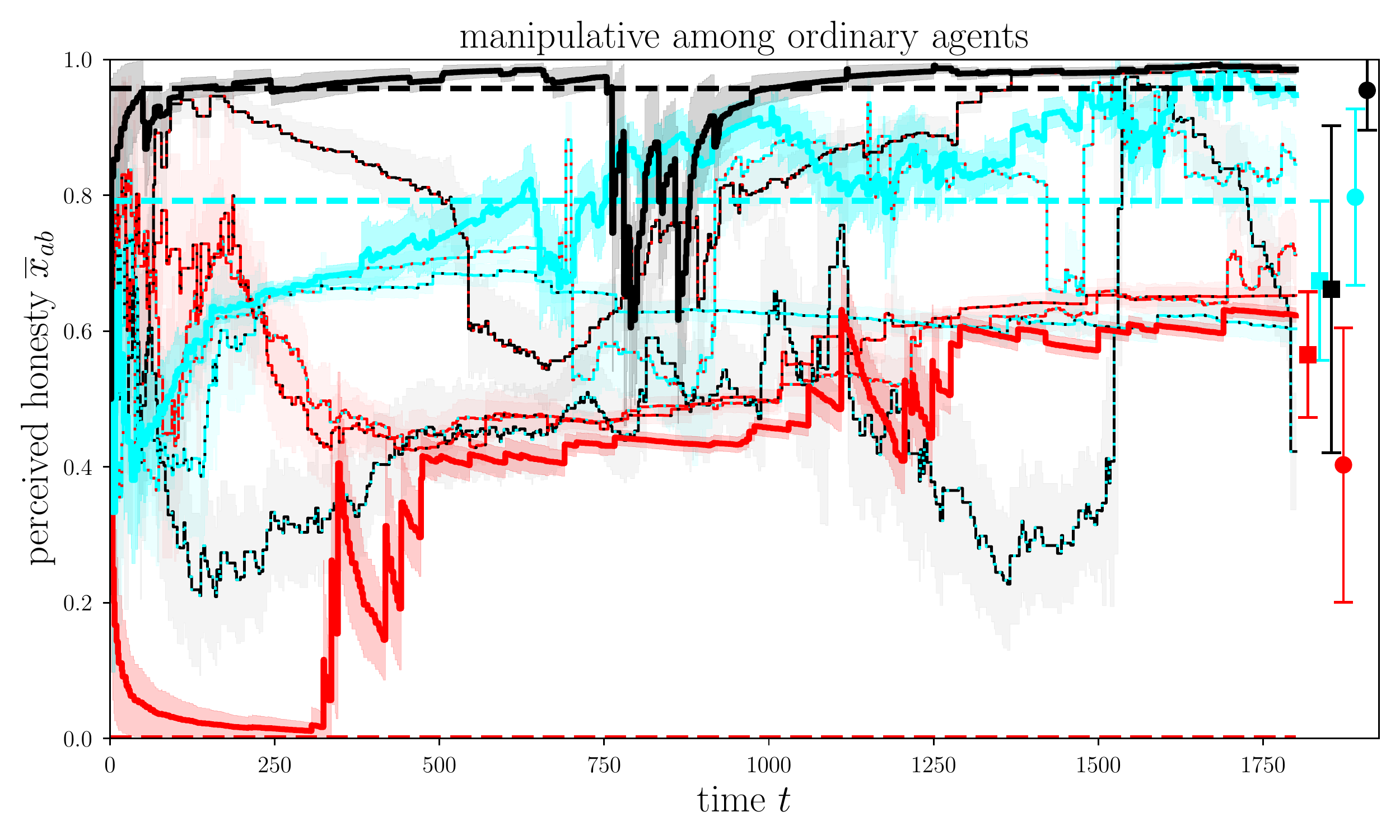}\includegraphics[width=0.5\textwidth]{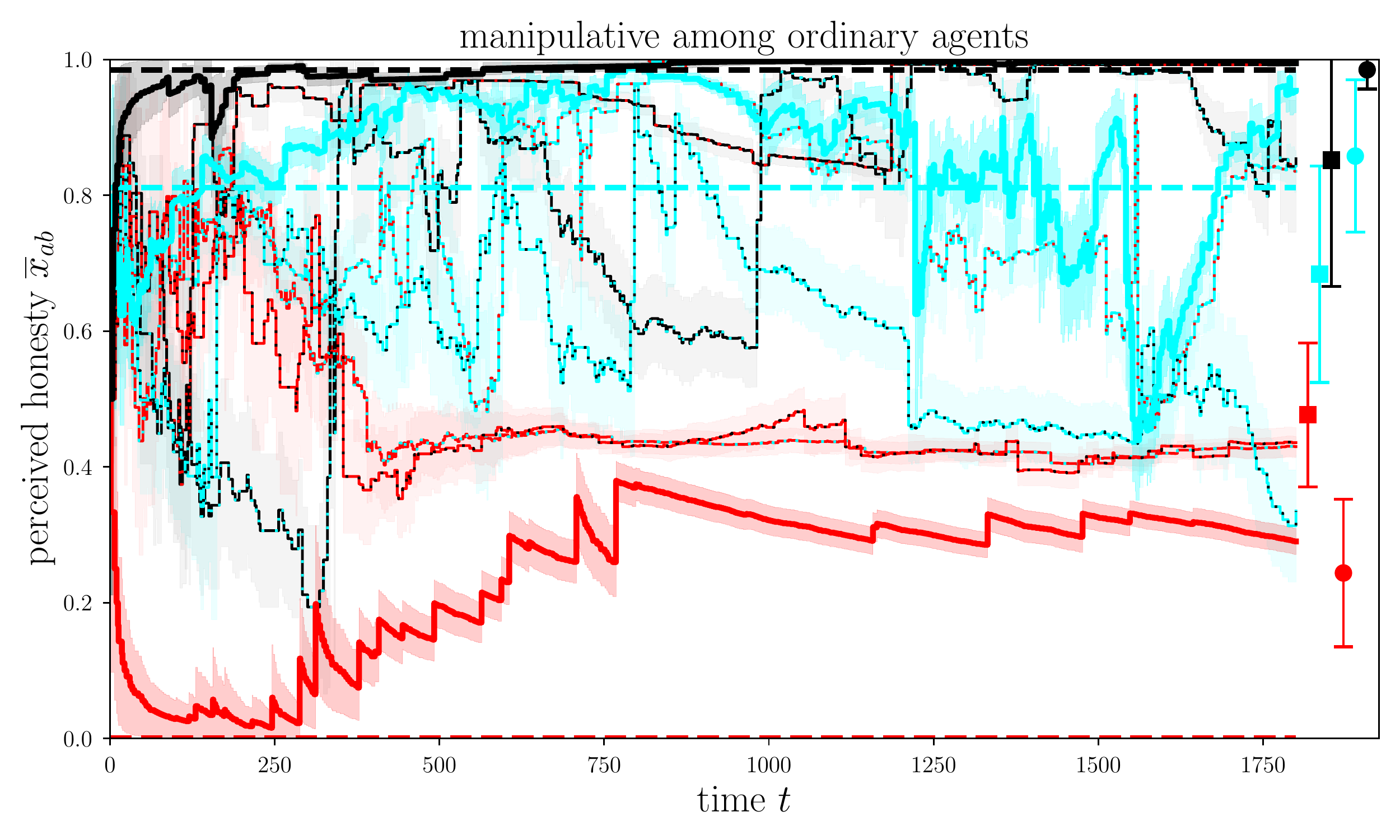}

\includegraphics[width=0.5\textwidth]{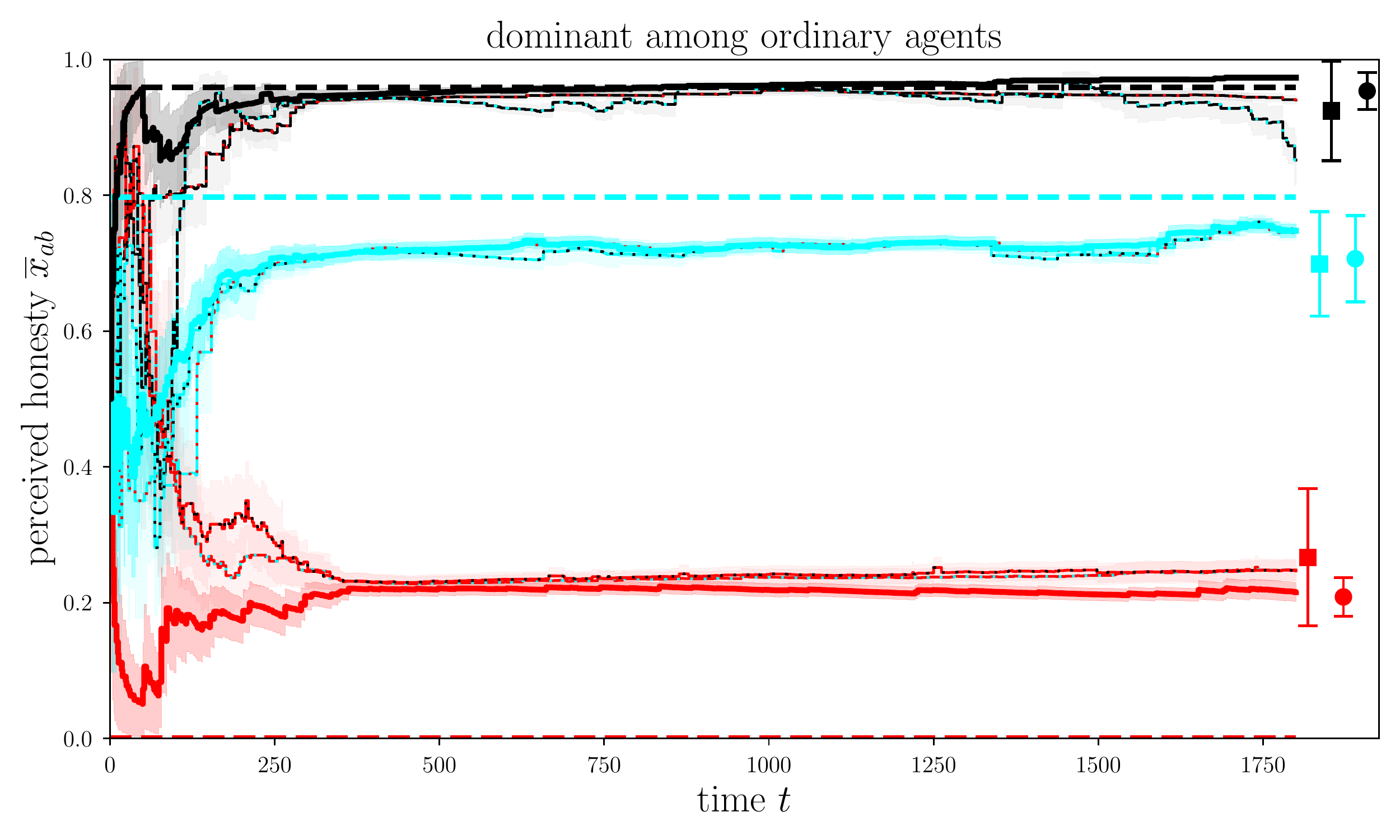}\includegraphics[width=0.5\textwidth]{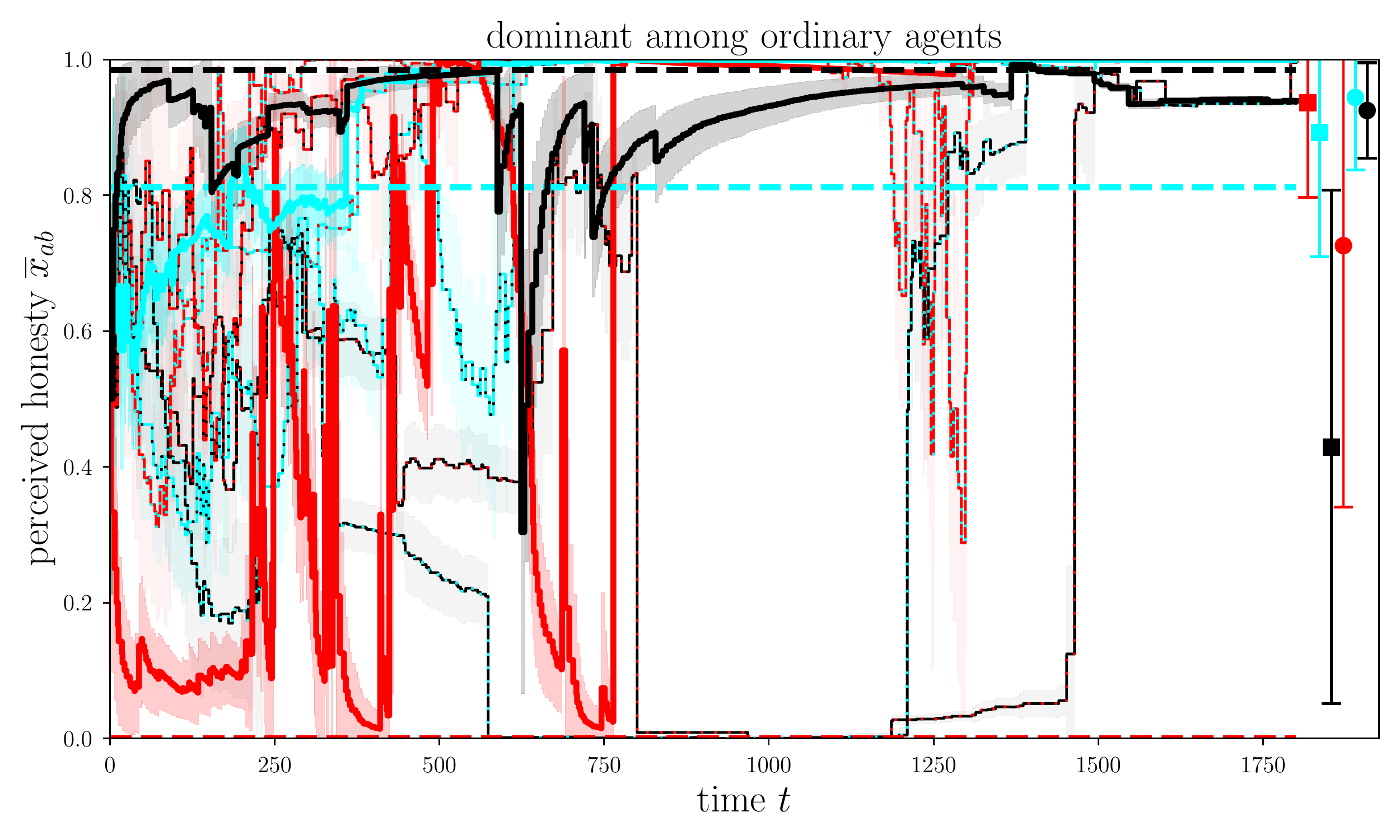}

\includegraphics[width=0.5\textwidth]{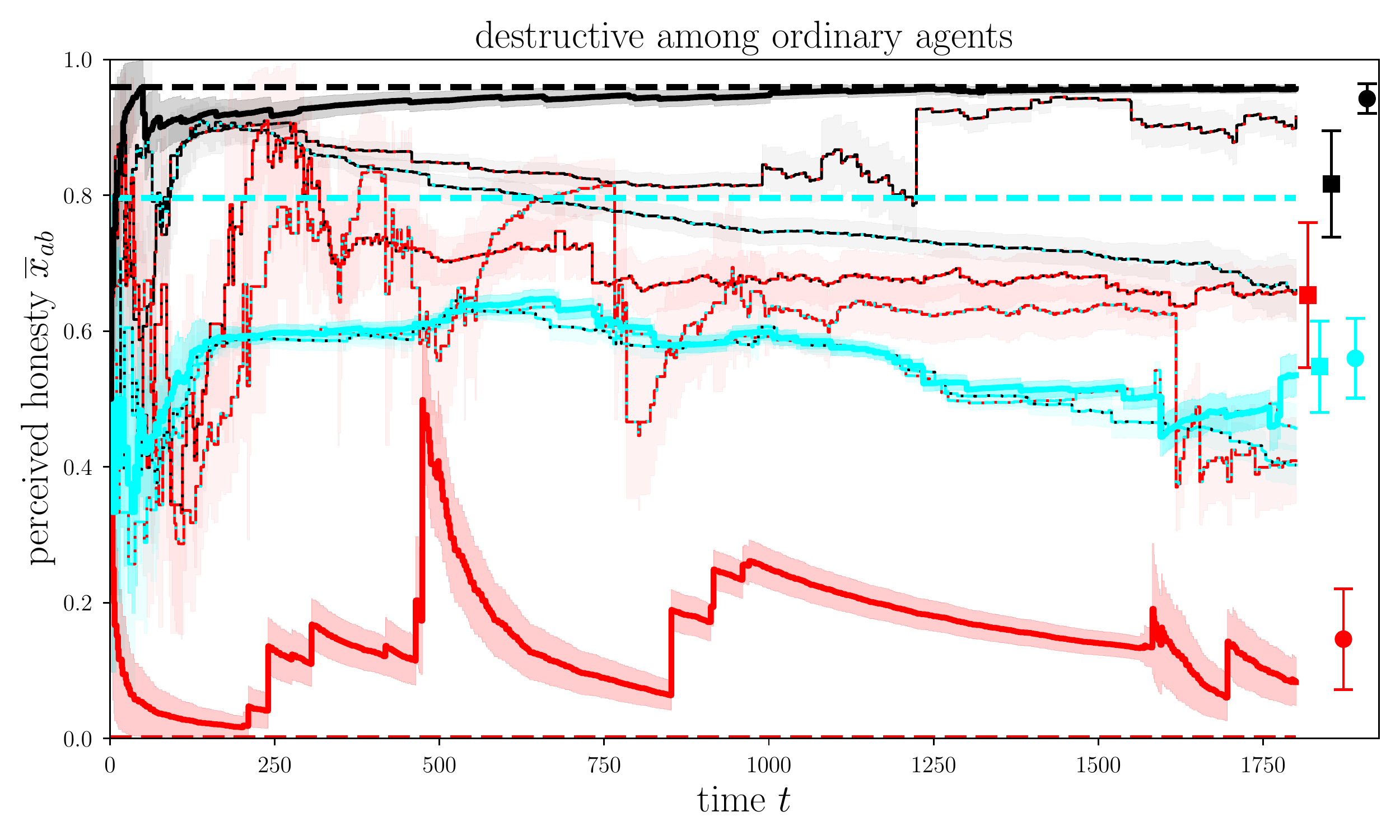}\includegraphics[width=0.5\textwidth]{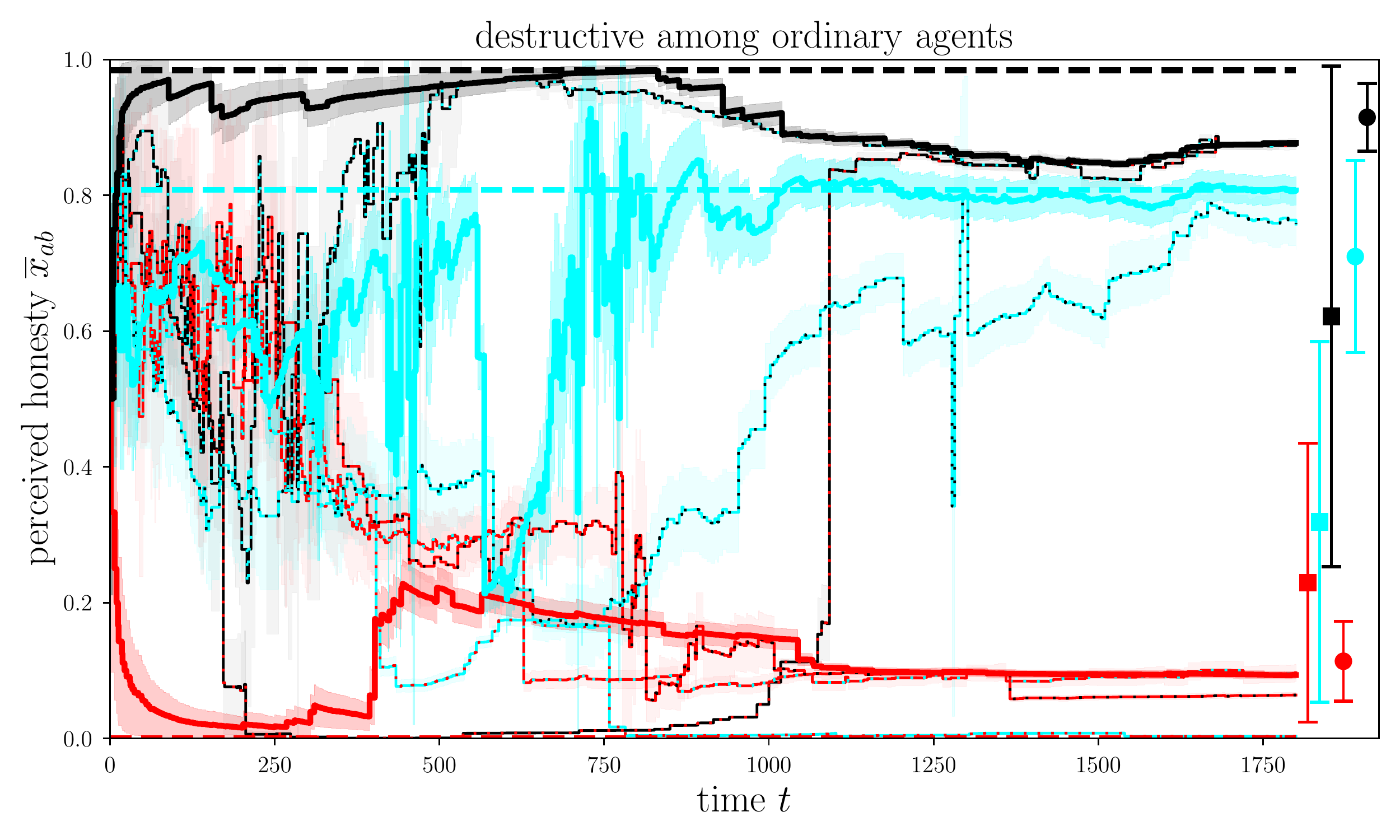}

\caption{As Fig.\ \ref{fig:Basic-communication-strategies}, but for agent
red being deceptive (first row), manipulative (second row), dominant
(third row), and destructive (fourth row). The left column shows simualtions
using random sequences No.\ 1 and the right using No.\ 2. \label{fig:Special-communication-strategies}}
\end{figure*}

\begin{figure*}[!t]
\vspace{-1.5em}
\includegraphics[width=0.5\textwidth]{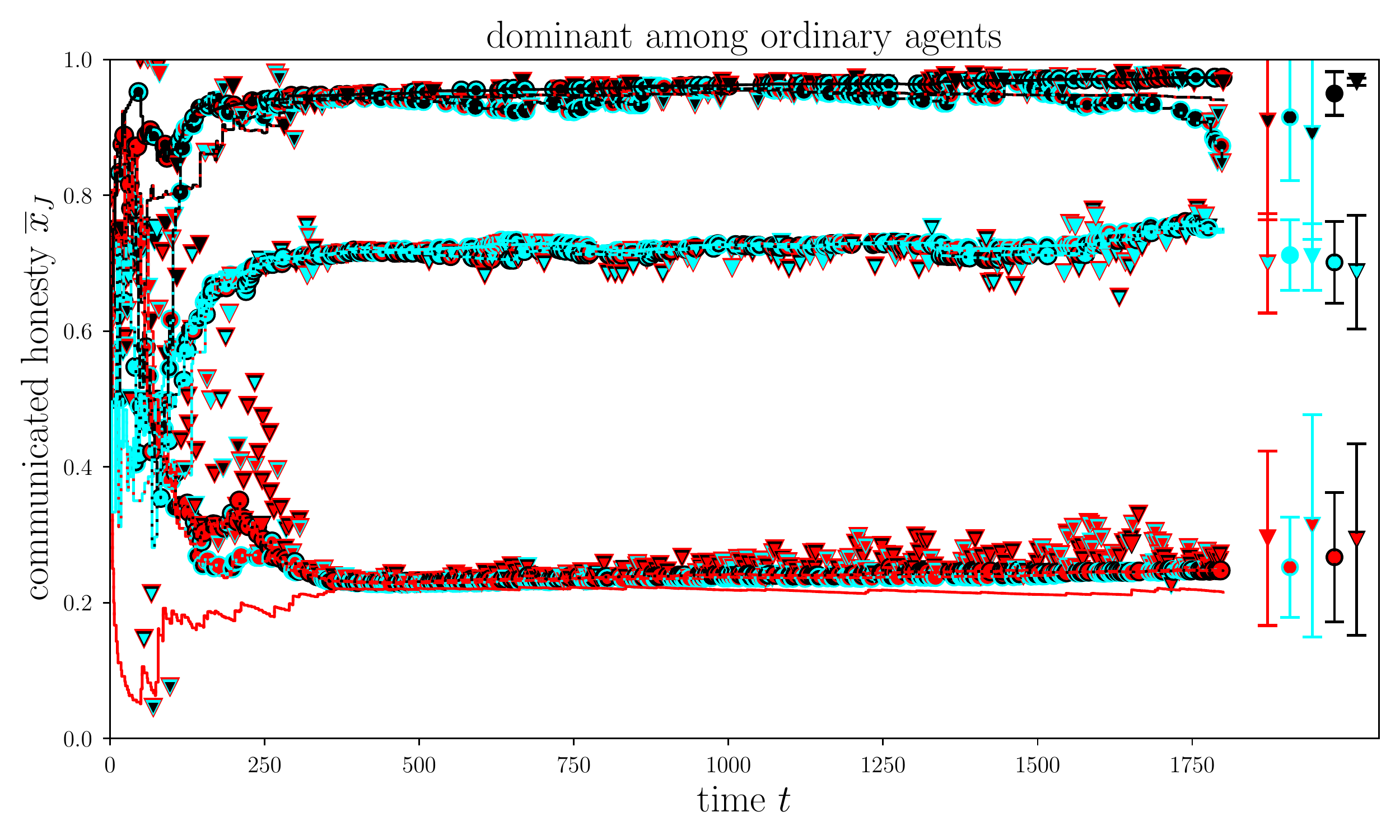}\includegraphics[width=0.5\textwidth]{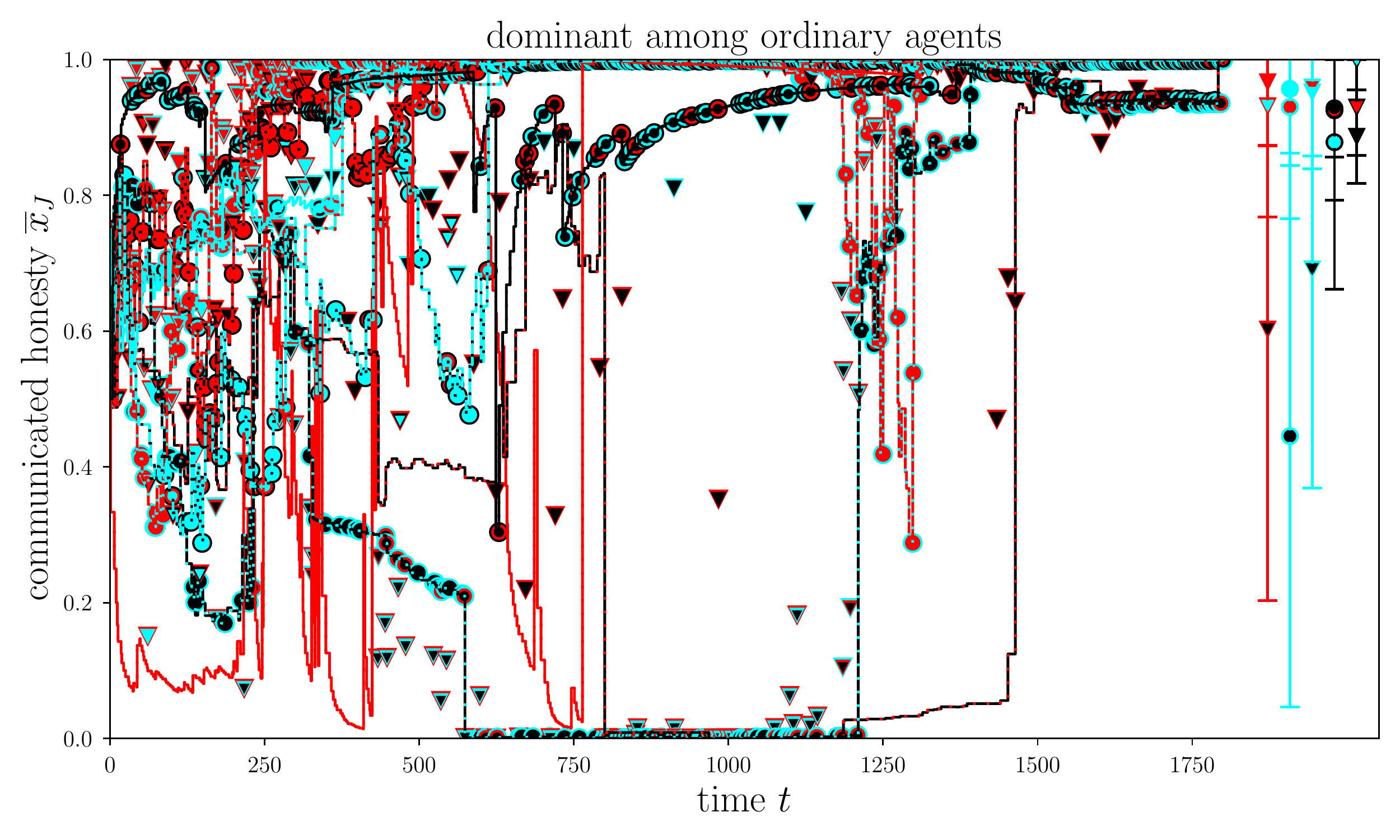}

\caption{Communication dynamics like in Fig.\ \ref{fig:Communication-patterns}
here for the dominant agent in the simulation with random sequences
No.\ 1 (left) and No.\ 2 (right) to highlight the typical social
atmospheres created by such agents. \label{fig:Dominant-communication-patterns}}
\end{figure*}

\subsection{Communication strategies\label{subsec:Communication-strategies}}

We now discuss the impact of the basic and special communication strategies.
The setup will be as in Sect.\ \ref{subsec:Game-simulations}, but
now agent red uses basic or special communication strategies. The
intrinsic honesty of the three agents and the random sequences determining
the course of simulation events will again be identical to what they
were in the simulations shown in Figs.\ \ref{fig:Reputation-game-simulations}
and \ref{fig:Communication-patterns} for some of the runs. These
are the runs with random number sequence No.\ 1 from our statistical
set of one hundred simulations to be discussed later. For the special
agents, we will also show runs with random sequence No.\ 2 to illustrate
the variance in the dynamics with otherwise identical setup.

\subsubsection{Basic communication strategies\label{subsec:Basic-communication-strategies-1}}

Fig.\ \ref{fig:Basic-communication-strategies} shows simulation
runs, with the setup of Fig.\ \ref{fig:Reputation-game-simulations},
but here agent red is either strategic, egocentric, flattering, shameless,
aggressive, or deceptive. The corresponding communication patterns
can be found in Appendix \ref{sec:Detailed-communication-strategies}.
None of the basic strategies adapted by agent red appears to be efficient
in boosting red's reputation, except for the flattering and the deceptive
strategies, which both let red lie more often.

The \textbf{strategic agent} red concentrates opinion exchanges on
the most reputed agent black, and thereby manages to convince black
that cyan is untrustworthy. This lets cyan, who actually is trustworthy,
doubt black's honesty as a reaction to black's opinion on them. However,
black's and cyan's reputations still stay well above that of red,
as red's reputation suffers from red's occasional confessions.

The \textbf{egocentric agent} red speaks mostly about themselves.
This has two visible consequences: Firstly, the others' opinions on
red converge faster, due to the larger number of confessions made
by red. Secondly, the lies of red are not able to follow the development
of the other agent's opinions on other agents that well.\footnote{See the horizontally aligned lies of red on cyan in the period $t=200$
to $800$ in the communication record displayed in Fig.\ \ref{fig:Basic-communication-strategies-2}.<}

The \textbf{flattering agent} red is somehow successful in obtaining
an enhanced reputation. The key factor is that red is preferentially
talking about others, thereby avoiding giving information about themselves
away via confessions. This helps red to establish a slightly higher
reputation than in the other scenarios discussed so far. The feedback
to red by the other agents lets red's self-esteem grow to this enlarged
value until $t=1000$. Thereafter, confessions by red are based on
this enhanced value and do not let the other agents' opinions fall
below it. We witness here a successful and advantageous self-deception
of an agent.

The \textbf{shameless agent} red lies without blushing, and therefore
is more convincing. As a result red's reputation grows slightly higher
than in the ordinary agent scenario. Red's reputation is held back
by red's confessions and the inertia the converging group opinion
generates against the pull of red's self-appraisal. We note, however,
the significantly reduced reputations of black owning to the more
convincing lies of a shameless agent red.

The \textbf{aggressive agent} red attacks preferentially cyan, who's
reputation and self-esteem suffer significantly from red's vilification.

Finally, the \textbf{deceptive agent} red manages to get the highest
reputation and self-esteem of red in all the scenarios discussed so
far, since red does not make a single confession, and self-promotes
with a high frequency.

\subsubsection{Special communication strategies\label{subsec:Special-communication-strategies-1}}

Figs.\ \ref{fig:Special-communication-strategies} shows runs for
agent red being clever (smart and deceptive), manipulative (clever,
anti-strategic, and flattering), dominant (clever, strategic, and
egocentric), and destructive (clever, strategic, aggressive, and shameless).
On the left panels of Fig.\ \ref{fig:Special-communication-strategies}
the random sequences are chosen as before (and like runs No.\ 1 of
the statistics ensemble), whereas on the right panels different sequences
(runs No.\ 2) were chosen. The latter highlight how different dynamical
regimes can appear in otherwise identical setups. The corresponding
communication patterns can be found in Appendix \ref{sec:Detailed-communication-strategies}.
For the dominant agent we display them also in Fig.\ \ref{fig:Dominant-communication-patterns}
for a more detailed discussion.

\begin{figure*}[!t]
\vspace{-1.5em}
\includegraphics[width=0.333\textwidth]{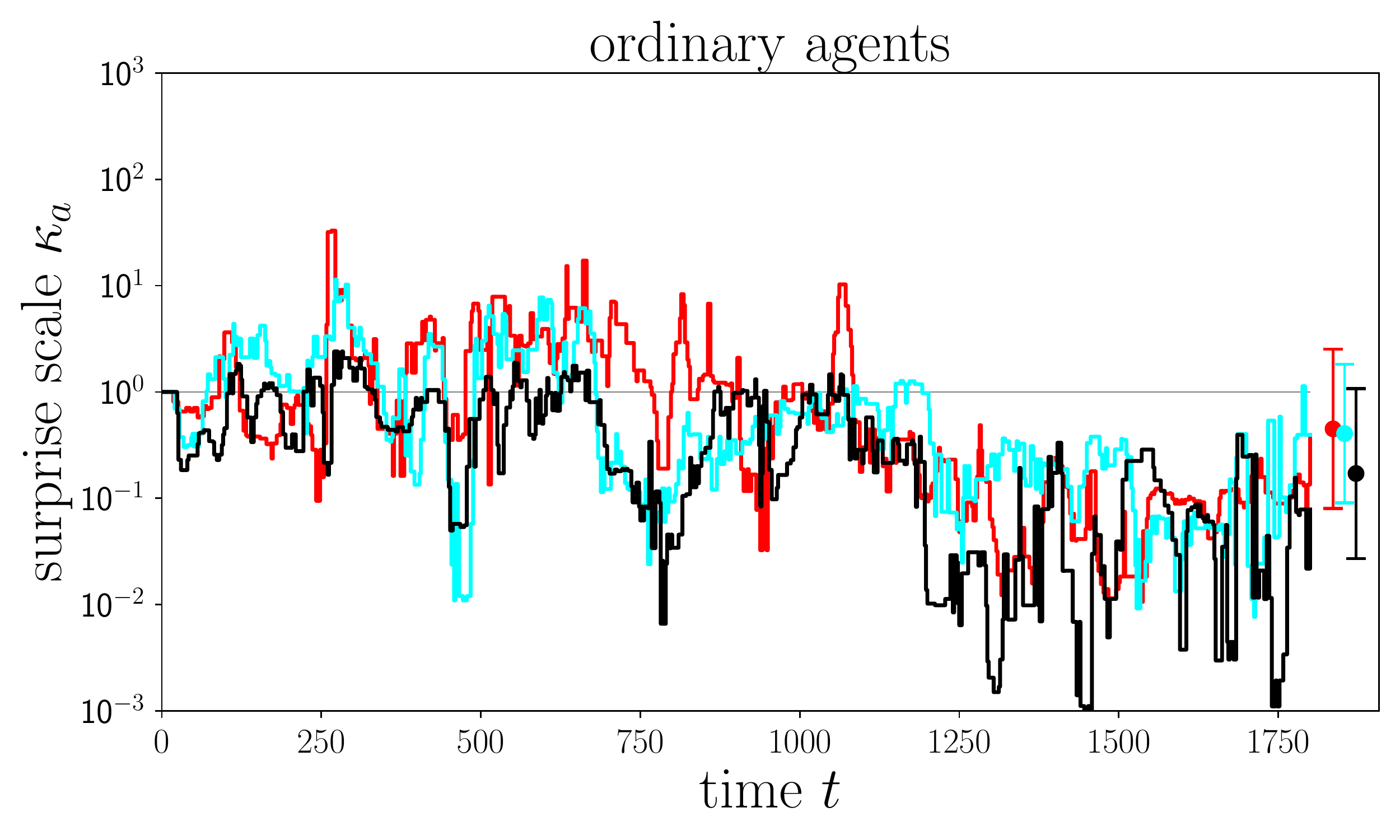}\includegraphics[width=0.333\textwidth]{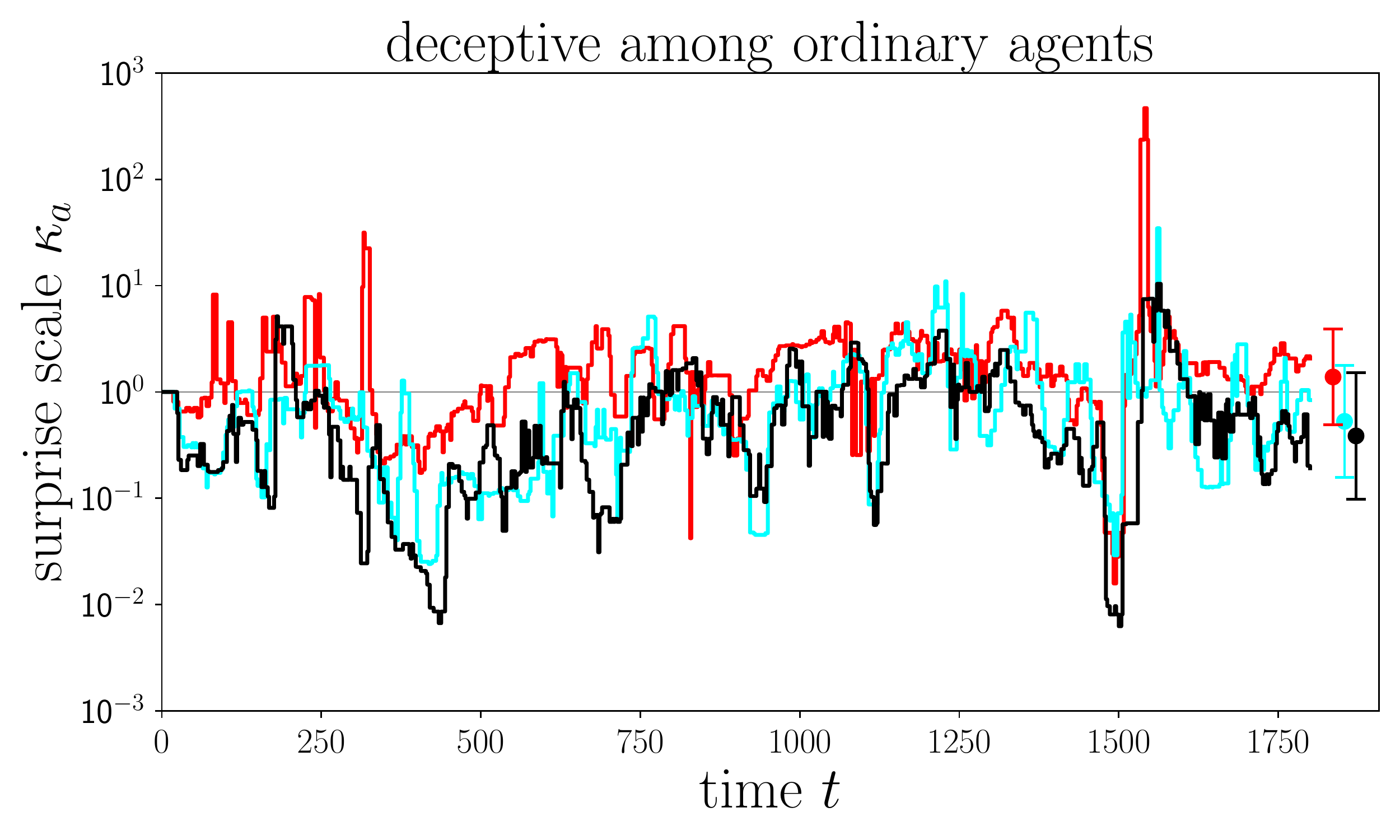}\includegraphics[width=0.333\textwidth]{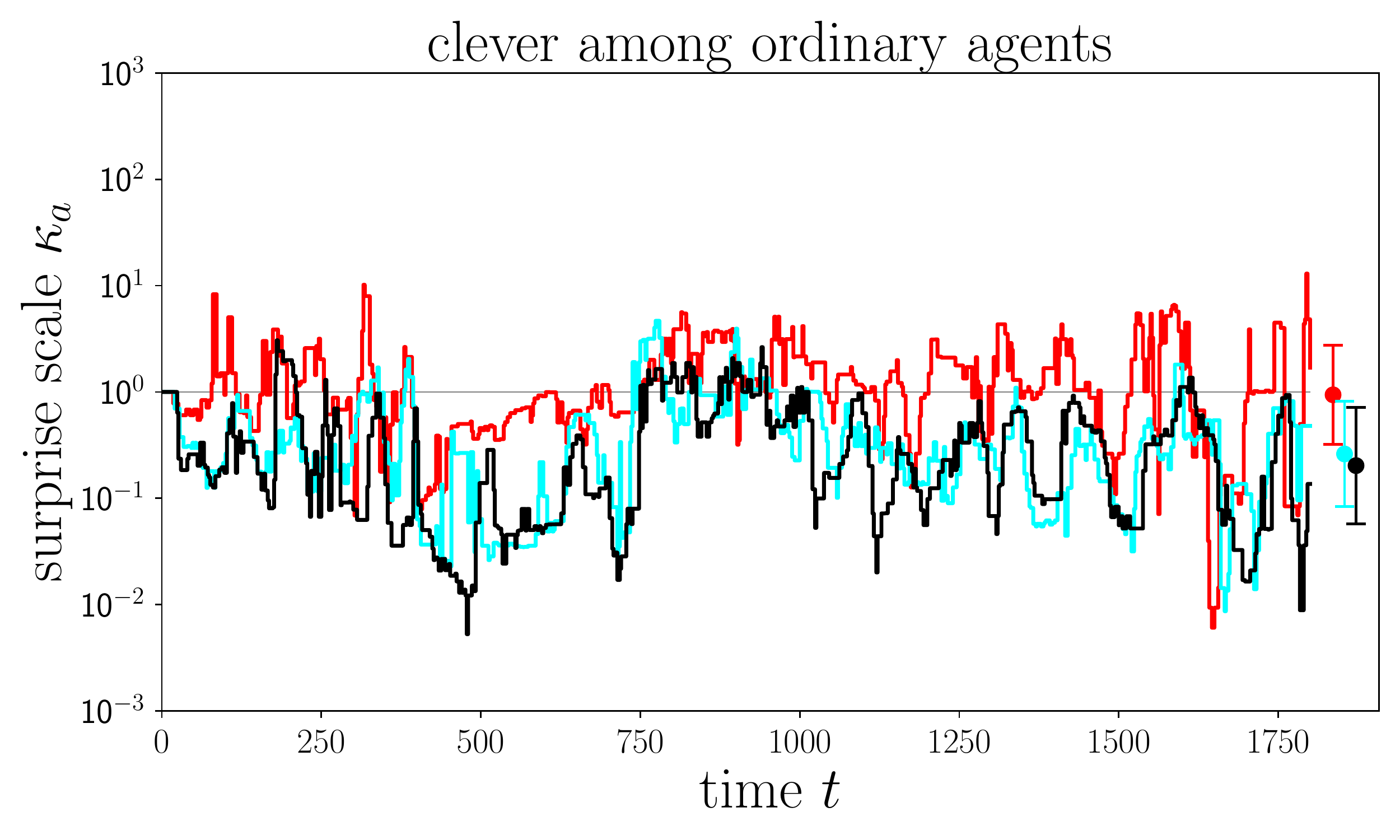}

\includegraphics[width=0.333\textwidth]{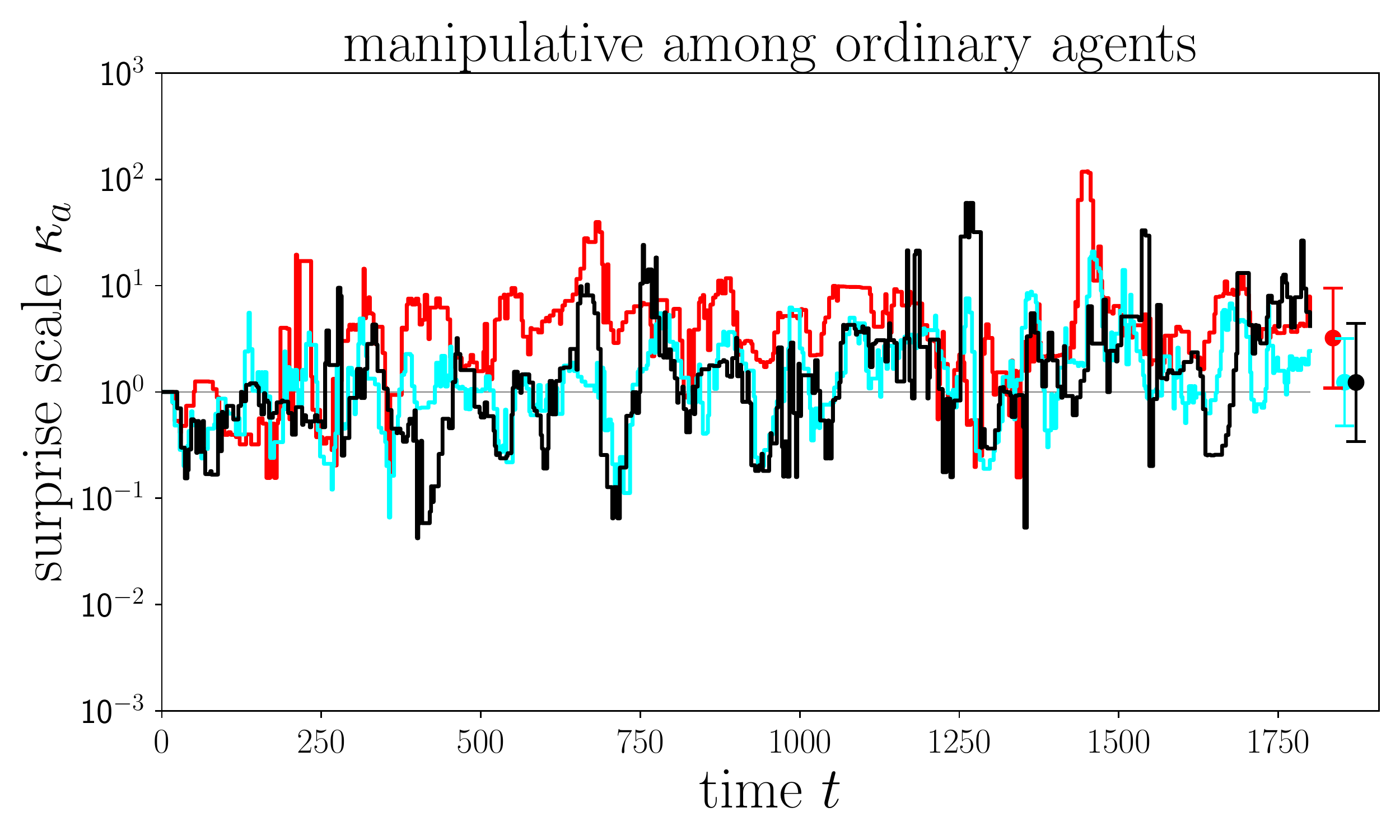}\includegraphics[width=0.333\textwidth]{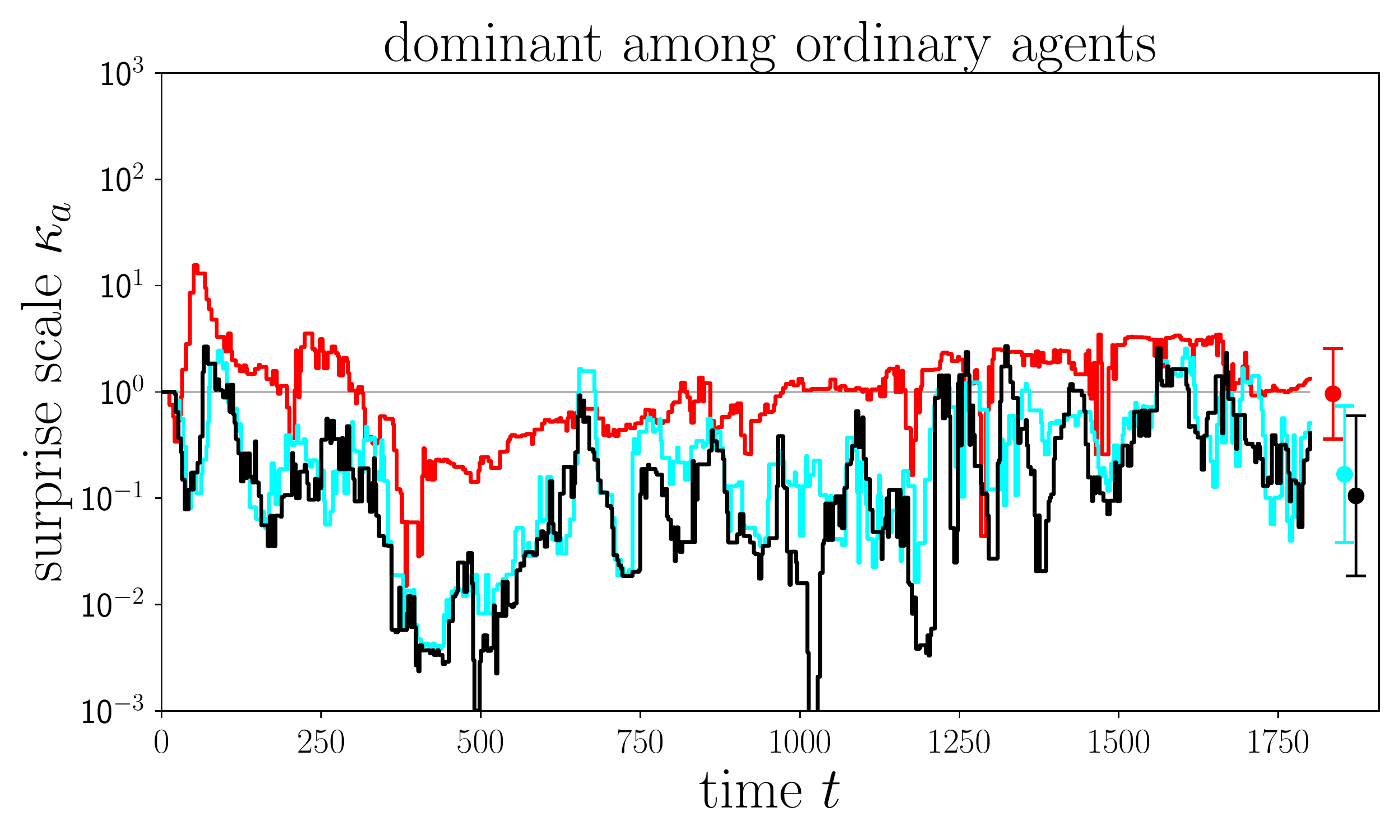}\includegraphics[width=0.333\textwidth]{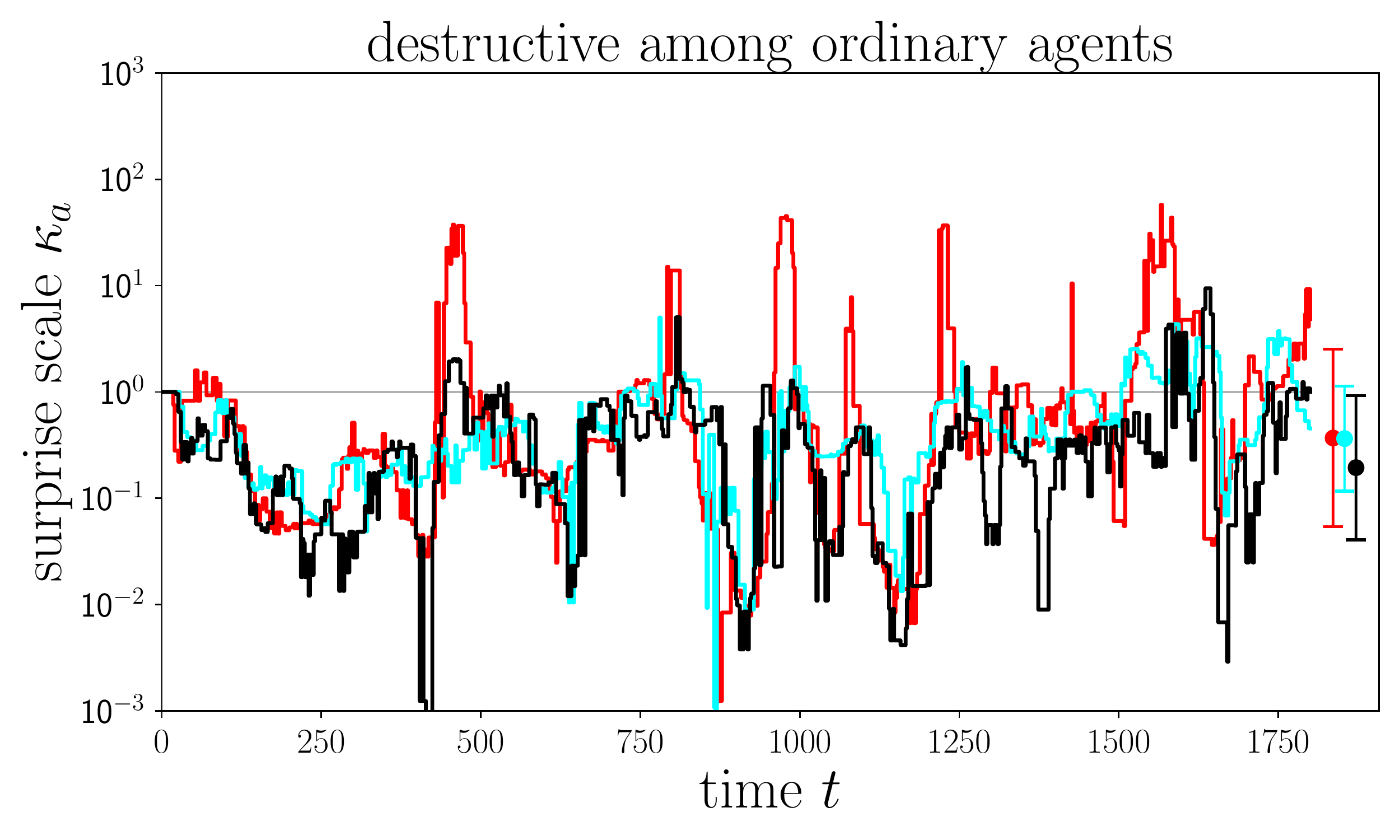}

\includegraphics[width=0.333\textwidth]{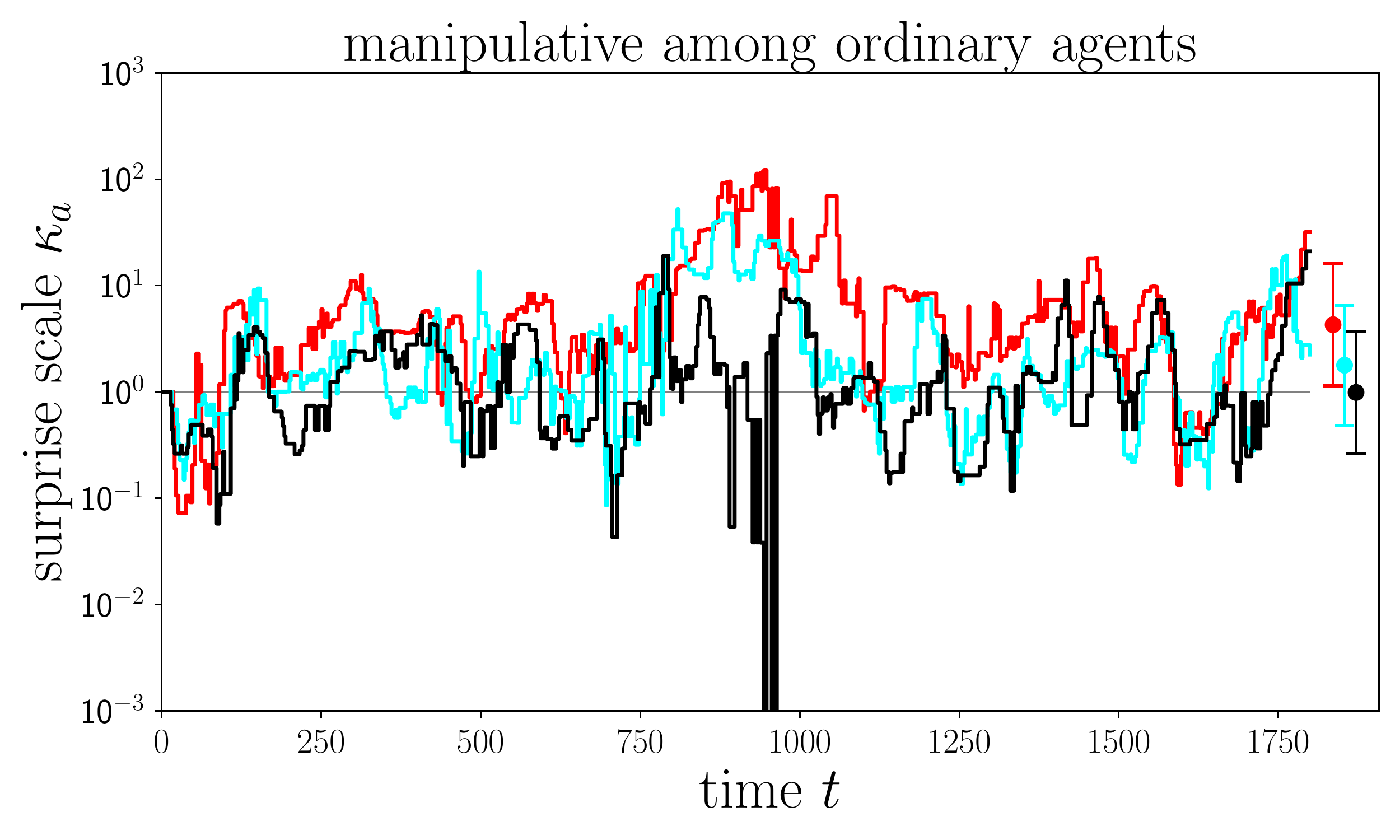}\includegraphics[width=0.333\textwidth]{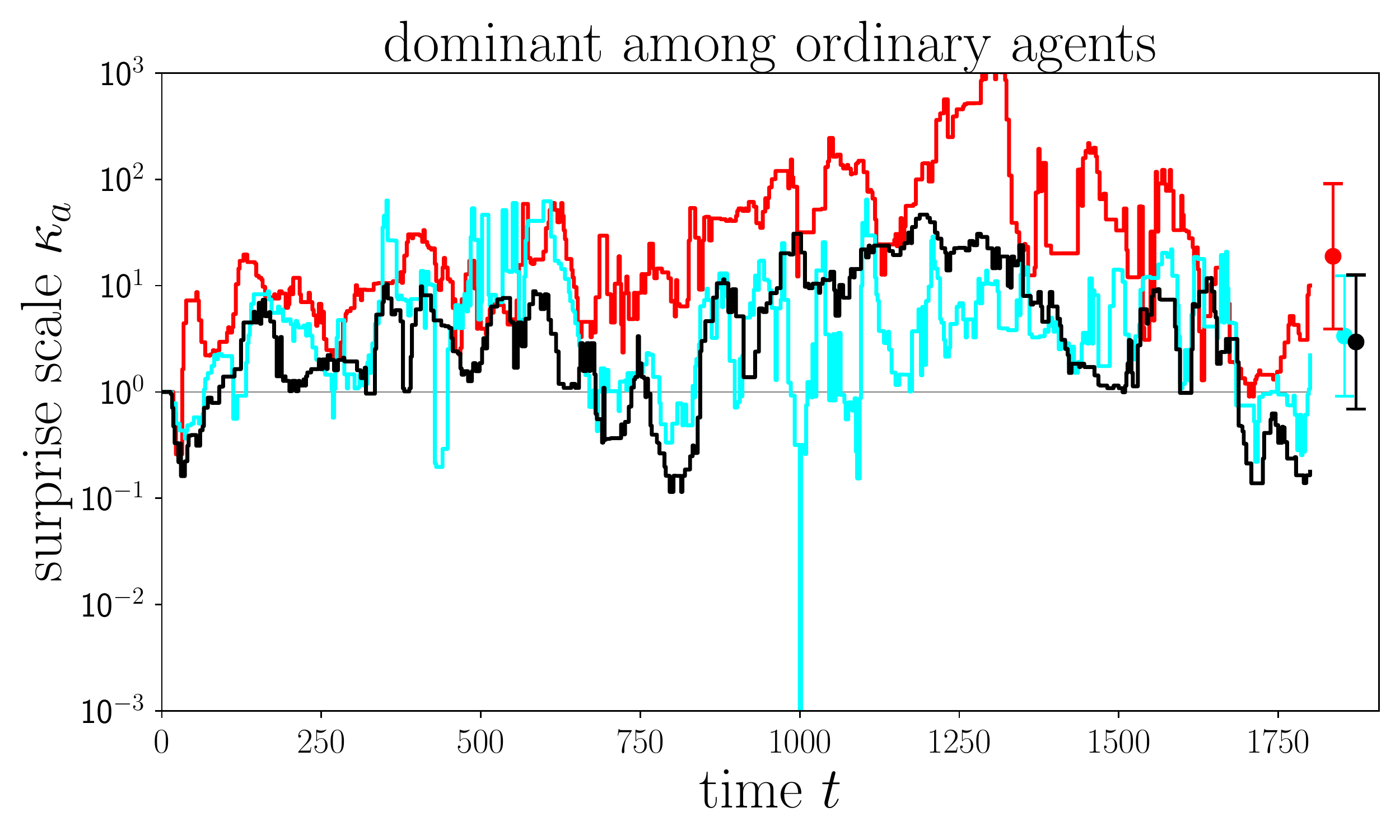}\includegraphics[width=0.333\textwidth]{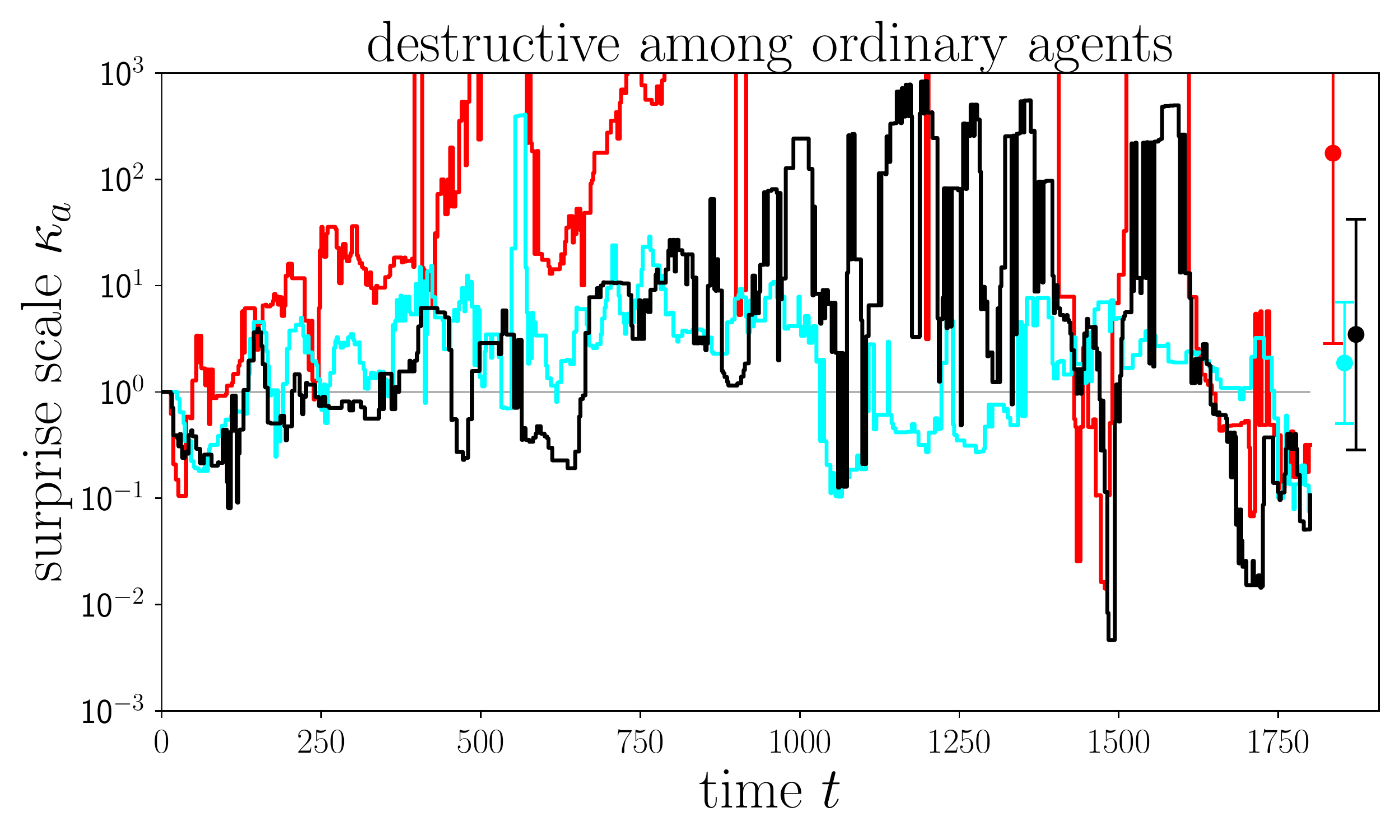}

\caption{Time evolution of $\kappa_{i}$ with $i\in\{\text{black, cyan, red}\}$
in different scenarios. The strategy used by agent red is specified
in each panel's titles. The top and middle rows are for the random
sequences of runs No.\ 1 and the bottom row for No.\ 2. The data
points with bars at the end of the evolutionary tracks show the mean
and standard deviation of the evolution of $\log_{10}\,\kappa_{i}$
for each agent $i$ in their corresponding color. Hereby, any averaging
was performed on a logarithmic scale. \label{fig:kappa-comparison}}
\end{figure*}

Furthermore, Fig.\ \ref{fig:kappa-comparison} shows the evolution
of the lie detection scale $\kappa_{a}$ for an instructive selection
of simulation runs. A larger $\kappa_{a}$ of agent $a$ implies that
this agent is used to receive messages that diverge more from the
own opinions. This can make the agent blind for smaller lies.

The runs shown there with red being an \textbf{ordinary agent} shows
that usually $\kappa_{a}$ varies on a logarithmic scale around unity,
with a typical variance of one order of magnitude up or down.

The \textbf{clever agent} red performs slightly worse in terms of
reputation than in the run where red is deceptive (see discussion
before). The lie detection scale $\kappa_{\text{red}}$ of the clever
agent red is significantly larger than that of the other two agents
in the same run. However, thanks to being clever, red's lies match
the believes of the other agents better and these experience therefore
reduced surprises compared to what they experience in the deceptive
scenario. This will also be the case for many of the runs with the
other special agents, to which the clever agent scenario is a reference
for comparison.

Compared to the case when agent red is clever, the \textbf{manipulative
agent} red is much more successful. As red is mostly flattering cyan,
the latter gets a significant self-esteem boost in the simulation
No.\ 1 (second row, left panel of Fig.\ \ref{fig:Special-communication-strategies}),
and partly also in No.\ 2 (second row, right panel).

By focusing on their reputation, the \textbf{dominant agent }red freezes
the group opinion on red (third row, left panel of Fig.\ \ref{fig:Special-communication-strategies}),
preventing red to obtain a high reputation in about half of the simulations
with red being dominant. This is accompanied by strongly reduced variances
in opinions and therefore in $\kappa_{i}$ for every agent $i$ in
such runs (see central panel of Fig.\ \ref{fig:kappa-comparison}).
The other half of the runs show much more volatility in red's reputation
with about one fifth of these runs leading to a top reputation and
self-esteem for red. In the third row, right panel of Fig.\ \ref{fig:Special-communication-strategies}
shows run No.\ 2 for the dominant agent red, which illustrates this
latter case, and seems to be typical for this outcome. Before red's
dominance is established, a period of high opinion volatility and
large uncertainty seems to be necessary. Red's lies in this scenario
are often on the extremes (see right panel of Fig.\ \ref{fig:Dominant-communication-patterns}
for the period 600 to 1200), creating a social atmosphere that might
be characterized as \emph{toxic}, as any enemy of red is often blamed
to be a complete liar. The reason for this is that red's self-esteem
does not manage to catch up with red's inflated reputation due to
red knowing their lies. Therefore, the many conversations of red about
red lead to a high level of cognitive dissonance, which inflates $\kappa_{\text{red}}$
by two orders of magnitude above the usual $\kappa_{\cdot}$ values
(see bottom middle panel of Fig.\ \ref{fig:kappa-comparison}). As
$\kappa_{\text{red}}$ is also used by red in lie construction, red's
expressed opinions tend to be largely on the extreme, either very
positive (about red and friends) or very negative (about enemies).
Only after red's self-esteem manages to become as high as red's reputation,
does $\kappa_{\text{red}}$ fall to a more normal level.

The \textbf{destructive agent} red manages to establish a high reputation
in run No.\ 1, but not in run No.\ 2. In the latter red largely
destroys cyan's reputation during the initial period with a concentrated
attack, though. Red's surprise scale $\kappa_{\text{red}}$ in this
run takes very extreme values, mostly due to the large difference
between red's and the others' opinions about them. As a consequence,
red speaks extremely negative about them, however, without being believed.

\begin{figure*}[!t]
\includegraphics[width=0.5\textwidth]{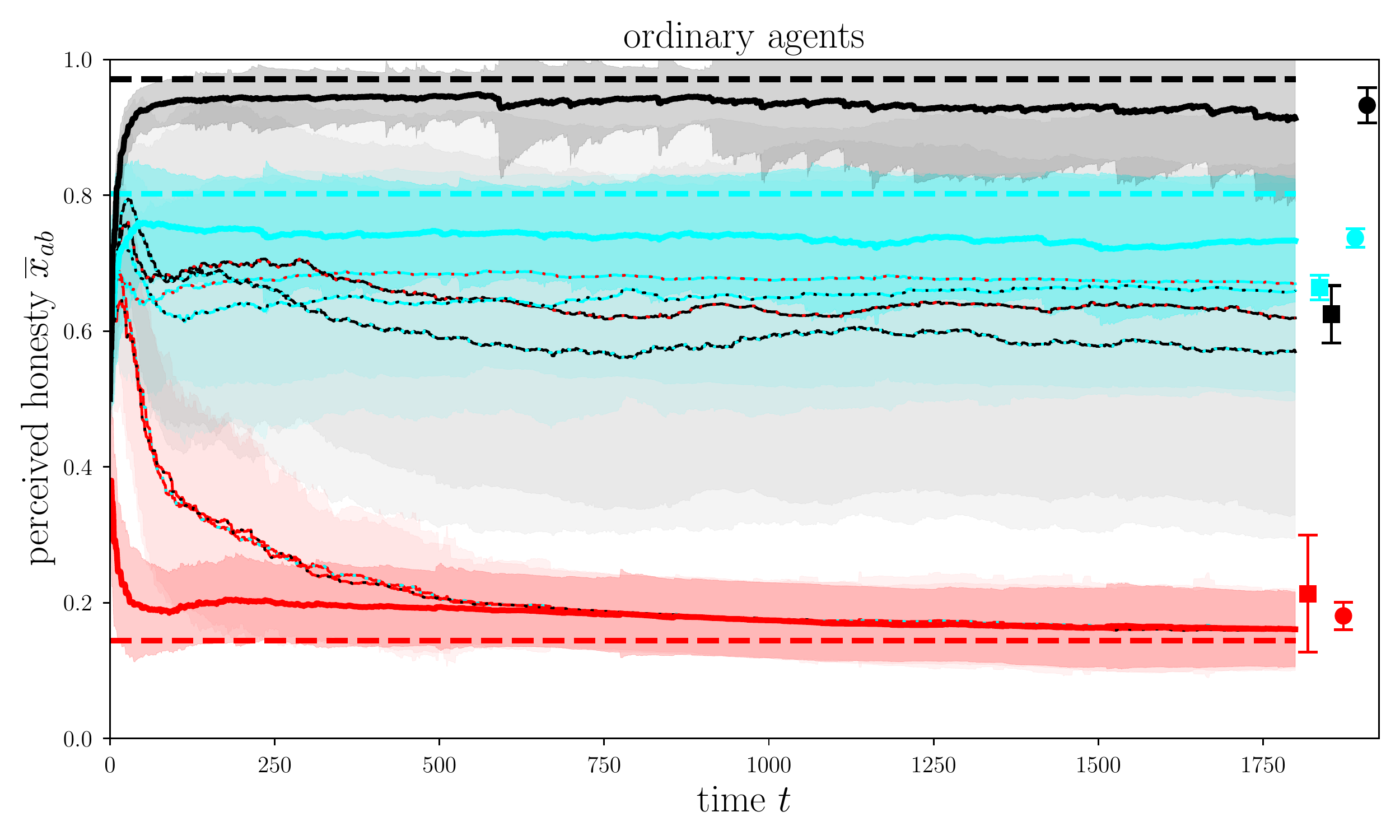}\includegraphics[width=0.5\textwidth]{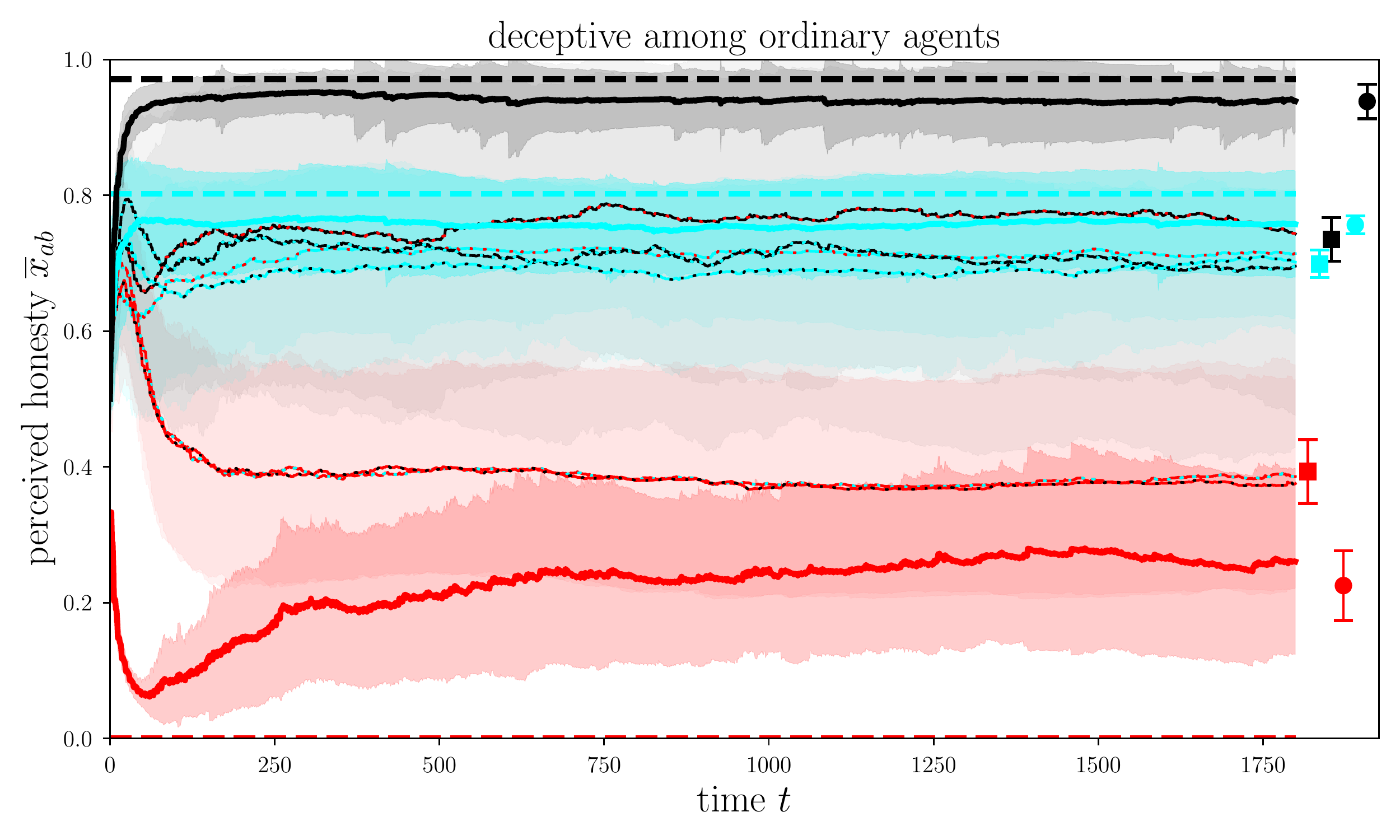}

\includegraphics[width=0.5\textwidth]{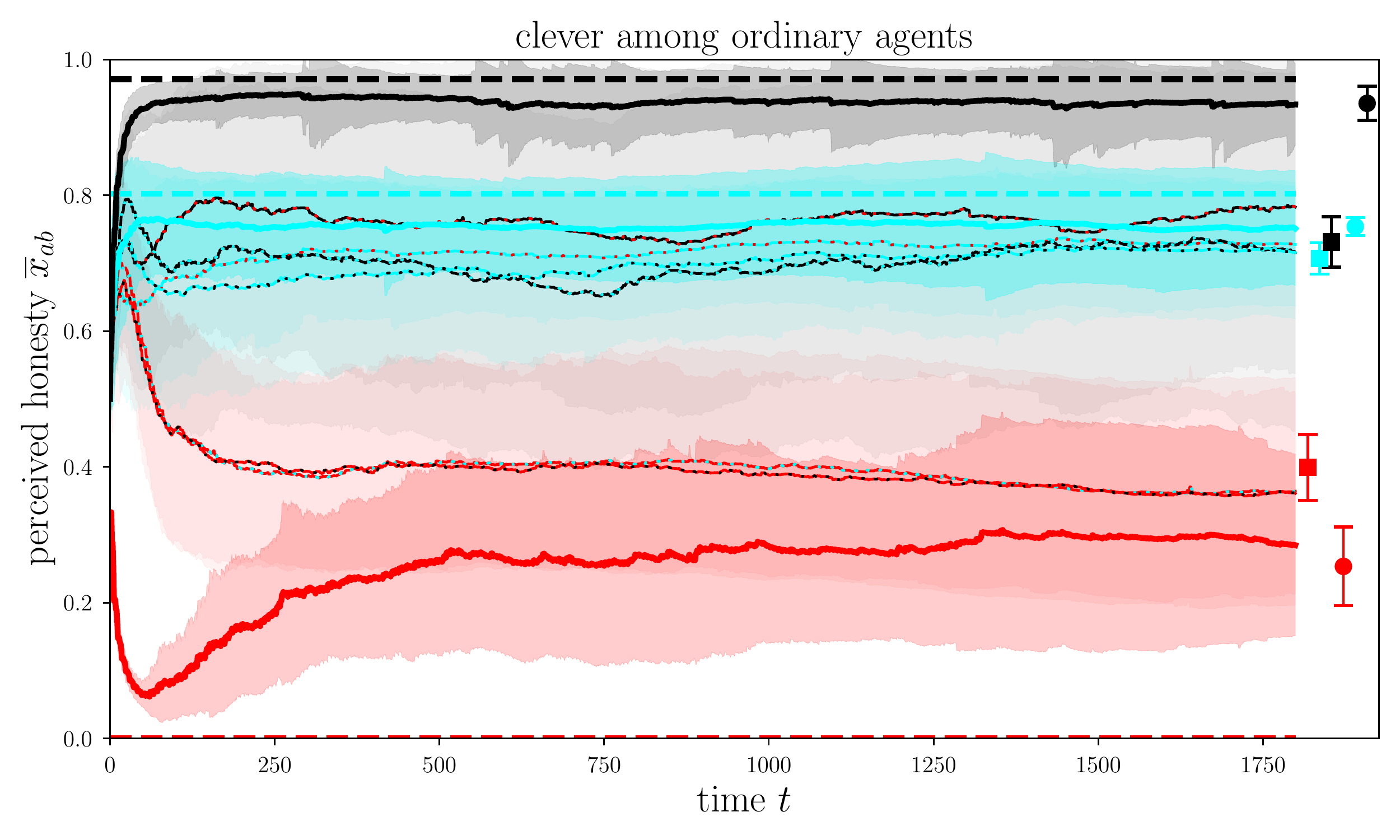}\includegraphics[width=0.5\textwidth]{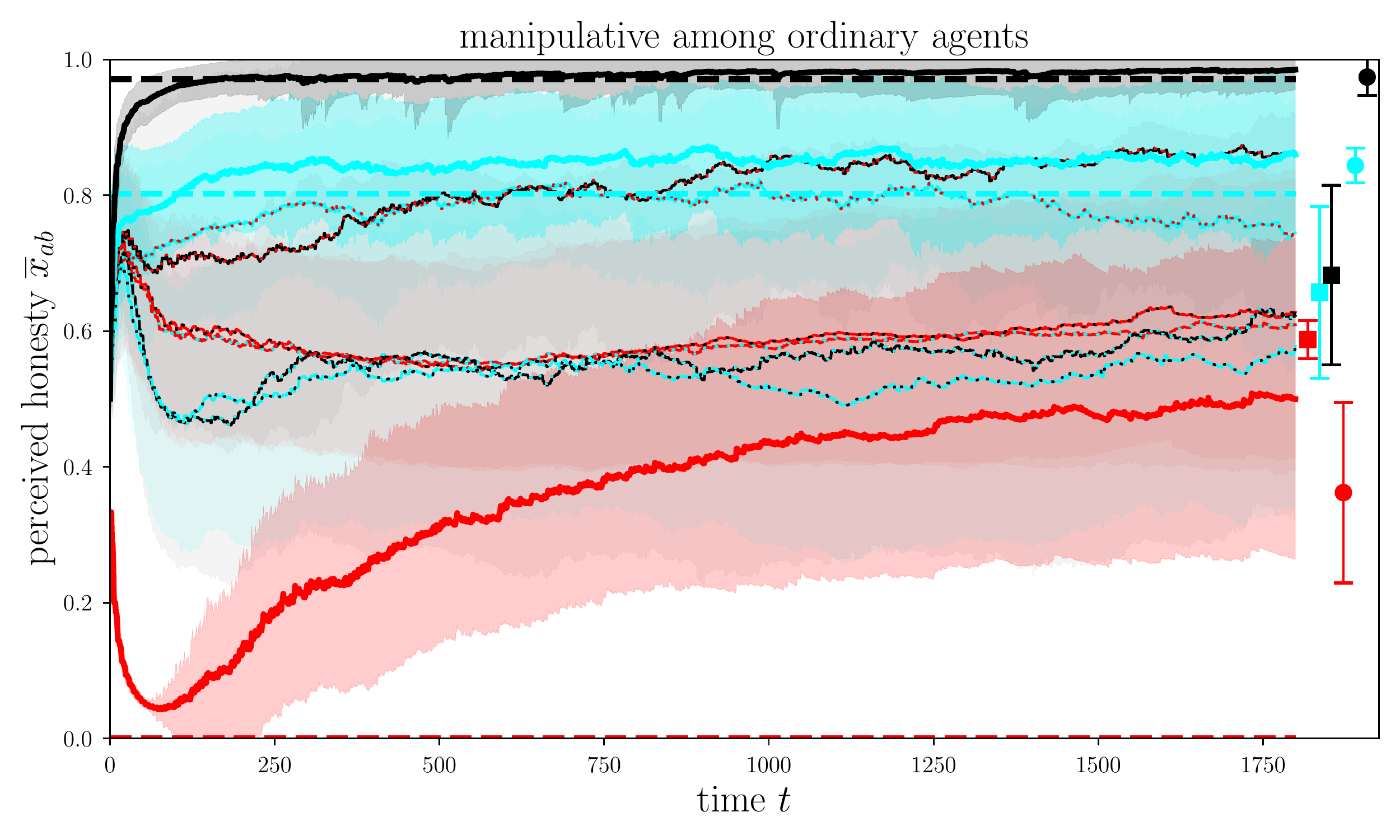}

\includegraphics[width=0.5\textwidth]{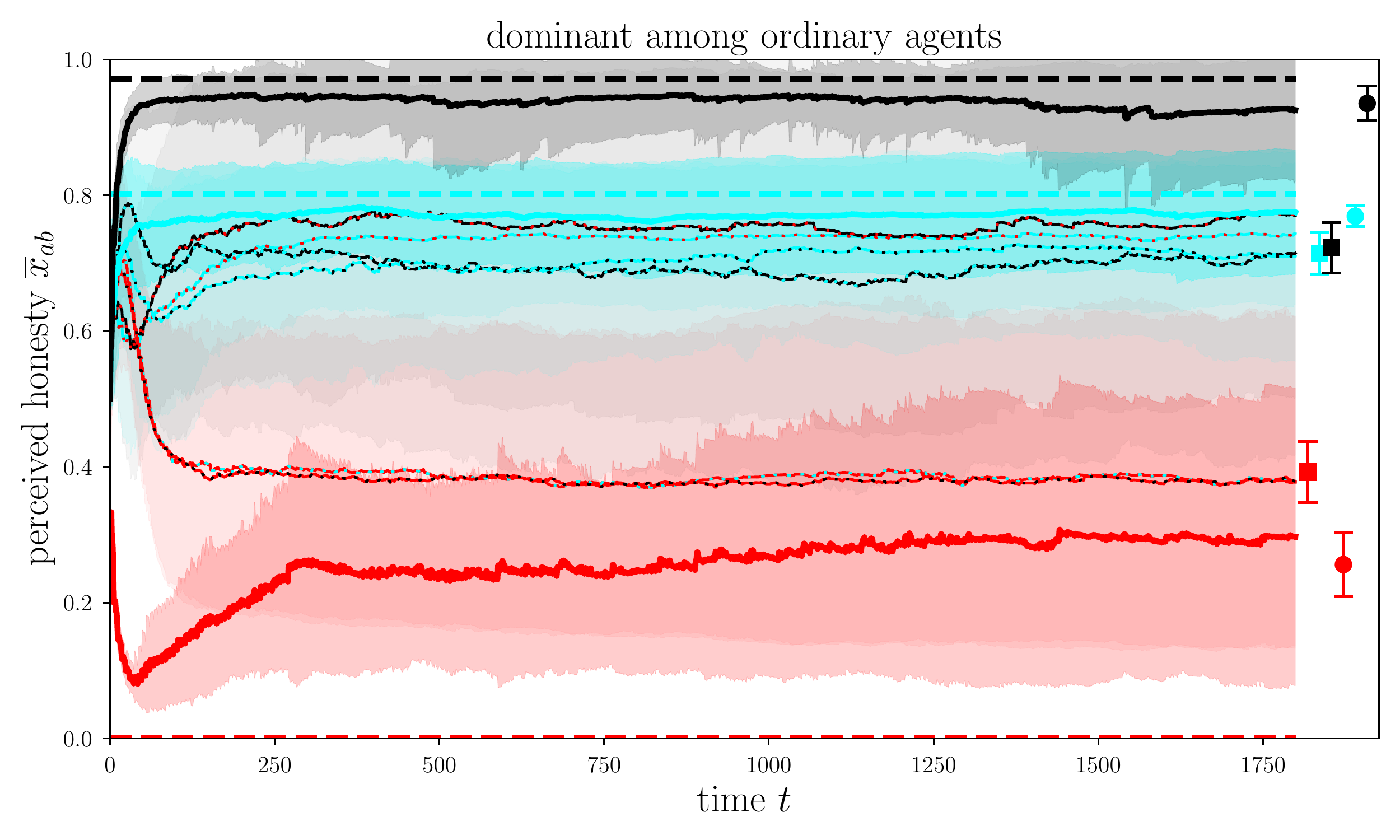}\includegraphics[width=0.5\textwidth]{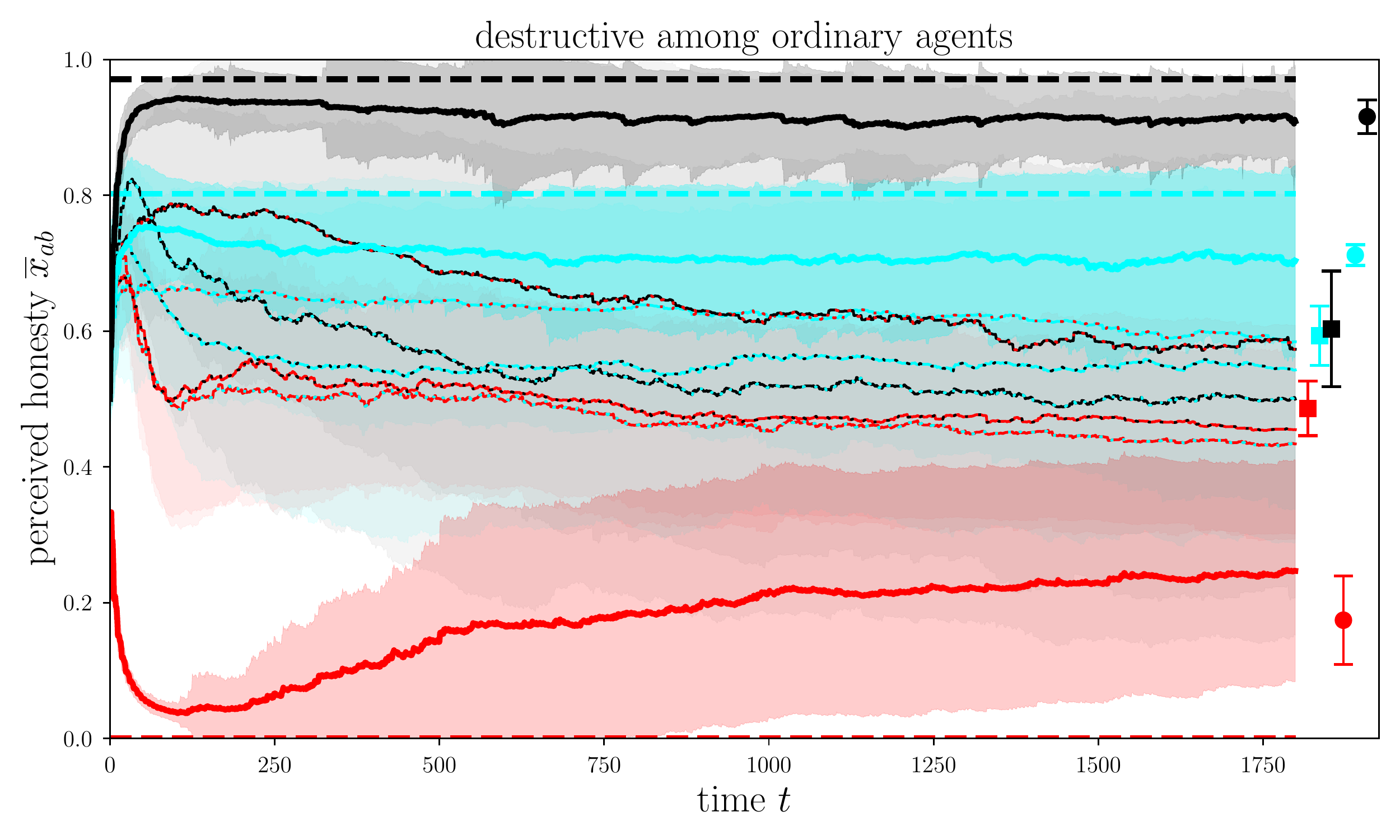}

\caption{Statistical summaries of 100 simulation runs with differing random
sequences. Shown are the mean (lines) and dispersion (shaded areas)
of reputations (thin lines) and self-esteems (thick lines) averaged
over the 100 runs for the same moments. The colors code agents in
the same way as in the other figures. The points and bars on the right
indicate the mean and dispersion of the displayed temporal mean curves
of reputations (via squares) and self-esteems (via circles). The bars
do not take into account the dispersion of the individual runs (which
is indicated by the shaded areas). \label{fig:Statistics-dynamics}}
\end{figure*}
\begin{figure*}[!t]
\includegraphics[width=0.333\textwidth]{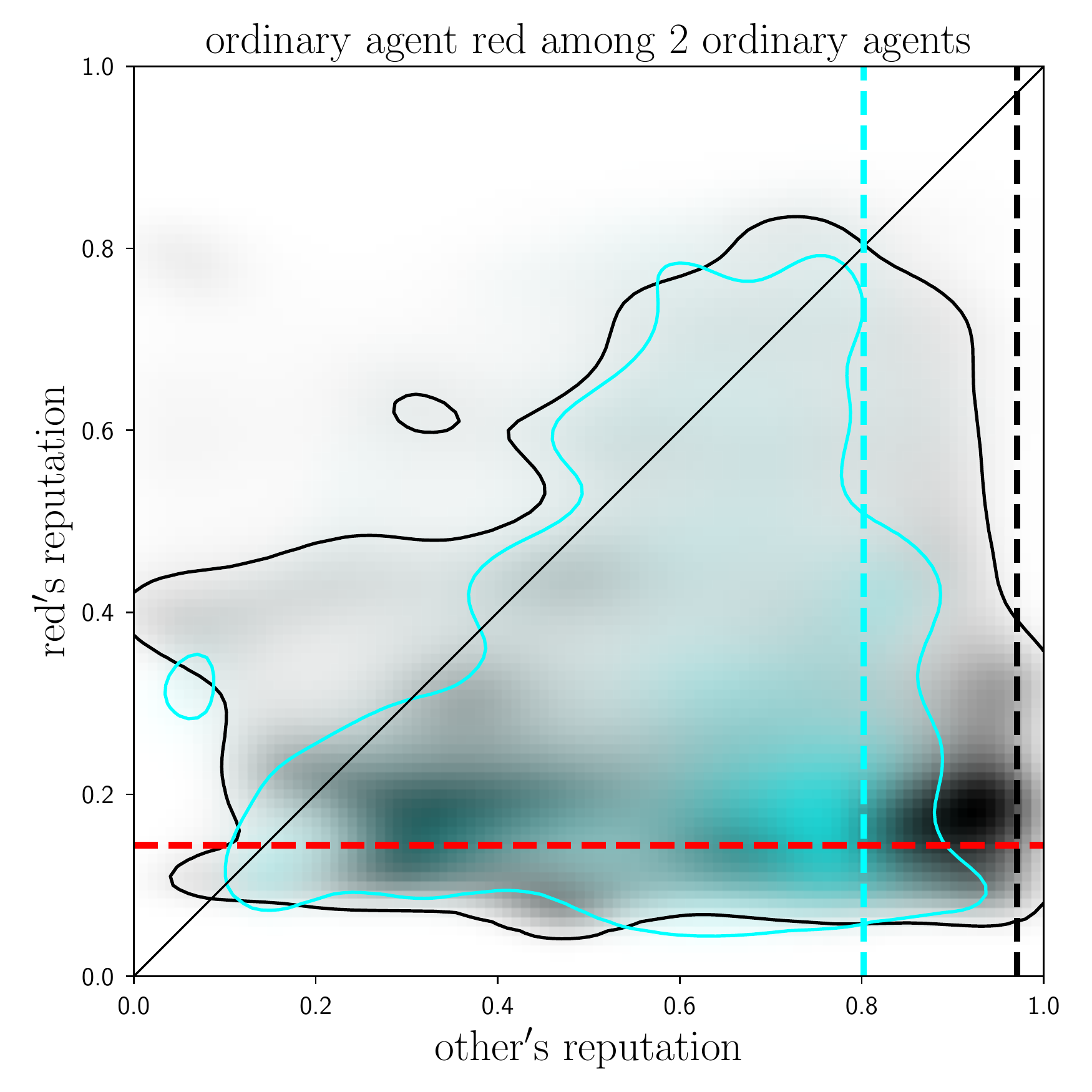}\includegraphics[width=0.333\textwidth]{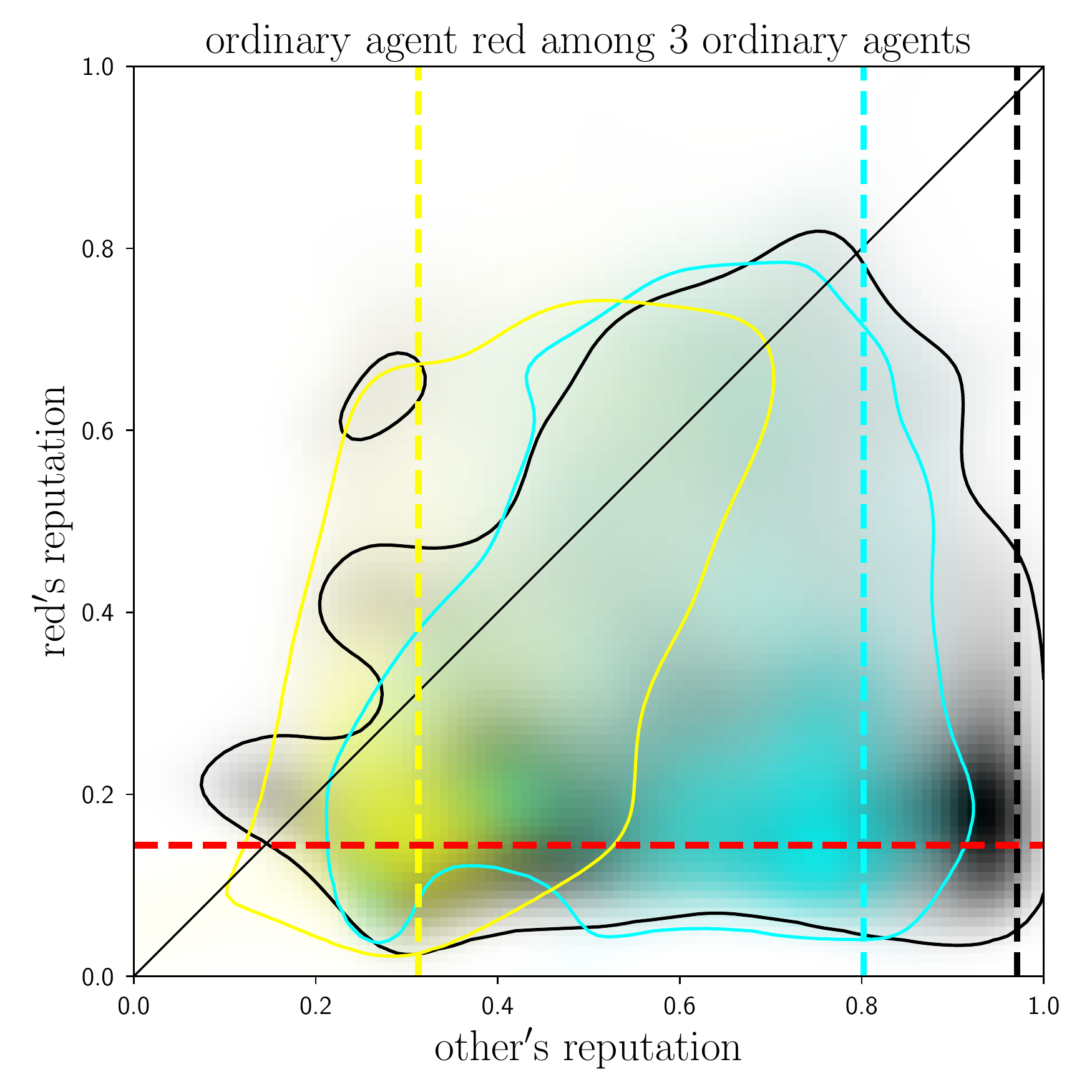}\includegraphics[width=0.333\textwidth]{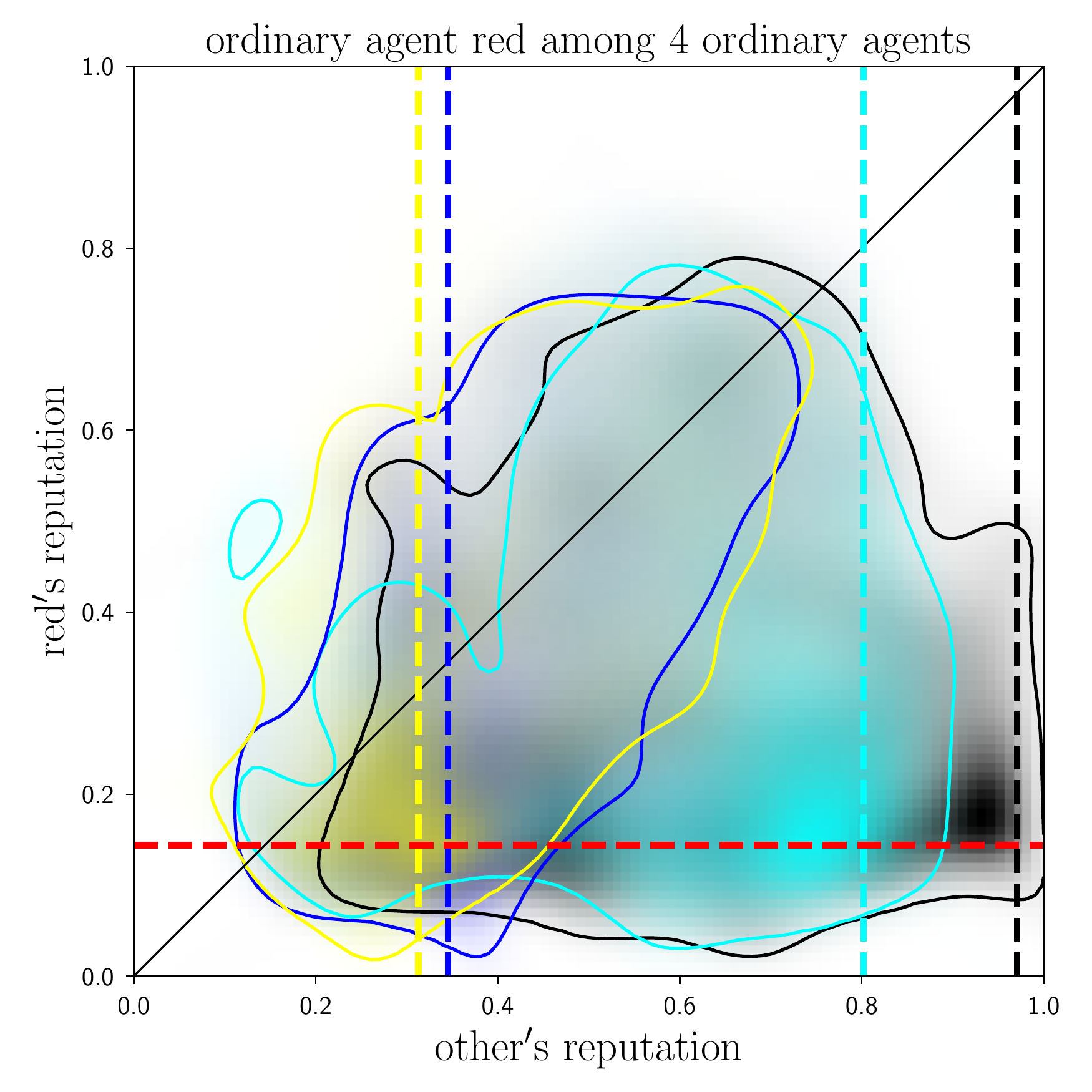}

\includegraphics[width=0.333\textwidth]{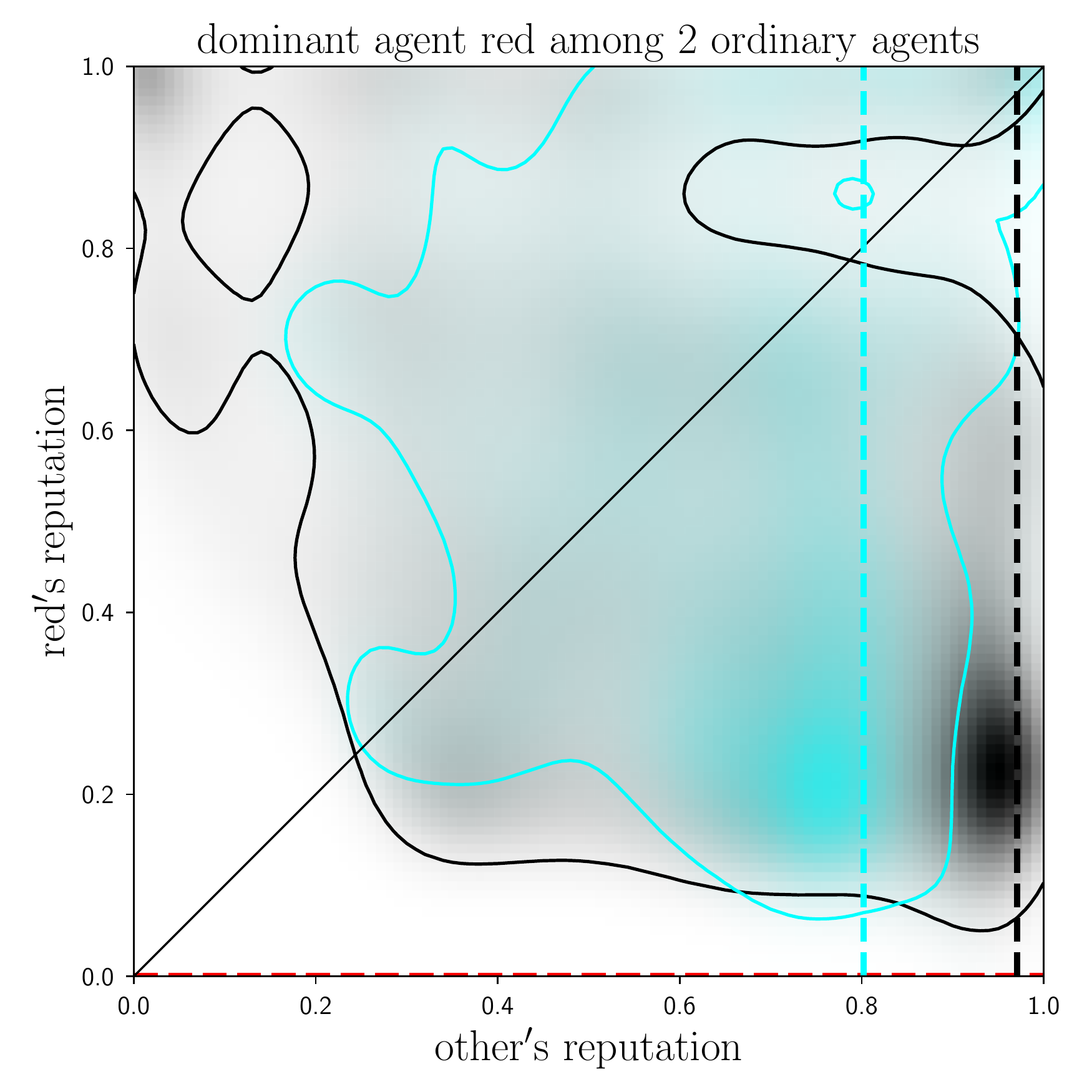}\includegraphics[width=0.333\textwidth]{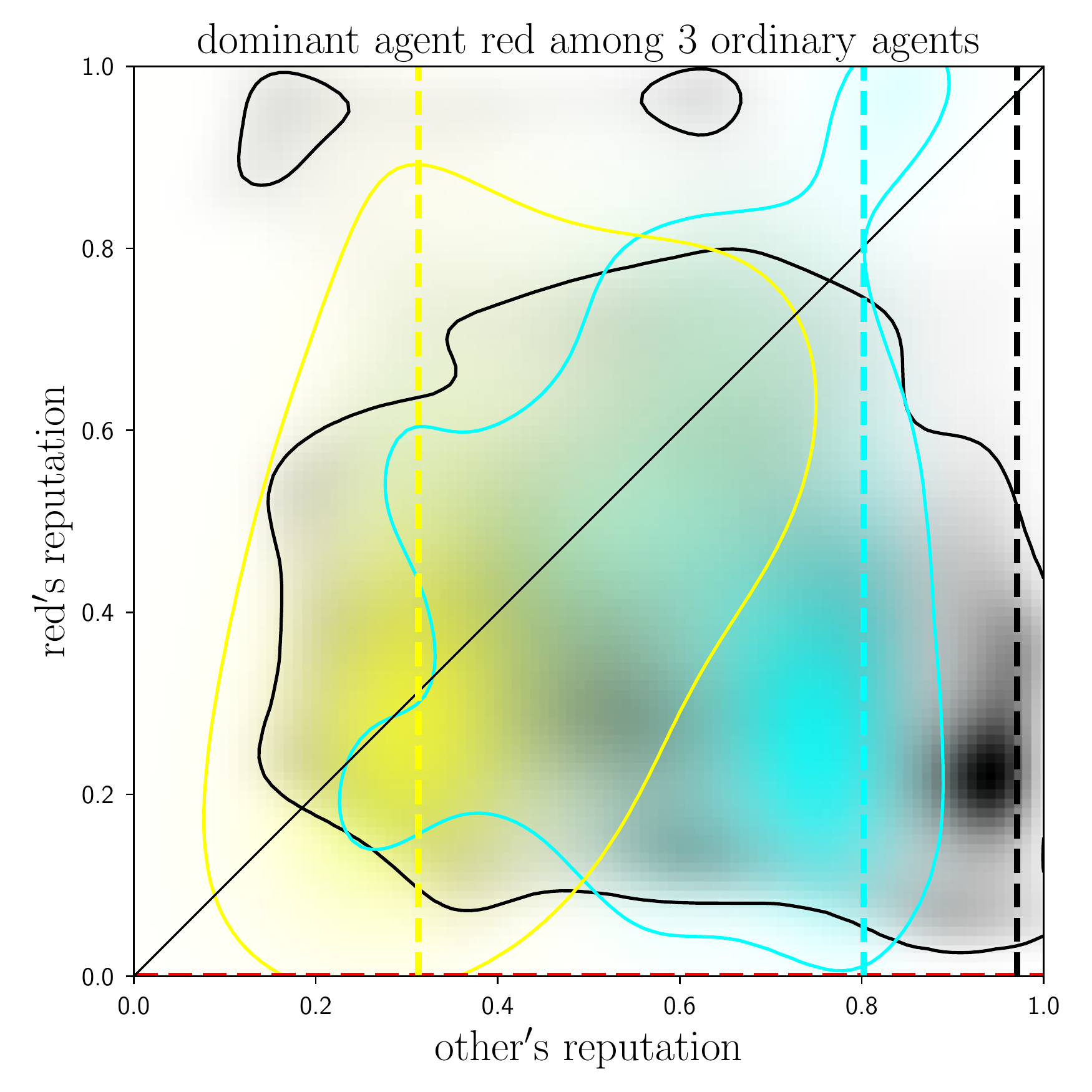}\includegraphics[width=0.333\textwidth]{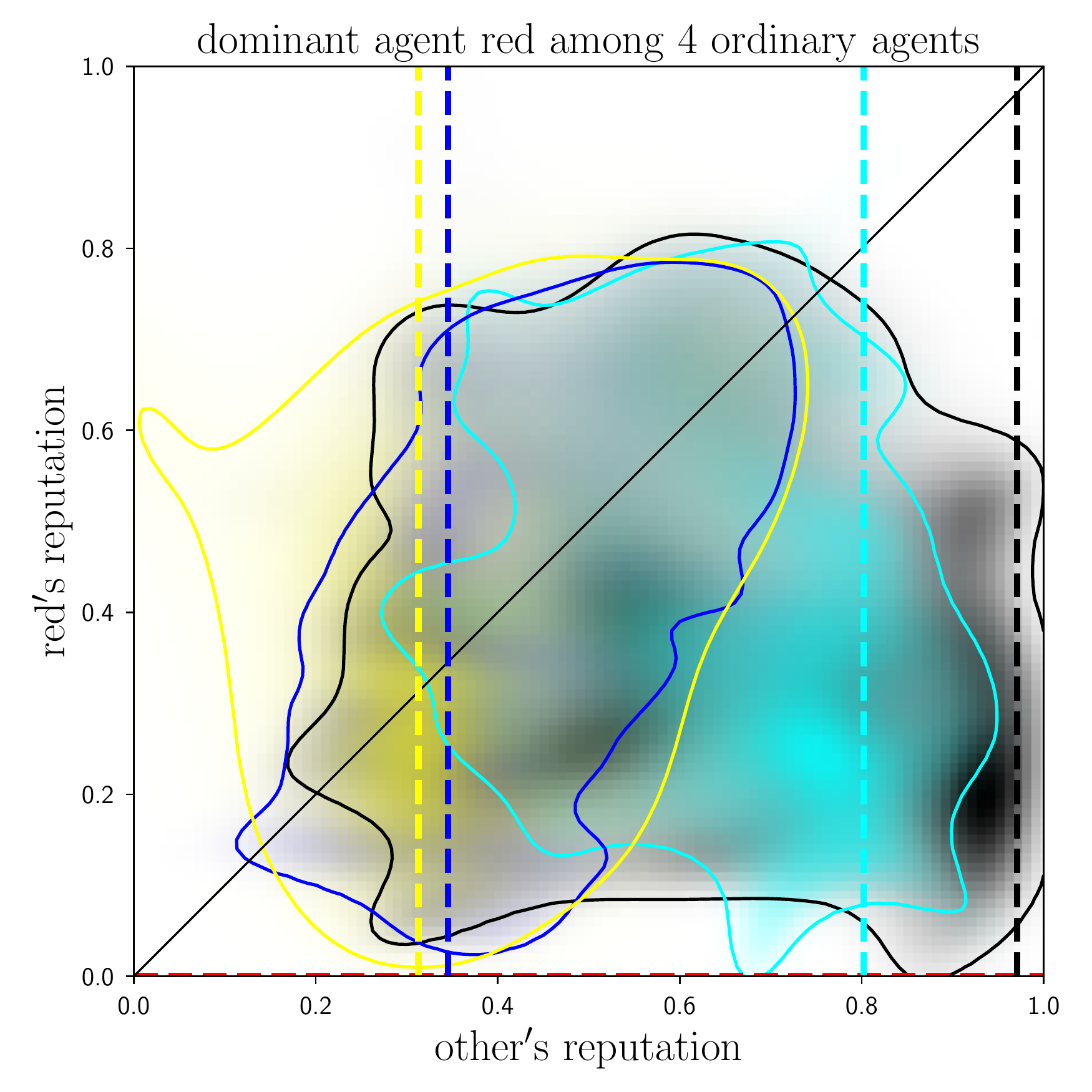}

\caption{Two dimensional density of the reputation values for pairs of agents.
The vertical coordinate is the reputation agent red has in the eyes
of black and cyan (averaged), the horizontal one is the reputation
of either agent cyan or black with the corresponding other two (also
averaged). The contour lines are at level of 0.5\% of the peak values
of the corresponding densities. Both, the densities and lines, are
in color of the other agent to which red is compared. The intrinsic
honesty of an agent is marked as the dashed line in the agent's color.
If agents were not confused, the densities should peak where the dashed
lines cross. The diagonal line indicates equal reputation of red and
the corresponding other agent. For locations above the diagonal line
agent red has a higher reputation than the corresponding other agent.
Shown are runs for ordinary (top row) and dominant agents (bottom
row) for scenarios with three (left), four (middle), and five (right)
agents. The corrsponding plots for other types of agent red are provided
in App.\ \ref{sec:Detailed-figures}.\label{fig:Statistics-scatter-3A}}
\end{figure*}
\begin{figure*}[!t]
\includegraphics[width=0.5\textwidth]{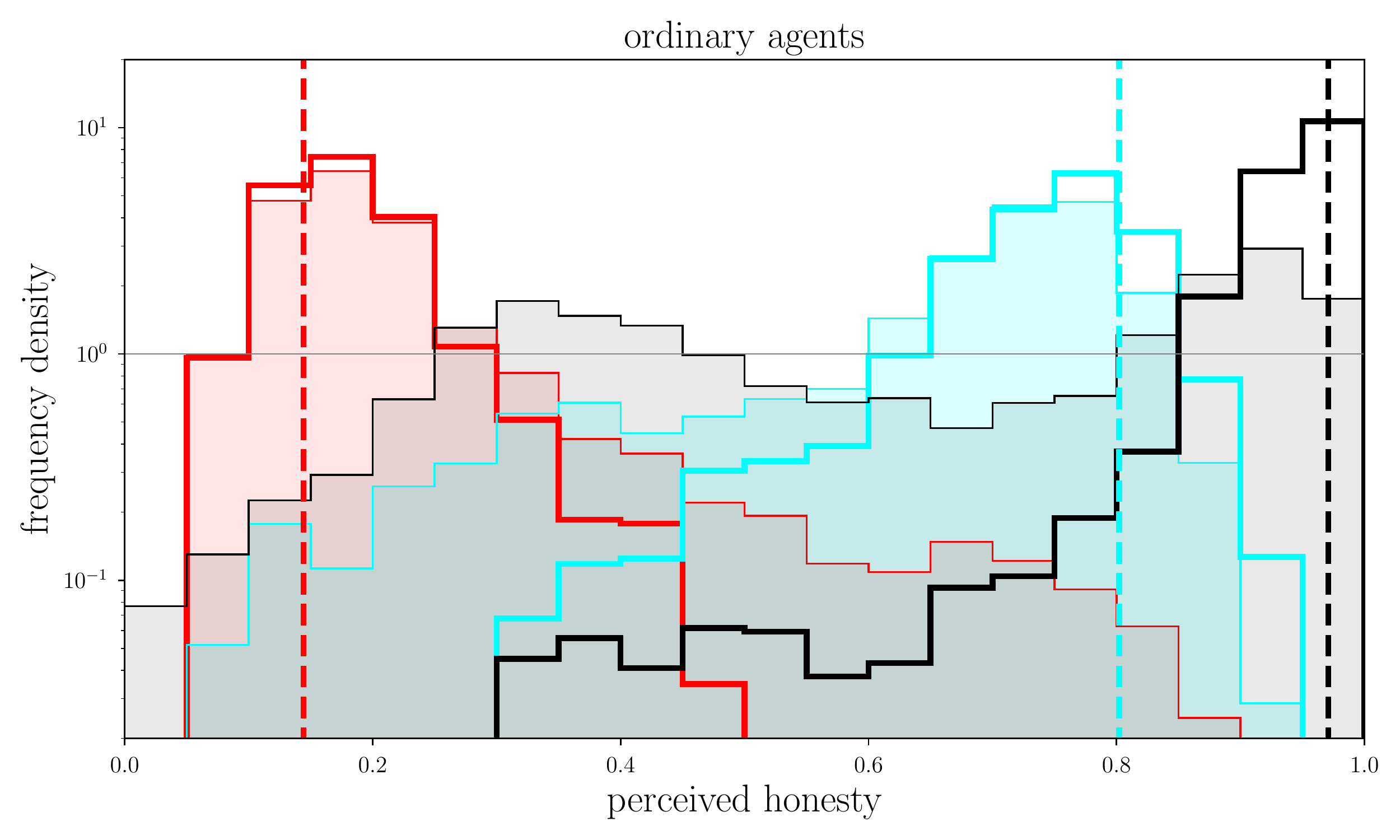}\includegraphics[width=0.5\textwidth]{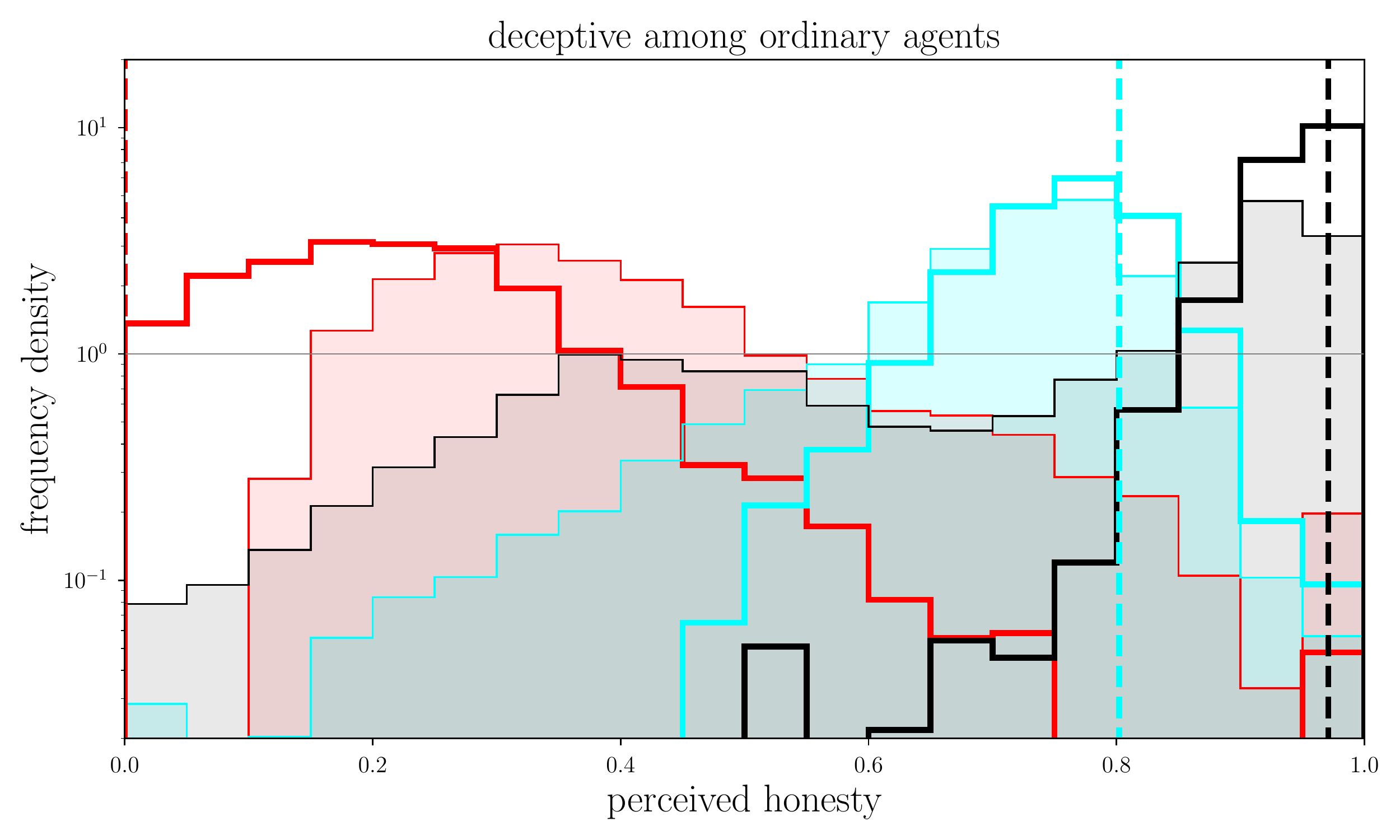}

\includegraphics[width=0.5\textwidth]{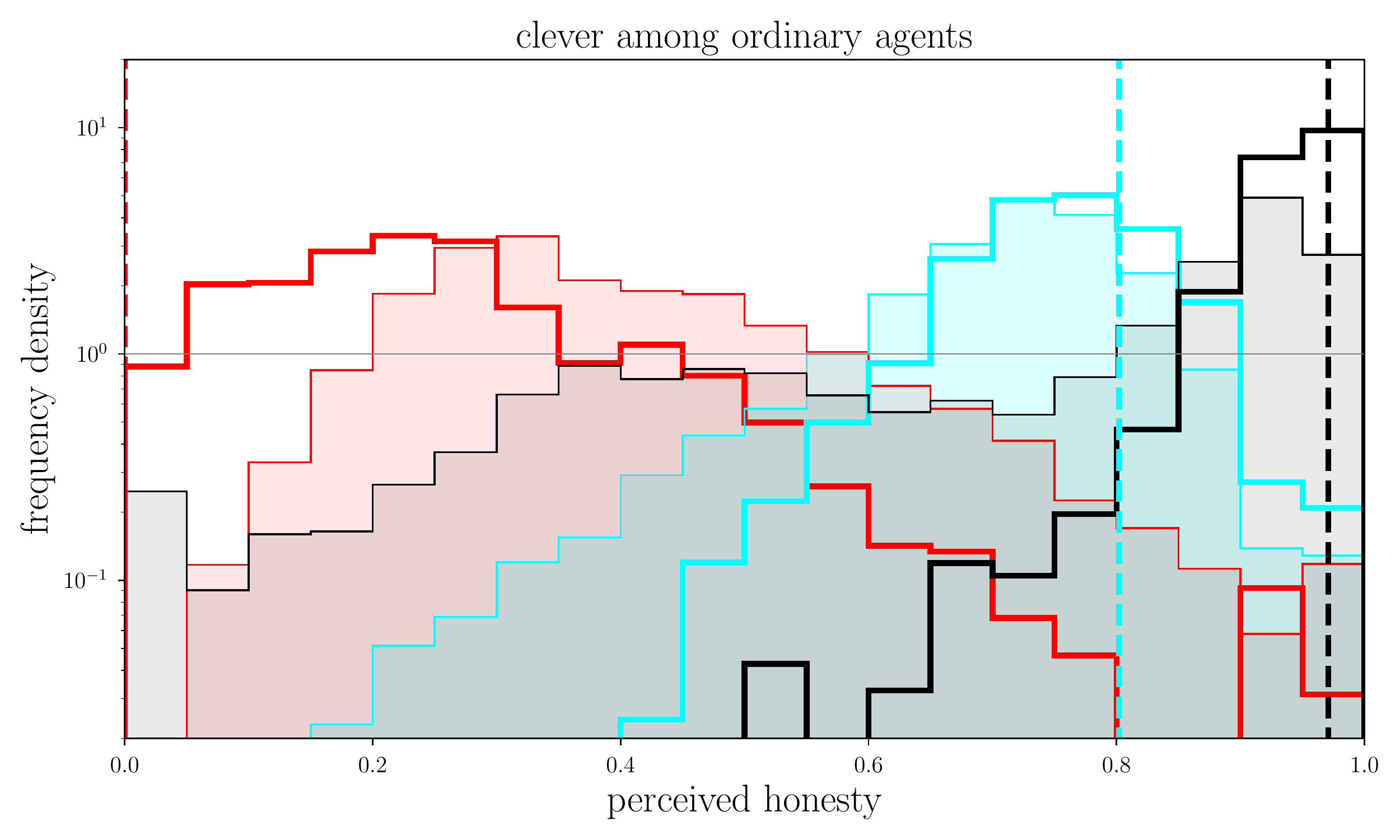}\includegraphics[width=0.5\textwidth]{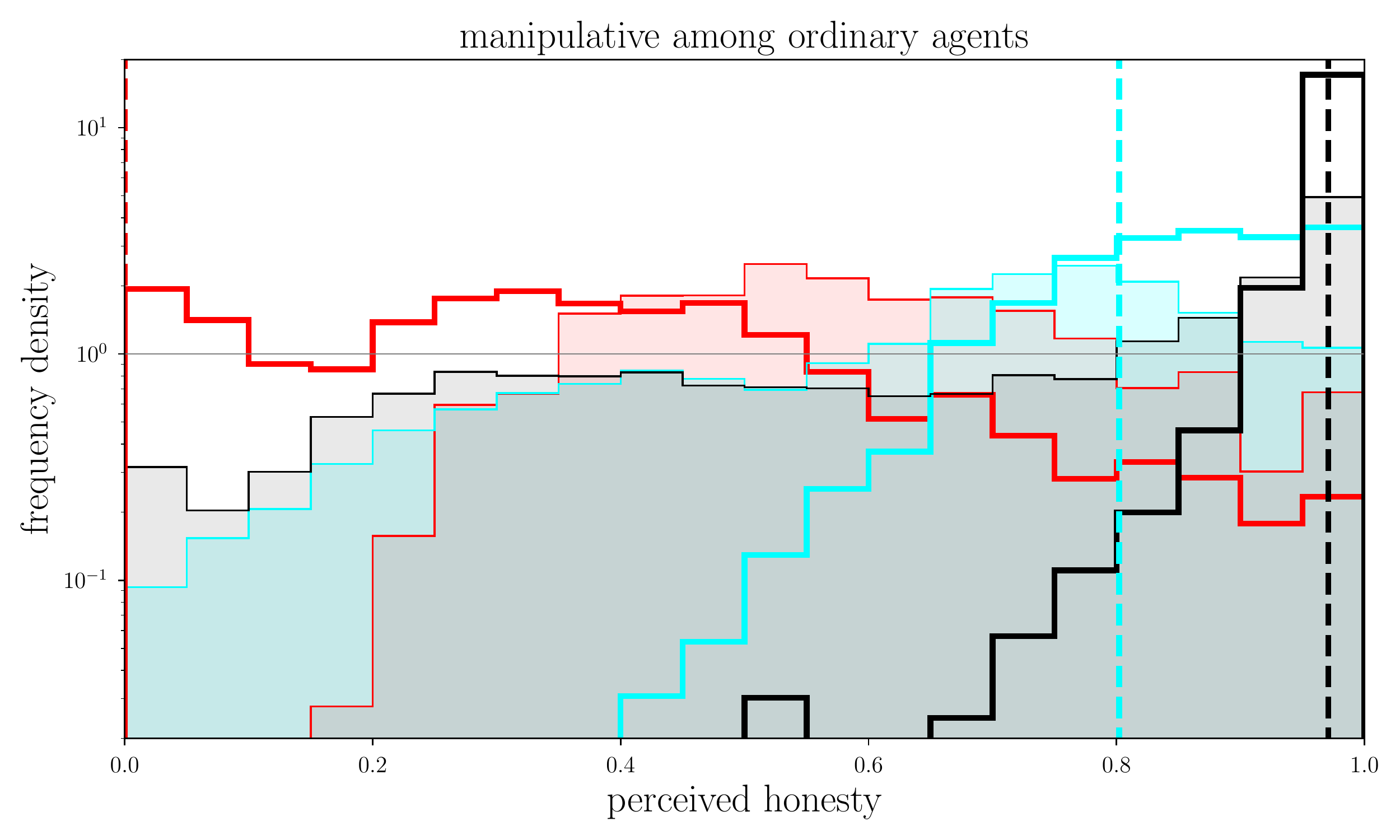}

\includegraphics[width=0.5\textwidth]{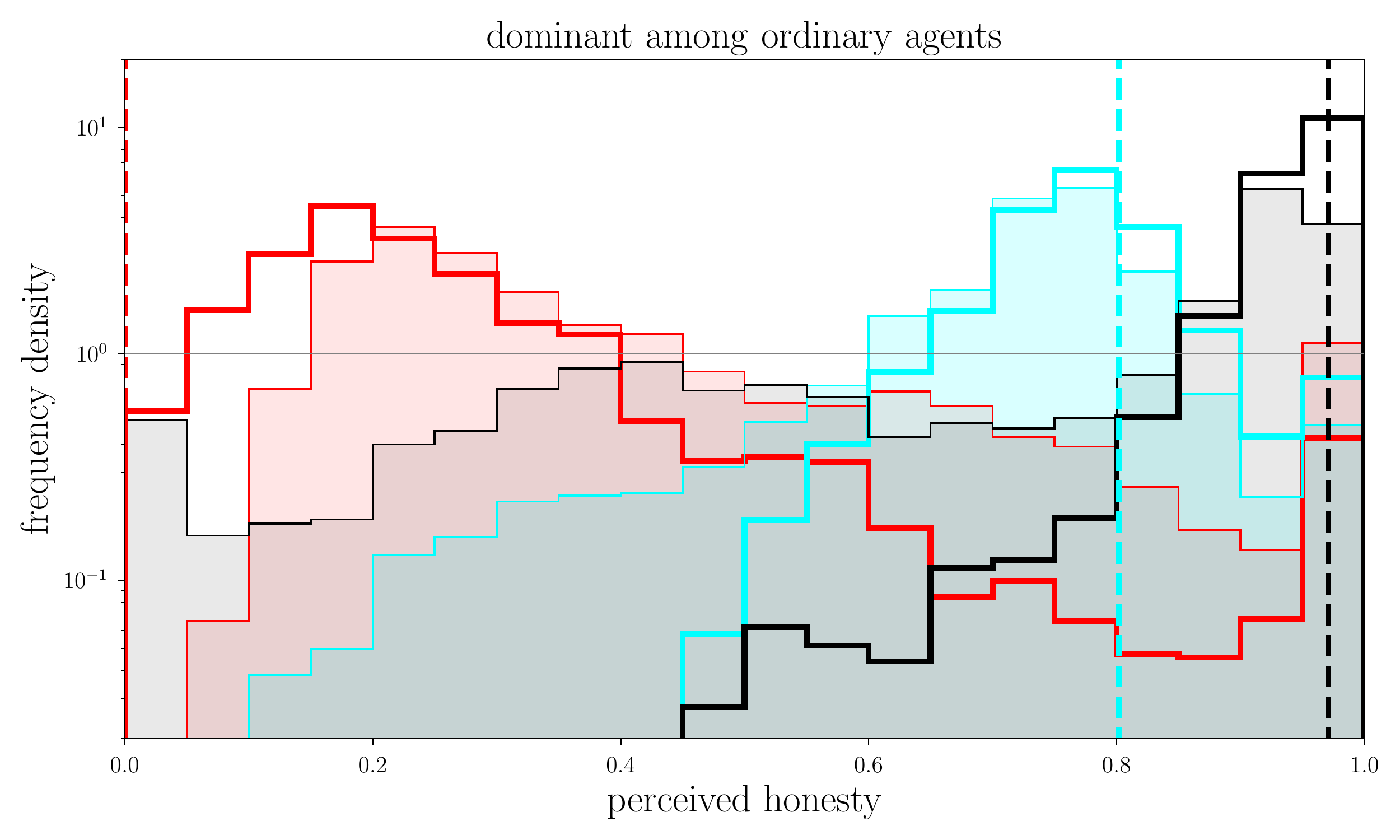}\includegraphics[width=0.5\textwidth]{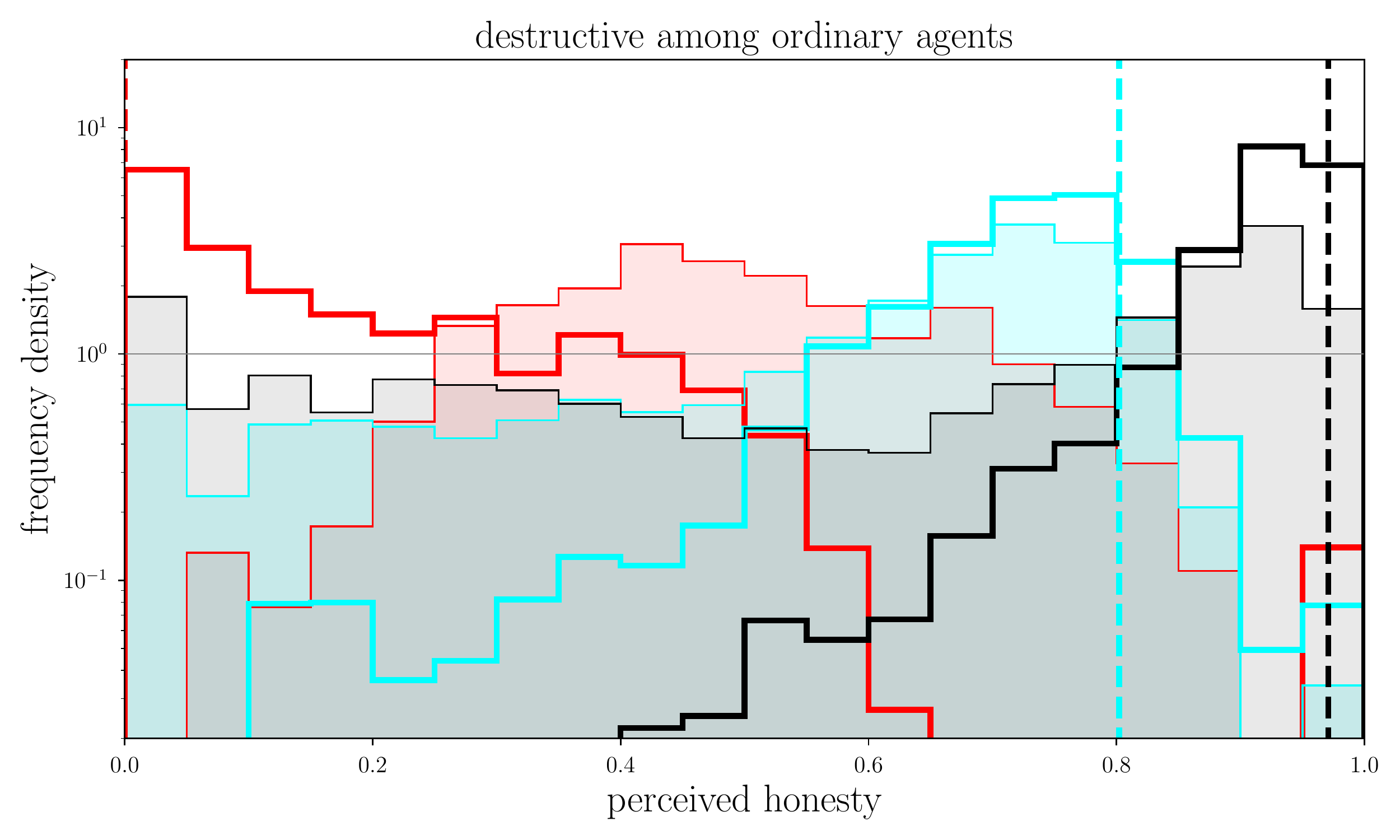}

\caption{Frequency densities of agents (as indicated by color) to have a certain
reputation (thin lines with shading below) or self-esteem (thick lines)
based on the runs underlying also Fig.\ \ref{fig:Statistics-dynamics}.
An uniform distribution would appear as marked by the thin horizontal
gray lines. The true honesty of an agent is marked as a vertical dashed
line in the agent's color. \label{fig:Statistics-histogram}}
\end{figure*}

\subsection{Statistics\label{subsec:Statistics}}

\subsubsection{Reputation statistics}

\begin{figure}[!t]
\includegraphics[width=0.5\textwidth]{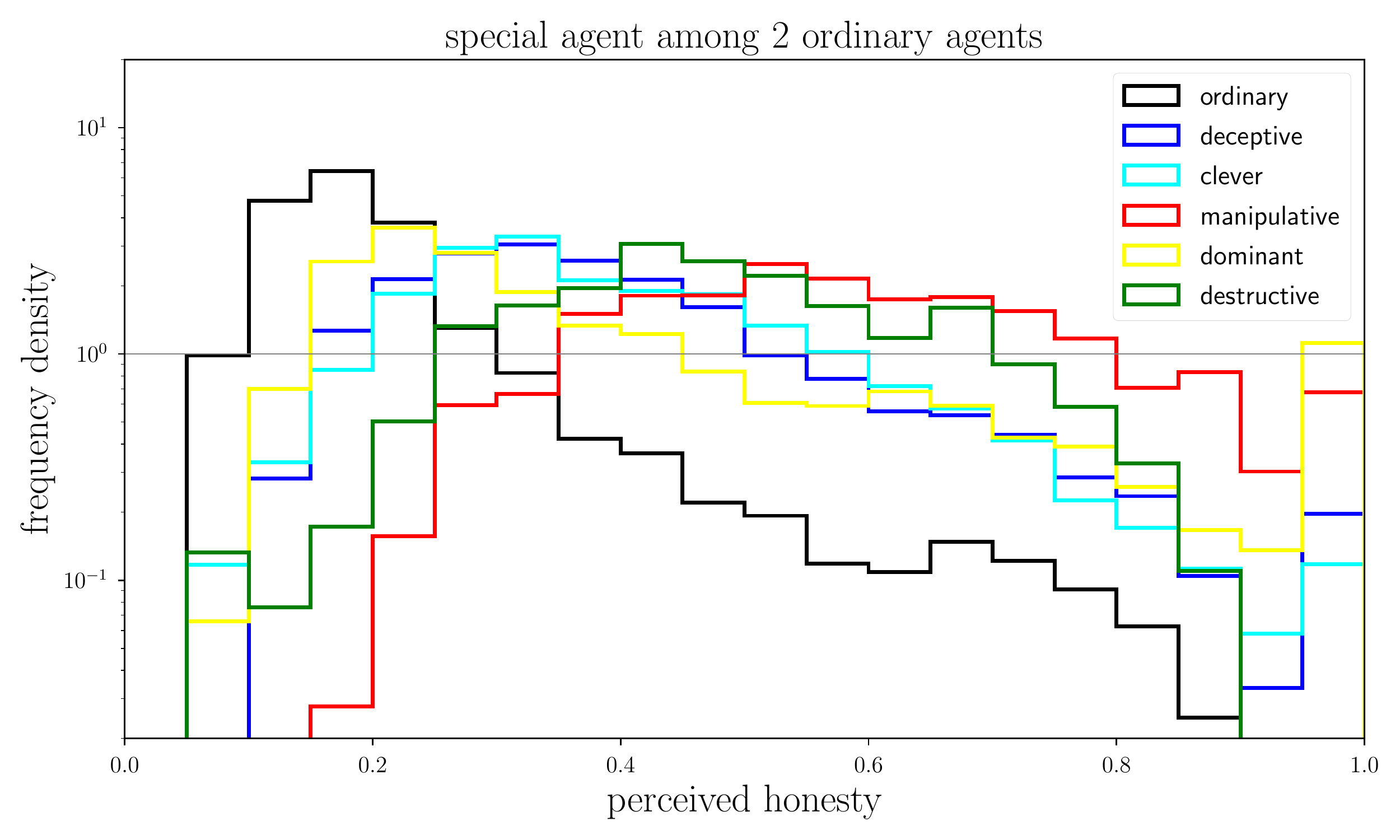}

\includegraphics[width=0.5\textwidth]{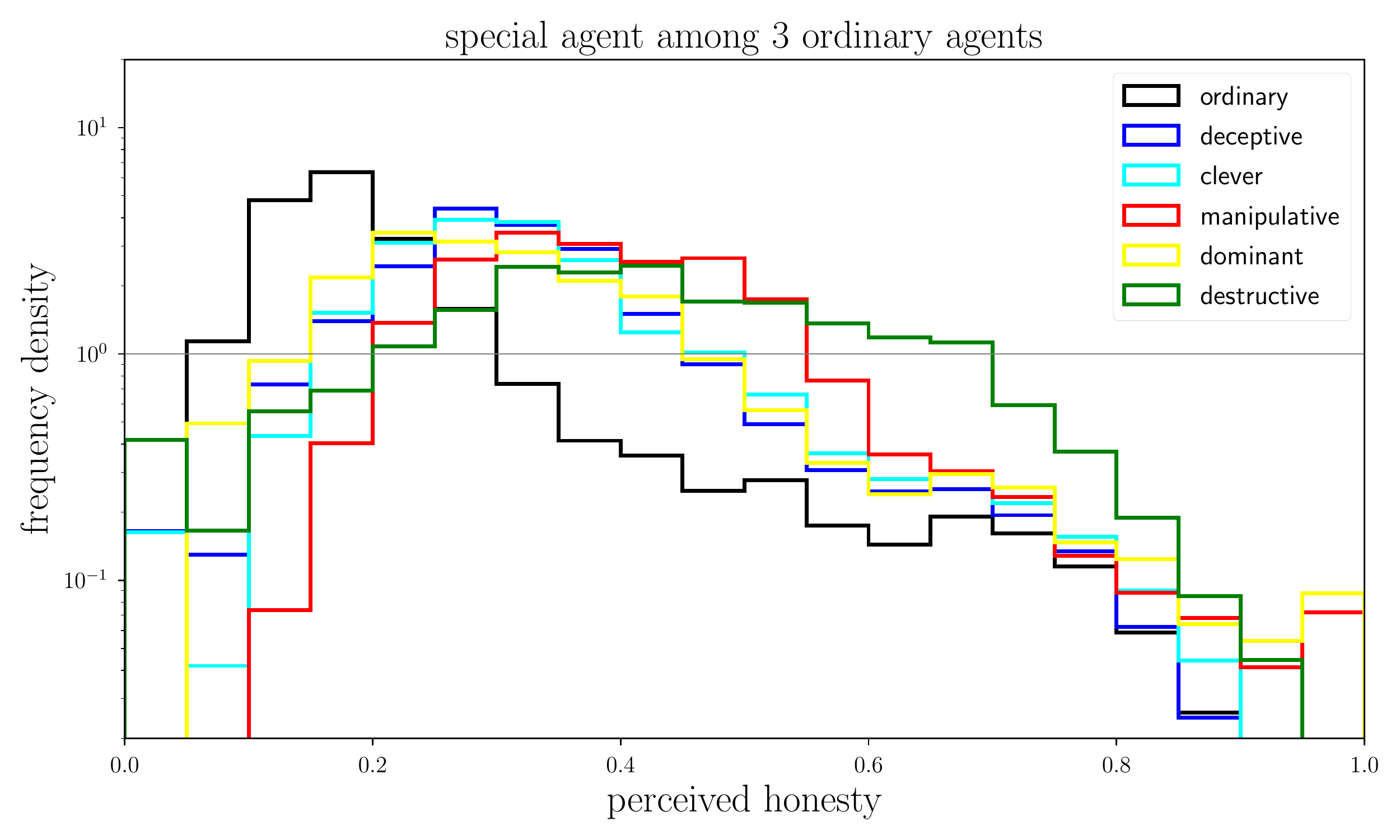}

\includegraphics[width=0.5\textwidth]{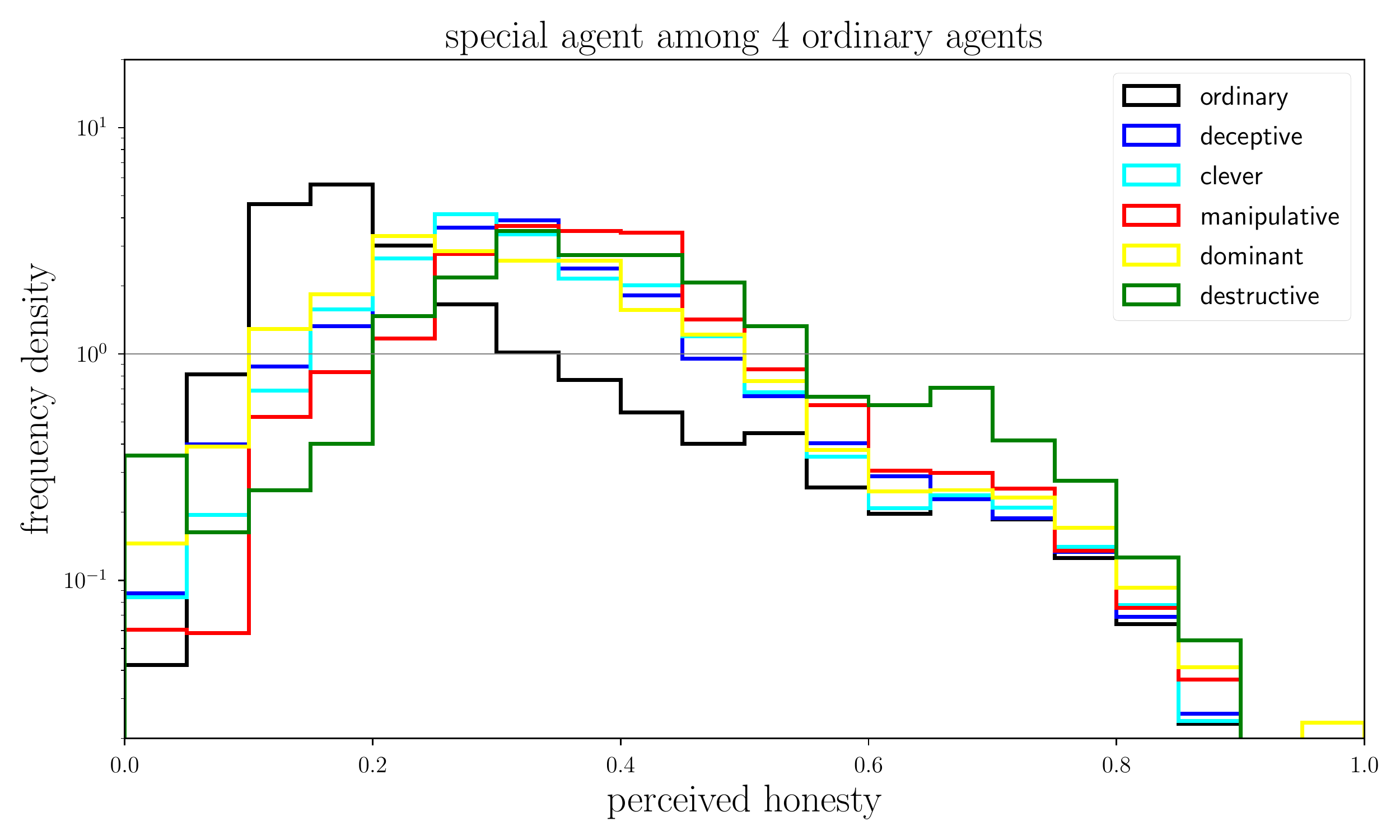}

\caption{Top: Frequency densities of the red agent being an ordinary or special
agent (as indicated by the color of the lines) to have a certain reputation
based on the runs underlying also Figs.\ \ref{fig:Statistics-dynamics}
and \ref{fig:Statistics-histogram}. Middle: The same as top panel,
now just with an additional ordinary agent with honesty $x_{\text{yellow}}=0.31$.
Bottom: The same as middle panel, just with a further ordinary agent
with honesty $x_{\text{blue}}=0.35$. An uniform distribution would
appear as marked by the thin horizontal gray lines. \label{fig:Statistics-histogram-1}}
\end{figure}

In order to see how robust the observed impact of special agents on
the individual runs are, an ensemble of one hundred simulations with
differing random sequences was performed for each of the setups in
which red is ordinary, deceptive, clever, manipulative, dominant,
and destructive. All other configuration parameters are kept identical.
Fig.\ \ref{fig:Statistics-dynamics} shows the time evolution of
the ensemble mean and dispersion of the agent's reputations and self-esteems
for the different scenarios. Fig.\ \ref{fig:Statistics-scatter-3A-1-2}
displays the correlations between the reputation of agent red (the
least honest one) and that of agents cyan and black (the two more
honest ones) for the different scenarios.\footnote{Fig.\ \ref{fig:Statistics-scatter-3A-1-2} shows a reconstructed
density based on the simulation data points as given by the various
time snapshots in the hundred simulations. The density is w.r.t.\ the
plane spanned horizontally by black's (or cyan's) average reputation
(averaged over the other two agents) and vertically by that of red.
Details of the used density reconstruction technique can be found
in \cite{2021arXiv210513483G}.\label{fn:density-reconstruction}} Fig.\ \ref{fig:Statistics-histogram} shows histograms of the agents'
reputations and their self-esteems occurring during the simulations
for the different scenarios. We name these scenarios after the strategy
agent red uses in them.

In the ordinary scenario, with \textbf{ordinary agent} red, reputation
and self-esteem values of agents roughly reflect their honesty. Red
is not able to significantly increase their reputation or self-esteem
beyond red's honesty ($x_{\text{red}}=0.14$) in most of the runs,
only in a few cases high values are reached (Fig.\ \ref{fig:Statistics-histogram}
shows a peak in the histogram of red's reputation at $0.2,$ with
a fat tail towards larger reputations up to $0.9$). Agent cyan's
reputation ($\approx0.7$) and self-esteem ($\approx0.75$) are only
slightly lower than they should be ($x_{\text{cyan}}=0.80$). The
largest disparity between reputation and honesty happens typically
for agent black, who is too honest to defend their reputation (black's
reputation is on average at $\approx0.65$, but shows large variance,
whereas $x_{\text{black}}=0.97$). Black's reputation shows even a
bimodal distribution, with a high reputation peak (at $0.925$) shortly
below black's honesty ($x_{\text{black}}=0.97$) and a low reputation
peak at a much lower value ($0.325$). The reason for this low reputation
peak is again black's high honesty, which lets black more often express
positions that are in contradiction to those of the other agents,
letting black appear as untrustworthy. This can be regarded as a reputation
game manifestation of the Cassandra syndrome: the most honest agent
may appear less trustworthy than more dishonest agents. We note that
agent cyan's reputation, who is also mostly honest, is slightly bimodal
as well.

In the scenario with the \textbf{deceptive agent} red, red reaches
on average a significantly higher reputation ($0.4$) and self-esteem
($0.26$) than in the ordinary scenario ($0.2$ and $0.16$, respectively).
Red's reputation distribution histogram shows now a broad plateau
(from $0.05$ to $0.3$), a fat tail towards higher reputations (up
to $0.95$), and a distinct peak at highest reputations ($>0.95$).
This peak indicates that once accepted as being very honest, red can
defend this position thanks to the higher influence a reputed agent
has.

We note that cyan's and black's self-esteems are higher than in the
ordinary scenario and more focused on their intrinsic honesty values.
Red's more frequent lies in this scenario exhibit a stabilizing force
to the other's self-esteems. As lies mostly mirror the other's beliefs,
they can strengthen those beliefs if they are not too biased.

The scenario with the \textbf{clever agent} red looks nearly indistinguishable
to the previous one, except for the self-esteem of agent red, which
is now slightly higher ($0.3$) near the end of the simulation time.
Being smart, red realizes that black and cyan are mostly honest when
they speak about red's honesty (which appears to them to be $0.4$).
Therefore, red's self-opinion is more strongly drawn towards this
value than in the previous scenario.

The scenario with the \textbf{manipulative agent} red shows that the
manipulative strategy is the most successful in allowing red to reach
on average the highest reputation and self-esteem among all scenarios
investigated. Both quantities show also the strongest rising trends
at the end of the simulated period. Red's chance of being regarded
as very reputed ($>0.95$) is nearly five times higher in the manipulative
scenario compared to the clever one. Compared to the clever scenario,
cyan and black's reputations are lower and show more variance. Cyan's
reputation is now reaching lowest values nearly as frequently as black's,
thanks to cyan's higher exposure to red's confusing lies (red is anti-strategical
here, thus mostly talking to cyan). It is noteworthy that the self-esteems
of black and cyan are enhanced not only w.r.t.\ the clever scenario,
but also w.r.t.\ black and cyan's intrinsic honesty. This is due
to the flattering they get from red, which boosts their self-esteem.
Although all agent's reputations are generally higher here compared
to the clever scenario, the number of cases in which red's reputation
surpasses the ones of the others is strongly increased (see Fig.\ \ref{fig:Statistics-scatter-3A-1-2}).

The \textbf{dominant agent} red does not reach a higher average reputation
than the clever agent red, but red's reputation displays a larger
dispersion in the dominant scenario than in the clever one or any
of the others. Red's chances to be regarded as very reputed ($>0.95$)
is the largest in the dominant scenario, being ten and two times higher
than in the clever and manipulative scenarios, respectively. Red's
self-esteem is higher on average by being dominant than being only
clever, despite the lower average reputation of red in the dominant
scenario. The higher frequency of conversations about red in the dominant
scenario couples red's self-esteem more strongly to red's reputation.
This effect outweighs the lower average reputation of red in this
scenario. Being strategic, red targets predominantly black with self-promotion
lies and thereby drives black's opinion away from the other's. As
a consequence, black gets often confused and this lets black's reputation
reach lowest values ($<0.05$) so frequently that black's reputation
distribution histogram exhibits a distinct peak there. Fig.\ \ref{fig:Statistics-scatter-3A}
confirms this interpretation, with exhibiting the lowest reputations
for black for moments when red reaches highest reputation values.

In the scenario with the \textbf{destructive agent} red, red reaches
on average a reputation significantly higher ($0.45$) than in the
clever and dominant scenarios ($0.4$). However, the destructive red's
reputation exhibits a slowly declining temporal trend, whereas the
ones of them being manipulative or dominant are increasing or constant,
respectively. Destructive red's reputation is uni-modal (with a broad
peak centered on $\approx0.5$) and reaches neither the highest nor
the lowest reputation values. Red's self-esteem evolution is initially
low but constantly raising during the further simulated period. Their
self-esteem distribution function, however, peaks strongly at lowest
values ($<0.05$). This stronger detaching of red's self-esteem from
their reputation in the destructive scenario is caused by them avoiding
themself as a topic; red mostly talks about enemies, not about red.
The impact of red's destructive strategy on red's enemies is also
clearly visible: Both other agents, black and cyan, experience now
a high chance to be without any reputation. Furthermore, their reputations
with red show a declining temporal trend, meaning that red believes
more and more red's own lies. Despite having a low reputation on average
in the destructive scenario, agent red surpasses the other agent's
reputations frequently by destroying those. If the goal is to be highly
deceptive, but still more reputed than other agents, the destructive
strategy seems to be a choice as good as the manipulative one. 

A comparison of red's reputation histogram for the different strategies
used is given by Fig.\ \ref{fig:Statistics-histogram-1}. This shows
that among the strategies investigated here of deceptive agents, on
average, the manipulative one seems to be the most successful, followed
by the destructive one. If, however, success is defined as reaching
the highest reputation values, the dominant strategy seems to be most
favorable.

Fig.\ \ref{fig:Statistics-histogram-1} also shows the reputation
histogram results for runs with four or five agents. The three agents
of the previous simulations were kept, just one or two additional
agents are introduced there, who have a low honesty of $x_{\text{yellow}}=0.31$
and $x_{\text{blue}}=0.35$. These simulations can be regarded to
be statistically independent of the simulation with three agents and
w.r.t.\ each other for most practical purposes.

The corresponding reputation density plots for the four and five agent
simulations for ordinary and dominant agents are shown in Fig.\ \ref{fig:Statistics-scatter-3A}.\footnote{Those for the other agents can be found in Fig.\ \ref{fig:Statistics-scatter-3A-1-2}
in App.\ \ref{sec:Detailed-figures}.} One sees that the reputations of these additional, mostly dishonest
agents are correlated with that of red, and the correlation gets stronger
the more dishonest agents are present. This indicates that there is
some synergy between these least honest agents. There are two effects
that can cause this. First, less honest agents are better in befriending
each other. Second, there is a generic benefit for liars to draw from
an atmosphere of general confusion that a larger number of dishonest
agents creates. Their lies fly easier there.

These plots show further that the special strategies still pay off
within larger groups, but with a reduced reputation gain compared
to the three agent scenario. Now, the destructive agent red manages
to reach higher average reputation values than by being manipulative
or dominant. The latter are still more efficient in reaching the highest
reputation values.

It seems safe to claim that these simulations show that the introduced
special deceptive strategies are more successful than just being deceptive
or clever. The details of which strategy is best with respect to the
different success metrics might also depend on the precise composition
of the social group. This was not varied much here, as we kept agent
black very honest and agent cyan mostly honest in all runs. We leave
the investigation of such effects to future research.

\subsubsection{Friendship statistics\label{subsec:Friendship-statistics}}

\begin{figure}
\includegraphics[width=0.1665\textwidth]{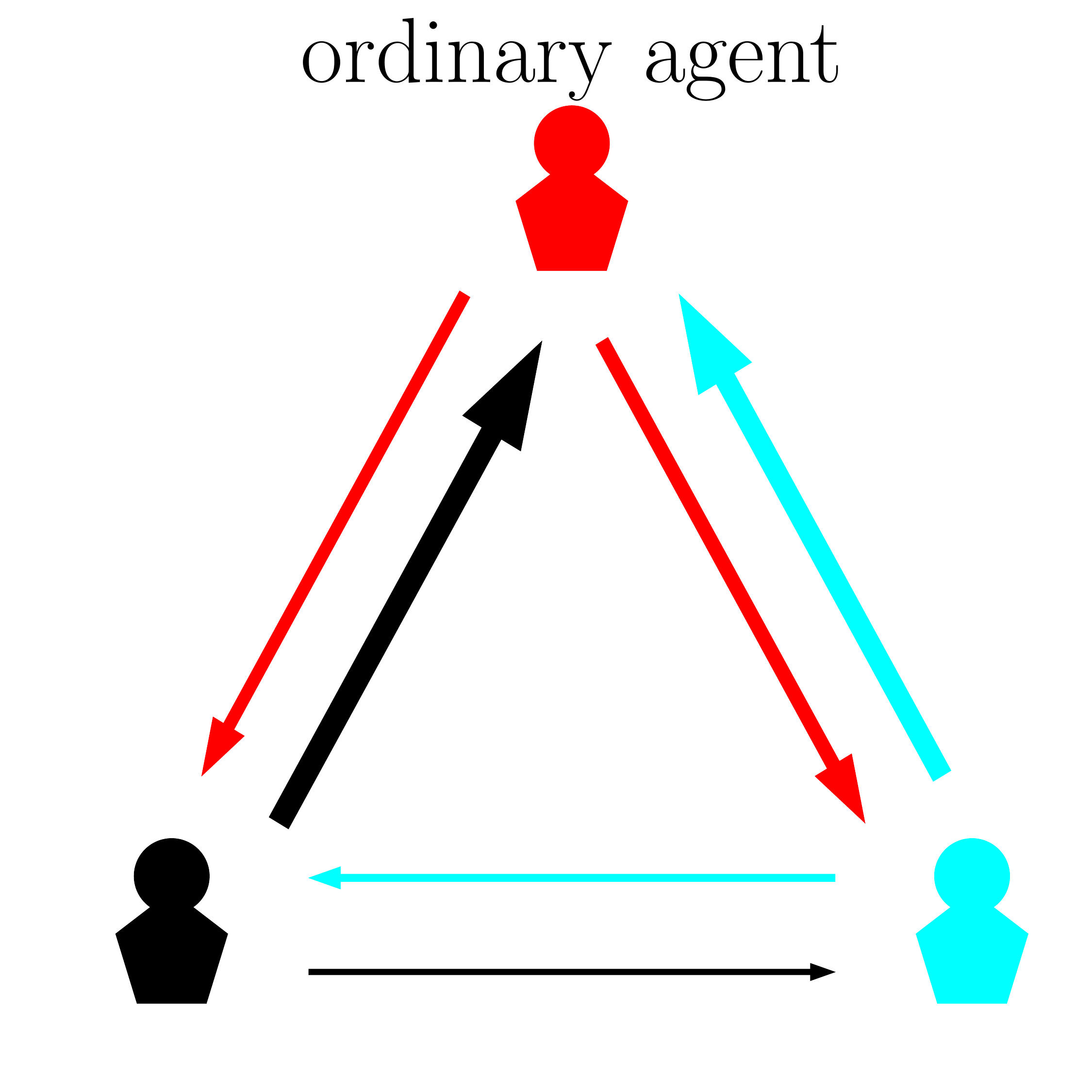}\includegraphics[width=0.1665\textwidth]{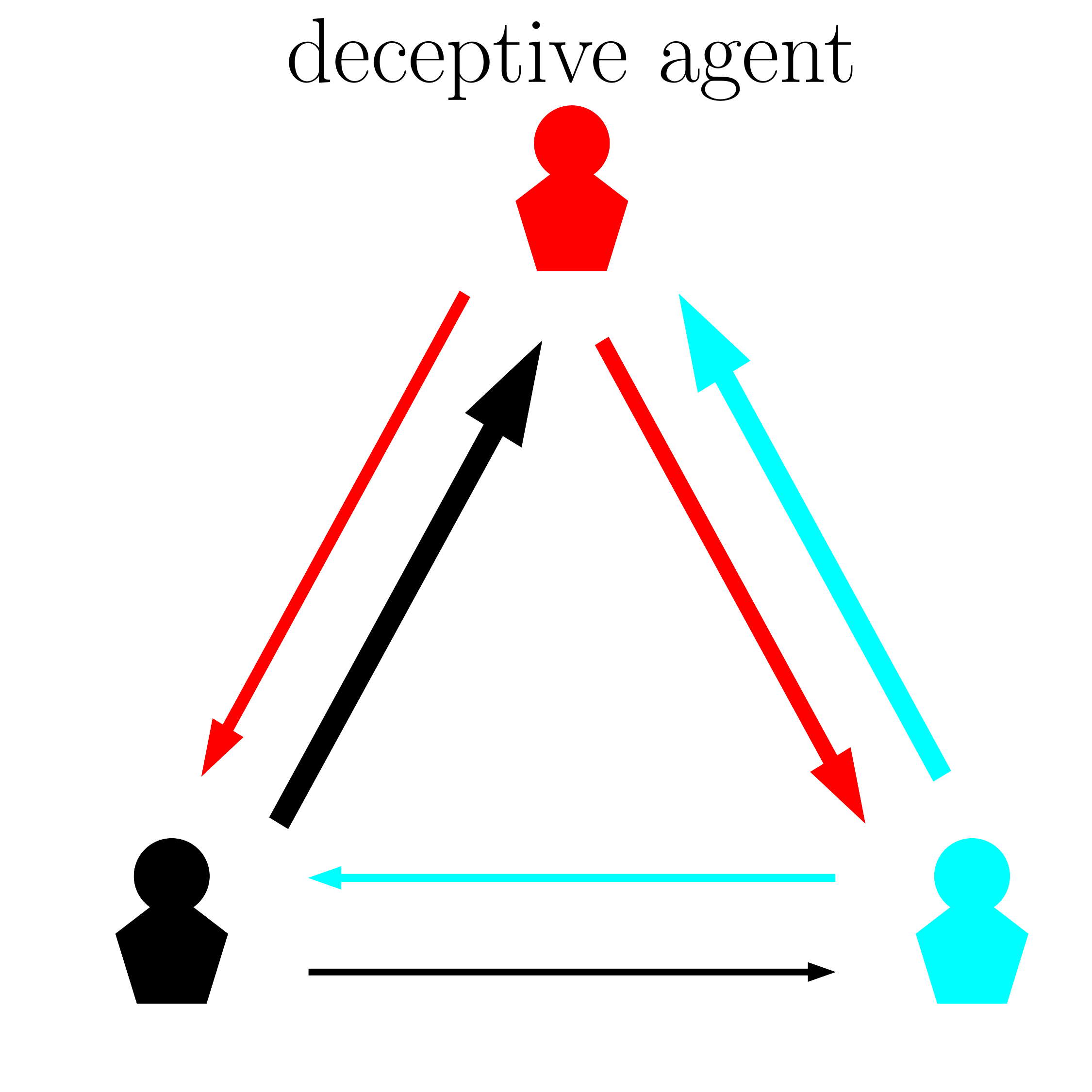}\includegraphics[width=0.1665\textwidth]{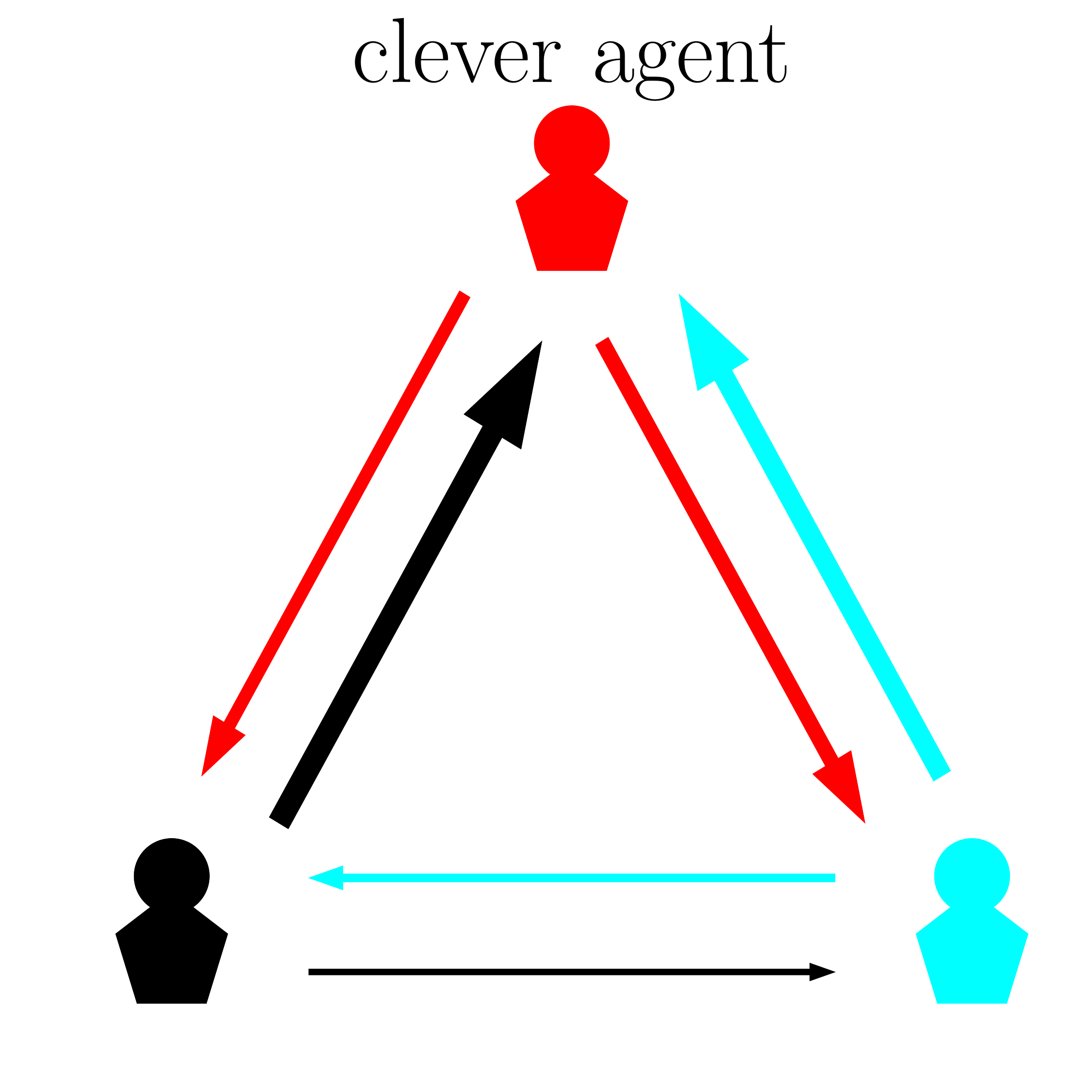}

\includegraphics[width=0.1665\textwidth]{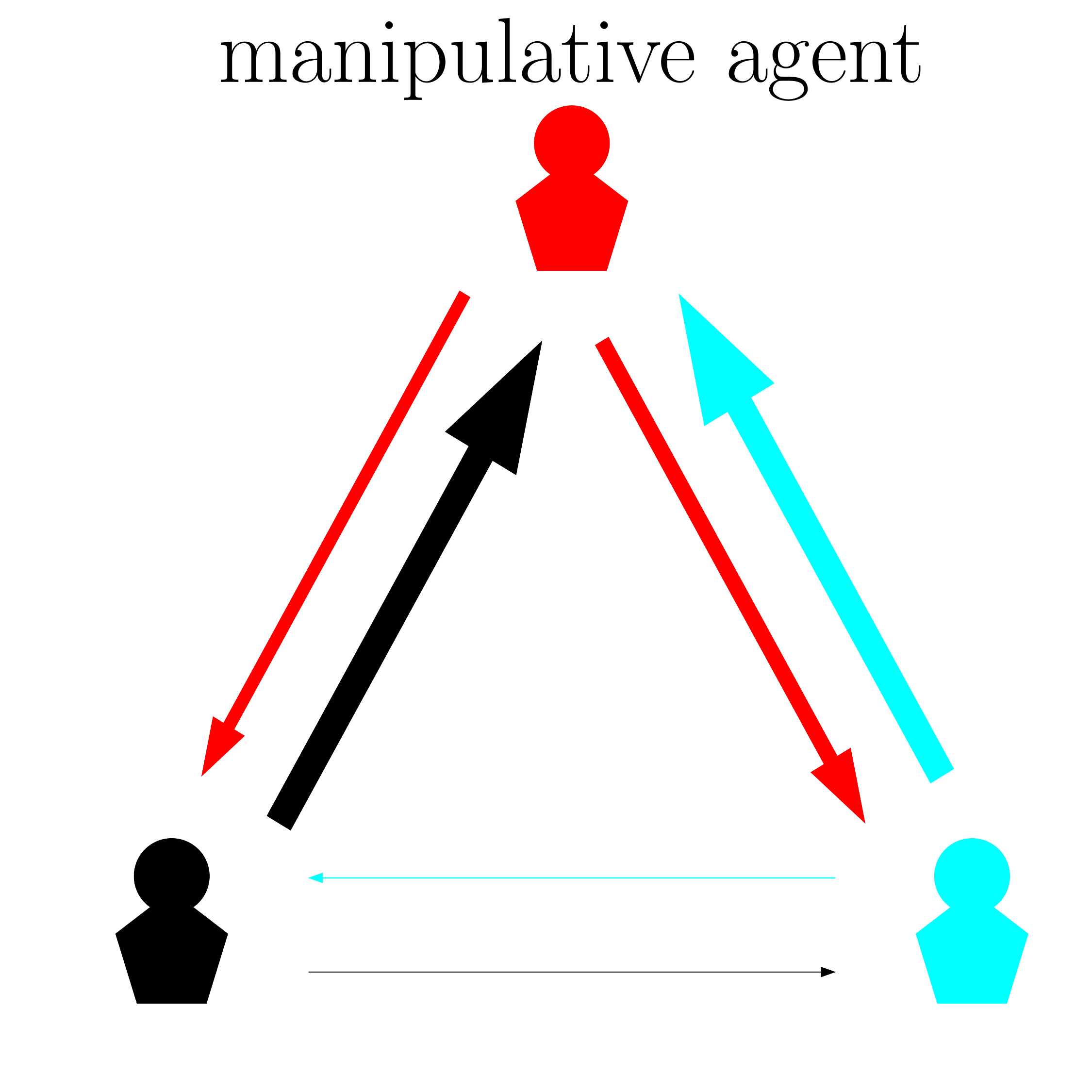}\includegraphics[width=0.1665\textwidth]{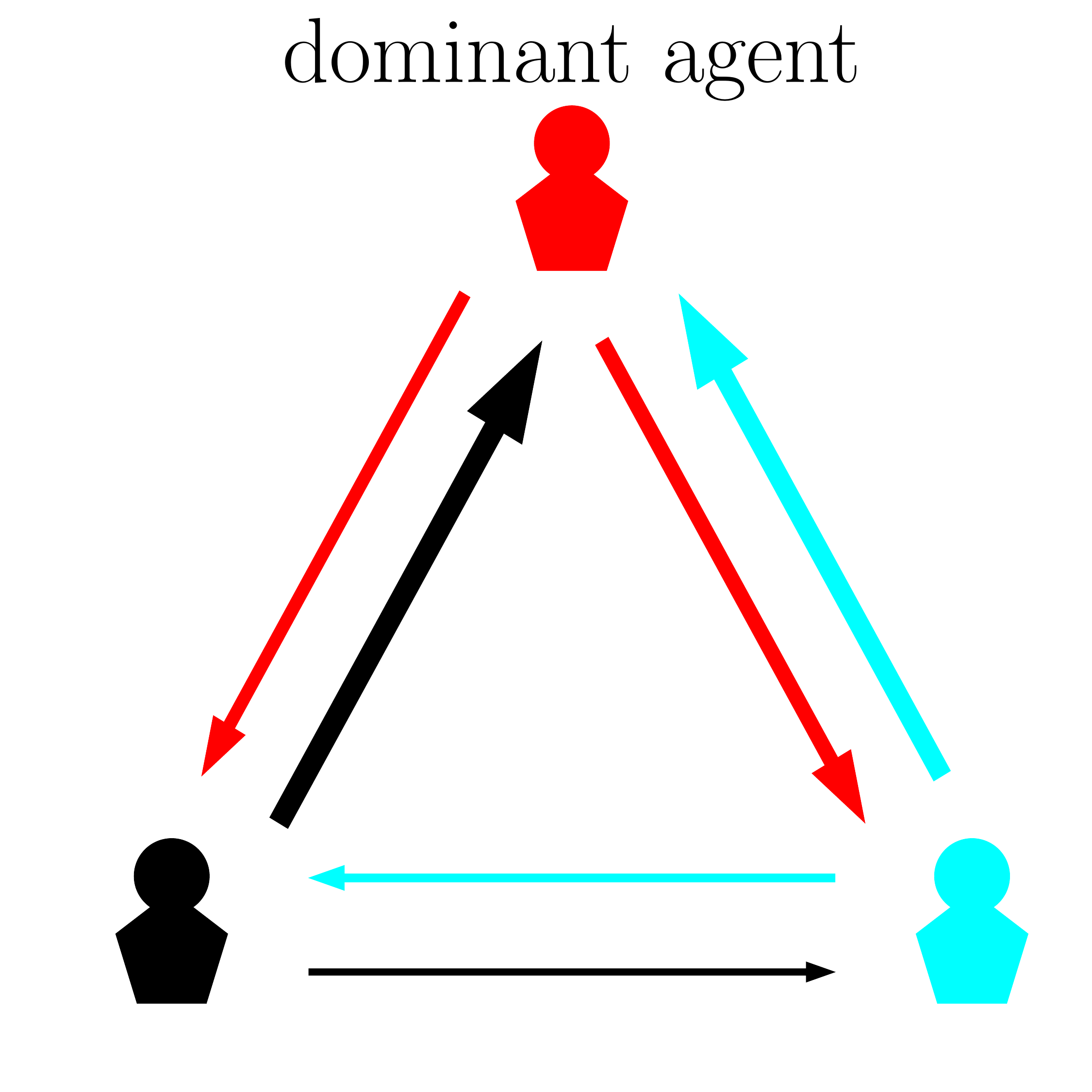}\includegraphics[width=0.1665\textwidth]{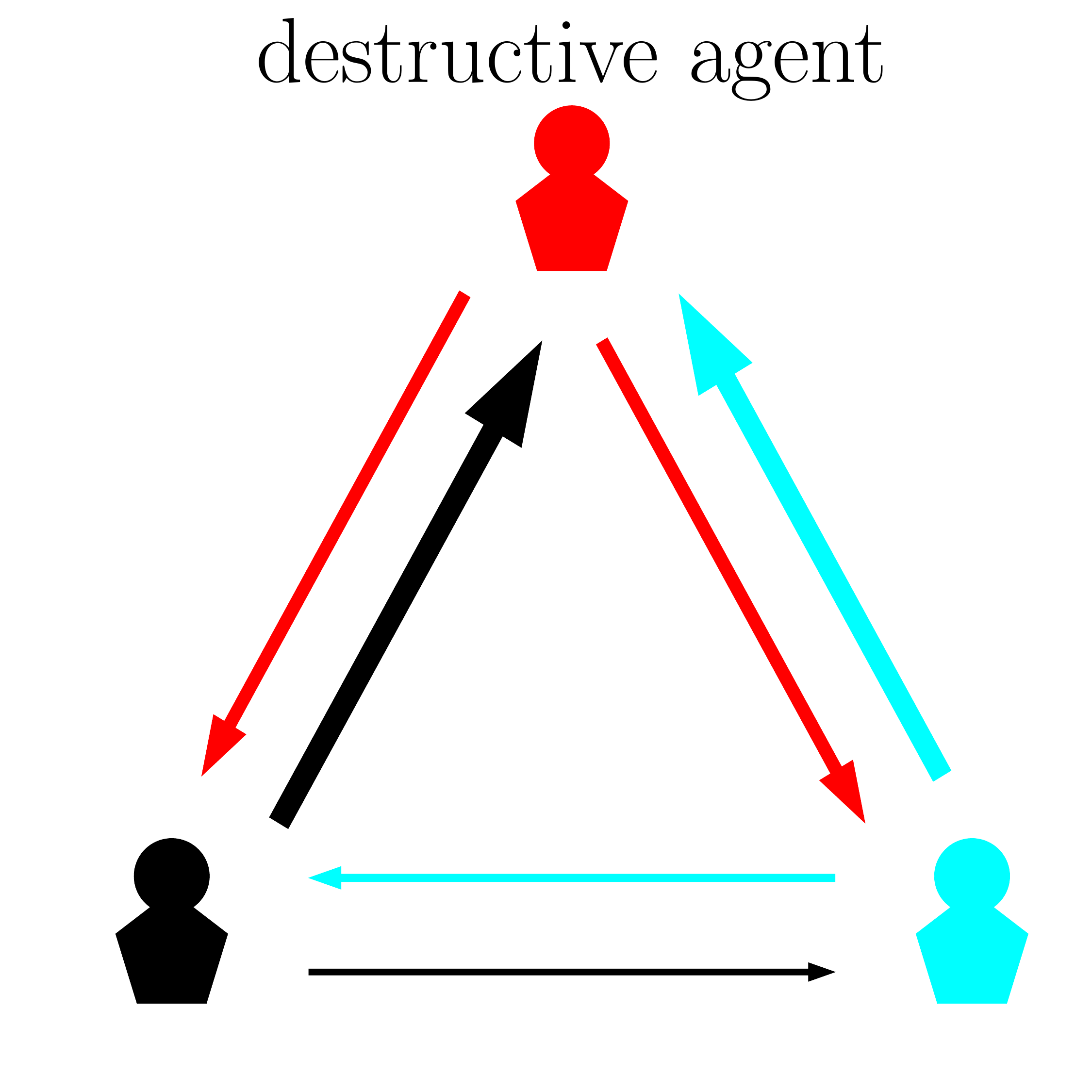}

\caption{Friendship network, where the thickness of an arrow $a\rightarrowtriangle b$
from agent $a$ to $b$ (as indicated by their colors) expresses how
often $a$ regards $b$ as a friend within the hundred runs. \label{fig:Friendship-network}}
\end{figure}

In the following, we want to investigate the friendship relational
network of agents in the different setups. For the simulations with
three agents, these are displayed in Fig.\ \ref{fig:Friendship-network}
and show that the most dishonest agent (red) manages to befriend best
the others, in particular when being manipulative (bottom left). Red's
own friendship budget is nearly equally distributed among the other
two agents, with a slight preference for cyan, who, also being a bit
dishonest, is slightly better in maintaining friendships than black.

\begin{figure*}[t]
\includegraphics[width=0.333\textwidth]{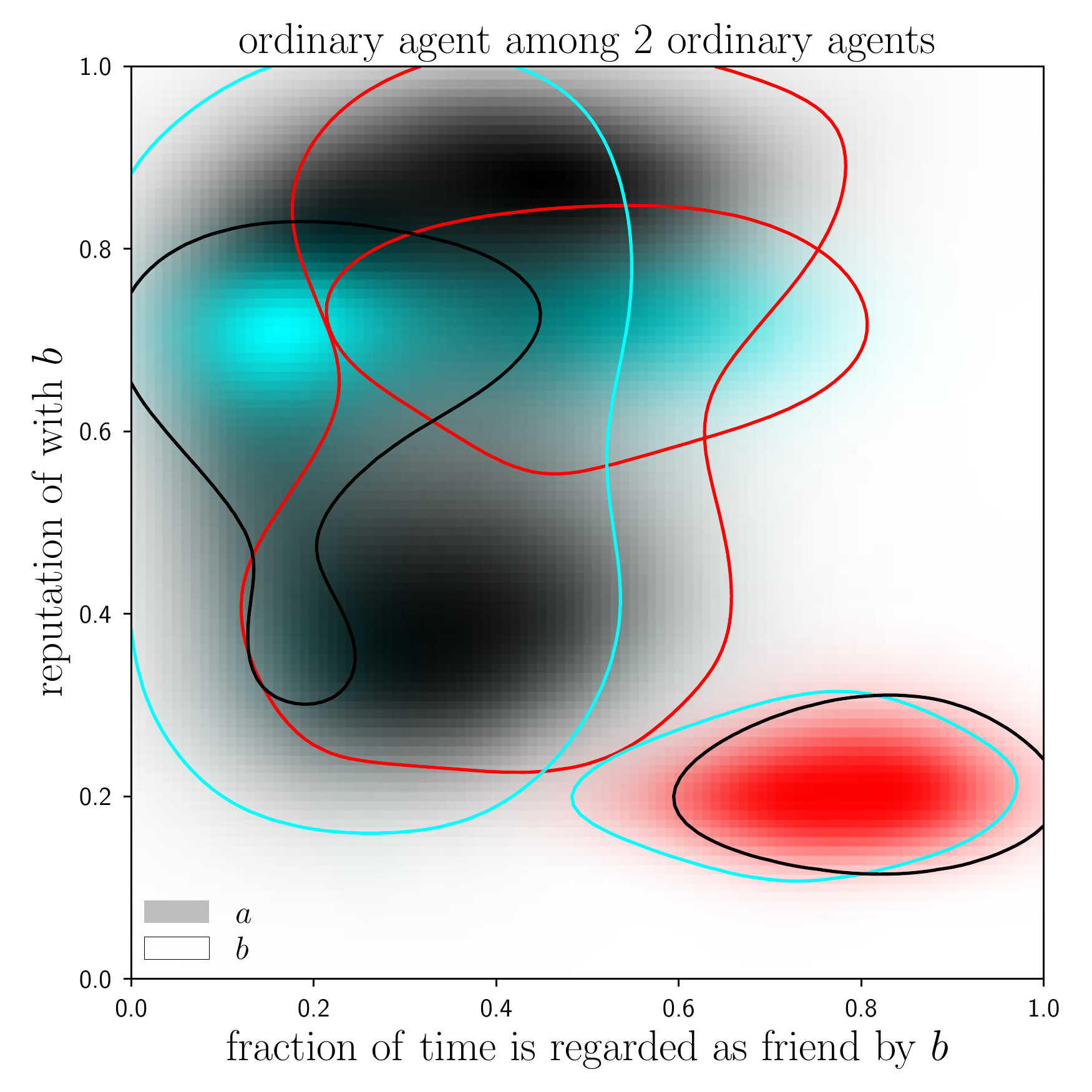}\includegraphics[width=0.333\textwidth]{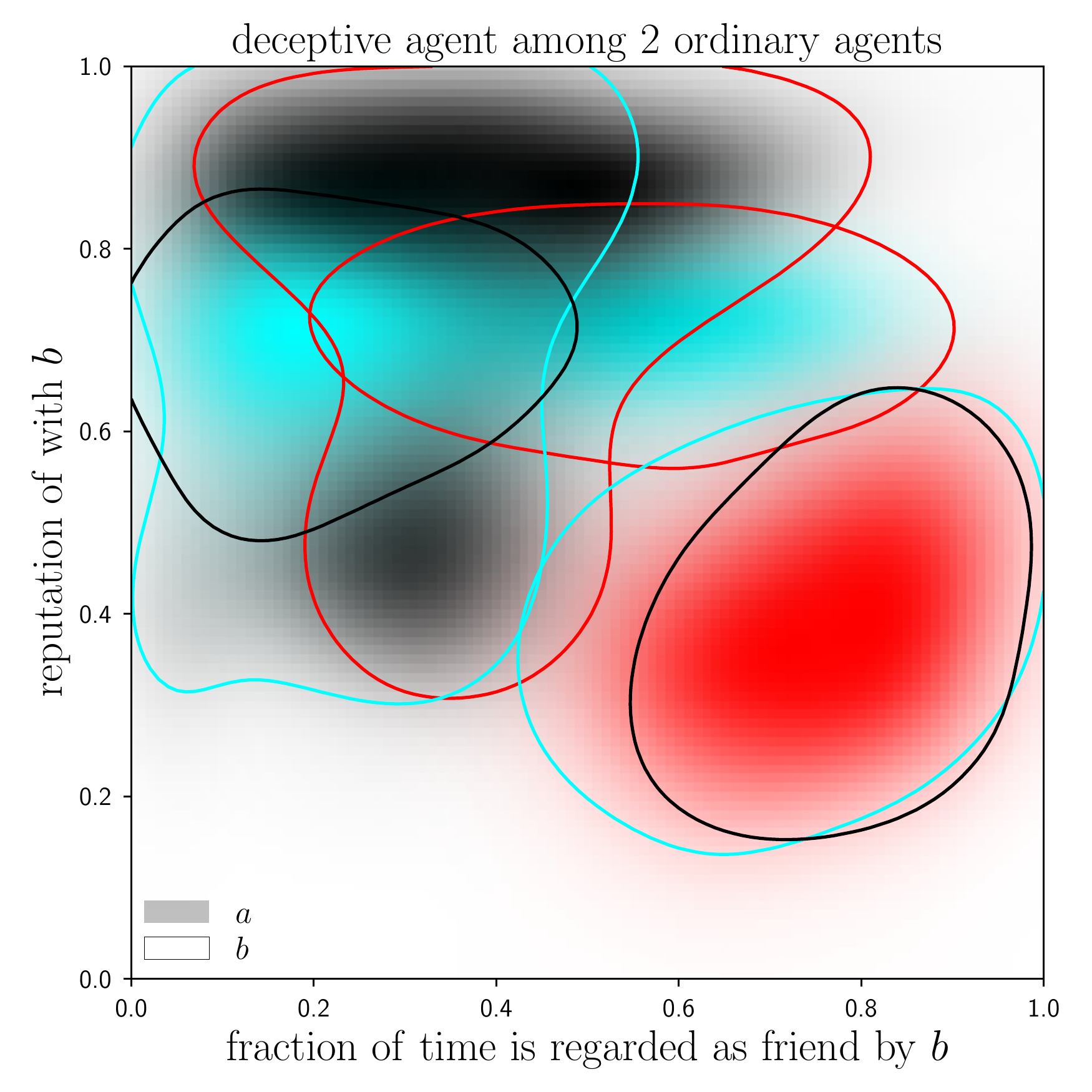}\includegraphics[width=0.333\textwidth]{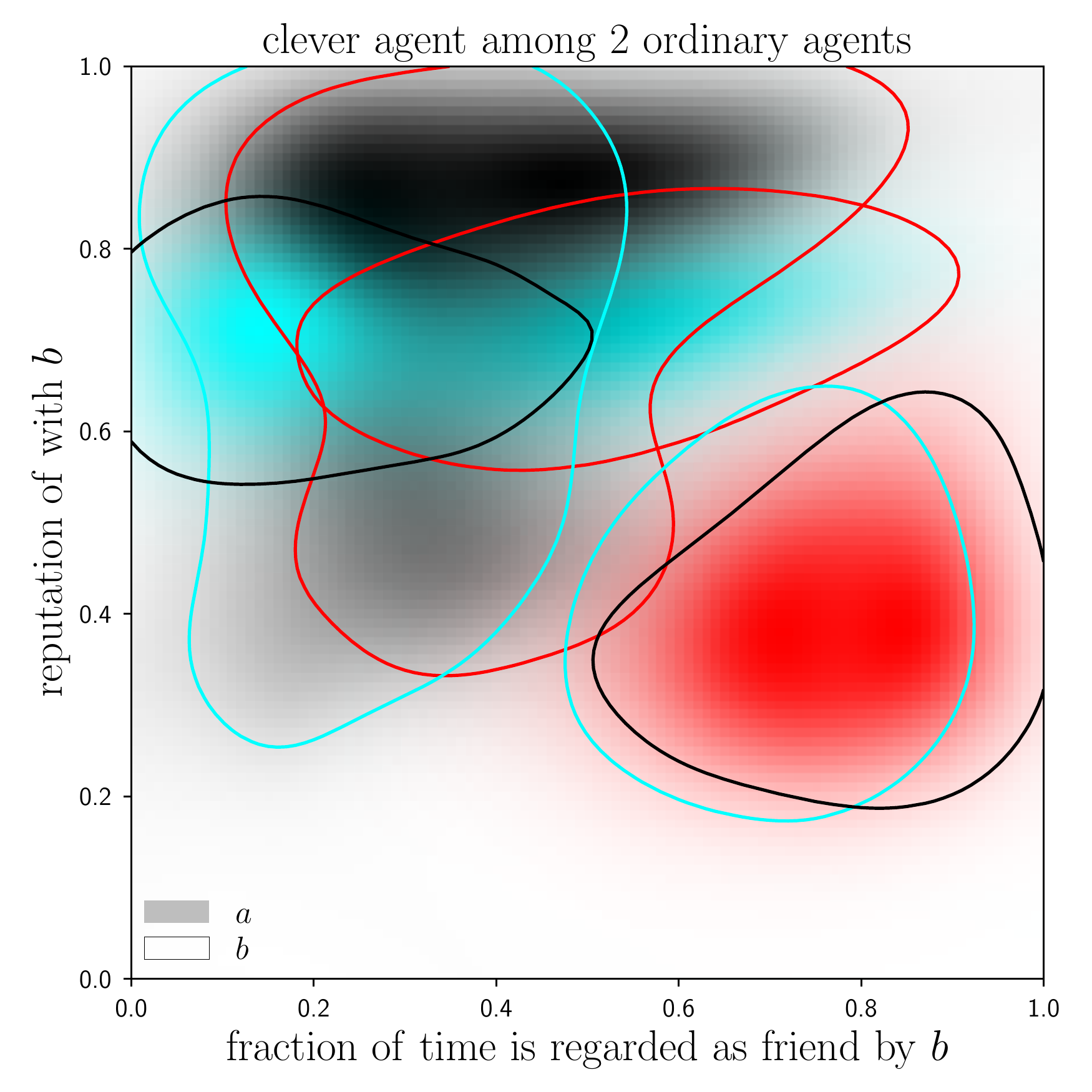}

\includegraphics[width=0.333\textwidth]{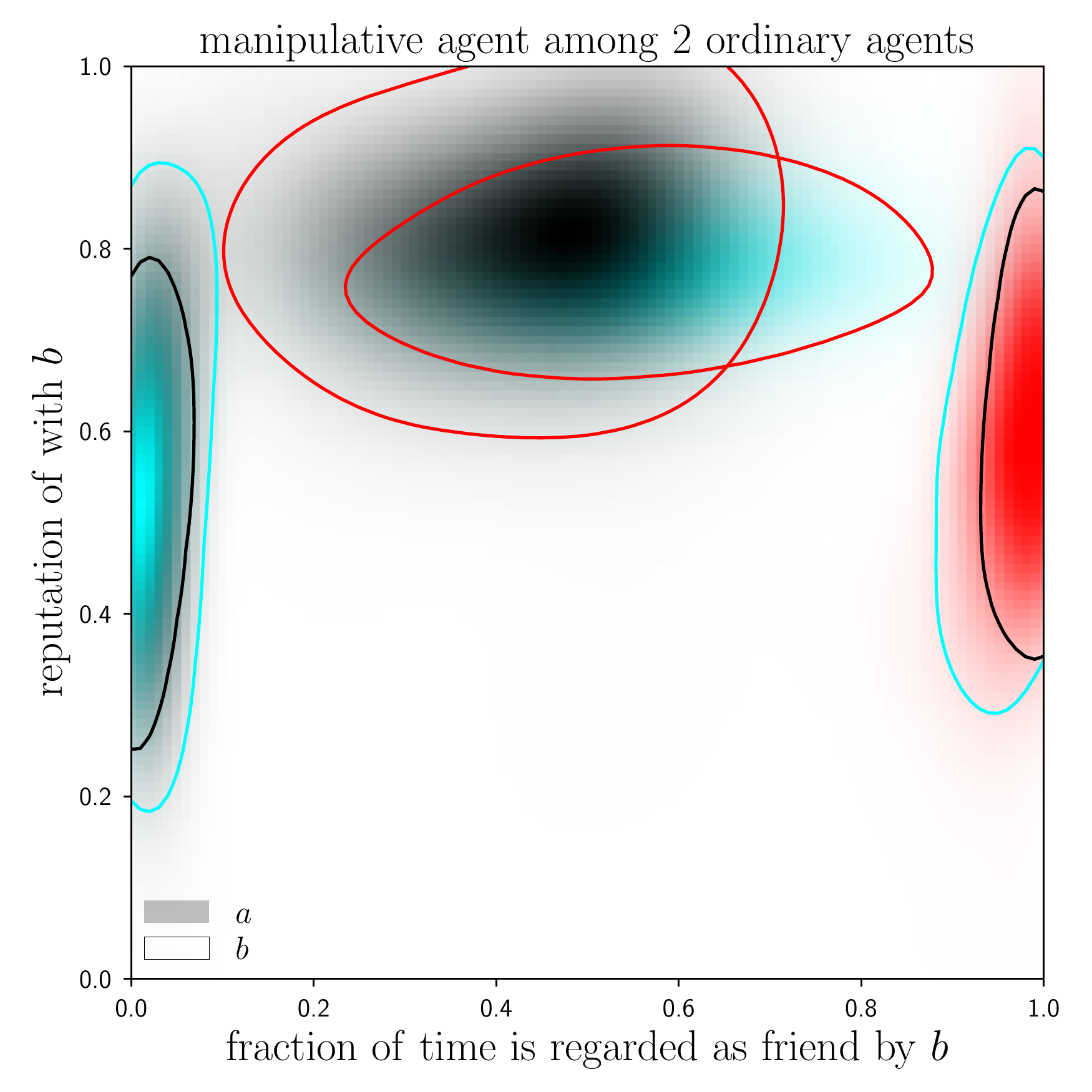}\includegraphics[width=0.333\textwidth]{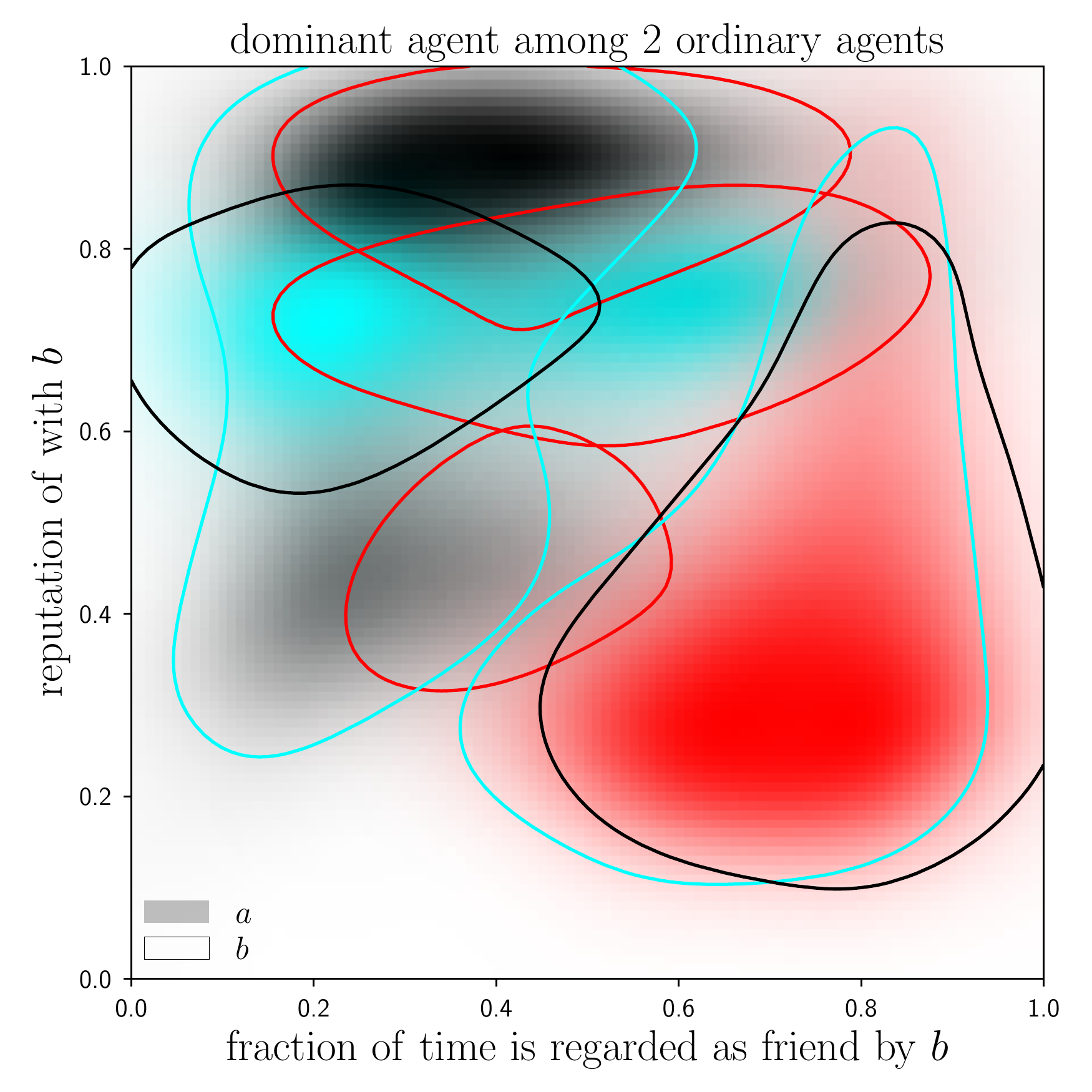}\includegraphics[width=0.333\textwidth]{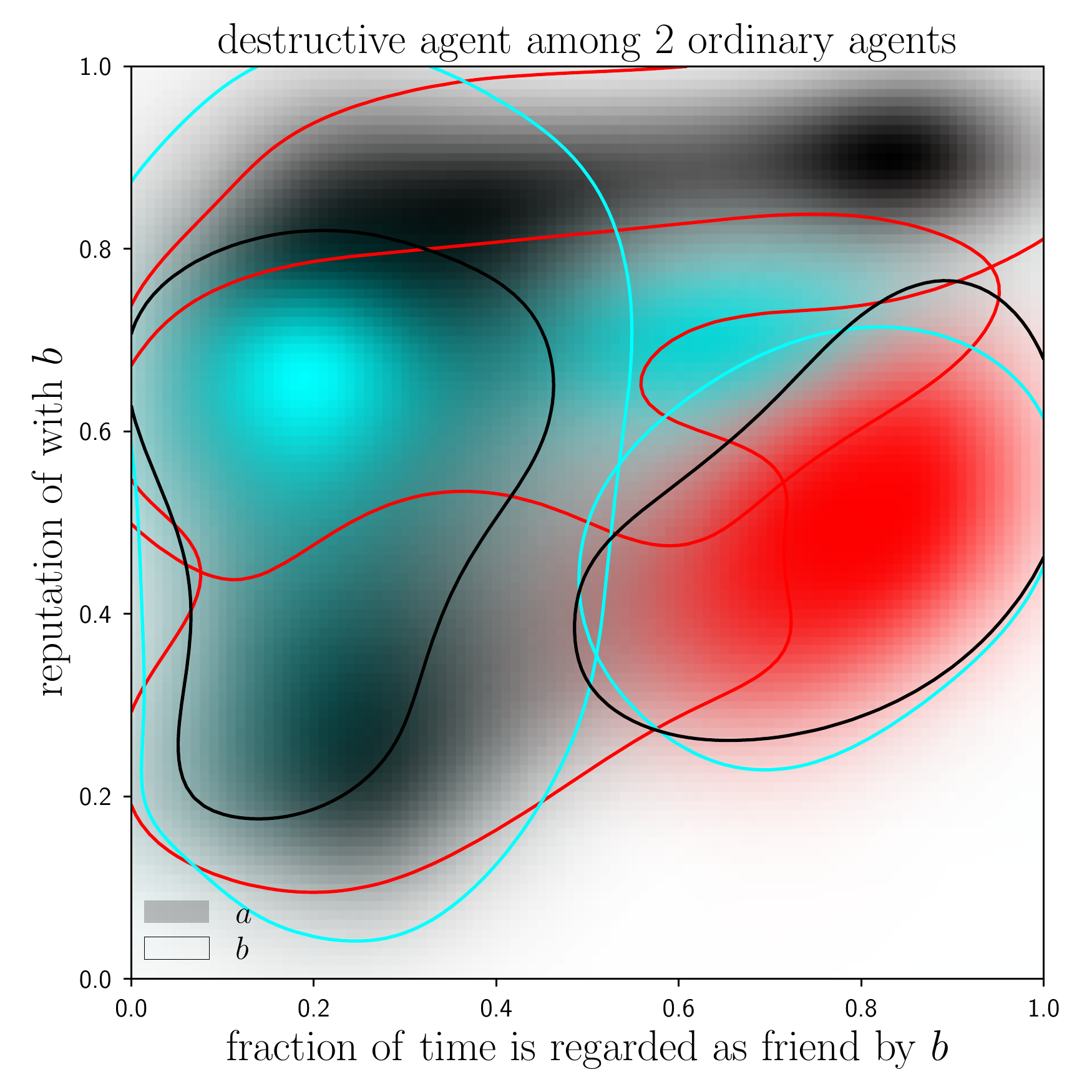}

\caption{Distribution of reputation and friendship relations between pairs
of agents in the hundred runs. For the reputation of agent $a$ with
$b$ ($\overline{x}_{ba}$, on the vertical axis) and for the time
fraction $a$ is regarded as a friend by $b$ (on the horizontal axis)
the density distribution color indicates agent $a$ and the color
of the contour line, which is at $10\%$ of the distribution's peak
value, indicates agent $b$. Displayed are the run-averaged friendship
and reputation distributions. Thus, the cyan distribution with red
contour expresses how cyan (agent $a)$ is seen by red (agent $b)$
on average within each of the hundred runs.\label{fig:Distribution-of-reputation}}
\end{figure*}

The correlation of friendship and reputation relations can be studied
in Fig.\ \ref{fig:Distribution-of-reputation}. For each of the hundred
runs time-average $a\rightarrowtriangle b$ reputation relation values
(with $a\rightarrowtriangle b$ meaning agent $b$'s reputation with
$a$) and the time-fraction of $a\rightarrowtriangle b$ friendships
(meaning agent $a$ regards $b$ as friend) were calculated and displayed.
For visual clarity of the plot, the hundred points in the friendship-reputation
plane of each $a\rightarrowtriangle b$ relation were converted into
a density.\footnote{The technique described in footnote \ref{fn:density-reconstruction}
was used for this.} Fig.\ \ref{fig:Distribution-of-reputation} confirms the observation
made with Fig.\ \ref{fig:Friendship-network} that the most dishonest
agents are preferentially regarded as friends. No distinct correlation
between the friendship strengths and reputation values within the
same $a\rightarrowtriangle b$ relation is seen, with two exceptions,
the dominant and the destructive agents. The density distributions
show different levels of dispersion in the friendship and reputation
dimensions, but not much (linear) correlation between these variables.

The different strategies of agent red manifest themselves by clearly
distinct friendship-reputation relation patterns. The ordinary agent
red (top left panel of Fig.\ \ref{fig:Distribution-of-reputation})
manages to become both other agents' preferred friend, at a moderate
time averaged reputation of about $0.2$. Becoming deceptive (top
middle panel) increases red's reputation to typically $0.4$ without
changing the friendship network much. The other agents' reputations
increase thereby also by a comparable margin. Red becoming clever
(deceptive and smart, top right panel) lets the other agents' reputations
increase further on average, as red's higher smartness now less often
classifies them incorrectly as dishonest. The manipulative agent red
(bottom left panel) manages to nearly monopolize black and cyan's
friendship, which turns them thereby into permanent mutual enemies.
As the manipulative agent red interviews the others frequently about
their self-images, red is well informed about their honesty. In contrast
to this, the dominant agent red, who mostly speaks about red and less
about others, therefore often incorrectly classifies black as less
reliable (see distinct lower red contour in bottom middle panel).
The dominant red's own reputation can occasionally become very large,
but usually stays below of that of the other two agents and that of
the manipulative red agent. Finally, the destructive agent red (bottom
right panel) creates the largest dispersion in other agents' reputation
and friendship values.

For the destructive agent red a clear correlation exists between the
reputation and friendship red has for others, which is caused by the
destructive agent's tendency to heavily damage the reputations of
any enemy. This primarily destroys red's enemies reputation with red's
friends, but red's disrespectful opinions are mirrored by red's friend
and thereby imprints also onto red's own beliefs on red's enemies.
It is interesting that this friendship-reputation correlation effect
is stronger for red's view on black than on cyan. This is a consequence
of cyan's lower honesty, which allows cyan to participate in the destruction
of black's reputation whenever being in a mutual friendship with red.

The relation of run averaged reputations and friendship relations
for the simulations with four and five agents are shown in App.\ \ref{sec:Detailed-communication-strategies}.
The trends observed with three agents are less obvious there.

\subsection{Social atmospheres\label{subsec:Social-atmosphere}}

The visual inspection of the belief state and reputation dynamics
in Figs.\  \ref{fig:Reputation-game-simulations}-\ref{fig:Communication-patterns},
Figs.\  \ref{fig:Propaganda-simulation}-\ref{fig:Dominant-communication-patterns},
and App.\ \ref{sec:Detailed-communication-strategies} shows a variety
of social atmospheres, ranging from frozen situations, in which opinions
quickly converge to static values (e.g.\ dominant agent run shown
on the left of Figs.\ \ref{fig:Special-communication-strategies}-\ref{fig:Dominant-communication-patterns}),
over adaptive regimes, in which individual and collective learning
curves can be observed (e.g.\ ordinary agent run in Figs.\ \ref{fig:Reputation-game-simulations}-\ref{fig:Communication-patterns}),
to very chaotic situations, in which the beliefs and expressed opinions
change rapidly (e.g.\ dominant agent run shown on the right of Figs.\ \ref{fig:Special-communication-strategies}-\ref{fig:Dominant-communication-patterns}).
In order to classify these different regimes and to see how different
strategies are related to those we introduce a measure of social chaos
in a run as 
\begin{eqnarray}
\text{chaos} & := & \langle(\overline{x}_{ij}(t)-\overline{\overline{x}_{ij}})^{2}\rangle_{i,j\in\mathcal{A};t\in[0,T]}^{\nicefrac{1}{2}}\text{ where}\label{eq:chaos}\\
\overline{\overline{x}_{ij}} & := & \langle\overline{x}_{ij}(t)\rangle_{t\in[0,T]}.\label{eq:opinion-temporal-mean}
\end{eqnarray}
This characterizes the average volatility of all beliefs of a run.

Fig.\ \ref{fig:Statistics-chaos-3A} displays the relation of run
averaged reputations of agents and this measure of social chaos in
different scenarios (ordinary, deceptive, clever, manipulative, dominant,
and destructive). All fully deceptive agents (all agents red except
the ordinary agent red) seem to create and benefit from social chaos,
as higher chaos values are reached and the average reputation of agent
red correlates with this. The manipulative and dominant agents seem
to benefit most strongly from chaos, whereas the destructive agent
red shows the lowest level of a correlation between red's reputation
and the level of social chaos.

The reputation of the more honest agents black and cyan is mostly
anti-correlated with the level of social chaos, at least in the cases
where those exhibit a high reputation. Agent black sometimes gets
into the Cassandra-syndrome regime, where black's reputation is low,
despite black being very honest. Interestingly, in this low reputation
regime black's reputation is positively correlated with the level
of social chaos. The steeper reputation-chaos correlation of black
in black's low reputation regime compared to the corresponding correlation
for red indicates that a different mechanism is here at work for black
(in comparison to red). A plausible explanation is that the effect
generating the Cassandra syndrome for black becomes inefficient beyond
a certain level of chaos. Chaos increases the lie detection threshold
$\kappa_{i}$ of every agent $i$, and therefore makes them more tolerant
for deviating opinions and thus for those expressed by black when
being in the Cassandra syndrome mode.

The relation of run averaged reputations with social chaos for simulations
with four and five agents are displayed in App.\ \ref{sec:Detailed-communication-strategies}.
The correlations visible in the three agent case are not visible there.

\section{Discussion\label{sec:Discussion}}

\subsection{The game and its players}

We introduced a\emph{ reputation game simulation} to study emerging
social and psychological phenomena. The game illustrates the vulnerability
of individuals or groups to certain kinds of malicious communications.
The rules of the game were designed to study a number of effects witnessed
in group dynamics and can be summarized (and generalized) as follows:
\begin{quote}
\emph{A number of players exchange opinions on the other's reputations
(a partly shared reality) while aiming for orientation, reputation,
and power.}
\end{quote}
The terms opinions, reputation, and power should be briefly explained
in the game's context. Here, the exchanged opinions are messages that
can be honest or dishonest. Honesty is defined in the game as the
frequency in which the players communicate their beliefs. Orientation,
knowledge about the environment (or reality) \cite{festinger1954theory},
is necessary for the agents to reach their other two goals, reputation
and power. Reputation is defined as the beliefs of others about a
player's honesty. In the game, reputation is a prerequisite for power,
which here is the ability to influence the environment, as only the
statements of reputed players have a significant chance to impact
other's belief systems. Ultimately, reputation and power are helpful
in the real world to obtain other resources, which are, however, not
modeled explicitly in the game. Although a high reputation can be
reached by an agent by being honest, this typically does not imply
a large empowerment, as is shown by the fact that the most honest
agent often receives a low reputation in the presence of a deceptive
agent. An honest player has little ability to steer others' beliefs
in comparison to a frequent liar. Thus, the most powerful players
should be the ones that are least honest, but with a high reputation.
The increase of their reputation with respect to their intrinsic honesty
therefore seems to be a good measure of power. Honest agents might
become reputed, but are rarely powerful.

A number of decisions of agents in the game appear to be driven by
chance, but this does not need to be the case. In principle agents
could make decisions according to more sophisticated, deterministic
calculations. However, using randomness for now permits to set up
the game without having to discuss all principles behind decisions
in detail. Nevertheless, a number of behavior strategies were introduced
to understand their impact on the game.

These strategies were chosen to resemble to a certain degree deceptive
strategies used by humans. In particular, the manipulative, dominant,
and destructive strategies introduced here resemble real world strategies
that are used (neither necessarily nor exclusively) by members of
the dark triad, Machiavellian, narcissistic, and sociopathic personalities.

\subsection{The player's minds}

The agent's information processing is designed to follow information
theoretical principles, within some limits. The used cognitive model
tries to follow the optimal Bayesian logic, however, agents are unable
to memorize all fine details of the resulting high dimensional probability
distributions. We believe that such a bound rationality model roughly
captures how a human mind operates. Trying to maintain orientation
in a complex and changing world requires to follow information principles.
These principles, however, demand computational resources beyond what
is available to most finite physical systems, such as humans, our
agents, or other AI systems. Thus, compromises in the accuracy to
represent and process information are always necessary, and these
could be the basis of some of the \emph{cognitive biases} observed
in real world psychology \cite[e.g.][]{10.2307/1884852,Tversky&Kahneman,fiske1991social}
and AI \cite{Yu238,challen2019artificial}.

The limitations of the agents' knowledge representation, which is
only a direct product of one dimensional, parametrized probability
functions and not a multidimensional, non-parametric distribution
as required by Bayesian logic, can be exploited by adverse strategies
of other agents.\emph{} For example, a statement about some agent's
honesty that strongly disagrees with the receiver's belief implies
a bimodal posterior probability, with a peak associated with the possibility
of an honest message and a second peak associated with the possibility
of a lie. The relative height of these peaks depends on the clues
the receiver got about the message honesty. However, this bimodal
distribution cannot be stored in the agents' belief representation
and the information needs to be compressed into this form. As information
is inevitably lost in this compression, the resulting reasoning of
agents will be imperfect or \emph{irrational} to a certain degree.
This imperfection can be exploited by adversarial attacks, for example
in form of large scale propaganda.

To decide whether a message is reliable, agents use a number of signs.
Critical agents judge the trustworthiness of a message according to
how much it fits their own beliefs or how surprising it is. The surprise
of a message is measured in terms of the divergence of the belief
resulting from accepting the message in comparison to the present
belief. This divergence (or surprise) is measured in the number of
bits that would be obtained by this update.  The scale against which
this surprise is compared to decide about the trustworthiness of messages
needs to be learned and kept updated in a changing social environment.
This adaptability, however, opens the door to manipulative attacks.
Exposing an agent to a large number of strongly diverging opinions
inflates this scale, thereby reduces the ability to detect lies, and
thus makes manipulations easier. This seems to be the principle of
\emph{gas lighting} communication patterns used by dark triad personalities.
We simulated the case where agents are exposed to many propaganda
messages, which strongly diverge from their own beliefs, and observed
that even agents, which were initially getting more and more skeptical
about the trustworthiness of the propaganda, converted eventually
to the opinion expressed by the propaganda. The exposure to the propaganda
let their reference surprise scale inflate, and thereby their lie
detection break. Interestingly, the initially most skeptical agents
convert most strongly to the propaganda position, since the propaganda
causes the largest mental dissonance in the more skeptical minds within
our simulation.

In order to make agents more immune to propaganda we also introduced
a smart receiver strategy, which compares a message with what the
speaker seems to believe on a topic as well as what the speaker's
typical lies on a topic seem to be. These two reference points, but
also the need to construct credible lies, require agents to maintain
a mental representation of other's belief systems, i.e. a rudimentary
Theory of Mind. Here, we propose a simple description of the Theory
of Mind updates, which is certainly ad-hoc and should be revised in
future research. Smart agents, which are better in maintaining and
using their Theory of Mind, are indeed more immune against propaganda
and slightly better in discriminating lies from honest statements.
Our special agents are all smart as well as deceptive (= pathological
liars).

\subsection{The player's strategies}

\begin{figure*}[!t]
\includegraphics[width=0.333\textwidth]{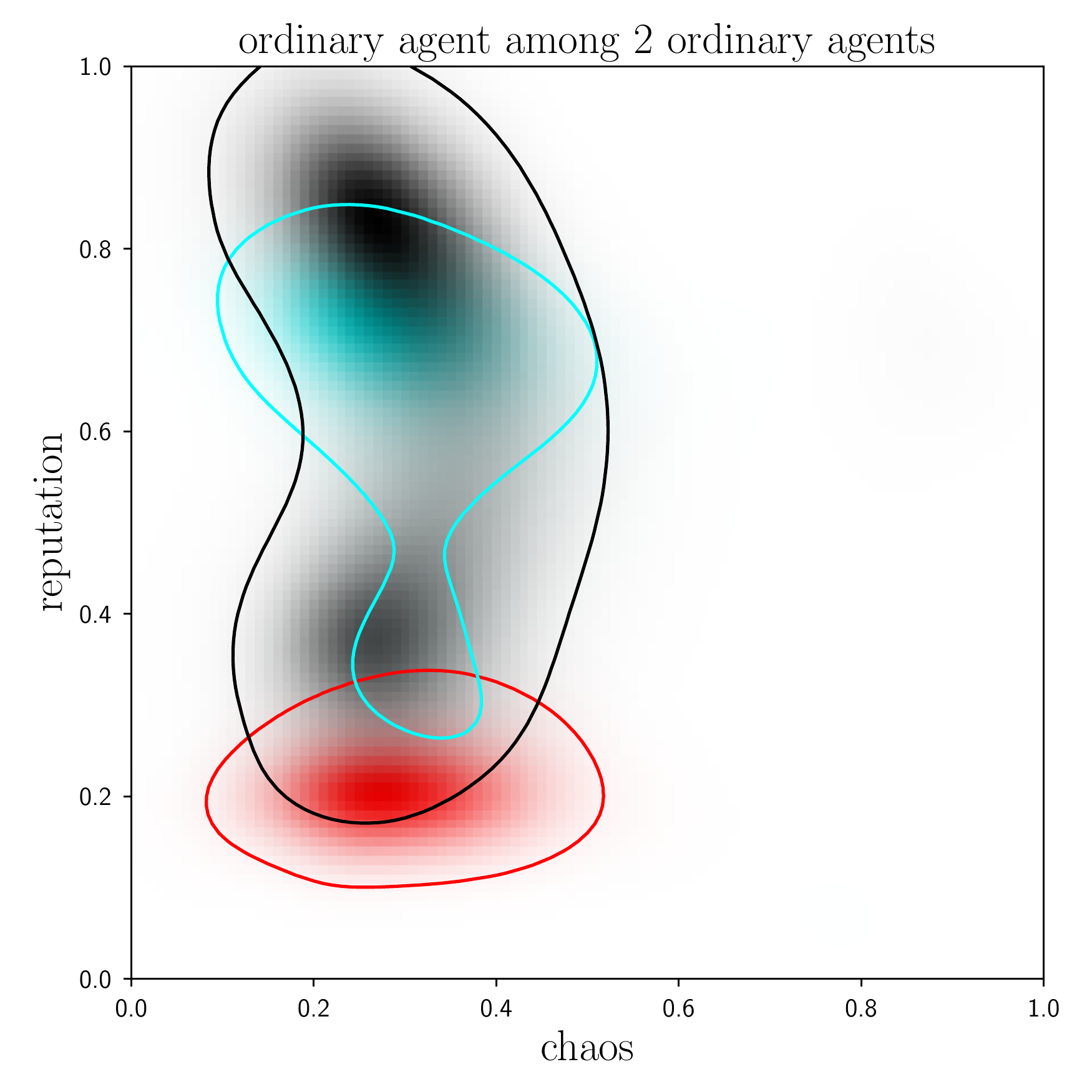}\includegraphics[width=0.333\textwidth]{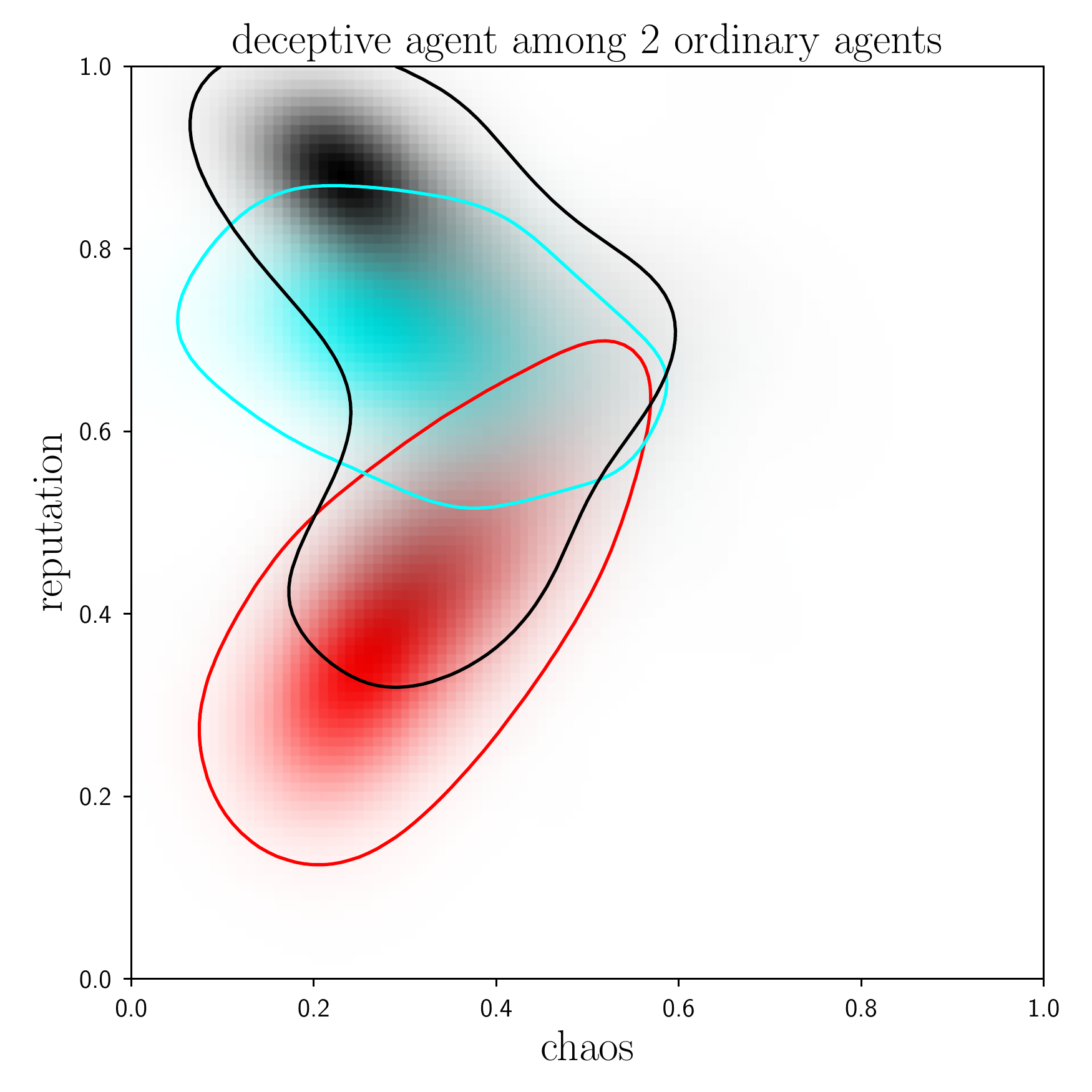}\includegraphics[width=0.333\textwidth]{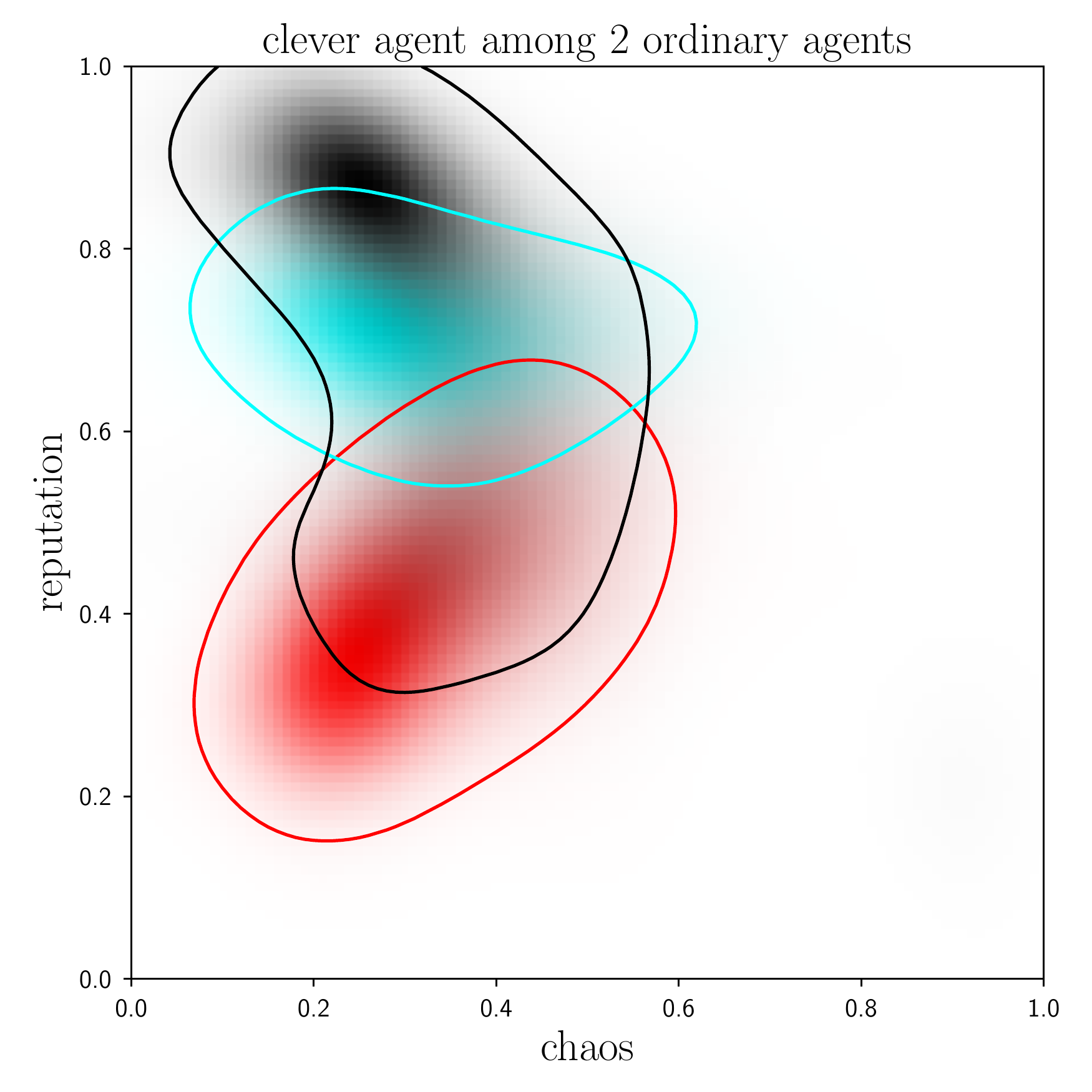}

\includegraphics[width=0.333\textwidth]{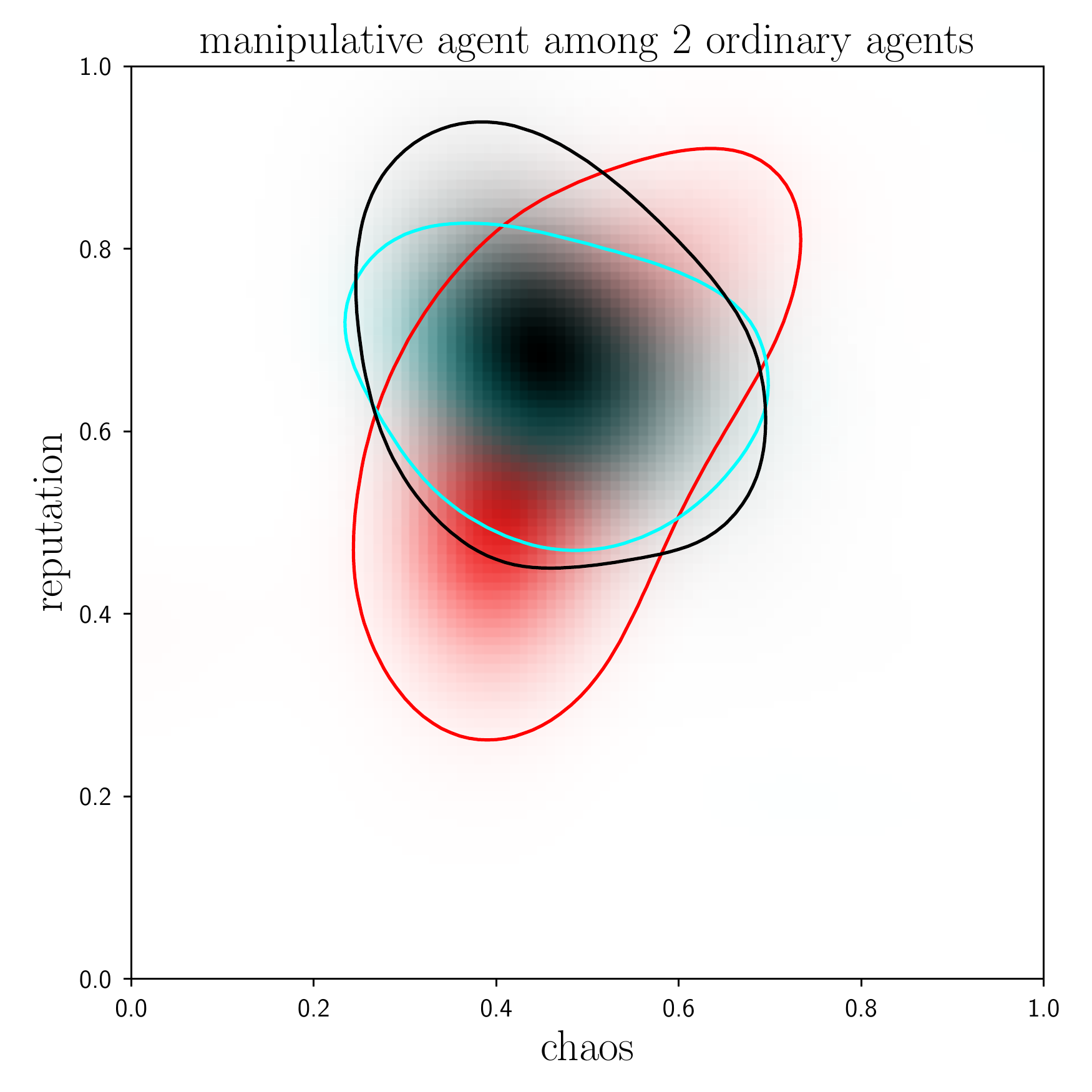}\includegraphics[width=0.333\textwidth]{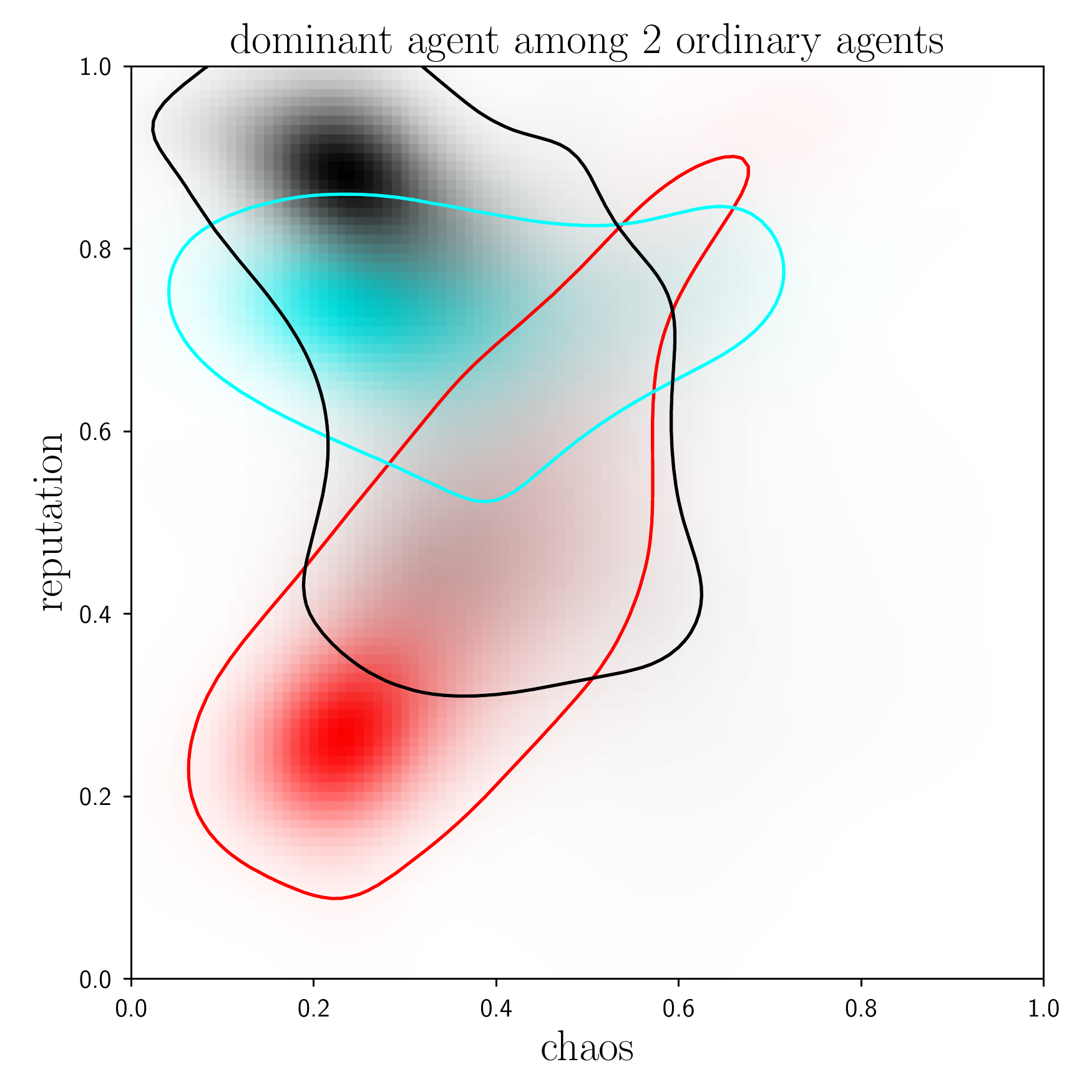}\includegraphics[width=0.333\textwidth]{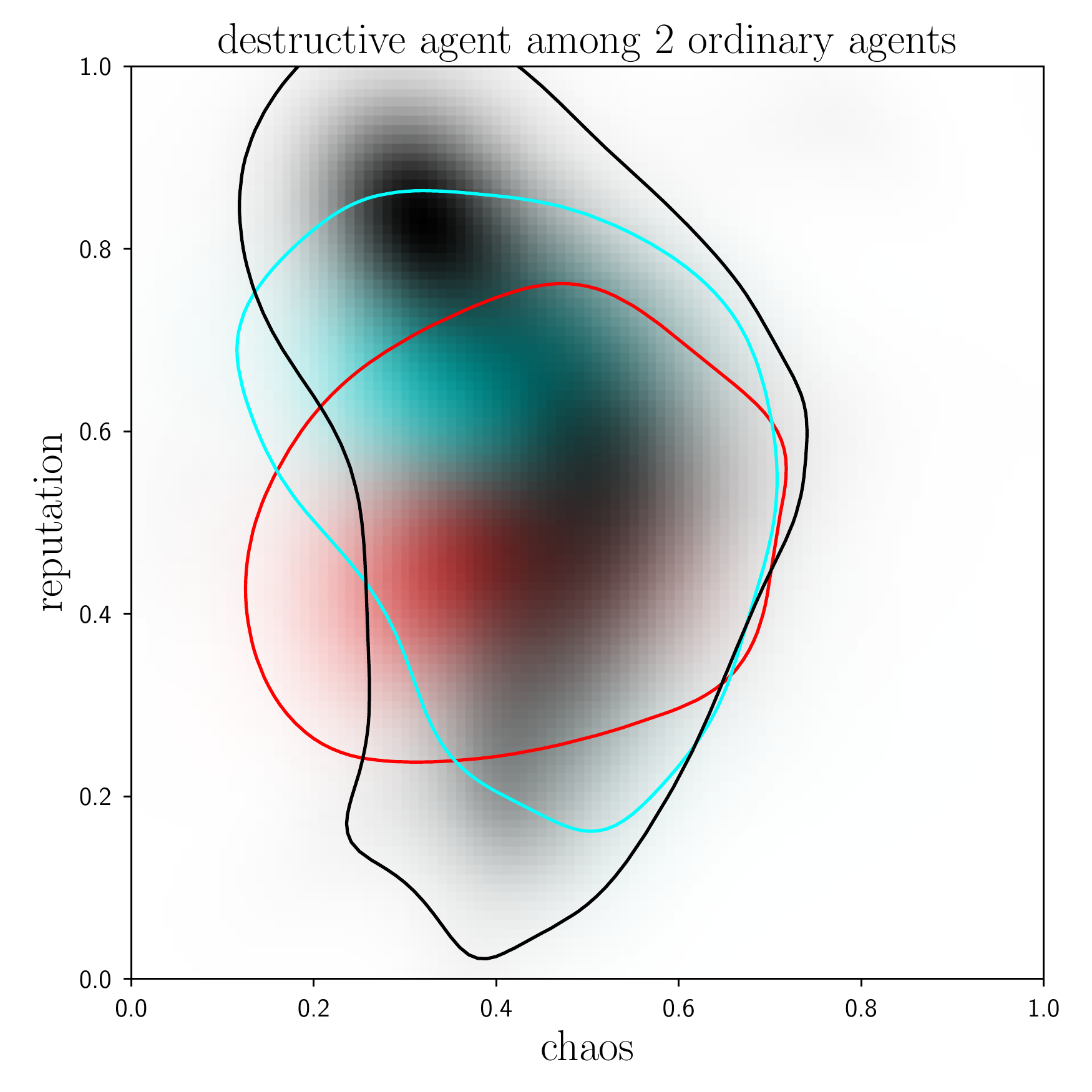}

\caption{Run averaged reputation of agents over the level of social \emph{chaos},
as measured by the temporal dispersion of all mean opinions within
each run calculated via Eq.\ \ref{eq:chaos}. The individual run
results are converted into a density as described in footnote \ref{fn:density-reconstruction}.
The lines mark the level of 5\% of the peak value of the shown density
of the same color.\label{fig:Statistics-chaos-3A}}
\end{figure*}

The basic strategies that agents can adopt are referred to as being
strategic, anti-strategic, egocentric, flattering, aggressive, shameless,
and deceptive. They can all be combined to form more complex, special
strategies such as the clever (deceptive and smart), manipulative
(clever, anti-strategic, and flattering), dominant (clever, strategic,
and egocentric), and destructive (clever, strategic, aggressive, and
shameless) strategies. The latter three are introduced to emulate
communication patterns frequently associated with Machiavellian, narcissistic,
and sociopathic personalities, respectively. Reputation game simulations
permit to investigate the effectiveness of such communication patterns
in achieving the goal of a high reputation and large power. Our simulations
verify that such strategies are indeed effective to achieve such goals,
at least in a statistical sense, not only in comparison to ordinary
agents, but also if we compare to clever agents (= deceptive and smart).

The manipulative strategy most often leads to the highest relative
reputation within small groups, the dominant strategy is able to reach
the absolute highest reputation values most frequently, and the destructive
strategy seems to become more efficient than the other two in larger
groups (see Sect.\ \ref{subsec:Statistics}). How many of these results
can be transferred to real human communication settings will require
more detailed investigations.

\subsection{Emergent phenomena}

The dynamic of the game is complex, stochastic, and chaotic. Nevertheless,
emergent trends and patterns can be observed that resemble real world
socio-psychological phenomena. Here, we list the ones we observed
in the simulations.

The game setup is an \textbf{echo-chamber}, with only a few agents
talking to each other, and who, thanks to the imperfect tracking of
other agents' information sources, do not realize when another agent's
apparent new information is in fact an echo of an earlier, own statement.
Emergent echo-chambers in sub-sets of agents can also be observed
in simulation runs (even though the size of the population is below
the size of real world echo chambers). The reputation network between
agents defines who really listens to each other, where \emph{``really
listening''} is meant in the sense of accepting a received message
as honest. The formation of an echo-chamber can for example clearly
be observed in the simulation of cross-communicating ordinary agents
under constant propaganda shown in Fig.\ \ref{fig:Propaganda-simulation}
and discussed in Sect.\ \ref{subsec:Propaganda-and-resilience}.
There, agents cyan and yellow form an echo-chamber, characterized
by a growing mutual trust and converging opinions on red, a process
in which agent black does not participate.

The occurrence of \textbf{group opinion building} is very manifest
in most simulations and is discussed in the context of the smart agent
in Sect.\ \ref{subsec:Receiver-strategies}, of the shameless agent
in Sect.\ \ref{subsec:Basic-communication-strategies-1}, and the
dominant agent in Sect.\ \ref{subsec:Special-communication-strategies-1}.
There the phenomena of a \textbf{freeze-in of group opinions} was
explicitly mentioned, which happens frequently thanks to the general
echo-chamber setup of our reputation game.

The echo-chamber effect allows also for \textbf{self-deception} of
agents, which can be observed in many simulation runs. For example
the run with the flattering among ordinary agents shown in Fig.\ \ref{fig:Basic-communication-strategies}
and discussed in Sect.\ \ref{subsec:Communication-strategies} shows
clearly self-deception of the mostly dishonest agent red. Despite
the better direct information from the own self-observations, red's
final self-esteem follows red's enhanced reputation (w.r.t.\ red's
honesty), despite the basis for this enhancement being red's own lies.

The largest self-deceptions in the simulations can be found in some
of the runs with dominant agents. There, the self-deception might
be classified as an\textbf{ self-esteem boost via narcissistic supply}.
By preferring as conversation partners the most reputed agents and
as topics the dominant agent themselves, dominant agents set up their
environment in a way that makes efficient self-deception most likely.
If the most reputed agent talks positively and frequently about the
dominant agent to that agent, the dominant agent will start to believe
in the echo of the own propaganda. The reputed agent has thereby become
the supplier of the self-esteem boost. The transition from a realistic
self-perception to the boosted self-esteem state can be seen in Fig.\ \ref{fig:Special-communication-strategies}
(third row, right panel, e.g.\ shortly after time $t=250$, $400,$
and $750$), and is a frequent phenomena for dominant agents as visible
from the statistics presented in Fig.\ \ref{fig:Statistics-histogram},
Fig.\ \ref{fig:Statistics-histogram-1}, and Sect.\ \ref{subsec:Statistics}.

Different \textbf{social atmospheres} can also be observed, for example
by comparing the runs with dominant agents displayed in in the left
and right panel of Fig.\ \ref{fig:Special-communication-strategies}
and Fig.\ \ref{fig:Dominant-communication-patterns}. In the first
of the two displayed runs, the group opinions quickly freeze in, thanks
to rapidly converging statements of the different agents. In the second
run, the dynamics of the opinions is highly volatile and the expressed
opinions scatter largely. Not only the opinions of the deceptive agent
red, but also that of the other agents show large variance, even in
case the latter are honest. This is because the focusing of group
opinions is less efficient in such a situation. Fully deceptive agents
create and most strongly benefit from chaotic social atmospheres,
see Fig.\ \ref{fig:Statistics-chaos-3A} and Sect.\ \ref{subsec:Social-atmosphere}.

This large diversity of opinions leads for all agents to an enlarged
surprise reference scale for identifying lies. It also leads to larger
lies, as the size of a lie is gauged against this scale during lie
construction. This again leads to an even larger scale, forming a
run away effect. As a consequence, the agent's critical lie detection
breaks down and the propaganda of the do\-minant agent can pull opinions
as strongly as if it would act on uncritical agents. See Fig.\ \ref{fig:Propaganda-simulation}
for the reduced resilience of uncritical agents against propaganda
and Fig.\ \ref{fig:kappa-comparison} (bottom middle and right panels)
and Sect.\ \ref{subsec:Social-atmosphere} for the run away effect
of the lie detection scale.

Thus, an attack on the lie detection system by exposing the victims
to a large quantity of strong lies or just statements that create
cognitive dissonances can be a successful strategy in the simulation,
in particular for strongly self-promoting agents. A real world counterpart
of such a strategy is \textbf{gaslighting} in which the victims are
exposed to statements designed to confuse the victim's belief system
\cite[e.g.][and references therein]{sweet2019sociology}. Gaslighting
is a strategy often associated to narcissistic personalities. It is
currently not explicitly implemented in the repertoire of strategies
used by the dominant agent. Nevertheless, a variant of gaslighting
seems to occur in our reputation game as a side product of the dominant
agent's strong focus on a single topic (the dominant agent) and a
single conversation partner (the most reputed agent). If dominant
agents become reputed, their self-esteem might stay low, for their
many lies. This leads to a large cognitive dissonance for them, as
in the frequent conversations they have about themselves, they are
confronted with opinions that largly divergence from their self-picture.
As a consequence, their reference scale for lies increases. Since
they use this scale for lie construction and all their communications
are lies, they \textbf{express extreme opinions} on any conversation
topic.\footnote{See triangles marking lies in the right panel of Fig.\ \ref{fig:Dominant-communication-patterns},
which are either extremely positive (being at the top of the reputation
range) or extremely negative (being at the bottom of the range) in
the period 600 to 1200.} Since the extreme statements made by such a dominant agent with diverging
reputation and self-image also affects the lie reference scales of
other agents, the lies of those also become more extreme as well.
A \textbf{toxic social atmosphere} can therefore result, which persists
until the dominant agent's self esteem and reputation agree, either
on a high or on a low level. If the reputation and self-esteem of
a dominant agent are both high, this agent has managed to manipulate
the others into providing \textbf{narcissistic supply}, i.e. helping
to maintain the inflated self-image of the dominant agent (see right
panel of Fig.\ \ref{fig:Dominant-communication-patterns}).

Too much \emph{cognitive dissonance}, which agents experience if exposed
to large scale propaganda, can lead to a \textbf{breakdown of the
mental defense} against lies, as shown in Fig.\ \ref{fig:Propaganda-simulation}.
Working countermeasures that agents can take are honest and trustful
exchanges with other propaganda victims and being smart in detecting
lies. Both measures make agents more resilient against propaganda,
as discussed in Sect.\ \ref{subsec:Propaganda-and-resilience}.

We also observed some form of \textbf{Cassandra syndrome} within the
simulations, in which the most honest agents experience the largest
chance to get the lowest reputation and are unlikely to be believed
anymore. The opinions expressed of an honest agent are bound to this
agent's beliefs and therefore do not follow as much an evolving group
opinion as the opinions expressed by a dishonest agent, who targets
other beliefs when lying. As a consequence, the expressed opinions
of an honest agent might detach from the group position, which then
lets the others perceive this agent as dishonest. These will then
discard the opinions expressed by the most honest agent. Such a Cassandra
syndrome situation can occur among ordinary agents, but becomes substantially
more frequent when a dominant agent is present and manages to dominate
the group. Interestingly, the Cassandra syndrome effect weakens with
increasing levels of social chaos, probably due to the general loss
of the other agents' ability to discriminate between honest and dishonest
messages.

Finally, we see a strong positive correlation of the reputations of
the least honest agents. The mechanism generating this are the more
easily maintained mutual friendships of dishonest agents, the general
liar's benefit from confusion, and the resulting inflation of the
lie detection surprise scale in the presence of more other dishonest
agents. This can lead to a \textbf{deception symbiosis}, in which
the confusion created by a pathological liar makes it easier for other
liars to plant their lies as well. This not only seems to hold for
our agents. The negative impact of confusing, extreme messages on
the ability of humans to discriminate correct and false statements
is a documented psychological effect \cite{10.3389/fpsyg.2013.00453}.

\subsection{Robustness and assumptions}

The dynamics of our simulations are highly chaotic, which raises the
question how robust the results are in particular w.r.t.\ the model
assumptions. In this initial study, we are unable to answer this question
fully and have to leave this open for future research.

However, a number of parameter studies were performed in order to
calibrate the model parameters such that a meaningful dynamics appeared.
For example, a scan of different values of the caution parameter used
in lie construction $f_{\text{caution}}$ revealed that having smaller
values helps deceptive agents to build up a higher reputation. However,
for the sake of being brief, we just picked the from this perspective
sub-optimal value of $f_{\text{caution}}=0.3$ and did not present
results for other values. Another robustness check performed was changing
the number of agents from three to five. The friendship and reputation
distributions observed for the different strategies with three agents
could be observed there as well, however, they were less pronounced.

For these reasons, the numerical results of our simulations should
not be regarded as proofs of certain relations, but rather as possible
scenarios.

\subsection{Future directions}

Our reputation game simulation, as introduced here, is intended as
a starting point for further developments and investigations. Probably
most of its ingredients need to be revised and extended. Here, we
want to discuss a few possible future directions.

Currently, the beliefs of agents about others' honesties and their
representation of other agents' own beliefs have disparate dynamics.
In principle, this could be unified by agents just emulating other
agents in their minds by using the same computational infrastructure
for this, which they use for their own thinking. With such an architecture
for the Theory of Mind, not only the description might become more
natural, it might also be possible to simulate phenomena like hallucination
as cross-talk between an emulated and the own personality of an agent.

The characters of agents are currently static, programmed strategies.
Agents could be enabled to discover and learn strategies on their
own, from trial and error, or by watching the actions of other agents.
The level of randomness of their actions could also become an adjustable
parameter. It would be interesting to see under which conditions for
example the malicious strategies introduced here would develop on
their own in an evolutionary scenario.

The language of agents can be enriched. More topics could be introduced,
as aspects of an outer reality, or additional properties of agents.
Also enabling agents to quote each other would be very interesting.

The mental representation agents used to memorize the learned can
be made more complex. Real humans are, to some degree, able to remember
an entanglement of statements. They can even remove information partly
if it turns out that its source was deceptive. Agents could be provided
with similar abilities.

Furthermore, the parameters of the used cognitive model might be calibrated
against real world data. Finally, the sizes of the simulated social
networks need to be increased significantly to mimic real social networks
or even social media interactions. For simulation of the latter, the
effects of attention steering AI systems should be included, in order
to emulate their impact on society.

\section{Conclusions\label{sec:Conclusions}}

To conclude, we have introduced a reputation game as a socio-psychological
simulation that is built on the premise that agents should process
information according to simplified information theoretical principles.
We showed that a large number of known sociological and psychological
effects naturally seem to emerge from this premise.

With sufficient care, a number of conclusions might be drawn from
our agent based model that can be of interest to different communities.
Most of these insights might not be new, and well known in the corresponding
fields, however, we believe that there is a value to having them confirmed
by a reputation game simulation.

For a \textbf{social scientist} our reputation game simulation might
indicate the minimal set of rules and parameters that are necessary
to reproduce known socio-psychological effects. The simulation shows
that despite being highly chaotic, the outcome of social dynamics
might depend in a stochastic, but statistically robust way on a small
number of key parameters. For example, the simulations show that a
single maliciously deceptive individuals can drastically change the
character of interactions in a small social group. For the \textbf{cognition
researcher}, the level to which the necessary information compression
of cognitive systems makes them more prone to manipulations might
be an interesting aspect of our model. A \textbf{psychologists} might
be interested in the regime of large cognitive dissonance that agents
using a dominant strategy often experience when they already managed
to build up a high reputation, but still have a low-self esteem. It
manifests in a very toxic behavior that only stops when self-image
and external image start to coincide, either due to the ``narcissistic
supply'' of the other agents having become strong enough to boost
the self-perception of the dominant agent, or when it becomes absent.
For social media \textbf{policy makers}, the simulation might illustrate
how toxic social atmospheres develop when the participants' belief
systems are challenged too much. Finally, we not only show how indoctrination
via propaganda might work on the individual mind, but also how one
can resist it. Not surprisingly, honest exchange with other critical
minds seems to be effective. We believe that this should be of interest
to basically \textbf{everyone}.
\begin{acknowledgement*}
We acknowledge detailed feedback on the manuscript by Prof.\ Dr.\ Matthias
Bartelmann on mathematical and sociological aspects, Vincent Eberle
on mathematical and graphical aspects, Philipp Frank on information
theoretical aspects, Dr.\ Julia Stadler on logical aspects, and Prof.\ Dr.\ Sonja
Utz on sociological and psychological aspects, by two constructive
referees that helped to streamline the article and to connect it better
to the existing sociological literature, as well as help by Jakob
Roth and Matteo Guardiani in plotting densities.
\end{acknowledgement*}
\printbibliography

\appendix

\section{Information representation}

\label{sec:Information-representation}

We first introduce probabilistic reasoning, before discussing the
agent's belief representation and updating in Sect.\ \ref{subsec:Belief-representation}
and\ \ref{subsec:Believe-update}, respectively. The optimal data
compression is introduced in Sect.\ \ref{subsec:Optimal-believe-approximation}.
\begin{table*}
\begin{tabular}{|c|c|c|c|}
\hline 
variable or symbol & ref. & range & meaning\tabularnewline
\hline 
\hline 
$P(A|B),$$\mathcal{P}(x|y)$ & \ref{subsec:Probabilistic-reasoning} & $[0,1]$,$\mathbb{R}_{0}^{+}$ & probability of $A$ given $B$, PDF of $x$ given $y$\tabularnewline
\hline 
$n$ & \ref{subsec:Basic-elements} & $\mathbb{N}$ & number of agents\tabularnewline
\hline 
$\mathcal{A}$ & \ref{subsec:Basic-elements} & $\{\text{red},\,\text{cyan},\,\text{...}\}$ & set of $n$ named agents\tabularnewline
\hline 
$a$, $b$, $c$, $i$ & \ref{subsec:Basic-elements} & $\mathcal{A}$ & some agents\tabularnewline
\hline 
$a$, $b$, $c$ & \ref{sec:The-rules-of-the-game} & $\mathcal{A}$ & usually sender, receiver, and topic of a communication\tabularnewline
\hline 
$x_{i}$ & \ref{sec:The-rules-of-the-game} & $[0,1]$ & honesty of agent $i$\tabularnewline
\hline 
$\underline{x}=(x_{i})_{i\in\mathcal{A}}$ & \ref{sec:The-rules-of-the-game} & $[0,1]^{n}$ & indexed set of honesty of all agents\tabularnewline
\hline 
$\text{Beta}(x|\alpha,\beta)$ & \eqref{eq:beta-distribution} & $\mathbb{R}_{0}^{+}$ & beta distribution\tabularnewline
\hline 
$\mathcal{B}(\alpha,\beta)$, $\Gamma(\alpha)$ & \eqref{eq:beta-function} & $\mathbb{R}_{0}^{+}$ & beta, gamma function\tabularnewline
\hline 
$\psi(\alpha)=d\ln\Gamma(\alpha)/d\alpha$ & \eqref{eq:digamma} & $\mathbb{R}_{0}^{+}$ & digamma function\tabularnewline
\hline 
$I=(\mu,\lambda)$ & \ref{sec:The-rules-of-the-game} & $(-1,\infty]^{2}$ & stored belief about honesty of an agent\tabularnewline
\hline 
$I'$ & \ref{subsec:Information-handling} &  & some other belief, not necessarily in the format of $I$\tabularnewline
\hline 
$I''=(\mu'',\lambda'')$ & \ref{subsec:Optimal-believe-approximation} & $(-1,\infty]^{2}$ & encoding of $I'$ into storage format\tabularnewline
\hline 
$\mu$ & \ref{sec:The-rules-of-the-game} & $(-1,\infty]$ & number of honest statements counted for an agent\tabularnewline
\hline 
$\lambda$ & \ref{sec:The-rules-of-the-game} & $(-1,\infty]$ & number of lies counted for an agent\tabularnewline
\hline 
$I_{0}$ & \ref{subsec:Belief-representation} & $(-1,\infty]^{2}$ & prior information, here $I_{0}=(0,0)$\tabularnewline
\hline 
$I_{ab}$ & \ref{subsec:Basic-elements} & $(-1,\infty]^{2}$ & belief of agent $a$ on honesty of agent $b$\tabularnewline
\hline 
$I_{a}=(I_{ai})_{i\in\mathcal{A}}$ & \ref{subsec:Basic-elements} & $(-1,\infty]^{2\times n}$ & beliefs of $a$ on honesty of all agents\tabularnewline
\hline 
$I_{abc}=(\mu{}_{abc},\lambda_{abc})$ & \eqref{eq:ToM-1} & $(-1,\infty]^{2}$ & $a$'s assumption about belief of $b$ about $c$\tabularnewline
\hline 
$\widetilde{I}_{abc}=(\widetilde{\mu}{}_{abc},\widetilde{\lambda}_{abc})$ & \eqref{eq:ToM-2} & $(-1,\infty]^{2}$ & $a$'s assumption about $b$'s intention for $c$\tabularnewline
\hline 
$\text{KL}_{x}(I',I'')$ & \eqref{eq:KL} & $\mathbb{R}_{0}^{+}$ & Kullback-Leibler divergence $\mathcal{D}_{\text{KL}}(\mathcal{P}(x|I')||\mathcal{P}(x|I''))$\tabularnewline
\hline 
$\overline{x}_{I}:=\langle x\rangle_{(x|I)}$ & \eqref{eq:x-mean} & $[0,1]$ & expected $x$ given information $I$\tabularnewline
\hline 
$\left(\sigma{}_{I}\right):=\langle(x-\overline{x}_{I})\rangle_{(x|I)}$ & \eqref{eq:x-dispersion} & $[0,\nicefrac{1}{\sqrt{2}})$ & uncertainty dispersion of $x$ given information $I$\tabularnewline
\hline 
$\overline{x}_{ab}:=\overline{x}_{I_{ab}}$ & \eqref{eq:x-mean} & $[0,1]$ & reputation of $b$ with $a$\tabularnewline
\hline 
$t$ & \ref{sec:The-rules-of-the-game} & $\mathbb{N}$ & time as measured in communication events\tabularnewline
\hline 
$a\overset{c}{\rightarrow}b$ & \ref{sec:The-rules-of-the-game} & $\mathcal{A}^{3}$ & communication of $a$ to $b$ about $c$\tabularnewline
\hline 
$J=J_{a\overset{c}{\rightarrow}b}(t)=(\mu_{J},\lambda_{J})$ & \ref{sec:The-rules-of-the-game} & $(-1,\infty]^{2}$ & message in communication $a\overset{c}{\rightarrow}b$ at time $t$\tabularnewline
\hline 
$\Delta J=J_{a\overset{c}{\rightarrow}b}-I_{abc}$ & \eqref{eq:DeltaJ} & $\mathbb{R}^{2}$ & apparent novel information in $J_{a\overset{c}{\rightarrow}b}$ on
$c$\tabularnewline
\hline 
$\text{h}=\text{honest}$ & \ref{subsec:Belief-representation} & $\{\text{true},\text{ false}\}$ & whether message was honest, meaning $J_{a\overset{c}{\rightarrow}b}=I_{ac}$\tabularnewline
\hline 
$\neg\text{h}=\text{lie}$ & \ref{subsec:Belief-representation} & $\{\text{true},\text{ false}\}$ & whether message was a lie, meaning $J_{a\overset{c}{\rightarrow}b}\neq I_{ac}$\tabularnewline
\hline 
$\text{state}$ & \ref{subsec:Basic-lie-detection} & $\{\text{h},\,\neg\text{h}\}$ & state of a message\tabularnewline
\hline 
$\text{b}=\text{blush}$ & \ref{subsec:Basic-lie-detection} & $\{\text{true},\text{ false}\}$ & whether speaker blushed because of lying\tabularnewline
\hline 
$o=o_{J}$ & \ref{subsec:Basic-lie-detection} & $\{\text{b},\neg\text{b}\}$ & blushing observation of comm. $J$, $\text{b}=\text{blush}$\tabularnewline
\hline 
$f_{\text{b}}$ & \eqref{eq:blush} & $0.1$ & frequency of blushing while lying\tabularnewline
\hline 
$d=(a\overset{c}{\rightarrow}b,J,o)$ & \eqref{eq:y_J} & $\mathcal{A}^{3}(-1,\infty]^{2}\{\text{b},\neg\text{b}\}$ & data: communication, message, blushing observation\tabularnewline
\hline 
$(\mu,\lambda)^{+}:=\begin{cases}
(\mu,\lambda) & \text{if }\mu,\lambda\ge0\\
I_{0} & \text{else}
\end{cases}$ & \eqref{eq:Jplus} & $(-1,\infty]^{2}$ & ensures convex PDFs, reduces confusing updates\tabularnewline
\hline 
$y_{J}$ & \eqref{eq:y_J} & $[0,1]$ & probability of received message being honest\tabularnewline
\hline 
$\text{KL}_{J}$ & \ref{subsec:Basic-lie-detection} & $\mathbb{R}_{0}^{+}$ & amount of new information in message $J$ if honest\tabularnewline
\hline 
$K_{b}$ & \ref{subsec:Basic-lie-detection} & $\left(\mathbb{R}_{0}^{+}\right)^{10}$ & last ten non-zero $\text{KL}{}_{J}$s encountered by agent $b$\tabularnewline
\hline 
$\kappa_{b}=\text{median}(K_{b})$ & \eqref{eq:kappa_b} & $\mathbb{R}_{0}^{+}$ & scale $b$ compares $\text{KL}_{J}$ against to judge honesty of $J$\tabularnewline
\hline 
$\mathcal{S}_{J}=\nicefrac{\text{KL}_{J}}{\kappa_{b}}$ & \eqref{eq:f_s} & $\mathbb{R}_{0}^{+}$ & relative surprise of message $J$ for agent $b$\tabularnewline
\hline 
$\mathcal{R}(d)$ & \eqref{eq:R(d)} & $\mathbb{R}_{0}^{+}$ & ratio of likelihoods for $J$ lie and for $J$ honest\tabularnewline
\hline 
$I^{\varoplus}=I+(1,0)$ & \ref{subsec:Believe-update} & $(-1,\infty]^{2}$ & belief $I$ on speaker, updated for being honest\tabularnewline
\hline 
$I^{\ominus}=I+(0,1)$ & \ref{subsec:Believe-update} & $(-1,\infty]^{2}$ & belief $I$ on speaker, updated for being dishonest\tabularnewline
\hline 
$\text{const}$, $\text{const}'$, ... & \eqref{eq:KL_decomposition} & $\mathbb{R}$ & irrelevant constants\tabularnewline
\hline 
\end{tabular}

\caption{Used variables and symbols, the Sec.\ or Eq.\ of their definition,
their ranges, and meanings.\label{tab:Used-variables-and}}
\end{table*}

\subsection{Probabilistic reasoning\label{subsec:Probabilistic-reasoning}}

Agents need to maintain a picture of their social environment, to
know who is honest and who is not. Since they do not have direct access
to the intrinsic honesty parameters of any other agent, nor even to
their own, they need to deduce these values from the information they
get. This information, however, is incomplete, noisy, and often biased,
with a noise level that depends on the evolving \textbf{social atmosphere}\footnote{Social atmosphere refers here to the ensemble of assumptions agents
base their decisions on and the statistical properties of the consequential
communications.}. Therefore, agents have to cope with significant amounts of uncertainty.

\begin{figure*}
\includegraphics[width=0.5\textwidth]{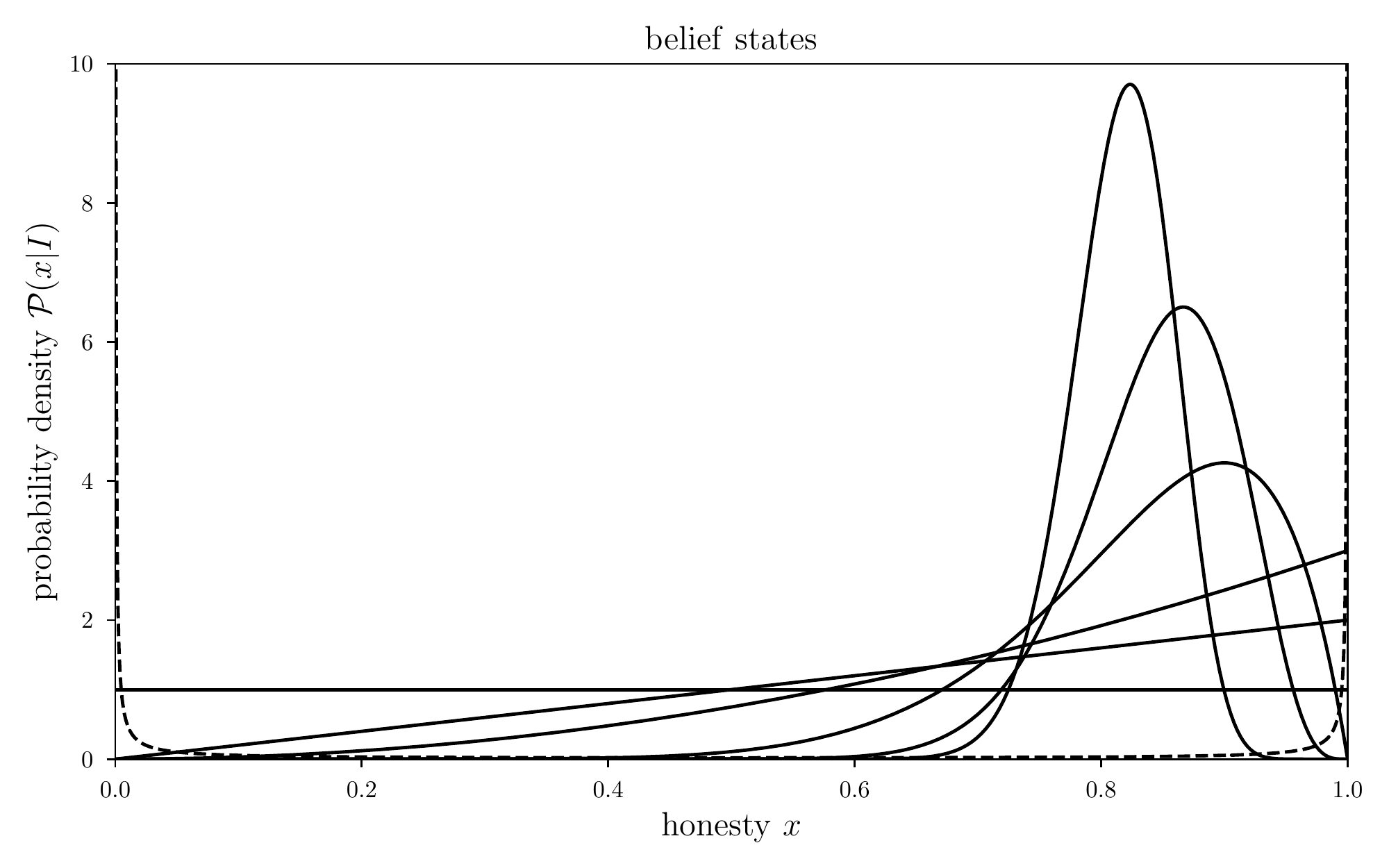}\includegraphics[width=0.5\textwidth]{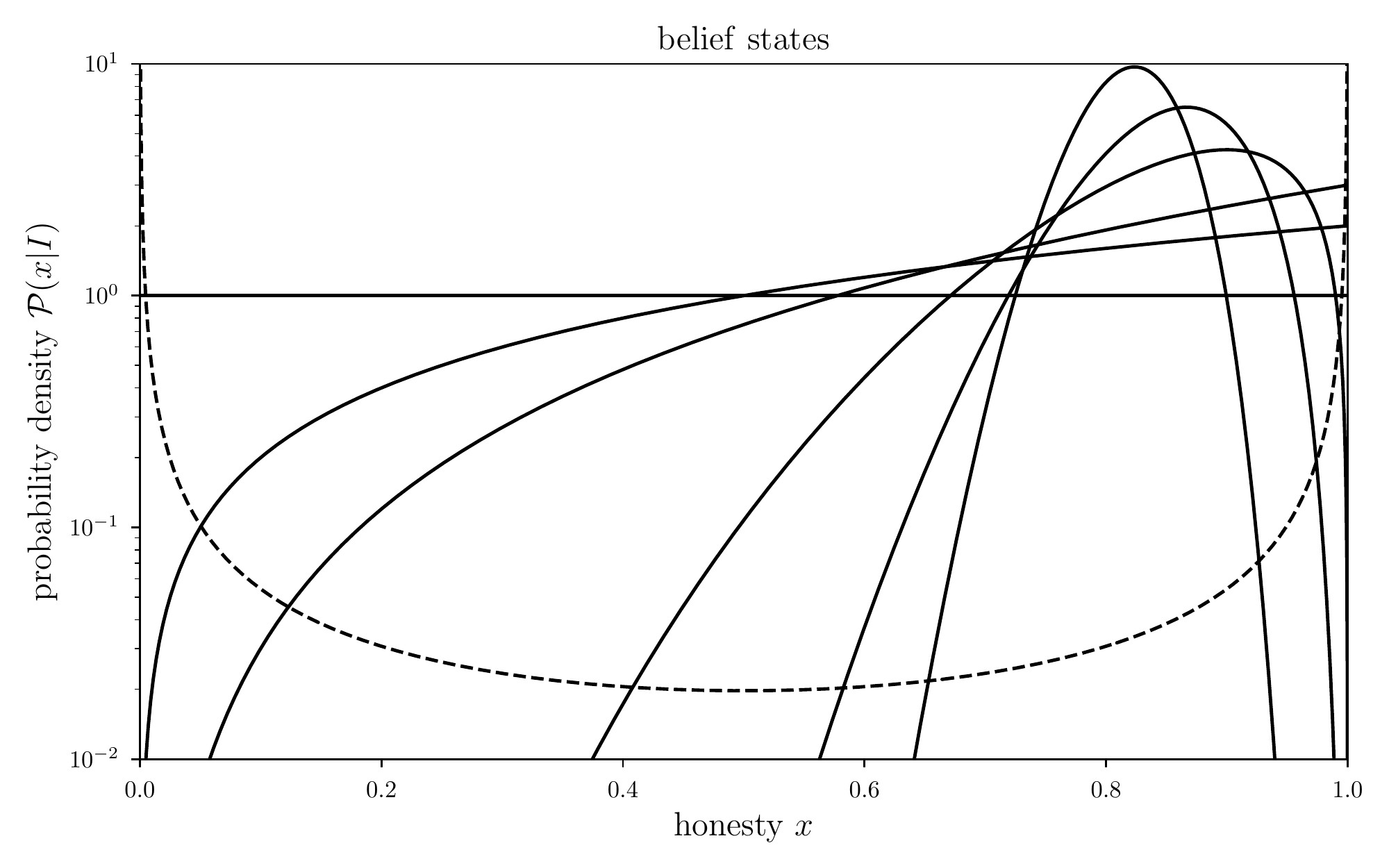}

\caption{PDFs corresponding to belief states about an agent's honesty $x$
as given by $I=(0,0)$, $(1,0)$, $(2,0)$, $(9,1)$, $(26,4)$, $(70,15)$
as marked by solid lines, and $(-0.99,-0.99)$ as marked by the dashed
line. The peaks of the PDFs increase in this order. The case shown
with the dashed line corresponds to believing the respective agent
to be either a very frequent liar or to be very honest, but not an
occasional liar. Left: Linear scale. Right: Logarithmic scale. \label{fig:PDFs-corrsponding-to-believes}}
\end{figure*}

Bayesian probabilities are ideal for logical reasoning under uncertainty
\cite{Cox1946,Cox1963}. Thereby, probabilities are regarded as a
device that keeps book of the plausibility of different possibilities
given some information $I$. Assigning a probability value $P(E|I)\in[0,1]$
to a possibility or an event $E$ therefore is not necessarily expressing
how often $E$ happens on average, $i.e.$ its frequency, but expresses
the strength of the belief in $E$ being the case. If, however, an
event $E$ has a frequency $f$, then the event's probability equals
this frequency if the latter is known, $P(E|I,f)=f$ with $I=\text{\textquotedblleft}f\text{ is the frequency of }E\text{\textquotedblright}$.

Probabilities are subjective, in the sense that different probability
values are assigned by agents with different knowledge. They are objective,
in the sense that given the same knowledge state, any ideal mind should
assign the same probability values. We use this in the following by
only labeling the belief state $I_{a}$ of an agent $a$ on some quantity
$x$, but not explicitly the induced probability $P(x|I_{a})$ used
by this agent. Any other agent $b$ with an identical belief state
$I_{b}=I_{a}$ would assign exactly the same probability to $x$,
$P(x|I_{b})=P(x|I_{a})$.

If there is a number of imperfectly known continuous quantities $x_{1},\ldots\,x_{n}$
then the PDF $\mathcal{P}(\underline{x}|I)$, with $\underline{x}=(x_{1},\ldots\,x_{n})$,
expresses their joint probability density. The probability (density)
of individual quantities is obtained from this by marginalization
over the other parameters,
\begin{equation}
\mathcal{P}(x_{i}|I)=\int dx_{1}\ldots dx_{i-1}\,dx_{i+1}\ldots dx_{n}\,\mathcal{P}(\underline{x}|I).
\end{equation}
In case the quantities are independent, the joint probability factorizes
into marginal ones,
\begin{equation}
\mathcal{P}(\underline{x}|I,\text{independence})=\mathcal{P}(x_{1}|I)\,\cdots\,\mathcal{P}(x_{n}|I).
\end{equation}

Often, probabilities do not factorize, $\mathcal{P}(\underline{x}|I)\neq\mathcal{P}(\underline{x}|I,\text{independence})$.
This expresses the entanglement between quantities, like that certain
combination of two variables are particularly probable. Complicated
entanglements can arise in the setting of a reputation game, since
agents make statements about the trustworthiness of each other that
are only believed in case they appear trustworthy themselves.

Here, $x_{a}$ will denote the honesty of agent $a$, with $x_{a}=0$
stating that agent $a$ always lies and $x_{a}=1$ that $a$ is always
honest. These honesty values are denoted by the tuple $\underline{x}=(x_{i})_{i\in\mathcal{A}}$
with $\mathcal{A}$ the set of agents. Any agent $b$ will maintain
a belief state $I_{b}$ about these honesty values in form of the
PDF $\mathcal{P}(\underline{x}|I_{b})$. This is updated when new
data $d$ becomes available according to Bayes' theorem,
\begin{equation}
\mathcal{P}(\underline{x}|d,I_{b})=\frac{\mathcal{P}(d,\underline{x}|I_{b})}{\int d\underline{x}\,\mathcal{P}(d,\underline{x}|I_{b})}=\frac{\mathcal{P}(d|\underline{x},I_{b})\,\mathcal{P}(\underline{x}|I_{b})}{\mathcal{P}(d|I_{b})}.\label{eq:Bayes}
\end{equation}
Here, $\mathcal{P}(d|\underline{x},I_{b})$ is the likelihood, the
probability to have obtained the data $d$ given $\underline{x}$
and $I_{b}$. $\mathcal{P}(\underline{x}|d,I_{b})$ is the posterior,
the probability for $\underline{x}$ given $d$ and $I_{b}$. The
latter PDF is the knowledge about $\underline{x}$ updated by the
data.

\subsection{Belief representation\label{subsec:Belief-representation}}

Ideally, after receiving new data $d$, agent $b$ would update the
knowledge by just memorizing it, i.e. $I_{b}\rightarrow I_{b}'=(d,I_{b})$,
and use all recorded statements and Bayes' theorem to construct their
current beliefs. However, this would be computational expensive, as
then all reasoning has to be repeated over and over again whenever
new information arrives or an action has to be chosen. Therefore,
our agents will follow the design of many cognitive systems, which
only store and update some compressed information. This will be the
tuple $I_{b}=(I_{bi})_{i\in\mathcal{A}}$ consisting of $n$ parameter
tuples $I_{bi}$ that describe agent $b$'s honesty impression of
agent $i$, as well as some auxiliary information $A_{b}$. As we
will not use probabilistic updates for the auxiliary information to
limit the complexity of the simulation, we will omit $A_{b}$ in our
equations in the following. Thus, we write $\mathcal{P}(d|\underline{x},I_{b})$
instead of the more accurate $\mathcal{P}(d|\underline{x},I_{b},A_{b})$.

We will assume that agents do not store information on parameter entanglements,
but simply keep track of the individual marginal probabilities about
the honesty of each other agent and themselves. The knowledge of agent
$b$ about the honesty of all agents is then given by the direct product
of individual marginal probabilities,
\begin{equation}
\mathcal{P}(\underline{x}|I_{b})=\prod_{i\in\mathcal{A}}\mathcal{P}(x_{i}|I_{bi}).\label{eq:product-assumption}
\end{equation}

The functional form of the belief about the honesty of a single agent
$c$, $\mathcal{P}(x_{c}|I_{ac})$, should be derived here from the
case where agent $a$ makes unambiguous observations, namely the self-observation
of their own actions. To investigate this, let us first concentrate
on the case agent $a$ communicates honestly, the message is in the
state $\text{\textquotedblleft honest\textquotedblright}=\text{h}$.
This happens with the frequency $x_{a}$. The update of the self-belief
state $I_{aa}$ of agent $a$ should then be according to Eq.\ \ref{eq:Bayes}
\begin{eqnarray}
\mathcal{P}(x_{a}|I_{aa},\text{ h}) & = & \frac{P(\text{h}|x_{a},I_{aa})\,\mathcal{P}(x_{a}|I_{aa})}{P(\text{h}|I_{aa})}\\
 & \propto & x_{a}\,\mathcal{P}(x_{a}|I_{aa}),
\end{eqnarray}
since $P(\text{h}|x_{a},I_{aa})=x_{a}$. Thus, whenever agent $a$
communicates honestly, the probability expressing the self-perception
should be multiplied with $x_{a}$ and then normalized.

Now, let us investigate the case of agent $a$ lying, the message
state is $\text{\textquotedblleft lie\textquotedblright}=\text{\ensuremath{\neg\text{h}}}$,
which happens with frequency $1-x_{a}$. Then we have
\begin{eqnarray}
\mathcal{P}(x_{a}|I_{aa},\neg\text{h}) & = & \frac{P(\neg\text{h}|x_{a},I_{aa})\,\mathcal{P}(x_{a}|I_{aa})}{\mathcal{P}(\neg\text{h}|I_{aa})}\\
 & \propto & (1-x_{a})\,\mathcal{P}(x_{a}|I_{aa}),
\end{eqnarray}
since $P(\neg\text{h}|x_{a},I_{aa})=1-x_{a}$. Thus, whenever lying,
the self-perception probability should be multiplied with $(1-x_{a})$.

It is therefore reasonable to represent the self-perception via numbers
of honest and dishonest statements, $\mu_{aa}$ and $\lambda_{aa}$,
respectively. The corresponding probability is then 
\begin{equation}
\mathcal{P}(x_{a}|I_{aa})=\frac{(\mu_{aa}+\lambda_{aa}+1)!}{\mu_{aa}!\,\lambda_{aa}!}\,x_{a}^{\mu_{aa}}(1-x_{a})^{\lambda_{aa}},
\end{equation}
with $I_{aa}=(\mu_{aa},\lambda_{aa})$. Here, it is assumed that the
prior distribution in absence of further information is flat, $\mathcal{P}(x_{a}|I_{0})=1$
with $I_{0}=(0,0)$.

We adopt this functional form for the honesty information representation
for all agents. We drop agent indices for a moment and the requirement
of integer parameters $\mu$ and $\lambda$ by allowing $\mu,\lambda\in(-1,10^{6}]$
in the following, where the lower limit ensures proper (integrable)
PDFs and the upper limit numerical stability. With this, the corresponding
probability generalizes to
\begin{equation}
\mathcal{P}(x|I):=\frac{x^{\mu}(1-x)^{\lambda}}{\mathcal{B}(\mu+1,\lambda+1)}=\text{Beta}(x|\mu+1,\lambda+1),\label{eq:single-agent-belief}
\end{equation}
with $I=(\mu,\lambda)$,
\begin{equation}
\mathcal{B}(\alpha,\beta):=\int_{0}^{1}dx\,x^{\alpha-1}(1-x)^{\beta-1}=\frac{\Gamma(\alpha)\Gamma(\beta)}{\Gamma(\alpha+\beta)}\label{eq:beta-function}
\end{equation}
 being the beta function, and 
\begin{align}
\text{Beta}(x|\alpha,\beta) & :=\frac{x^{\alpha-1}(1-x)^{\beta-1}}{\mathcal{B}(\alpha,\beta)}\label{eq:beta-distribution}
\end{align}
the beta distribution. This provides a bit more flexibility compared
to the case of $\alpha,\beta\in\mathbb{N}$ to express small information
gains, which is needed in case the obtained data contains ambiguous
information. Such probabilities $\mathcal{P}(x|I)$ for a number of
belief states for an agent's honesty are shown in Fig.\ \ref{fig:PDFs-corrsponding-to-believes}.

We note that $\mathcal{P}(x|I)$ defined this way has a mean and variance
of
\begin{eqnarray}
\overline{x}_{I} & := & \langle x\rangle_{(x|I)}=\int_{0}^{1}\!\!\!\!dx\,x\,\mathcal{P}(x|I)=\frac{\mu+1}{\mu+\lambda+2},\label{eq:x-mean}\\
\sigma_{I}^{2} & := & \langle(x-\overline{x}_{I})\rangle_{(x|I)}=\frac{\overline{x}_{I}(1-\overline{x}_{I})}{\mu+\lambda+3},\label{eq:x-dispersion}
\end{eqnarray}
and denote with $\overline{x}_{ab}:=\overline{x}_{I_{ab}}$ the reputation
$b$ has in the eyes of $a$.

\subsection{Belief update\label{subsec:Believe-update}}

When receiving a statement $J=J_{a\overset{c}{\rightarrow}b}(t)=(\mu_{J},\lambda_{J})_{a\overset{c}{\rightarrow}b}(t)$
from agent\textbf{ $a$} at time $t$, agent $b$ assesses the reliability
of the statement by assigning the probability 
\begin{equation}
y_{J}=P(\text{h}|d,I_{b})
\end{equation}
to the possibility that $a$ communicated honestly depending on $b$'s
receiver strategy as will be discussed in App.\ \ref{sec:detailed-receiver-strategies}.
This assignment is based on the prior belief $I_{ba}$ on $a$'s honesty
(and some auxiliary information $A_{b}$) and the data $d=d(t)=(a\overset{c}{\rightarrow}b,J,o)(t)$,
which consist of the message $J$ and the observation $o$ whether
agent $a$ blushed or not (accidentally revealed a lie or not).

\subsubsection{Untrustworthy message}

\begin{figure*}[t]
\includegraphics[width=0.5\textwidth]{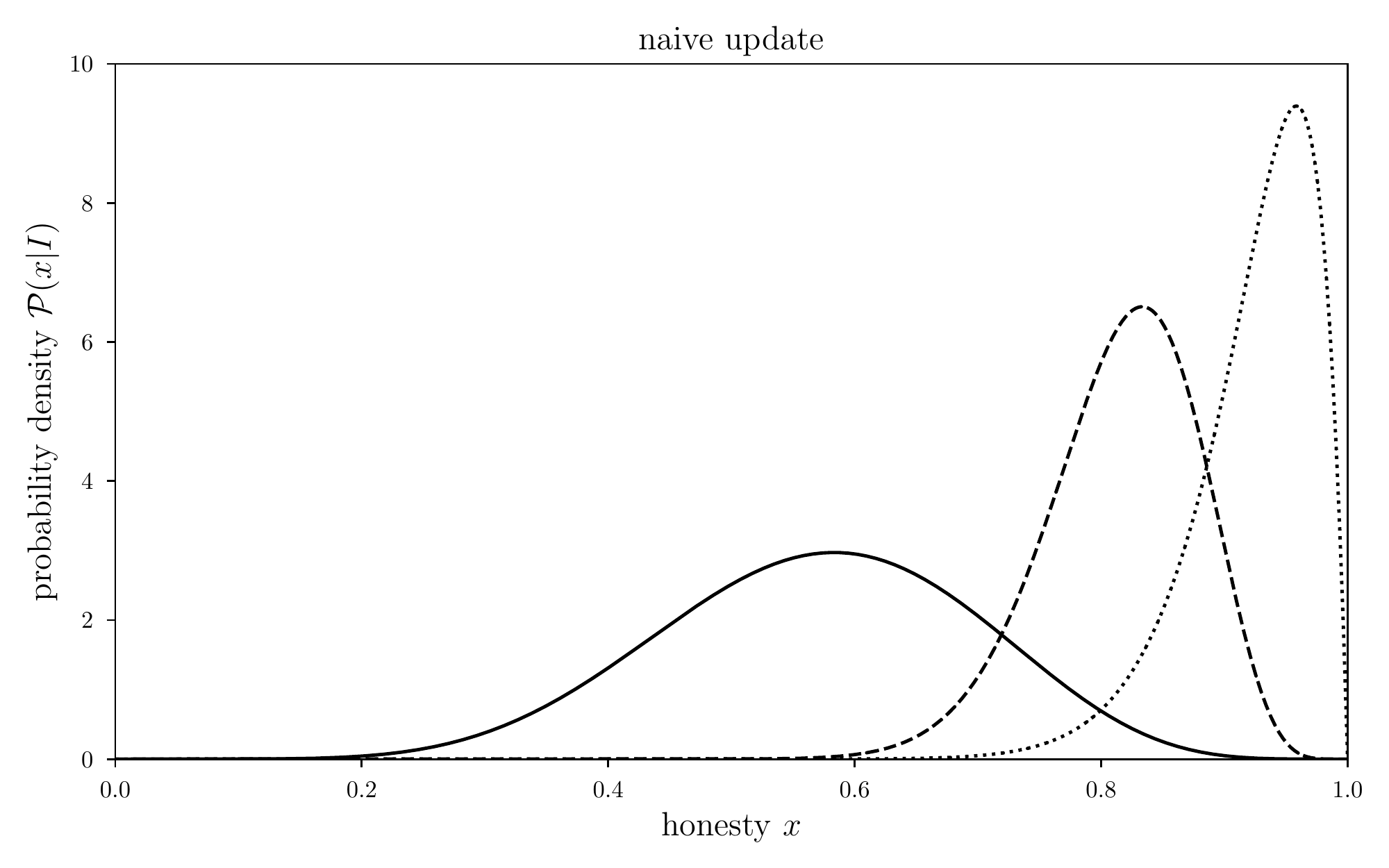}\includegraphics[width=0.5\textwidth]{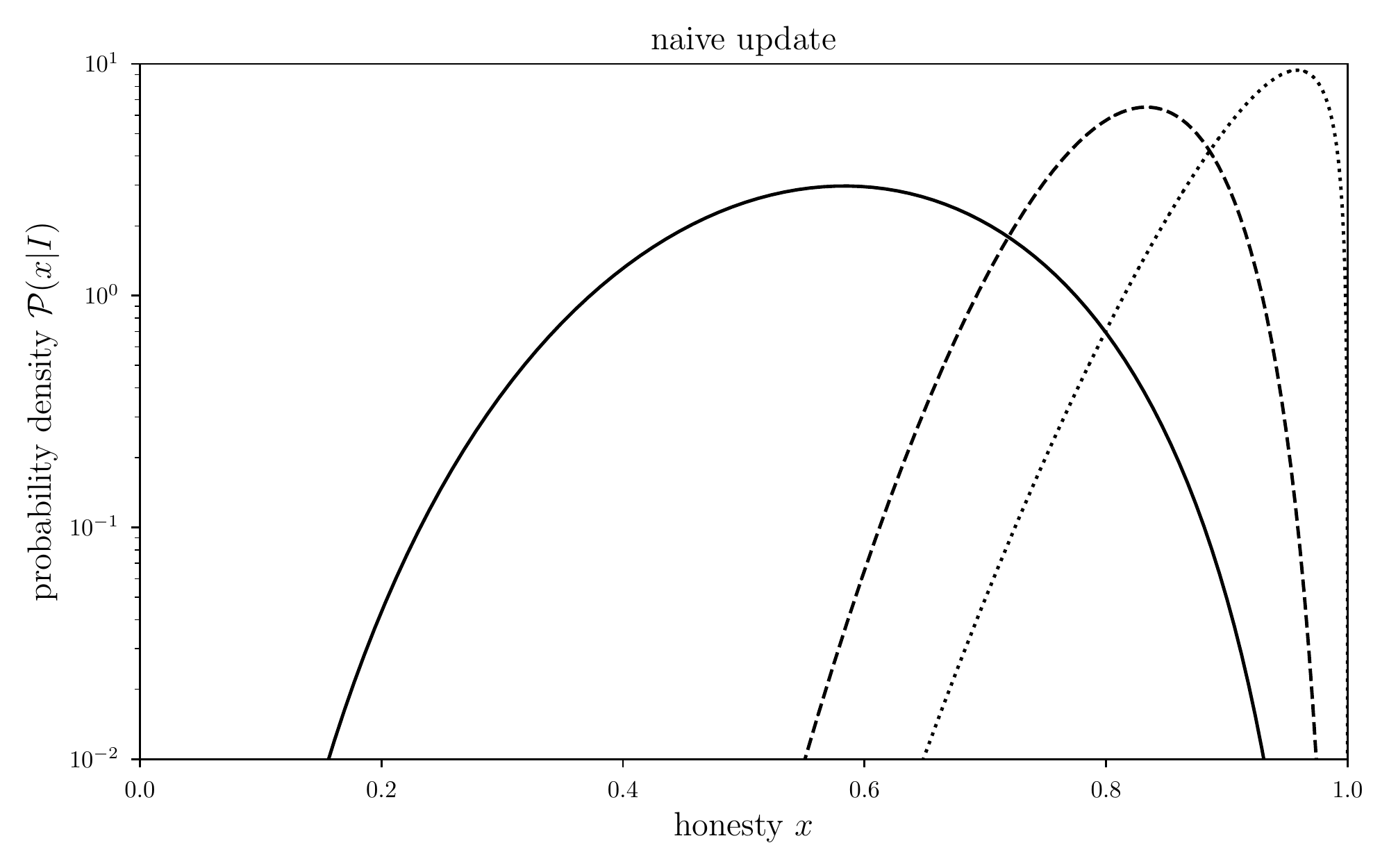}

\caption{Trustful belief update of agent $b$ hearing for the first time from
$a$ about $c$, thus assuming $I_{bac}=I_{0}=(0,0)$. The initial
belief state $I_{bc}=(7,5)$ (solid line) changes after receiving
the communication $J_{a\protect\overset{c}{\rightarrow}b}=(23,1)$
(dotted line) to $I'_{bc}=I_{bc}+J_{a\protect\overset{c}{\rightarrow}b}-I_{bac}=(30,6)$
(dashed line). The information gain of the update is $2.4\text{ nits}=3.4\text{ bits}$.
Left: Linear scale. Right: Logarithmic scale.\label{fig:Believe-update-of}}
\end{figure*}
If a communication $J$ appears completely untrustworthy to the receiver
$b$ they will set $y_{J}=0$ and ignore the statement made about
the conversation topic. However, $b$'s opinion about the speaker
$a$ will be updated. Agent $b$'s posterior about $a$ will change
according to
\begin{eqnarray}
\mathcal{P}(x_{a}|\neg\text{h},I_{ba}) & = & \frac{P(\neg\text{h}|x_{a},I_{ba})\,\mathcal{P}(x_{a}|I_{ba})}{P(\neg\text{h}|I_{ba})}\\
 & \propto & x_{a}^{\mu_{ba}}\,(1-x_{a})^{1+\lambda_{ba}},
\end{eqnarray}
since $P(\neg\text{h}|x_{a},I_{ba})=1-x_{a}$ irrespective of $I_{ba}$.
This new belief state is represented by increasing $b$'s lie-counter
$\lambda_{ba}$ for $a$ by one, 
\begin{equation}
I_{ba}(t)\rightarrow I_{ba}(t+1)=(\mu_{ba}(t),\lambda_{ba}(t)+1)=I_{ba}^{\varominus}(t).\label{eq:lie}
\end{equation}
All other beliefs of agent $b$ stay unchanged, $I_{bi}(t+1)=I_{bi}(t)$
for all $i\in\mathcal{A}\backslash\{a\}$. For later usage we introduced
the notation $I^{\varominus}:=I+(0,1)$ for a belief state $I$ updated
by one lie. Similarly, $I^{\varoplus}:=I+(1,0)$ should denote a belief
state $I$ updated by one observed honest statement.

\subsubsection{Trustful update}

If, however, agent $b$ is convinced that the statement received from
$a$ is honest, then agent $b$ assigns $y_{J}=1$. Let's assume that
$a$ does not make a statement about themselves, $a\neq c$. If agent
$b$ believes the statement is honest ($b$ thinks ``$J_{a\overset{c}{\rightarrow}b}(t)=I_{ac}(t)$'')
then $b$ only needs to identify the new information in it.\footnote{To limit the complexity of the simulation, our agents ignore the possibility
that other agents might be misinformed.} For spotting the news in an expressed opinion, the receiver needs
to know the opinion of the speaker at the time of their last conversation.
Agents maintain guesses on each other's previous beliefs for this
purpose. The guess of agent $b$ at time $t$ what $a$ believed about
$c$ at their last conversation is denoted as $I_{bac}(t)=I_{bac}=(\mu_{bac},\lambda_{bac})$.
How this is maintained is explained later in Sect.\ \ref{subsec:Auxiliary-parameters-update}.
The new information in an honest statement of $a$ on $c$ is then
\begin{equation}
\Delta J=\Delta J(t)=(\Delta\mu,\Delta\lambda)(t)=J_{a\overset{c}{\rightarrow}b}(t)-I_{bac}(t).\label{eq:DeltaJ}
\end{equation}
If all accounting was correct, $\Delta\mu,\Delta\lambda\ge0$ should
be the case. If this is not the case, something went wrong and agent
$b$ better assumes not to have received any new information, as expressed
in $\Delta J\rightarrow I_{0}=(0,0)$. We denote this by
\begin{equation}
\Delta J^{+}=(\Delta\mu^{+},\Delta\lambda^{+}):=\begin{cases}
(\Delta\mu,\Delta\lambda) & \text{if \ensuremath{\Delta}\ensuremath{\mu},\ensuremath{\Delta}\ensuremath{\lambda\ge}0}\\
I_{0} & \text{else}.
\end{cases}\label{eq:Jplus}
\end{equation}
Agent $b$ therefore realizes that agent $a$ is reporting agent $c$
to have made $\Delta\mu^{+}$ new honest and $\Delta\lambda^{+}$
new dishonest statements since they spoke last about $c$. The belief
update on $c$ should then be\footnote{In case $\Delta\mu^{+},\Delta\lambda^{+}$ are not integer numbers,
a slightly more sophisticated argumentation leads to exactly the same
result: Agent $b$ assumes that $a$'s past belief, which has lead
to $\Delta J^{+}$, was based on some effective, accumulated data
$d'$ and happened according to Bayes' rule. This implies $\mathcal{P}(x_{c}|d',I_{bac})=\mathcal{P}(d'|x_{c},I_{bac})\mathcal{P}(x_{c}|I_{bac})/\mathcal{P}(d'|I_{bac})$
and means that the data $d'$ must have been such that $\mathcal{P}(d'|x_{c},I_{bac})\propto x_{c}^{\Delta\mu^{+}}(1-x_{c})^{\Delta\lambda^{+}}$.
This should therefore also be the likelihood $\mathcal{P}(\Delta J^{+}|x_{c},I_{bc})$
that agent $b$ uses for the update. This leads to Eq.\ \ref{eq:trustworthy-update}
as well.}
\begin{eqnarray}
\mathcal{P}(x_{c}|\Delta J^{+},\text{h},I_{b}) & \!\!\!=\!\!\! & \frac{\mathcal{P}(\Delta J^{+}|x_{c},I_{bc})\,\mathcal{P}(x_{c}|I_{bc})}{\mathcal{P}(\Delta J^{+}|I_{bc})}\\
 & \!\!\!\propto\!\!\! & x_{c}^{\mu_{bc}+\Delta\mu^{+}}\,(1-x_{c})^{\lambda_{bc}+\Delta\lambda^{+}}\!.\ \ \ \label{eq:trustworthy-update}
\end{eqnarray}
This can be represented by agent $b$ just increasing the counts for
assumed honest and dishonest statements of $c$, i.e.
\begin{eqnarray}
I_{bc}(t) & \rightarrow & I_{bc}(t+1)=I_{bc}(t)+\Delta J^{+}(t).\label{eq:truth}
\end{eqnarray}
Such a trustful update is illustrated in Fig.\ \ref{fig:Believe-update-of}.

Since agent $b$ assumes that agent $a$ has said the truth, $b$
registers 
\begin{equation}
I_{ba}(t)\rightarrow I_{ba}(t+1)=(\mu_{ba}(t)+1,\lambda_{ba}(t))=I_{ba}^{\varoplus}(t).\label{eq:truth-speaker}
\end{equation}
All other beliefs of agent $b$ stay unchanged, $I_{bi}(t+1)=I_{bi}(t)$
for all $i\in\mathcal{A}\backslash\{a,c\}$.

Finally, we need to deal with the case that agent $a$ made a self-statement
that agent $b$ regards as absolutely honest. Then, the two above
update rules for $c$ and $a$ just need to be merged into a single
one for $a$, 
\begin{equation}
I_{ba}(t)\rightarrow I_{ba}(t+1)=I_{ba}^{\varoplus}(t)+\Delta J^{+}(t)\label{eq:self-thruth-speaker}
\end{equation}
and $I_{bi}(t+1)=I_{bi}(t)$ for all $i\in\mathcal{A}\backslash\{a\}$.

\subsubsection{Skeptical update}

The two cases of updates discussed above lead to joint belief states
on $a$ and $c$ for agent $b$ that again are represented by product
states without any entanglement,
\begin{equation}
\mathcal{P}(x_{a},x_{c}|d(t),I_{b}(t))=\mathcal{P}(x_{a}|I_{ba}(t+1))\,\mathcal{P}(x_{c}|I_{bc}(t+1)).
\end{equation}
When agent $b$ is unsure whether $a$ was honest or lied, the resulting
belief state should be a superposition of the state after an assumed
honest communication and a perceived lie. The former is given by Eq.\ \ref{eq:truth}
(Eq.\ \ref{eq:truth-speaker} for $a\neq c$ or Eq.\ \ref{eq:self-thruth-speaker}
for $a=c$) and the latter by Eq.\ \ref{eq:lie}. The superimposed
states should have weights according to their probabilities. Thus,
$y_{J}=P(\text{h}|d,I_{b})$ is the weight of the honest message state
and $1-y_{J}=P(\neg\text{h}|d,I_{b})$ the weight of the dishonest
message state.
\begin{figure*}[t]
\includegraphics[width=0.5\textwidth]{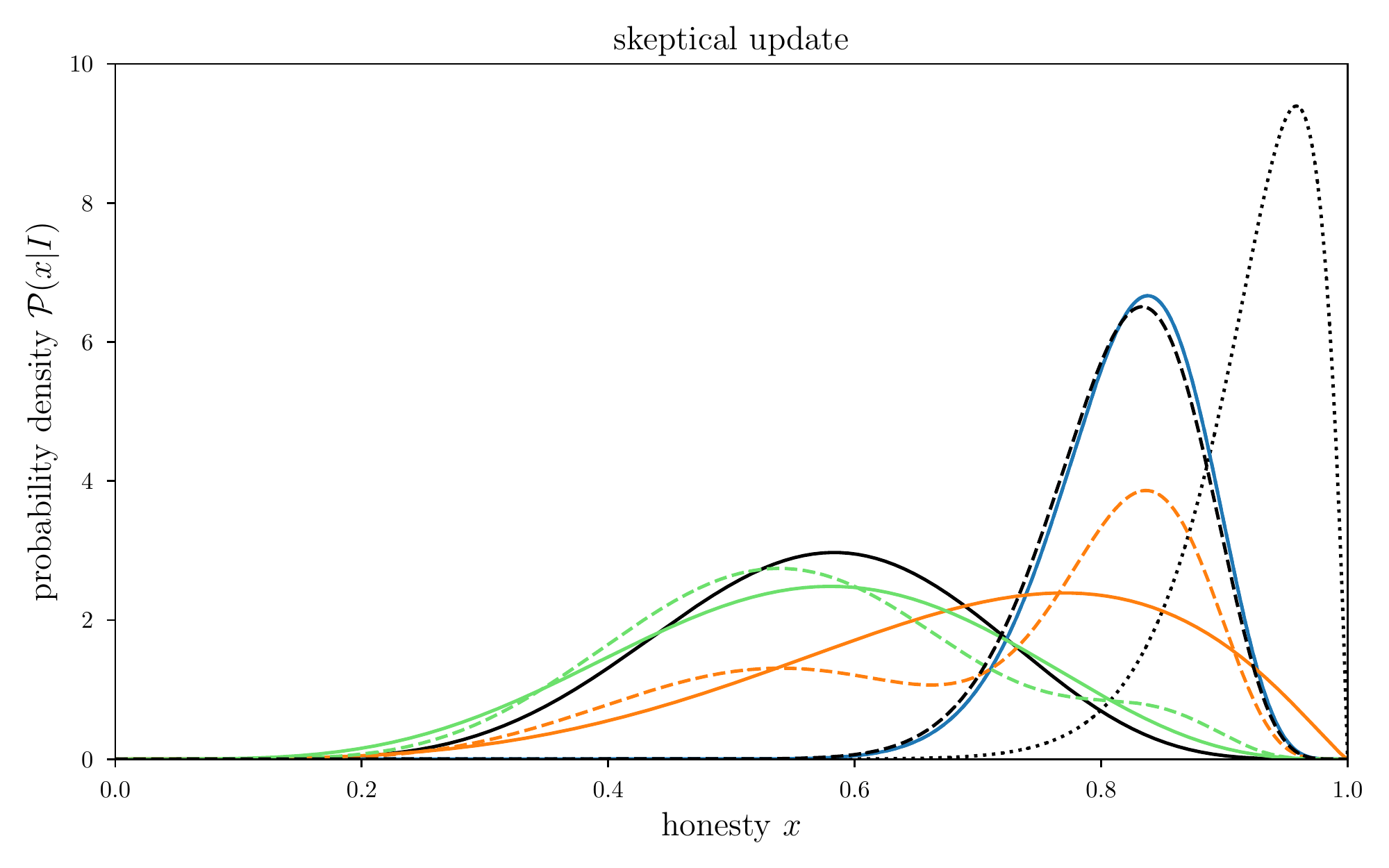}\includegraphics[width=0.5\textwidth]{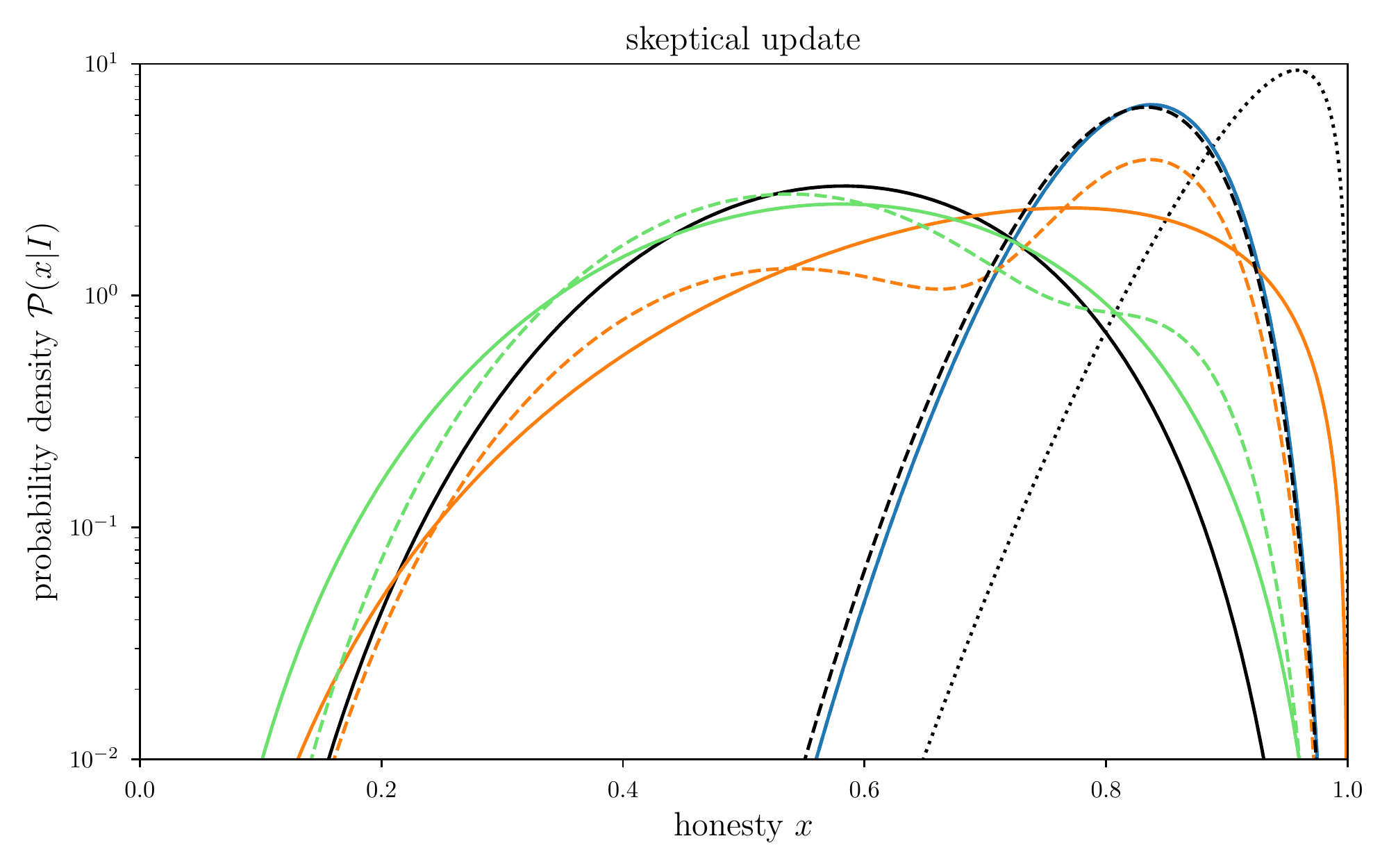}

\caption{Belief updates of an agent $b$ listening for the first time to the
self-appraisal of agent $a$. The initial belief state $I_{ba}=(7,5)$
(black solid line) changes after receiving the message $J_{a\protect\overset{a}{\rightarrow}b}=(23,1)$
(black dotted line) by an amount that depends on whether agent $b$
trusts the message fully ($y_{c}=1\Rightarrow I''=(31,6)$, blue),
with the sender's reputation ($y_{J}=\overline{x}_{ba}=0.57\Rightarrow I''=(4.0,1.2)$,
orange) or only a little ($y_{J}=0.1\Rightarrow I''=(4.6,3.3)$, green).
For these cases, the correct posteriors are shown with dashed lines
and the memorized PDFs as solid lines in the corresponding colors.
The perceived honesty of the message is included in the updates shown
in color, but not in the naive update (black dashed line). Left: Linear
scale. Right: Logarithmic scale.\label{fig:Believe-update-of-sceptical}}
\end{figure*}

Let us first assume that $a\neq c$. We then have
\begin{eqnarray}
\mathcal{P}(x_{a},x_{c}|d,I_{b}) & = & y_{J}\mathcal{P}(x_{a}|I_{ba}^{\varoplus})\,\mathcal{P}(x_{c}|\Delta J^{+},\text{h},I_{bc})\nonumber \\
 &  & +(1-y_{J})\,\mathcal{P}(x_{a}|I_{ba}^{\varominus})\,\mathcal{P}(x_{c}|I_{bc})\\
 & = & y_{J}\text{Beta}(x_{a}|I_{ba}^{\varoplus})\text{Beta}(x_{c}|I_{bc}+\Delta J^{+})\nonumber \\
 &  & +(1-y_{J})\,\text{Beta}(x_{a}|I_{ba}^{\varominus})\,\text{Beta}(x_{c}|I_{bc}).\nonumber \\
\label{eq:skeptical-update}
\end{eqnarray}
We note that this is not a direct product of marginal distributions
any more used in the agent's memories since $b$'s knowledge on the
honesty of $a$ and $c$ got entangled.

When $a$ speaks about themselves, we have $c=a$ and assign\footnote{Actually, since a self-statement is self-referential with respect
to its truth value, a logically fully consistent update would require
to solve an implicit self-consistent relation. This can be seen for
example in case agent $a$ makes the statement to be a notorious liar,
which if true is contradicted by just have made an honest confession.
We, as well as our agents, do not invest mental energy in such philosophical
calculations, but just use the pragmatic Eq.\ \ref{eq:dubious-self-statement}.} 
\begin{equation}
\mathcal{P}(x_{a}|d,I_{b})=y_{J}\text{Beta}(x_{a}|I_{ba}^{\varoplus}+\Delta J^{+})+(1-y_{J})\,\text{Beta}(x_{a}|I_{ba}^{\varominus}).\label{eq:dubious-self-statement}
\end{equation}
In general, this is also not in the format used by agent $b$ to memorize
beliefs, $\mathcal{P}(x_{a}|I_{ba}(t+1))\equiv\text{Beta}(x_{a}|I_{ba}(t+1))$,
which raises the need for a compression of the correct new belief
state into a memorizable, simpler form.

Since the cases of a certainly honest and a certainly dishonest message
are enclosed in Eqs.\ \ref{eq:skeptical-update} and \ref{eq:dubious-self-statement}
by setting $y_{J}=1$ and $y_{J}=0$, respectively, we only have to
consider skeptical updates in the following.

\subsection{Optimal belief approximation\label{subsec:Optimal-believe-approximation}}

Usually the honesty of a message is unclear to the receiver $b$.
In this case, the belief state $\mathcal{P}(\underline{x}|I')$ with
$I'=I'(t):=(d(t),I_{b}(t))$ as given by Eq.\ \ref{eq:skeptical-update}
is a superposition of the two belief states that would arise if the
message is known to be honest and to be dishonest. In order to cast
$\mathcal{P}(\underline{x}|I')$ into the functional form of $\mathcal{P}(\underline{x}|I)$
a new $I''=I_{b}(t+1)$ has to be found that captures as much as possible
the information of $I'$. The information loss in this approximation
of $I'$ by $I''$ is measured by the Kullback-Leibler (KL) divergence
\begin{eqnarray}
\text{KL}_{\underline{x}}(I',I'') & := & \mathcal{D}_{\text{KL}}(\mathcal{P}(\underline{x}|I')||\mathcal{P}(\underline{x}|I''))\label{eq:KL}\\
 & := & \int d\underline{x}\,\mathcal{P}(\underline{x}|I')\,\ln\left(\frac{\mathcal{P}(\underline{x}|I')}{\mathcal{P}(\underline{x}|I'')}\right)
\end{eqnarray}
in units of nits (=1.44 bits) \cite{2017Entrp..19..402L}. Thus, $\text{KL}_{\underline{x}}(I',I'')$
should be minimized with respect to $I''$, the parameters of the
approximate belief state.\footnote{Note that we regard the KL here to be a function of the information
sets $I'$ and $I''$ on the quantity $\underline{x}$, in contrast
to the standard convention to define $\mathcal{D}_{\text{KL}}$ as
a functional of the PDFs implied by those.} These then form the next information state $I_{b}(t+1)=I''$ of $b$.

Since the update concerns only the knowledge about agents $a$ and
$c$, the sender and topic of a message, only the beliefs about those
need updating. Side effects do not occur here as agents do not track
entanglements. Learning that $a$'s honesty is different from what
$b$ has previously assumed is not letting $b$ reevaluate $a$'s
past statements as $b$ neither memorizes those precisely, nor the
entanglements these imply.
\begin{figure*}[t]
\includegraphics[width=0.5\textwidth]{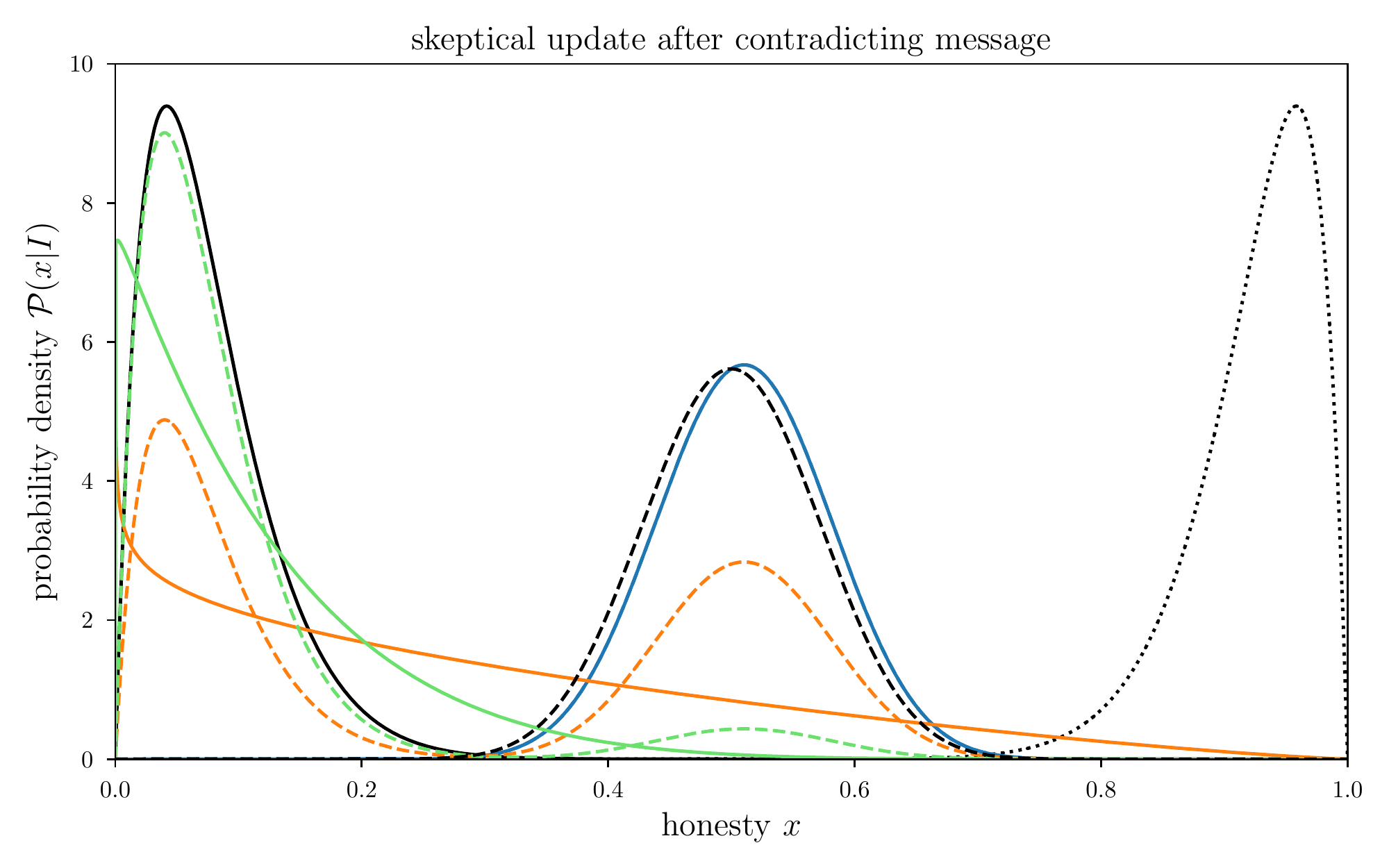}\includegraphics[width=0.5\textwidth]{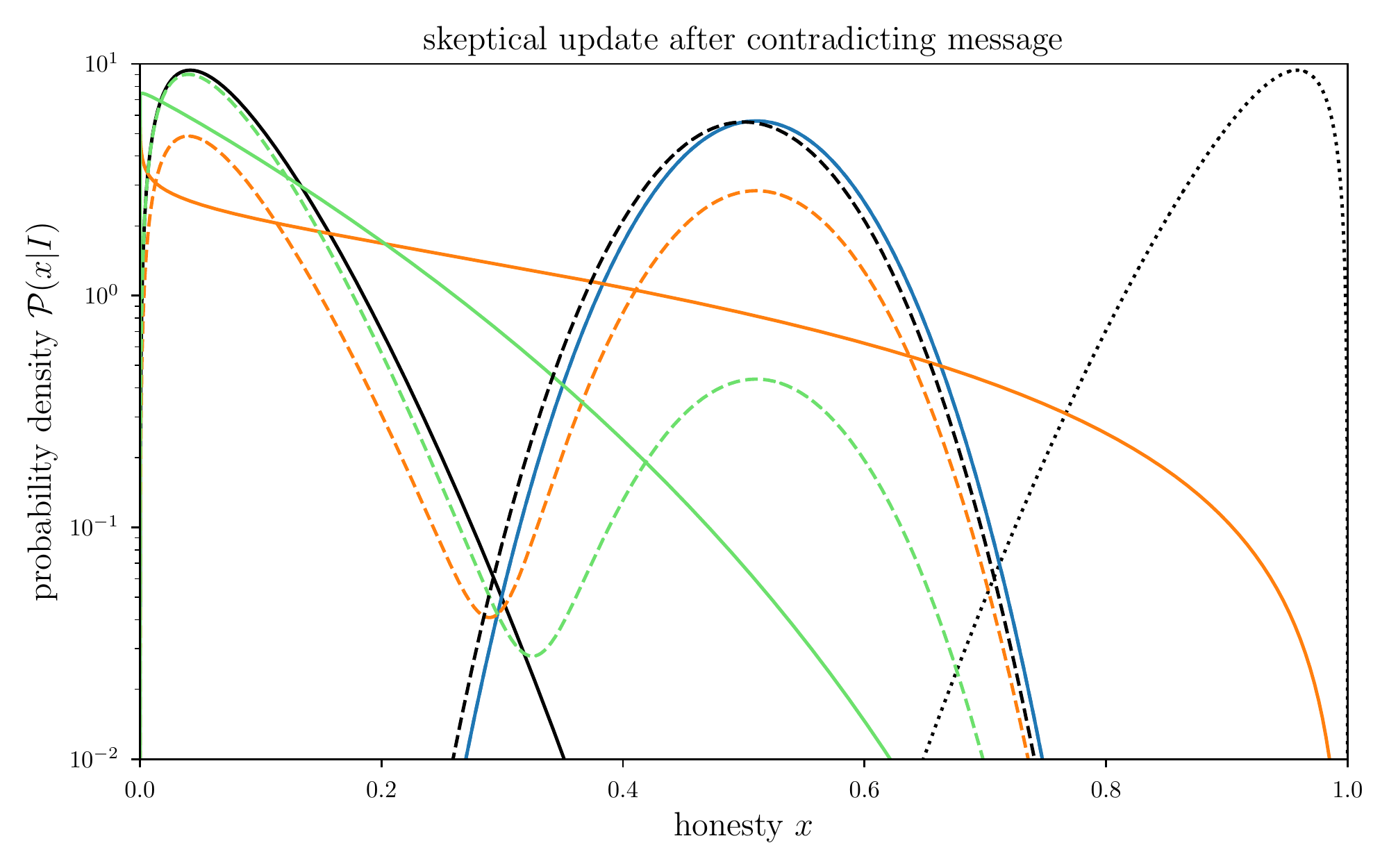}

\caption{Like Fig.\ \ref{fig:Believe-update-of-sceptical}, just with the
initial belief state of the receiver being $I_{ba}=(1,23)$ (solid
black line) and the trusts in the message being full ($y_{J}=1\Rightarrow I''_{ba}=(25,24)$,
blue line), undecided ($y_{J}=0.5\Rightarrow I''_{ba}=(0,1.5)$, orange
line), or poor ($y_{J}=\overline{x}_{ba}=0.08\Rightarrow I''_{ba}=(0,6.8)$,
green). As the initial belief and the message contradict each other
strongly, the last two updates (with $y_{J}<1$) can only coarsely
capture the bimodal posterior for the price of getting closer to the
uninformative state $I_{0}=(0,0)$. The displayed updates should only
be expected for naive (blue line) to uncritical (green line) agents,
as critical and smart agents would put far less trust in a so strongly
diverging opinion, as detailed in Sect.\ \ref{sec:detailed-receiver-strategies}.
\label{fig:Contradicting-update}}
\end{figure*}

Thus, the relevant KL for agent $b$'s belief update after receiving
information $d$ from $a$ about $c$ is $\text{KL}_{(x_{a},x_{c})}(I',I'')$
with $\mathcal{P}(x_{a},x_{c}|I')=\mathcal{P}(x_{a},x_{c}|d,I_{ba},I_{bc})$
being the accurate, potentially entangled belief state and 
\begin{eqnarray}
\mathcal{P}(x_{a},x_{c}|I'') & = & \mathcal{P}(x_{a}|I''_{ba})\,\mathcal{P}(x_{c}|I''_{bc})\\
 & = & \text{Beta}(x_{a}|I''_{ba})\,\mathcal{\text{Beta}}(x_{c}|I''_{bc})
\end{eqnarray}
being the simplified state over the relevant subspace of $x_{a}$
and $x_{c}$ that will be memorized. As the latter is a direct product
of one dimensional PDFs, it turns out that it is sufficient to perform
only two one dimensional updates based on the two marginals 
\begin{eqnarray}
\mathcal{P}(x_{a}|I') & = & \int_{0}^{1}dx_{c}\,\mathcal{P}(x_{a},x_{c}|I')\text{ and}\label{eq:update-speaker}\\
\mathcal{P}(x_{c}|I') & = & \int_{0}^{1}dx_{a}\,\mathcal{P}(x_{a},x_{c}|I').\label{eq:update-topic}
\end{eqnarray}

This is because the two dimensional $\text{KL}{}_{(x_{a},x_{c})}$
of the joint update on agents $a$ and $c$ separates into two one
dimensional KLs for the marginal distributions of $x_{a}$ and $x_{c}$,
\begin{eqnarray}
\text{KL}_{(x_{a},x_{c})}(I',I'') & \!\!\!\!=\!\!\!\! & \int\!\!dx_{a}\int\!\!dx_{c}\,\mathcal{P}(x_{a},x_{c}|I')\,\ln\frac{\mathcal{P}(x_{a},x_{c}|I')}{\mathcal{P}(x_{a},x_{c}|I'')}\nonumber \\
 & \!\!\!\!=\!\!\!\! & \int\!\!dx_{a}\int\!\!dx_{c}\,\mathcal{P}(x_{a},x_{c}|I')\,\times\nonumber \\
 &  & \left[-\ln\mathcal{P}(x_{a}|I''_{ba})-\ln\mathcal{P}(x_{c}|I''_{bc})\right]+\text{const}\nonumber \\
 & \!\!\!\!=\!\!\!\! & -\int\!\!dx_{a}\,\mathcal{P}(x_{a}|I')\,\ln\mathcal{P}(x_{a}|I''_{ba})\nonumber \\
 &  & -\int\!\!dx_{c}\,\mathcal{P}(x_{c}|I')\,\ln\mathcal{P}(x_{c}|I''_{bc})+\text{const}\nonumber \\
 & \!\!\!\!=\!\!\!\! & \text{KL}_{x_{a}}(I',I''_{ba})+\text{KL}_{x_{c}}(I',I''_{bc})+\text{const}',\nonumber \\
\label{eq:KL_decomposition}
\end{eqnarray}
and these can be minimized individually with respect to $I''_{ba}$
and $I''_{bc}$. Constant terms w.r.t.\ $I''$ are denoted $\text{const}$
and $\text{const}'$.

For calculating these single agent marginal KLs, $\text{KL}_{x_{a}}(I',I''_{ba})$
and $\text{KL}_{x_{c}}(I',I''_{bc})$, we need expressions for the
marginal updates on speaker and topic, $\mathcal{P}(x_{a}|I')$ and
$\mathcal{P}(x_{c}|I')$ as given by Eqs.\ \ref{eq:update-speaker}
and \ref{eq:update-topic}. The involved integrals can be calculated
analytically and the results for the different cases unify and generalize
to a single expression of the marginal update for any agent $i\in\mathcal{A}$,
\begin{equation}
\mathcal{P}(x_{i}|I')=y_{J}\text{Beta}(x_{i}|I_{bi}^{\text{h}})+(1-y_{J})\,\text{Beta}(x_{i}|I_{bi}^{\neg\text{h}}),\label{eq:generalized-marginal-update}
\end{equation}
with
\begin{eqnarray}
I_{bi}^{\text{h}} & := & I_{bi}+\left.(1,0)\right|_{i\text{ speaker}}+\left.\Delta J^{+}\right|_{i\text{ topic}},\\
I_{bi}^{\neg\text{h}} & := & I_{bi}+\left.(0,1)\right|_{i\text{ speaker}},\text{ and}\\
\left.I\right|_{\text{condition}} & := & \begin{cases}
I & \text{condition is true}\\
I_{0} & \text{condition is false}
\end{cases}
\end{eqnarray}
an information that only is taken into account in case the condition
is true. Eq.\ \ref{eq:generalized-marginal-update} is valid for
all agents $i\in\mathcal{A}$, including the topic $c$, the speaker
$a$, the receiver $b$, or anybody else. In case $i\notin\{a,c\}$,
Eq.\ \ref{eq:generalized-marginal-update} states that for agent
$i$ the initial belief is to be kept, $\mathcal{P}(x_{i}|I')=\text{Beta}(x_{i}|I_{bi})=\mathcal{P}(x_{i}|I_{b})$,
as no information about $i$ was revealed.

The single agent's marginal KLs are then 
\begin{eqnarray}
\text{KL}_{x_{i}}(I',I'') & = & y_{J}\text{KL}_{x_{i}}(I_{bi}^{\text{h}},I''_{bi})+\\
 &  & (1-y_{J})\,\text{KL}_{x_{i}}(I_{bi}^{\neg\text{h}},I''_{bi})+\text{const}\text{, with}\nonumber \\
\text{KL}_{x}(I,I'') & = & (\mu-\mu'')\,\left[\psi(\mu+1)-\psi(\mu+\lambda+2)\right]+\nonumber \\
 &  & (\lambda-\lambda'')\,\left[\psi(\lambda+1)-\psi(\mu+\lambda+2)\right]+\nonumber \\
 &  & \ln\frac{\mathcal{B}(\mu''+1,\lambda''+1)}{\mathcal{B}(\mu+1,\lambda+1)},\text{ and}\label{eq:KLdigamma}\\
\psi(\alpha) & = & \frac{d\ln\Gamma(\alpha)}{d\alpha}\label{eq:digamma}
\end{eqnarray}
the digamma function and $\text{const}$ an $I''$ independent constant.\footnote{The expression in Eq.\ \ref{eq:KLdigamma} can be derived using the
expectation value $\langle x\,\ln x\rangle_{\text{Beta}(x|\alpha,\beta)}=\frac{\alpha}{\alpha+\beta}\left[\psi(\alpha+1)-\psi(\alpha+\beta+1)\right]$
\cite{beta-wikipedia}.} These KLs, $\text{KL}_{x_{a}}$ for speaker $a$ and $\text{KL}_{x_{c}}$
for topic $c$, then have to be minimized numerically with respect
to $I''=I''_{bi}=(\ensuremath{\mu''_{bi},}\ensuremath{\lambda''_{bi})}$
for $i\in\{a,c\}$. Details of the numerical implementation are given
in Appendix \ref{sec:Numerical-Implementation-Details}. The parameters
obtained by minimizing $I''$ are stored as the updated belief $I_{bi}(t+1)$
of agent $b$ about agent $i$. Examples of such updates are shown
in Figs.\ \ref{fig:Believe-update-of-sceptical} and \ref{fig:Contradicting-update}.

\subsection{Numerical details\label{sec:Numerical-Implementation-Details}}

Now, we detail how the KL minimization introduced in Sec.\ \ref{subsec:Optimal-believe-approximation}
is performed numerically. We use the Python package \texttt{scipy}
\cite{2020SciPy} to implement and minimize the KLs with the second
order schemes \texttt{trust-exact} and \texttt{trust-ncg} \cite{doi:10.1137/1.9780898719857,nocedal2006conjugate}
in this sequence. In our experience, the former scheme seems to be
more robust, and therefore provides the starting point for the latter
scheme, which seems to be more accurate. Furthermore, we use the machine
learning package \texttt{jax} \cite{JAX2019} configured for 64 bit
calculations to auto-differentiate the KLs to obtain their required
Jacobians and Hessians as well as to speed up all KL-related computations
via just-in-time compilation, which accelerates them substantially.
Unfortunately, we found that the numerical results of the KL minimization
do not exactly agree if executed on different computers. Since the
game dynamics is chaotic, such tiny numerical differences can grow
and result in differing game evolution in the later parts of some
runs (visible to the eye typically after $t=1000$ in some of the
runs). We verified that the statistical results are not significantly
affected by this. Furthermore, to ensure $\mu'',\lambda''>-1$ in
every optimization step the $\text{KL}_{x}(I',I'')$ is modified to
the objective function 
\begin{eqnarray}
\text{KL}_{x}(I',(\gamma(\mu''),\gamma(\lambda''))+(\mu''-\gamma(\mu''))^{2}+(\lambda''-\gamma(\lambda''))^{2}=\!\!\!\!\!\!\!\!\!\!\!\!\!\nonumber \\
\text{KL}_{x}(I',(\gamma(\mu''),\gamma(\lambda''))+\text{ReLu}(\mu''-\gamma_{0})^{2}+\text{ReLu}(\lambda''-\gamma_{0})^{2}\!\!\!\!\!\!\!\!\!\!\!\!\nonumber \\
\end{eqnarray}
 with $\gamma(x)=\text{max}(x,\gamma_{0})$, $\gamma_{0}=-1+10^{-10}$,
and 
\begin{equation}
\text{ReLu}(x)=\begin{cases}
0 & x<0\\
x & x\ge0
\end{cases}
\end{equation}
the rectified linear unit function. This way, the correct minimum
is found as long it has coordinates $\mu'',\lambda''\ge\gamma_{0}>-1$.
The additional terms gently push the calculations back to this boundary
as soon it is violated. The upper limits of $\mu'',\lambda''\le10^{6}$
are enforced after the minimization via re-scaling both variables
by $10^{6}/\text{max(\ensuremath{\mu,\lambda})}$ in case one of them
exceeds this range.

\section{Detailed receiver strategies\label{sec:detailed-receiver-strategies}}

\begin{table*}[!t]
\centering{}%
\begin{tabular}{|c|c|c|c|c|c|c|c|}
\hline 
\multirow{2}{*}{receiver $b$} & listening & naive & speaker $a$ & blushing & confession & message & expectation\tabularnewline
 &  & trust & reputation &  &  & surprise & matching\tabularnewline
\hline 
\hline 
naive & $\checkmark$ & $\checkmark$ &  &  &  &  & \tabularnewline
\hline 
deaf &  &  & $\checkmark$ & $\checkmark$ &  &  & \tabularnewline
\hline 
uncritical & $\checkmark$ &  & $\checkmark$ & $\checkmark$ & $\checkmark$ &  & \tabularnewline
\hline 
critical & $\checkmark$ &  & $\checkmark$ & $\checkmark$ & $\checkmark$ & $\checkmark$ & \tabularnewline
\hline 
smart & $\checkmark$ &  & $\checkmark$ & $\checkmark$ & $\checkmark$ & $\checkmark$ & $\checkmark$\tabularnewline
\hline 
\hline 
$\mathcal{R_{\mathfrak{f}}}=\frac{\mathcal{P}(\mathfrak{f}|\neg\text{h})}{\mathcal{P}(\mathfrak{f}|\text{h})}$ &  & $0$ & $\overline{x}_{ba}^{-1}-1$ & $\frac{1-f_{\text{b}}}{P(\text{\ensuremath{\neg}b}|o)}$ & $P(\neg\text{c}\lor\text{b}|J,o)$ & $\frac{\mathcal{S}_{J}^{2}}{2}$ & $e^{\mathcal{S}_{\text{h}}-\mathcal{S}_{\neg\text{h}}}$\tabularnewline
\hline 
reference & Sec.\ \ref{subsec:Basic-lie-detection} & Eq.\ \ref{eq:naive} & Eq.\ \ref{eq:yJ} & Eq.\ \ref{eq:R_blush} & Eq.\ \ref{eq:R_confess} & Eq.\ \ref{eq:R_crit} & Eq.\ \ref{eq:R_em}\tabularnewline
\hline 
\end{tabular}\caption{Summary of agents' receiver strategies in terms of the used features
$\mathfrak{f}$, their lie-honest likelihood ratios $\mathcal{R_{\mathfrak{f}}}$,
and the references where these are specified. \label{tab:receiver-strategies}}
\end{table*}

In our reputation game, a speaker tries to construct effective lies
when deceiving. An effective lie should on the one hand be as big
as possible (as measured in bits) to pursue the speaker's agenda,
and on the other hand sufficiently small to go unnoticed by the receiver.
These are opposite requirements and the optimal scale depends on the
lie detection abilities of the receiver. It can therefore be assumed
that lie construction and detection strategies should be the result
of an antagonistic co-evolution. Here, we follow some imagined first
steps of such an evolution by first constructing some basic lie detection
strategies in Sect.\ \ref{subsec:Basic-lie-detection}, then introduce
an adapted lie construction strategy in Sect.\ \ref{subsec:Lie-construction},
and finally a \emph{smart }lie detection strategy adapted to this
in Sect.\ \ref{subsec:Smart-lie-detection}. Finally, we explain
how the Theory of Mind (or auxiliary) variables used by agents in
lie construction and detection are maintained in Sect.\ \ref{subsec:Auxiliary-parameters-update}.
An overview on the different receiver strategies is given in Tab.\ \ref{tab:receiver-strategies}.

\subsection{Basic lie detection\label{subsec:Basic-lie-detection}}

A lie detection strategy of an agent is a recipe for how to choose
the weight $y_{J}:=\mathcal{P}(\text{h}|d)$ of a message $J$ in
a communication $a\overset{c}{\rightarrow}b$, i.e. how to judge the
trustworthiness of a received message. For example, \textbf{naive
agents} always assign $y_{J}=1$ irrespectively of the data. This
is obviously a poor strategy. It already is problematic in case of
non-deceptive agents\footnote{Non-deceptive agents would even communicate honest statements when
they should lie according to their lie-frequencies $1-x_{a}$.} for the strong echo chamber effect it allows, which leads to a too
rapid convergence of premature opinions.

The message weight $y_{J}$ should best be assigned according to Bayes
theorem, yielding
\begin{eqnarray}
y_{J} & = & \frac{\mathcal{P}(d|\text{h})\,\mathcal{P}(\text{h})}{\mathcal{P}(d)}\label{eq:y_J}\\
 & = & \frac{\mathcal{P}(d|\text{h})\,\overline{x}_{ba}}{\mathcal{P}(d|\text{h})\,\overline{x}_{ba}+\mathcal{P}(d|\neg\text{h})\,(1-\overline{x}_{ba})},\\
 & = & \left[1+\mathcal{R}(d)\left(\overline{x}_{ba}^{-1}-1\right)\right]^{-1}\text{ with}\label{eq:yJ}\\
\mathcal{R}(d) & = & \frac{\mathcal{P}(d|\neg\text{h})}{\mathcal{P}(d|\text{h})}\label{eq:R(d)}
\end{eqnarray}
the likelihood ratio, $d=(a\overset{c}{\rightarrow}b,J,o)$ the data
available to $b$, and 
\begin{equation}
\mathcal{P}(\text{h})=\int dx\,x_{a}\mathcal{P}(x_{a}|I_{ba})=\langle x_{a}\rangle_{(x_{a}|I_{ba})}=\overline{x}_{ba}
\end{equation}
the prior probability that $b$ assigns to $a$ for being honest.
Thus, a receiver strategy is fully specified as soon as the likelihoods
$\mathcal{P}(d|\text{h})$ and $\mathcal{P}(d|\neg\text{h})$ are
given or even just their lie-to-honest likelihood ratio $\mathcal{R}(d)=\mathcal{P}(d|\neg\text{h})/\mathcal{P}(d|\text{h})$.

The reputation of a speaker has a strong influence on whether their
potentially outrageous statements will be believed or not. If we set
$y_{J}=\nicefrac{1}{2}$ to investigate which statements are at the
margin to being trustworthy and solve Eq.\ \ref{eq:yJ} for the likelihood
ratio
\begin{equation}
\mathcal{R}=\left.\frac{y_{J}^{-1}-1}{\overline{x}_{ba}^{-1}-1}\right|_{y_{J}=\nicefrac{1}{2}}=\frac{\overline{x}_{ba}}{1-\overline{x}_{ba}}\label{eq:reputation}
\end{equation}
we see that three agents with reputations $\overline{x}_{ba}=0.1$,
$0.5$, and $0.9$ reach $y_{J}=\nicefrac{1}{2}$ for $\mathcal{R}=\nicefrac{1}{9}$,
$1,$ or $9$, respectively. Thus, the lie-to-honest likelihood ratio
of a statement can be $81$ times larger for the most reputed agent
($\overline{x}_{ba}=0.9$) compared to that of a statement by the
least reputed of those agents ($\overline{x}_{ba}=0.1$) before it
is perceived as only half trustworthy. Statements of reputed agents
are much more trusted.

We assume in the following that these likelihoods are given by independent
probabilities for a number of data features $\mathfrak{f}_{j}(d)$
with $j$ labeling the different features. Thus, for the honesty $\text{state}\in\{\text{h},\,\neg\text{h}\}\text{ we have}$
\begin{eqnarray}
\mathcal{P}(d|\text{state}) & = & \prod_{j}\mathcal{P}(\mathfrak{f}_{j}(d)|\text{state}).\label{eq:data-likelihood}
\end{eqnarray}

The features used in basic lie detection are naive trust, speaker
reputation, blushing, confessions, and message surprise. Smart lie
detection will additionally use expectation matching. These features
will be explained in the following. The assumption of their independence
is not entirely realisitc, however, our aim is to set up a reasonably
functioning lie detection, but not necessarily the best possible.
The independence assumption permits to write
\begin{equation}
\mathcal{R}(d)=\prod_{j}\mathcal{R}_{j}(\mathfrak{f}_{j}(d))=\prod_{j}\frac{\mathcal{P}(\mathfrak{f}_{j}(d)|\neg\text{h})}{\mathcal{P}(\mathfrak{f}_{j}(d)|\text{h})}.
\end{equation}

To calculate the likelihood ratio, \textbf{naive agents} use only
naive trust, \textbf{uncritical agents} use additionally the speaker
reputation and blushing, \textbf{critical agents} use further confessions
and surprise information, whereas \textbf{smart agents} exploit expectation
matching in addition to the former features, which is whether a message
looks more like what the speaker seems to believe, or what the speaker
apparently wants them to believe. We also introduce \textbf{deaf agents},
who only use blushing information to learn about others, as an illustrative
reference. Deaf agents are also uncritical, as they do not inspect
the message content, neither for deciding about its honesty, not for
updating their beliefs.

\subsubsection{Naive trust}

\begin{figure*}
\includegraphics[clip,width=0.5\textwidth]{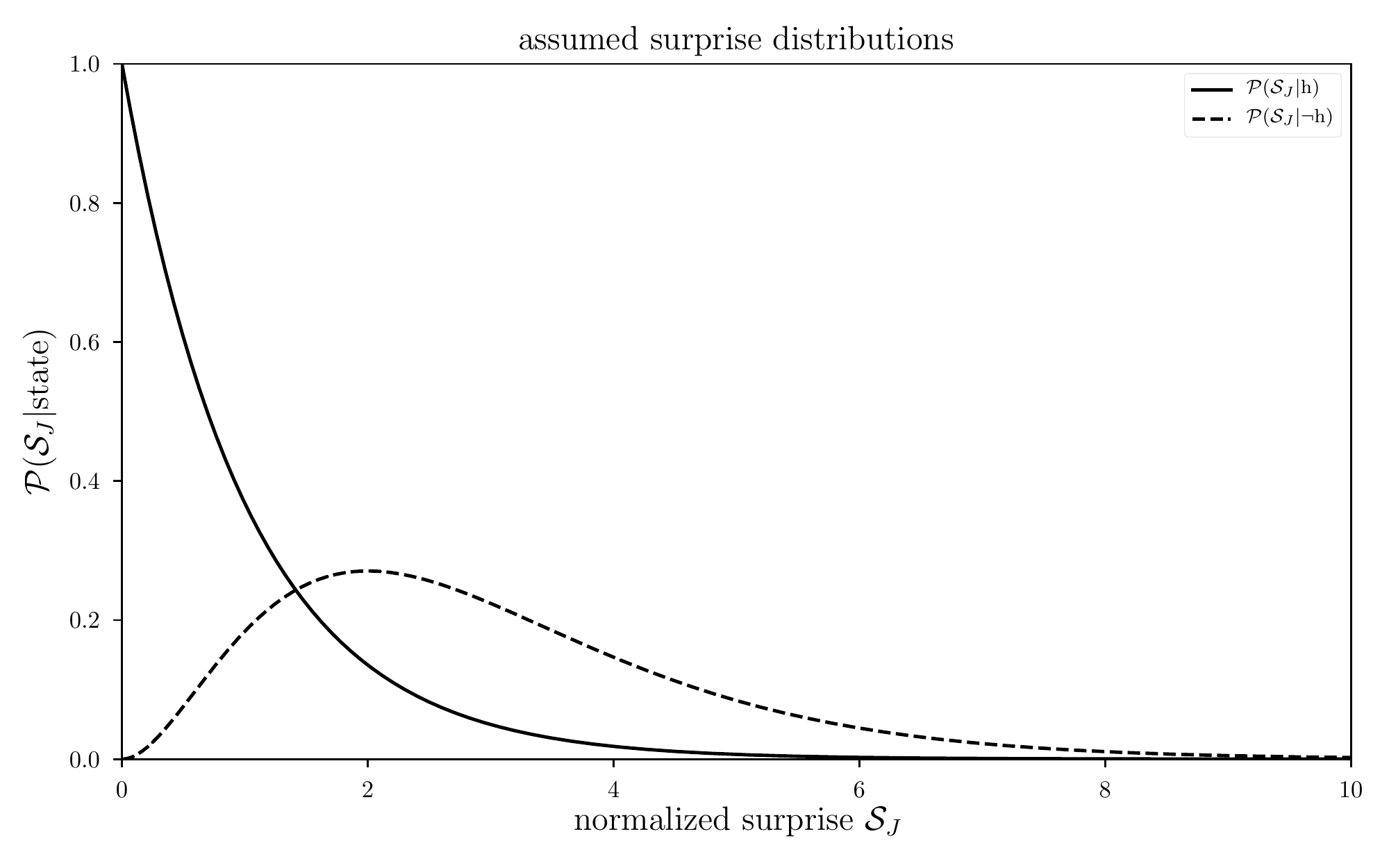}\includegraphics[clip,width=0.5\textwidth]{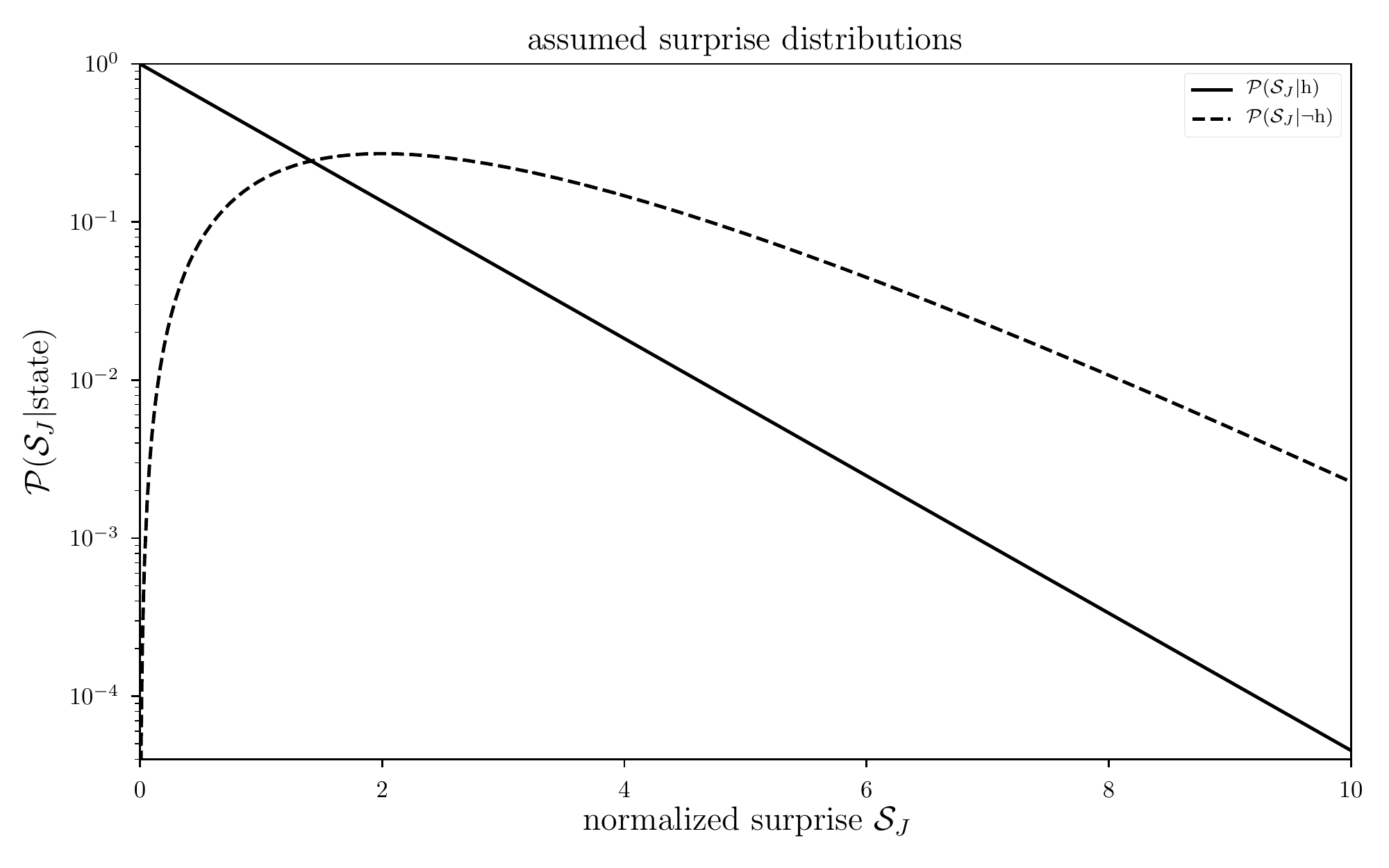}

\caption{Assumed likelihood for the surprise $\mathcal{S}_{J}$ of a received
message $J$ send from an honest agent (solid line) and a lying one
(dashed line) on linear scale (left panel) and logarithmic scale (right
panel).\label{fig:Assumed-PDF-for}}
\end{figure*}

Naive agents always trust the speaker and set $y_{J}=1$. This implies
that for them 
\begin{equation}
\mathcal{R}(d)=\mathcal{R}_{\text{naive}}(d)=0\label{eq:naive}
\end{equation}
 or $\mathcal{P}(d|\neg\text{h})=0$, meaning that they assume a lie
would never have reached them.

\subsubsection{Speaker reputation}

Let us first inspect the case that no feature beyond the message existence
is used at all, and that this existence does not imply any information
on the honesty, $\mathcal{P}(d|\text{h})=\mathcal{P}(d|\neg\text{h})$.
Therefore, $\mathcal{R}(d)=1$ and $y_{J}=\overline{x}_{ba}$. Thus,
without inspecting the message data agent $b$ assigns the prior average
belief on the honesty of $a$ to the message being honest. This already
provides some amount of defense against liars, since if identified
as such, they get their messages down weighted.

\subsubsection{Blushing}

The blushing feature $\mathfrak{f}_{\text{b}}(d)=o\in\{\text{blush},\text{no blush}\}=:\{\text{b},\neg\text{b}\}$
has the likelihood
\begin{eqnarray}
\mathcal{P}(\text{b}|\neg\text{h}) & = & f_{\text{b}},\label{eq:blush}\\
\mathcal{P}(\text{b}|\text{h}) & = & 0,\text{ and}\\
\mathcal{P}(\neg\text{b}|\text{state}) & = & 1-\mathcal{P}(\text{b}|\text{state}).
\end{eqnarray}
Therefore, $\mathcal{R}(\text{b})=\infty$ and $\mathcal{R}(\neg\text{b})=1-f_{\text{b}}$.
Thus, the uncritical agent assigns
\begin{equation}
\mathcal{R}_{\text{uncritical}}(d)=\mathcal{R}_{\text{b}}(o)=\frac{1-f_{\text{b}}}{P(\text{\ensuremath{\neg}b}|o)}=\begin{cases}
\infty & \text{b}\\
1-f_{\text{b}} & \neg\text{b}
\end{cases},\label{eq:R_blush}
\end{equation}
where $P(\text{\ensuremath{\neg}b}|o):=P(o=\neg b|o)\in\{0,1\}$ is
the logical theta function that is unity in case of no blushing, and
otherwise zero. The uncritical agent, who uses only blushing information,
assigns $y_{J}=0$ in case the speaker blushes, otherwise $y_{J}=\overline{x}_{ba}\left[(1-f_{\text{b}})+f_{\text{b}}\overline{x}_{ba}\right]^{-1}=\overline{x}_{ba}\left[0.9+0.1\overline{x}_{ba}\right]^{-1}\approx\overline{x}_{ba}$,
since $f_{\text{b}}=0.1$. The small enhancement of $y_{J}$ w.r.t.\ $\overline{x}_{ba}$
is due to the weak indication of honesty implied by non-blushing,
see Eq.\ \ref{eq:yJ} with $\mathcal{R}(d)=\mathcal{R}_{\text{b}}(\neg\text{b})=1-f_{\text{b}}=0.9$
inserted.

\subsubsection{Confession}

As agents rather overstate their honesty than understate it, a self-statement
of a currently non-blushing agent $a$ that indicates an honesty $\overline{x}_{J_{a\overset{a}{\rightarrow b}}}$
below the agent's reputation $\overline{x}_{ba}$ to $b$ must be
an honest confession. Whether a confession is present is given by
\begin{equation}
\mathfrak{f}_{\text{c}}(d):=\text{c}:=(\overline{x}_{J_{a\overset{a}{\rightarrow b}}}\!<\overline{x}_{ba})\in\{\text{true},\text{false}\}
\end{equation}
 and we have $P(\text{c}|\neg\text{h})=0,$ such that 
\begin{equation}
\mathcal{R_{\text{c}}}(\text{c},\neg\text{b})=\frac{P(\text{c}\land\neg\text{b}|\neg\text{h})}{P(\text{c}\land\neg\text{b}|\text{h})}=0\label{eq:R_c}
\end{equation}
and therefore $y_{J}=1$ if a confession is present. The absence of
a confession does not bear much information, as it could be caused
by a lie or by agent $b$ being misinformed about the true honesty
of $a$. The former has a probability of $1-x_{a}$, but the probability
of the latter is hard to estimate accurately. Thus, it is safer to
set the likelihood ratio for all other cases to be uninformative,
\begin{equation}
\mathcal{R}_{\text{c}}(\neg(\text{c}\land\neg\text{b}))=1,
\end{equation}
than to risk to get misleading hints. We collect all these cases in
\begin{equation}
\mathcal{R}_{\text{c}}(J,o)=P(\neg\text{c}\lor\text{b}|J,o),\label{eq:R_confess}
\end{equation}
again using the probability notation to express a logical theta function.

\subsubsection{Message surprise}

Critical agents use in addition to the blushing and confession information
the \emph{cognitive dissonance} the message generates if taken for
true, which we associate to the surprise of a message $J_{a\overset{c}{\rightarrow}b}$
with respect to their own beliefs. This surprise is $s_{J}=\text{KL}_{x_{c}}(J_{a\overset{c}{\rightarrow}b},I_{bc})$,
the number of nits a plain adaption of the message would cause in
$b$'s mind. This gets compared to an agent specific and learned reference
scale $\kappa_{b}$ to form the normalized surprise data feature
\begin{equation}
\mathfrak{f}_{\text{s}}(d):=\mathcal{S}_{J}:=\frac{s_{J}}{\kappa_{b}}=\frac{\text{KL}_{x_{c}}(J_{a\overset{c}{\rightarrow}b},I_{bc})}{\kappa_{b}}.\label{eq:f_s}
\end{equation}

Here, we make the ad-hoc assumption that critical agents implicitly
assume the distributions of $\mathcal{S}_{J}\in\mathbb{R}_{0}^{+}$
to be
\begin{eqnarray}
\mathcal{P}(\mathcal{S}_{J}|\text{h}) & := & e^{-\mathcal{S}_{J}}\text{ and}\label{eq:honest-surprise-distribution}\\
\mathcal{P}(\mathcal{S}_{J}|\neg\text{h}) & := & \frac{\mathcal{S}_{J}^{2}}{2}e^{-\mathcal{S}_{J}}\label{eq:dishonest-surprise-distribution}
\end{eqnarray}
such that for them
\begin{eqnarray}
\mathcal{R}_{\text{critical}}(d) & = & \mathcal{R}_{\text{s}}(\mathcal{S}_{J})\,\mathcal{R}_{\text{c}}(J,o)\,\mathcal{R}_{\text{b}}(o)\\
 & = & \frac{\mathcal{S}_{J}^{2}}{2}\,P(\neg\text{c}\lor\text{b})|J,o)\,\frac{1-f_{\text{b}}}{P(\text{\ensuremath{\neg}b}|o)}.\label{eq:R_crit}
\end{eqnarray}
This means critical agents assume the surprises of honest messages
to be distributed exponentially, with a clear peak at zero surprise,
whereas that of lies to have a typical surprise $s_{J}$ of at least
$\sqrt{2}\kappa_{b}$, the surprise level above which the lies should
dominate. The assumed surprise likelihood functions are depicted in
Fig.\ \ref{fig:Assumed-PDF-for}. They allow for a more critical
discrimination of lies from honest statements than blushing, confession,
and the speaker's reputation alone. Tuning their auxiliary parameters
$\kappa_{i}$ allows agents to adapt the absolute surprise distribution
functions $\mathcal{P}(s_{J}|\text{state})$ to the social situation
they find themselves in. This will be detailed later in Sect.\ \ref{subsec:Auxiliary-parameters-update}.

\subsection{Lie construction\label{subsec:Lie-construction}}

With the basic receiver strategies to detect lies in place, the question
can be addressed how agents should construct their lies.

Lies towards naive and uncritical agents can be arbitrarily big, as
these do not inspect the messages closely. Thus, these agents are
very vulnerable to propaganda.\footnote{In the current version of our game, this would not be fully exploited
by malicious agents, as agents do not infer the character of other
agents except for the level of their honesty.} Lies towards critical as well as smart agents need to balance the
push for the speaker's agenda, favoring larger lies, and the risk
to get caught, which increases with the size of the lie (where the
size is measured by the receiver in units of bits).

As a statement towards a critical or smart agent gets judged on the
basis of how much it diverges from the receiver's own opinion, it
better stays close to this opinion and deviates only so little in
the desired direction that it can go unnoticed. In order not to be
too predictable in this, lies are designed such that their surprises
approximately match the assumed surprise distribution of honest statements,
Eq.\ \ref{eq:honest-surprise-distribution}.

A liar $a$ proceeds in the following way when talking to an enemy
or a friend. The agent takes $I_{abc}$, the assumed belief of the
recipient $b$ on the topic $c$, decides on the direction $D\in\{(1,0),(0,1)\}$
of the bias to be applied according to whether $c$ is a friend to
$a$ or an enemy, respectively. The lie will be constructed as
\begin{equation}
J_{a\overset{c}{\rightarrow}b}=I_{abc}+\alpha\;D,
\end{equation}
with $\alpha\in\mathbb{R}_{0}^{+}$ such that the receiver is expected
to experience only a certain surprise by the lie. This is achieved
by drawing randomly a target normalized surprise $\mathcal{S}_{J}\hookleftarrow\mathcal{P}(\mathcal{S}_{J}|\text{h})$,
multiplying it with $\kappa_{a}\,f_{\text{caution}}$, where $\kappa_{a}$
is agent $a$'s substitute for $\kappa_{b}$ unknown by $a$ and $f_{\text{caution}}=0.3$
is a caution factor to compensate for the mistake thereby done, and
choose $\alpha$ via a numerical line search such that
\begin{equation}
\text{KL}_{x_{c}}(I_{abc}+\alpha\;D,I_{abc})=\kappa_{a}\,f_{\text{caution}}\mathcal{S}_{J}.
\end{equation}

For an agent being a topic, who is neither a friend or an enemy, a
\textbf{white lie} is used by setting $J_{a\overset{c}{\rightarrow}b}=I_{abc}$.
White lies do not necessarily bias the recipient's opinion on $c$,
however they let the speaker appear honest without revealing the speaker's
true opinion.

\subsection{Smart lie detection\label{subsec:Smart-lie-detection}}

A more efficient, \textbf{smart} lie detection takes into account
the way lies are constructed. The agent's lies are constructed as
biased copies of what the speaker $a$ thinks the receiver $b$ believes
on the topic. This opens the possibility for a smart agent $b$ to
discriminate messages by matching them up against expected honest
and dishonest statements of the speaker. For this $b$ needs an idea
of what $a$ believes on topic $c$, denoted as $I_{bac}$ (agent
$b$'s guess for $a$'s belief on $c$), as well as an idea of what
$a$ wants $b$ to believe on that topic, denoted as $\widetilde{I}_{bac}$
($b$'s guess for what $a$ wants $b$ to think about $c$). Which
of those matches better to the message $J_{a\overset{c}{\rightarrow}b}$
is then an indicator of the message's honesty. The data features used
by smart agents are the message surprises w.r.t.\ $I_{bac}$, and
$\widetilde{I}_{abc}$, $s_{\text{h}}:=\text{KL}_{x_{c}}(J_{a\overset{c}{\rightarrow}b},I_{bac})$
and $s_{\neg\text{h}}:=\text{KL}_{x_{c}}(J_{a\overset{c}{\rightarrow}b},\widetilde{I}_{bac})$,
respectively. The corresponding normalized surprises $\mathcal{S}_{\text{state}}:=s_{\text{state}}/\kappa{}_{b}$
(with $\text{state}\in\{\text{h},\,\neg\text{h}\}$) are again assumed
to be zero peaked exponential distributions,
\begin{eqnarray}
\mathcal{P}(\mathcal{S}_{\text{state}}|\text{state}) & := & e^{-\mathcal{S}_{\text{state}}},
\end{eqnarray}
with the lie detection scale parameter $\kappa{}_{b}$. This specifies
the distribution of $\mathcal{S}_{\text{h}}$ in case $\text{h}$,
as well as of $\mathcal{S}_{\neg\text{h}}$ in case $\neg\text{h}$.
The distribution of $\mathcal{S}_{\text{h}}$ in case $\neg\text{h}$
and that of $\mathcal{S}_{\neg\text{h}}$ in case $\text{h}$ are
not needed in detail, we only assume them to be identical,
\begin{equation}
\mathcal{P}(\mathcal{S}_{\text{h}}|\neg\text{h})=\mathcal{P}(\mathcal{S}_{\neg\text{h}}|\text{h}).
\end{equation}
Furthermore, we assume these two features to be independent of each
other, so that their lie-to-honest likelihood ratio becomes
\begin{eqnarray}
\mathcal{R}_{\text{em}}(\mathcal{S}_{\text{h}},\mathcal{S}_{\neg\text{h}}) & := & \frac{\mathcal{P}(\mathcal{S}_{\text{h}},\mathcal{S}_{\neg\text{h}}|\neg\text{h})}{\mathcal{P}(\mathcal{S}_{\text{h}},\mathcal{S}_{\neg\text{h}}|\text{h})}\\
 & = & \frac{\mathcal{P}(\mathcal{S}_{\text{h}}|\neg\text{h})}{\mathcal{P}(\mathcal{S}_{\text{h}}|\text{h})}\ \frac{\mathcal{P}(\mathcal{S}_{\neg\text{h}}|\neg\text{h})}{\mathcal{P}(\mathcal{S}_{\neg\text{h}}|\text{h})}\\
 & = & e^{\mathcal{S}_{\text{h}}-\mathcal{S}_{\neg\text{h}}}.\label{eq:R_em}
\end{eqnarray}
For the smart lie detection, this likelihood ratio is just multiplied
to the likelihood ratio critical agents use:
\begin{eqnarray}
\mathcal{R_{\text{smart}}}(d) & \!\!=\!\! & \mathcal{R}_{\text{em}}(\mathcal{S}_{\text{h}},\mathcal{S}_{\neg\text{h}})\,\mathcal{R}_{\text{s}}(\mathcal{S}_{J})\,\mathcal{R}_{\text{c}}(J,o)\,\mathcal{R}_{\text{b}}(o)\\
 & \!\!=\!\! & e^{\mathcal{S}_{\text{h}}-\mathcal{S}_{\neg\text{h}}}\frac{\mathcal{S}_{J}^{2}}{2}\,P(\neg\text{c}\lor\text{b})|J,o)\,\frac{1-f_{\text{b}}}{P(\text{\ensuremath{\neg}b}|o)}
\end{eqnarray}

Special deception strategies, which circumvent or even exploit smart
lie detection, can be imagined as well. These are beyond the scope
of this work. The above strategies are sufficient to illustrate what
kind of strategies might be used by real humans. Note that we do not
claim that the ones chosen here are exhaustive.

\subsection{Auxiliary parameters update\label{subsec:Auxiliary-parameters-update}}

We now summarize how all the auxiliary parameters forming the Theory
of Mind knowledge are maintained. After receiving the communication
$J=J_{a\overset{c}{\rightarrow}b}$ (and eventually having responded)
agent $b$ performs updates of the following parameters: Lists of
friends $\mathcal{F}_{b}$ and enemies $\mathcal{E}_{b}$, guesses
for agent $a$'s belief on and intention for $c$, $I_{abc}$ and
$\widetilde{I}_{abc}$, respectively, as well as the reference surprise
scale $\kappa_{b}$.

\subsubsection{Friends and enemies}

Agent $b$ updates the list of friends $\mathcal{F}_{b}$ and that
of enemies $\mathcal{E}_{b}$, where $\mathcal{F}_{b}=\{b\}$ and
$\mathcal{E}_{b}=\{\}$, initially. An agent in none of these lists
is regarded by $b$ as being neutral to $b$.

In case agent $a$ made a statement $J_{a\ensuremath{\overset{b}{\rightarrow}b}}$
about $b$ to $b$, agent $b$ memorizes how much \textbf{respect}
$r'_{bab}:=\overline{x}_{J_{a\ensuremath{\overset{b}{\rightarrow}b}}}$
agent $a$ thereby expresses for $b,$ where we define respect as
the by a communication stated honesty of an agent. Then the median
$\hat{r}_{b}=\text{median}(\{r'_{bib}\}_{i\in\mathcal{A}\backslash\{a,b\}})$
of the memorized respect values of all other agents is calculated
and compared to this updated one. If $r'_{bab}>\hat{r}_{b}$ agent
$a$ is added to the set $\mathcal{F}_{b}$ of $b$'s friends and
removed from $\mathcal{E}_{b}$, the list of $b$'s enemies (if listed
there). If $r'_{bab}<\hat{r}_{b}$ agent $a$ will be added to the
enemy list and removed from the friend list. In case $r'_{bab}=\hat{r}_{b}$,
these lists stay as they are.

In summary, an agent $a$ is regarded as a friend by $b$ whenever
$a$'s last statement about $b$ to $b$ was more positive than the
median of other agents' last statements at that point in time and
$a$ is regarded as an enemy, if this was less positive.

\subsubsection{Other's beliefs and intentions}

\begin{table*}[!t]
\centering{}%
\begin{tabular}{|c|c|c|c|c|c|c|}
\hline 
\multirow{2}{*}{agent $a$} & $P(a\overset{\cdot}{\rightleftarrows}b|a\overset{\cdot}{\rightleftarrows}\cdot)$ & $P(a\overset{c}{\rightleftarrows}b|a\overset{\cdot}{\rightleftarrows}b)$ & $P(\text{h}|a\overset{c}{\rightarrow}b)$ & $P(\text{b}|\neg\text{h})$ & \multicolumn{1}{c|}{deception} & \multicolumn{1}{c|}{receiver}\tabularnewline
 & $\propto(1-\delta_{ab})\times$ & $=$ & $=$ & $=$ & strategy & strategy\tabularnewline
\hline 
\hline 
deaf & $1$ & $\nicefrac{1}{n}$ & $x_{a}$ & $f_{\text{b}}$ & $a$, fr.$\uparrow$; en.$\downarrow$; n.$\circlearrowleft$ & \textcolor{blue}{deaf}\tabularnewline
\hline 
naive & $1$ & $\nicefrac{1}{n}$ & $x_{a}$ & $f_{\text{b}}$ & $a$, fr.$\uparrow$; en.$\downarrow$; n.$\circlearrowleft$ & \textcolor{blue}{naive}\tabularnewline
\hline 
uncritical & $1$ & $\nicefrac{1}{n}$ & $x_{a}$ & $f_{\text{b}}$ & $a$, fr.$\uparrow$; en.$\downarrow$; n.$\circlearrowleft$ & \textcolor{blue}{uncritical}\tabularnewline
\hline 
ordinary & $1$ & $\nicefrac{1}{n}$ & $x_{a}$ & $f_{\text{b}}$ & $a$, fr.$\uparrow$; en.$\downarrow$; n.$\circlearrowleft$ & critical\tabularnewline
\hline 
strategic & \textcolor{blue}{$\overline{x}_{ab}$} & $\nicefrac{1}{n}$ & $x_{a}$ & $f_{\text{b}}$ & $a$, fr.$\uparrow$; en.$\downarrow$; n.$\circlearrowleft$ & critical\tabularnewline
\hline 
anti-strategic & \textcolor{blue}{$(1-\overline{x}_{ab})$} & $\nicefrac{1}{n}$ & $x_{a}$ & $f_{\text{b}}$ & $a$, fr.$\uparrow$; en.$\downarrow$; n.$\circlearrowleft$ & critical\tabularnewline
\hline 
flattering & $1$ & \textcolor{blue}{$\delta_{bc}$} & \textcolor{blue}{$x_{a}(1-\delta_{bc})$} & $f_{\text{b}}$ & $a$,\textcolor{red}{{} }\textcolor{blue}{$b$,} fr.$\uparrow$; en.$\downarrow$;
n.$\circlearrowleft$ & critical\tabularnewline
\hline 
egocentric & $1$ & \textcolor{blue}{$\frac{1}{2}\,\left(\delta_{ac}+\nicefrac{1}{n}\right)$} & $x_{a}$ & $f_{\text{b}}$ & $a$, fr.$\uparrow$; en.$\downarrow$; n.$\circlearrowleft$ & critical\tabularnewline
\hline 
\multirow{2}{*}{aggressive} & \multirow{2}{*}{$1$} & $\nicefrac{1}{n}$\textcolor{red}{{} }\textcolor{blue}{if $\mathcal{E}_{a}=\{\}$} & \multirow{2}{*}{$x_{a}$} & \multirow{2}{*}{$f_{\text{b}}$} & \multirow{2}{*}{\textcolor{blue}{$a$, fr.$\circlearrowleft$;} en.$\downarrow$;n.$\circlearrowleft$} & \multirow{2}{*}{critical}\tabularnewline
 &  & \textcolor{blue}{$\nicefrac{\delta_{c\in\mathcal{E}_{a}}}{|\mathcal{E}_{a}|}$
else} &  &  &  & \tabularnewline
\hline 
shameless & $1$ & $\nicefrac{1}{n}$ & $x_{a}$ & \textcolor{blue}{$0$} & $a$, fr.$\uparrow$; en.$\downarrow$; n.$\circlearrowleft$ & critical\tabularnewline
\hline 
smart & $1$ & $\nicefrac{1}{n}$ & $x_{a}$ & $f_{\text{b}}$ & $a$, fr.$\uparrow$; en.$\downarrow$; n.$\circlearrowleft$ & \textcolor{blue}{smart}\tabularnewline
\hline 
deceptive & $1$ & $\nicefrac{1}{n}$ & \textcolor{blue}{$0$} & $f_{\text{b}}$ & $a$, fr.$\uparrow$; en.$\downarrow$; n.$\circlearrowleft$ & critical\tabularnewline
\hline 
clever & $1$ & $\nicefrac{1}{n}$ & \textcolor{blue}{$0$} & $f_{\text{b}}$ & $a$, fr.$\uparrow$; en.$\downarrow$; n.$\circlearrowleft$ & \textcolor{blue}{smart}\tabularnewline
\hline 
manipulative & \textcolor{blue}{$(1-\overline{x}_{ab})$} & \textcolor{blue}{$\delta_{bc}$} & \textcolor{blue}{$0$} & $f_{\text{b}}$ & $a$,\textcolor{red}{{} }\textcolor{blue}{$b$}, fr.$\uparrow$; en.$\downarrow$;
n.$\circlearrowleft$ & \textcolor{blue}{smart}\tabularnewline
\hline 
dominant & \textcolor{blue}{$\overline{x}_{ab}$} & \textcolor{blue}{$\frac{1}{2}\,\left(\delta_{ac}+\nicefrac{1}{n}\right)$} & \textcolor{blue}{$0$} & $f_{\text{b}}$ & $a$, fr.$\uparrow$; en.$\downarrow$; n.$\circlearrowleft$ & \textcolor{blue}{smart}\tabularnewline
\hline 
\multirow{2}{*}{destructive} & \multirow{2}{*}{\textcolor{blue}{$\overline{x}_{ab}$}} & $\nicefrac{1}{n}$\textcolor{red}{{} }\textcolor{blue}{if $\mathcal{E}_{a}=\{\}$} & \multirow{2}{*}{\textcolor{blue}{$0$}} & \multirow{2}{*}{\textcolor{blue}{$0$}} & \multirow{2}{*}{\textcolor{blue}{$a$, fr.$\circlearrowleft$;} en.$\downarrow$;n.$\circlearrowleft$} & \multirow{2}{*}{\textcolor{blue}{smart}}\tabularnewline
 &  & \textcolor{blue}{$\nicefrac{\delta_{c\in\mathcal{E}_{a}}}{|\mathcal{E}_{a}|}$
else} &  &  &  & \tabularnewline
\hline 
\end{tabular}\caption{Summary of agents' communication strategies, which determine how an
agent $a$ picks (if being initiator of a conversation $a\protect\overset{c}{\rightleftarrows}b$)
the partner $b$, topic $c$, whether (in any communication $a\protect\overset{c}{\rightarrow}b$)
$a$ is honest ($\text{h}$) or lies ($\neg\text{h}$), whether $a$
blushes ($\text{b}$), how $a$ lies about $a$, about a friend ($\text{fr.}\in\mathcal{F}_{a}$),
about an enemy ($\text{en.}\in\mathcal{E}_{a}$), and about a neutral
agent ($\text{n.}\in\mathcal{A}\backslash(\mathcal{F}_{a}\cup\mathcal{E}_{a}\cup a$)),
and how $a$ receives messages. ``$\uparrow$'' means that the testimony
in a lie is biased positively (with $(\delta\mu,\delta\lambda)=J_{a\protect\overset{c}{\rightarrow}b}-I_{abc}\propto(1,0)$)
and ``$\downarrow$'' means negatively ($(\delta\mu,\delta\lambda)\propto(0,1)$)
w.r.t.\ to $I_{abc}$, the by-$a$-assumed opinion of $b$ about
$c$. ``$\circlearrowleft$'' means white lies, in which $a$ tries
to tell $b$ exactly what $b$ beliefs about $c$, $J_{a\protect\overset{c}{\rightarrow}b}=I_{abc}$.
Differences in behavior w.r.t.\  an ordinary agent are marked in
\textcolor{blue}{blue}. \label{tab:Summary-of-agent's-strategies}}
\end{table*}

Agents maintain an image of the opinions and intentions of the other
agents. Agent $b$ does not have direct access to the beliefs of agent
$a$, but only to the received message $J=J_{a\overset{c}{\rightarrow}b}$.
This message has to be analyzed to determine $a$'s beliefs.

Agent $b$ extracts from the message what $a$ seems to believe on
topic $c$, whenever $a$ seems to be honest, and stores this as $I_{bac}$
(agent $b$'s best guess for $I_{ac}$). Similarly, agent $b$ can
also determine the intention of $a$ when $a$ is lying. In that case
the message contains what $a$ wants $b$ to believe about $c$. This
intention is stored by $b$ as $\widetilde{I}_{bac}$ (agent $b$'s
best guess for by lie distortions modified $I_{abc}$).

The updates for $I_{bac}$ and $\widetilde{I}_{bac}$ are done by
blending the message ($J$) into the present value of these variables
with a weight according to how much the message seems to be honest
(weight $y_{J}$) or dishonest (weight $1-y_{J}$), respectively:
\begin{eqnarray}
I_{bac}(t) & \rightarrow & I_{bac}(t+1)=y_{J}J+(1-y_{J})I_{bac}(t)\label{eq:ToM-1}\\
\widetilde{I}_{bac}(t) & \rightarrow & \widetilde{I}_{bac}(t+1)=(1-y_{J})J+y_{J}\widetilde{I}_{bac}(t)\label{eq:ToM-2}
\end{eqnarray}
These update rules can be regarded as modified DeGroot learning, which
is often treated as the counterpart to fully Bayesian updates that
have been used for the direct observation of others' opinions in this
work \cite{10.1257/mic.2.1.112,JADBABAIE2012210,MUELLERFRANK2014423}.
The only difference to classical DeGroot updates is the variable trust
matrix, which here always adapts to the presumed trustworthiness of
the message at hand. Similar rules are also used for agent based simulations
on trust networks \cite{10.2307/26395051}. These updates should provide
guesses of $b$ for $I_{ac}$ and $I_{abc}$ (modified by the bias
of the lie), respectively. The corresponding guesses, $I_{bac}$ and
$\widetilde{I}_{bac}$, become accurate whenever the speaker $a$
reveals to be honest ($y_{J}=1\Rightarrow I_{bac}=J=I_{ac}$) or to
be lying ($y_{J}=0\Rightarrow\widetilde{I}_{bac}=J\approx I_{abc}$),
respectively. Hopefully for $b$, these guesses should stay reasonably
accurate at other times.

\subsubsection{Typical surprises}

Agent $b$'s lie detection relies on the surprise reference scale
$\kappa_{b}$. This determines the assumed PDFs for message surprises
$s_{I}=\text{KL}_{x_{c}}(J,I)$ from various reference points I ($=I_{bc},\text{\ensuremath{I_{bac}}},$
and $\widetilde{I}_{bac}$). No static value can be assigned to $\kappa_{b}$,
as the surprise PDFs ($\mathcal{P}(s_{\cdot}|\text{h})$ and $\mathcal{P}(s_{\cdot}|\neg\text{h})$)
differ in different social situations and usually also evolve as a
function of time. A simple heuristic is used to update $\kappa_{b}$.

Initially, we set $\kappa_{b}=1$. For the update of $\kappa_{b}$,
it will be used that given the assumed surprise distributions for
honest and dishonest statements, Eqs.\ \ref{eq:honest-surprise-distribution}
and \ref{eq:dishonest-surprise-distribution}, respectively, and given
that half of the statements are a priori expected to be honest and
half to be dishonest (as implied by $\mathcal{P}(x_{a}|I_{0})=1$),
the median value for message surprises $s_{J}$ (with respect to $I_{bc}$)
should be located at $\sqrt{\pi}\kappa_{b}$. This is the expected
median of the assumed surprise distribution and marks the expected
transition from mostly honest to mostly dishonest statements. Thus,
agent $b$ just maintains a tuple $K_{b}$ with the $N_{\kappa}$
last non-zero surprises received and sets 
\begin{equation}
\kappa_{b}=\frac{\text{median}(K_{b})}{\sqrt{\pi}}\label{eq:kappa_b}
\end{equation}
whenever a new message arrives. The size of $N_{\kappa}$ determines
how quickly or slowly agent $b$ adapts to a changing social atmosphere,
and is set to $N_{\kappa}=10$ in our simulations. We initialize $K_{b}$
with $(\sqrt{\pi},\ldots\sqrt{\pi})$ to be consistent with the initial
$\kappa_{b}=1$.

\begin{figure*}[!t]
\vspace{-1.5em}
\includegraphics[width=0.5\textwidth]{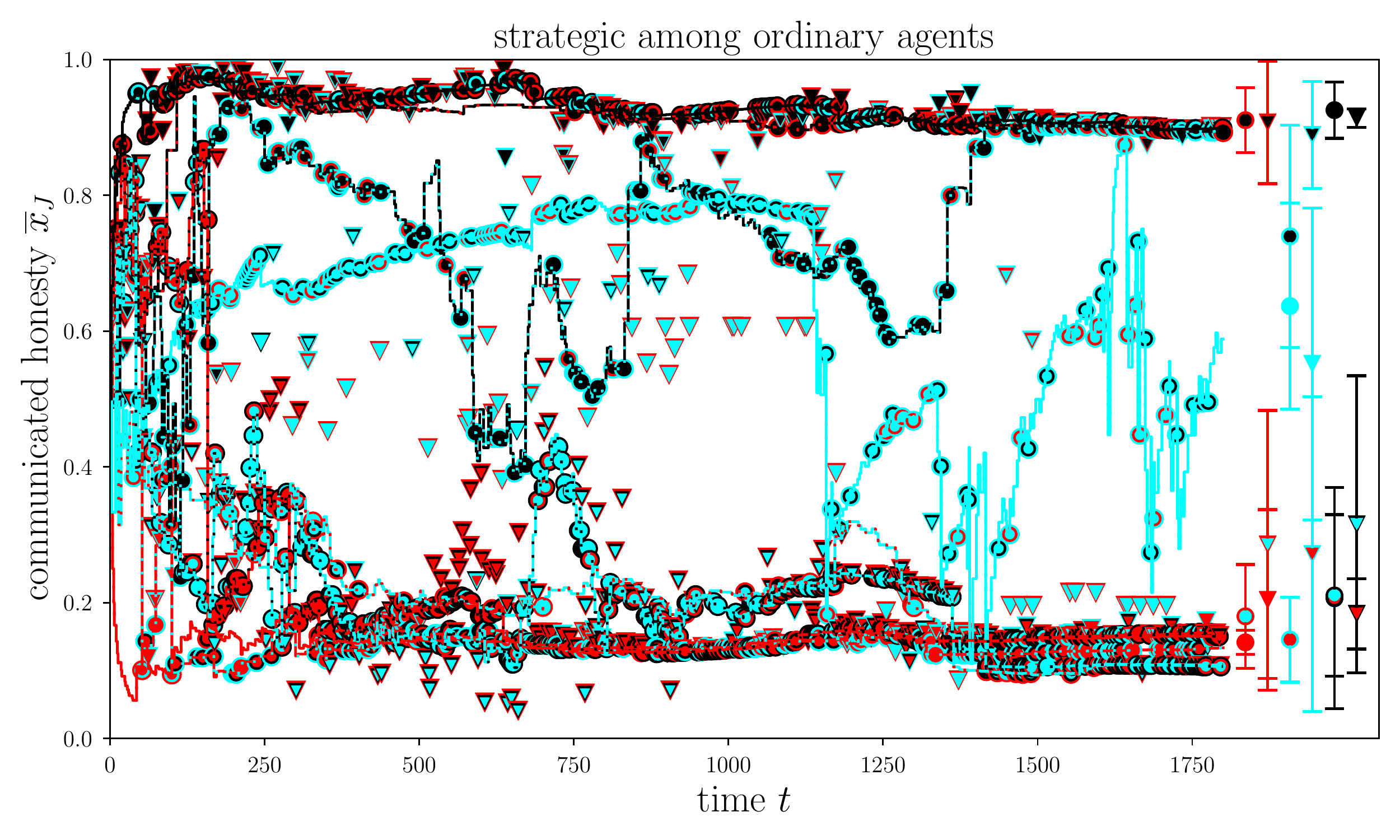}\includegraphics[width=0.5\textwidth]{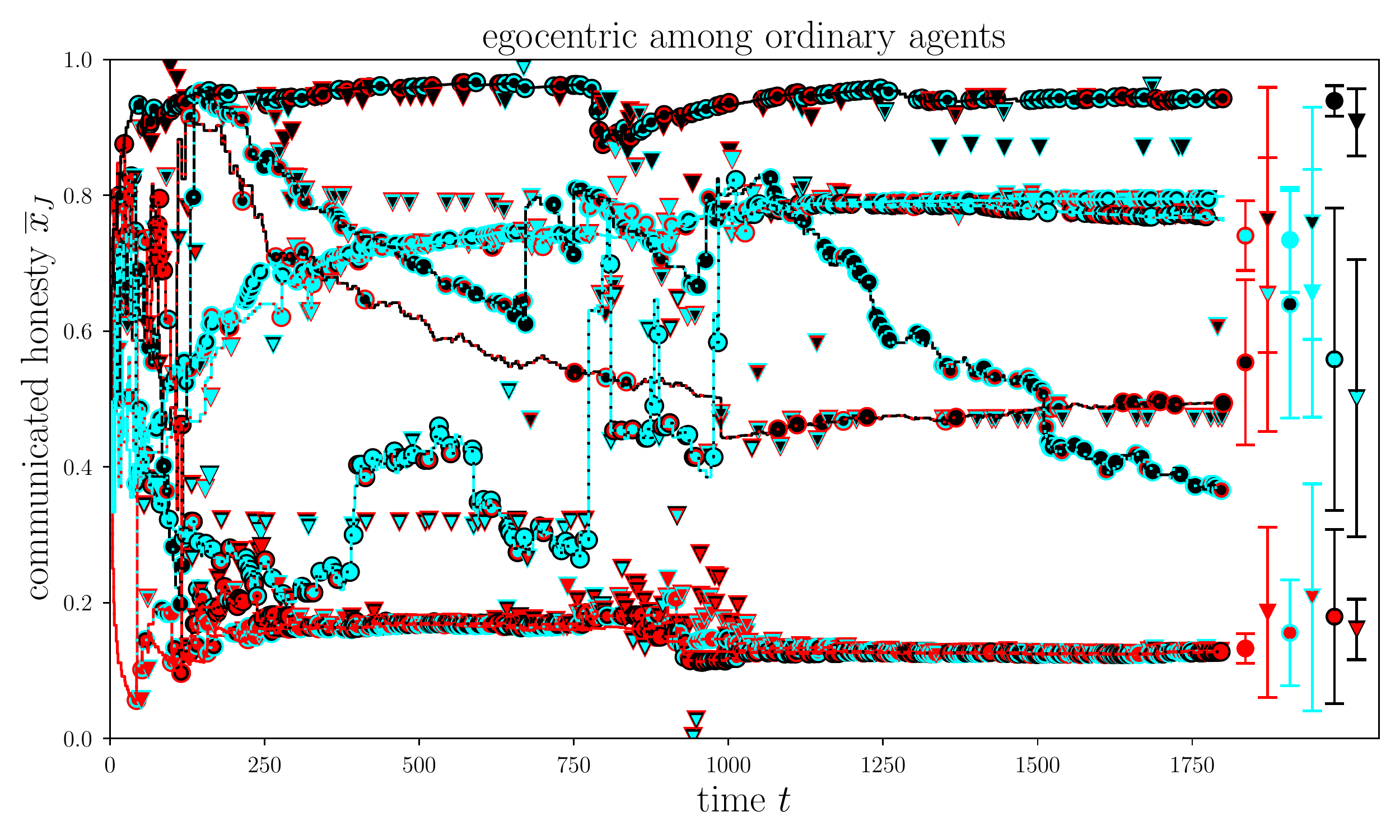}

\includegraphics[width=0.5\textwidth]{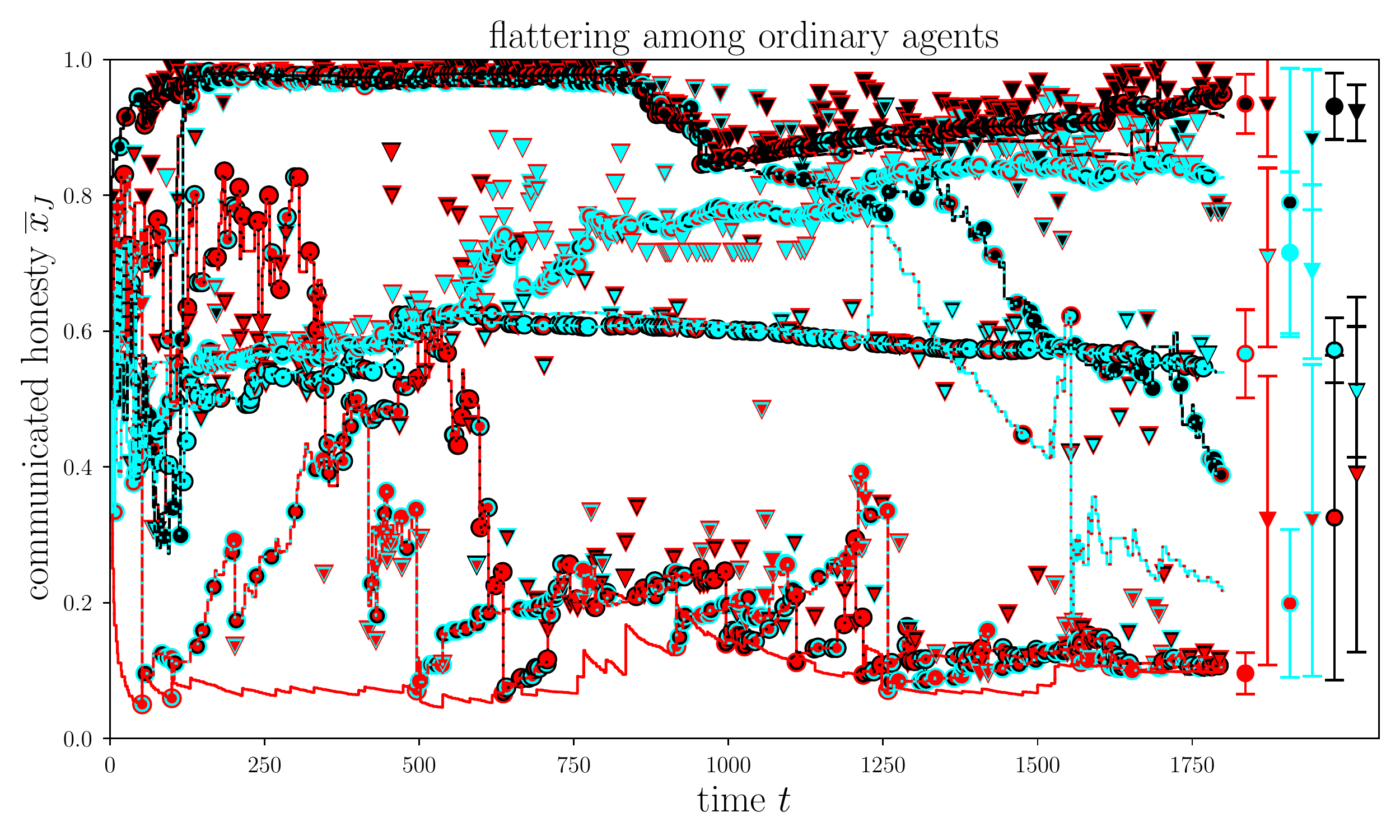}\includegraphics[width=0.5\textwidth]{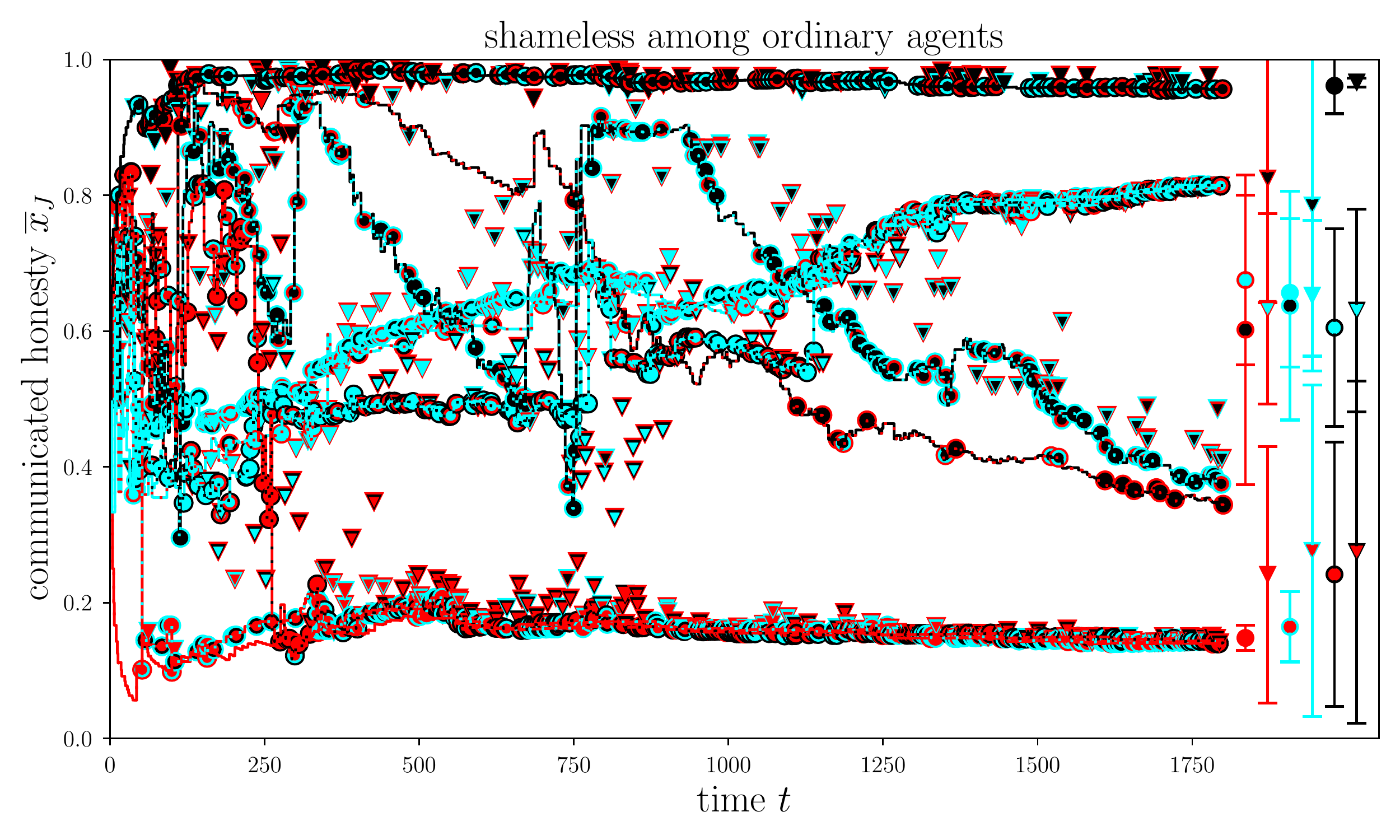}

\includegraphics[width=0.5\textwidth]{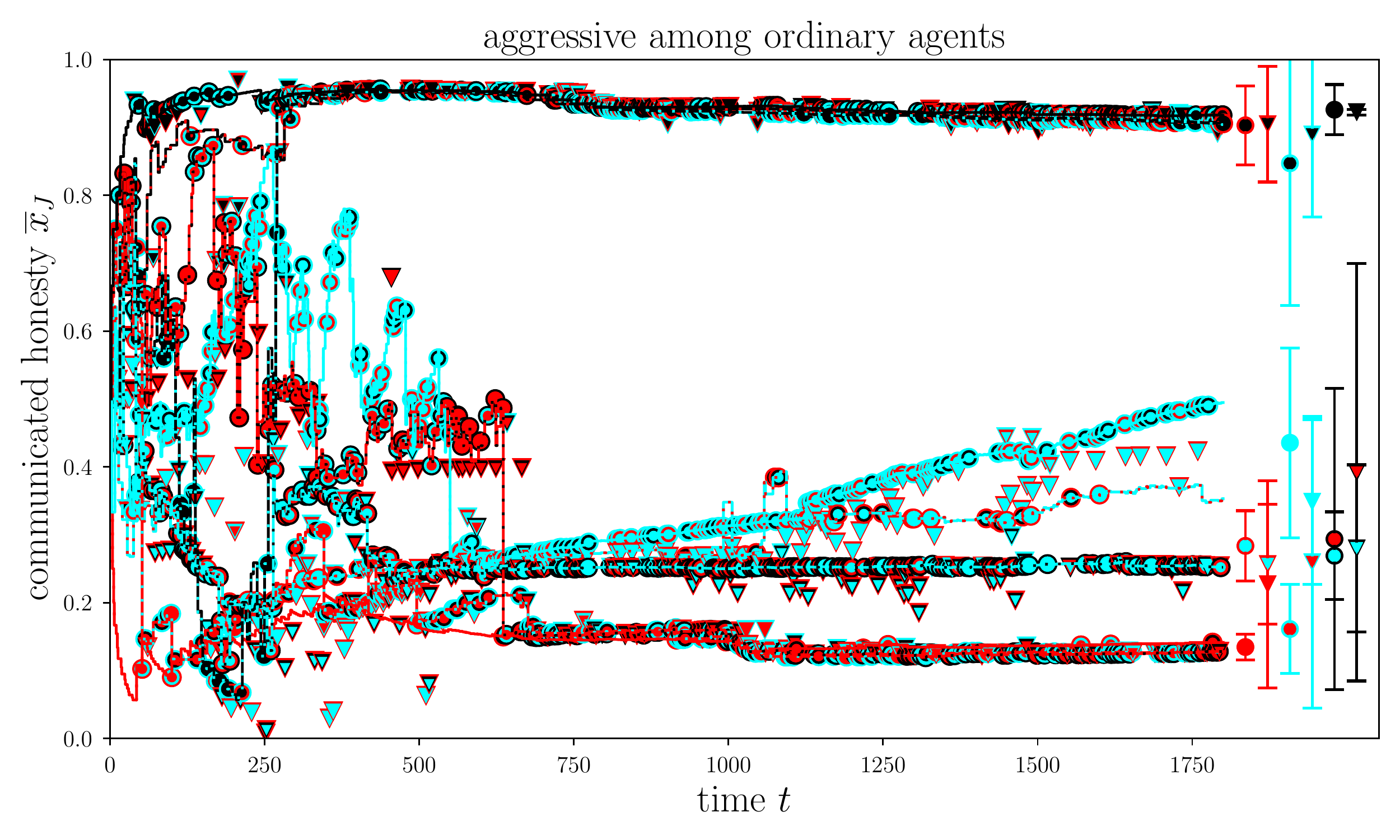}\includegraphics[width=0.5\textwidth]{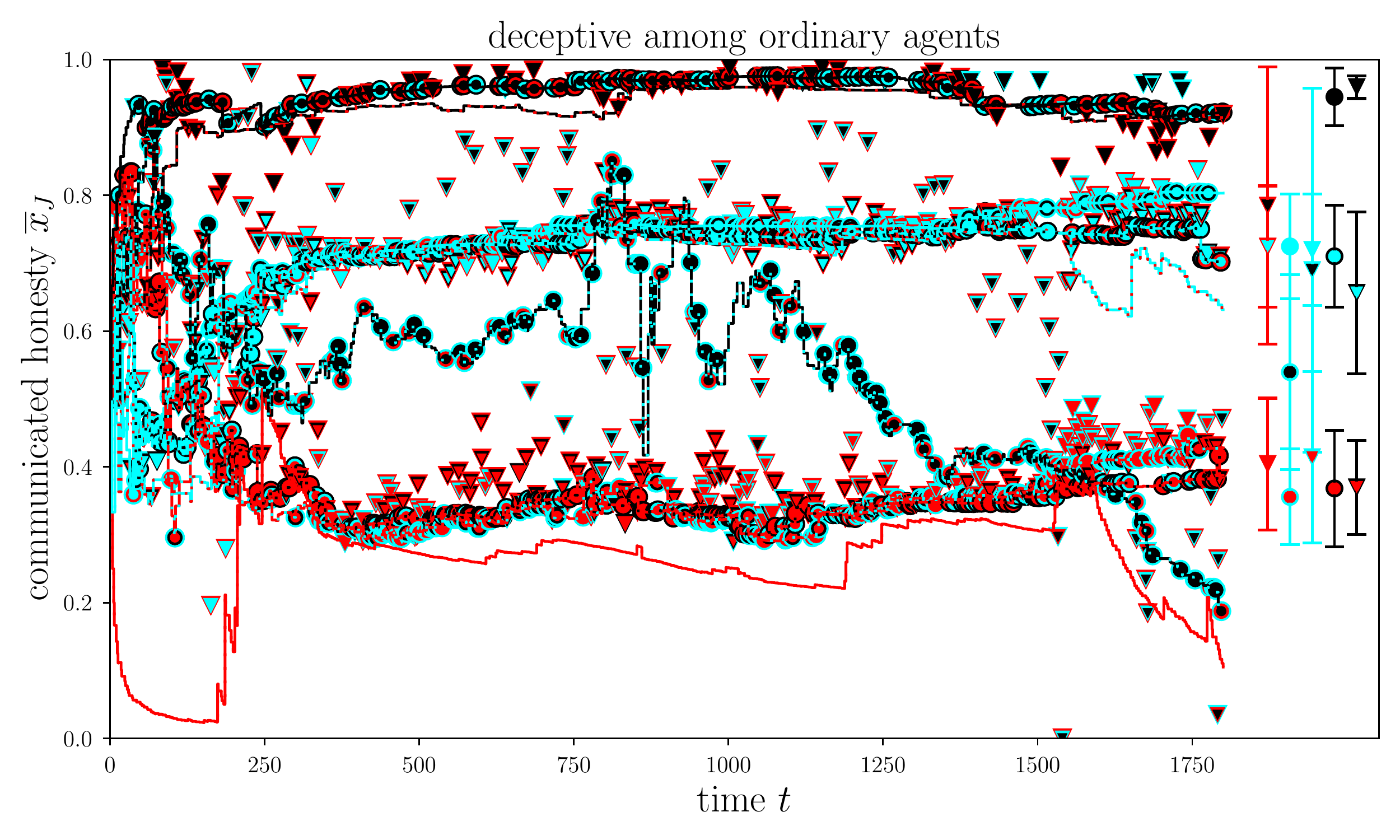}

\caption{Communication patterns as in Fig.\ \ref{fig:Communication-patterns}
of the simulations of basic communication strategies with agent red
being here strategic (top left), egocentric (top right), and flattering
(middle left), shameless (middle right), aggressive (bottom left),
and deceptive (fourth row). The used random sequences No.\ 1 are
also identical to the simulations shown in Fig.\ \ref{fig:Communication-patterns}.
\label{fig:Basic-communication-strategies-2}}
\end{figure*}

\section{Detailed communication strategies\label{sec:Detailed-communication-strategies}}

As a supplement to Sec.\ \ref{sec:Communication-strategies}, the
basic communication strategies are explained more rigorously here
once again.

The \textbf{ordinary agent} $a$ picks the communication partner $b$
randomly and uniformly from all other agents, $b\hookleftarrow\mathcal{A}\backslash\{a\}$,
the topic $c$ randomly and uniformly from all agents, $c\hookleftarrow\mathcal{A}$,
communicates honestly with the frequency $x_{a}$, promotes friends
and demotes enemies when lying, and uses a critical receiver strategy.

The \textbf{strategic agent} $a$, however, picks communication partners
according to their reputation, by setting 
\begin{equation}
P(a\overset{\cdot}{\rightleftarrows}b|a\,\text{strategic})=\frac{\overline{x}_{ab}}{\sum_{b\in\mathcal{A}\backslash\{a\}}\overline{x}_{ab}}\,(1-\delta_{ab}).
\end{equation}
By concentrating communications on presumably reputed, if not even
really honest agents, the strategic agent's opinions, if adapted by
$b$, might propagate more efficiently into third agents. This is
because the communicated opinion benefits from the reputed agents
being more influential and the higher frequency with which honest
agents express their true beliefs. Strategic agents therefore target
optimal multipliers for their communications. Being strategic will
be part of the dominant and the destructive strategies.

We call an agent preferring low reputed agents as communication partners
an \textbf{anti-strategic}\footnote{An \emph{anti-strategic} agent is also ``strategic'' in the original
sense of the word, similar to an \emph{anti-particle}, which actually
is a \emph{particle}, or an \emph{anti-correlation}, which is also
a \emph{correlation.} } agent: 
\begin{equation}
P(a\overset{\cdot}{\rightleftarrows}b|\,a\,\text{anti-strategic})=\frac{(1-\overline{x}_{ab})}{\sum_{b\in\mathcal{A}\backslash\{a\}}(1-\overline{x}_{ab})}\,(1-\delta_{ab}).
\end{equation}

Being anti-strategic may pay off for \textbf{flattering agents}, who
always lie positively when their conversation partner is the topic,
and pick the conversation partner as topic whenever they have the
opportunity to initiate a conversation,
\begin{equation}
P(a\overset{c}{\rightleftarrows}b|a\overset{\cdot}{\rightleftarrows}b,\,a\text{ flattering})=\delta_{bc}.
\end{equation}
Flattering agents should be efficient in befriending others. Being
an agent $b$'s friend pays off for $a$ whenever $b$ lies about
$a$. Thus, flattering agents are best advised to be anti-strategic
as well, in order to ensure the friendship of the most frequent liars
they can identify. Being flattering and anti-strategic will therefore
be part of the manipulative strategy.

\textbf{Egocentric agents} prefer to speak about themselves. In half
of the cases in which they initiate a conversation, they directly
pick themselves as topic, in the other half, they pick randomly and
uniformly from the set of all agents $\mathcal{A}$,
\begin{equation}
P(a\overset{c}{\rightleftarrows}b|a\overset{\cdot}{\rightleftarrows}b,\,a\text{ egocentric})=\frac{1}{2}\left(\delta_{ac}+\frac{1}{n}\right).
\end{equation}
Thus, they present themselves in more than half of the conversations
they initiate. Egocentric agents can benefit from being strategic,
as this should increase their reach. For this reason, being also egocentric
in addition to being strategic will be part of the dominant strategy.

\textbf{Aggressive agents} only speak about enemies when initiating
a conversation,
\begin{equation}
P(a\overset{c}{\rightleftarrows}b|a\overset{\cdot}{\rightleftarrows}b,\,a\text{ aggressive})=\frac{\delta_{c\in\mathcal{E}_{a}}}{|\mathcal{E}_{a}|},
\end{equation}
and neither praise friends nor themselves. The aggressive agent's
destructiveness w.r.t.\ other agents' reputations can unfold best
if the agent is also strategic and therefore the destructive agent
will be both, aggressive and strategic.

A\textbf{ shameless agent} $a$ lies without blushing, which gives
a clear advantage if lying frequently and we will assume the destructive
agent also to be shameless.

Finally, a (fully) \textbf{deceptive agent} $a$ lies without exception,
$x_{a}=0$, and therefore does not risk to make any confession or
to be caught lying due to contradictions between expressed true beliefs
and lies.

All special agents, the clever, the manipulative, the dominant, and
the destructive agent, are combinations of different basic strategies
as explained in Sect. \ref{sec:Communication-strategies}. Thus, since
the single basic strategies that have been put together never contradict
each other, the basic characteristics presented above can simply be
combined for all remaining special strategies. For an overview, see
Tab.\ \ref{tab:Summary-of-agent's-strategies}.

\section{Detailed figures\label{sec:Detailed-figures}}

We show here a number of figures that permit the inspection of further
details of the simulation runs, but which are too crowded to be discussed
in the main text.

\begin{figure*}[!t]
\vspace{-1.5em}
\includegraphics[width=0.5\textwidth]{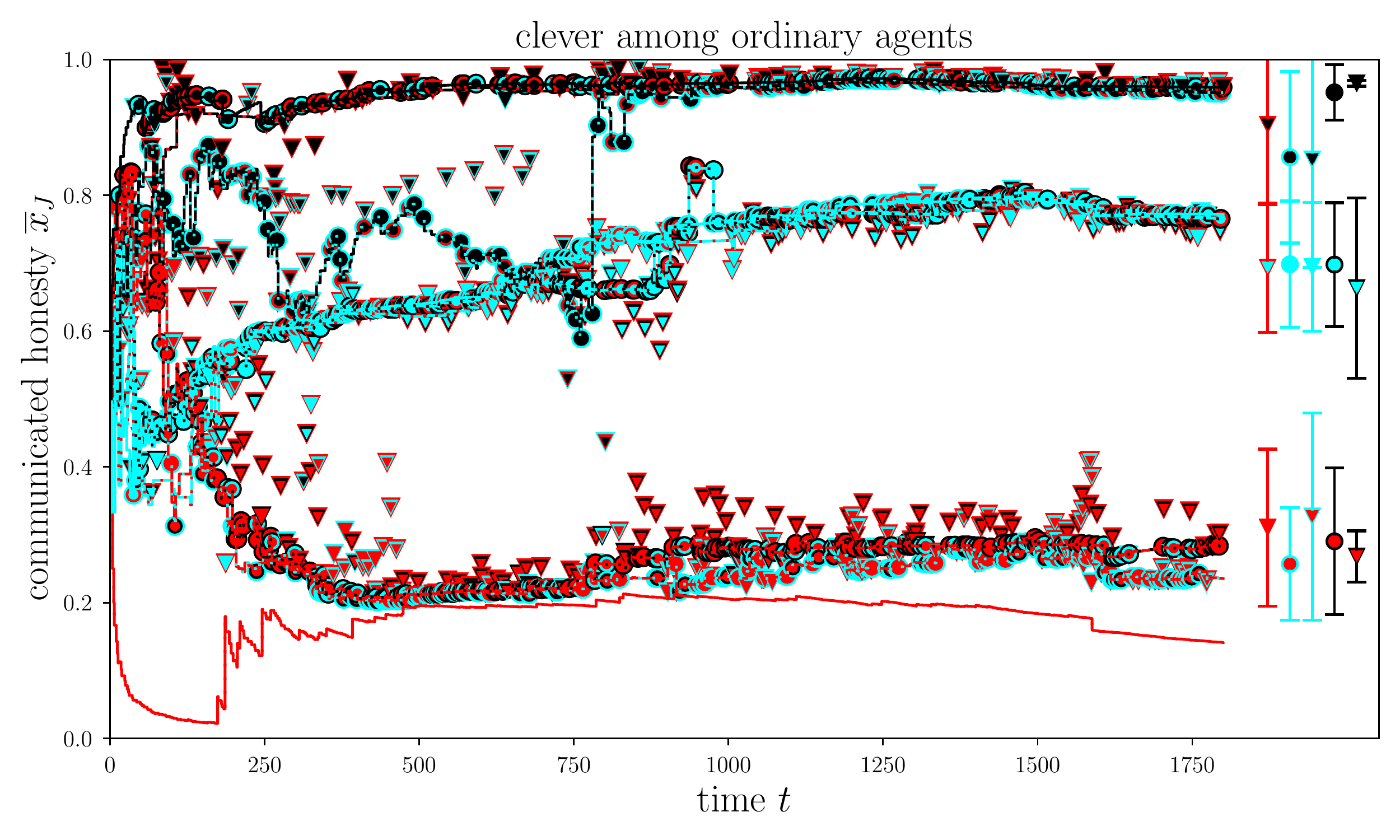}\includegraphics[width=0.5\textwidth]{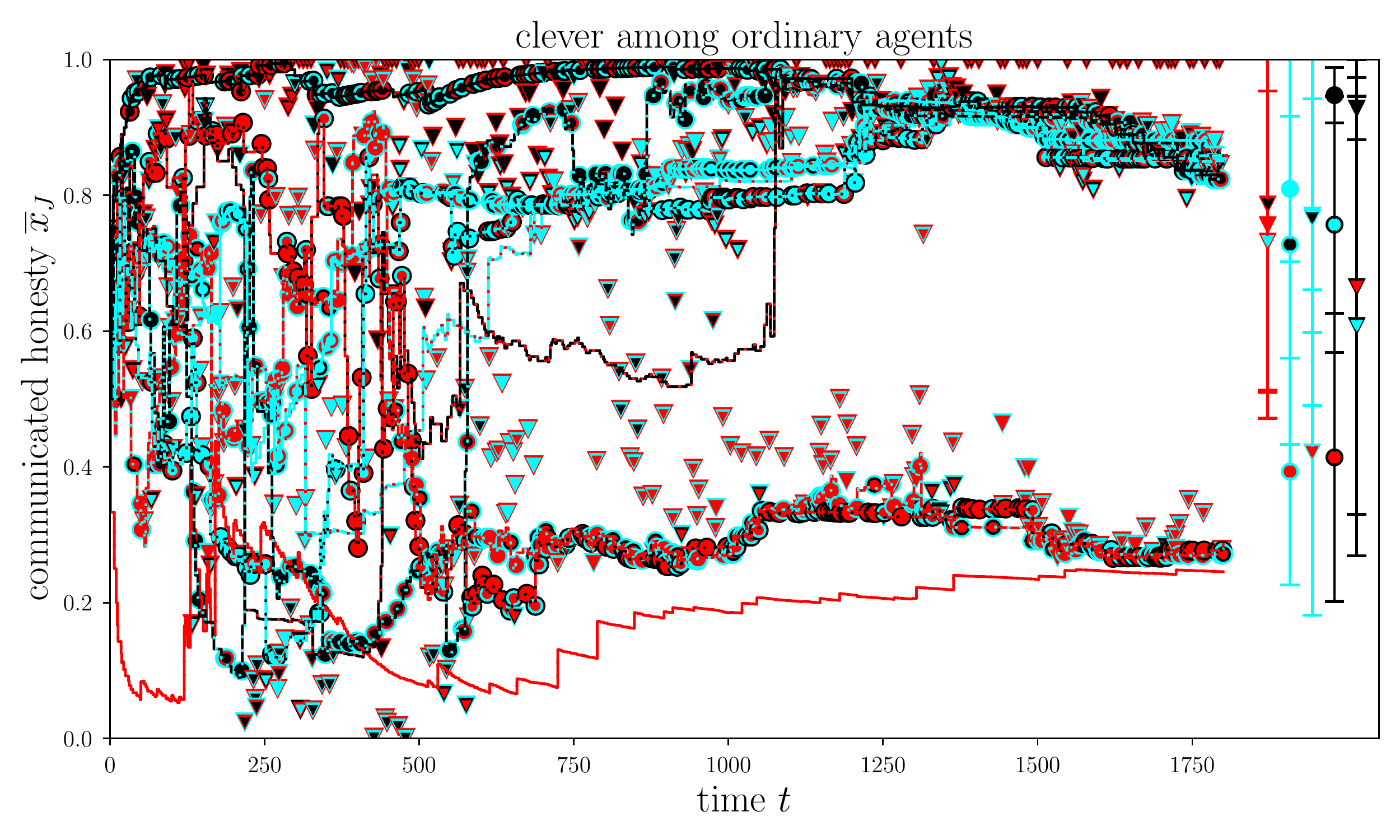}

\includegraphics[width=0.5\textwidth]{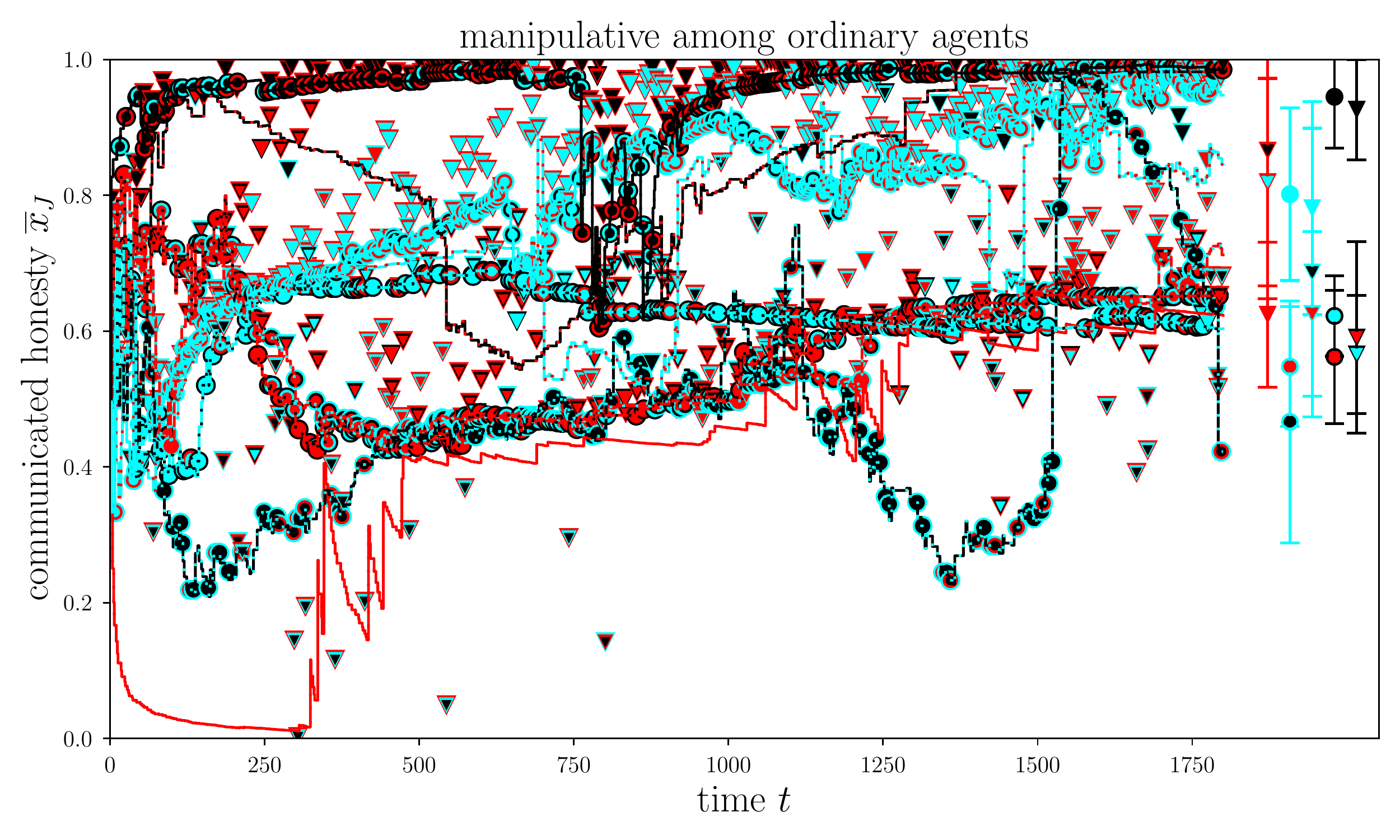}\includegraphics[width=0.5\textwidth]{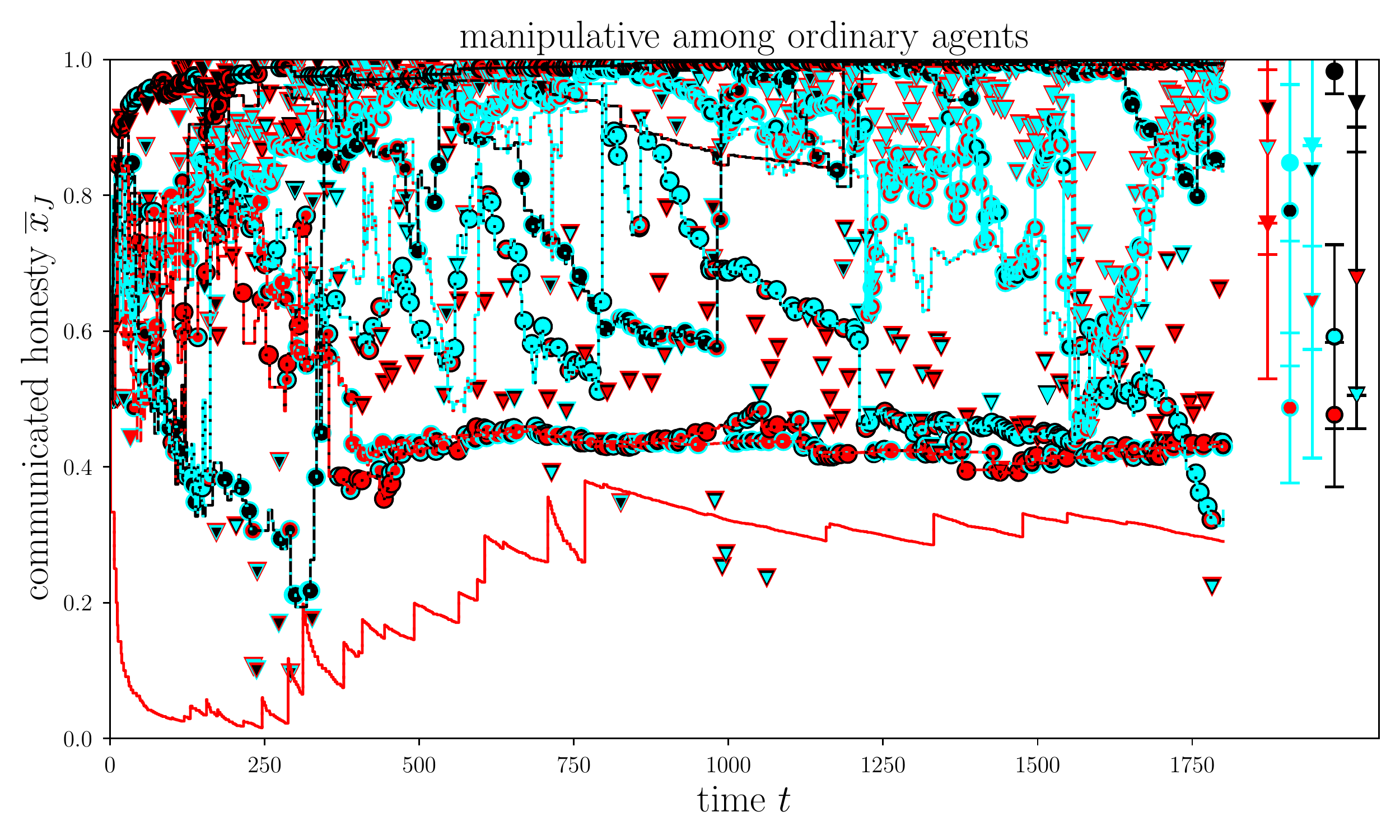}

\includegraphics[width=0.5\textwidth]{figures/14_dominant_NA3_RS0_comL}\includegraphics[width=0.5\textwidth]{figures/14_dominant_NA3_RS1_comL}

\includegraphics[width=0.5\textwidth]{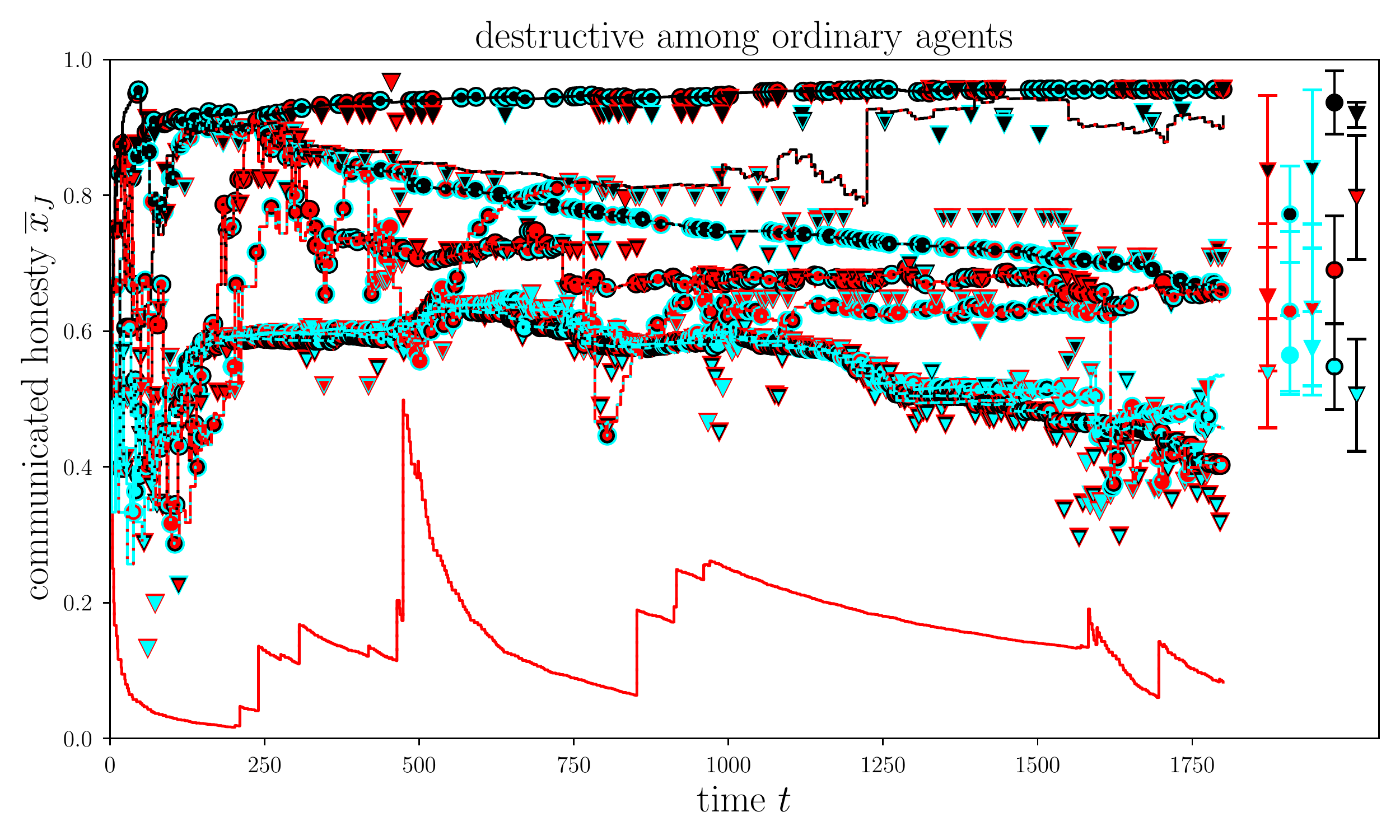}\includegraphics[width=0.5\textwidth]{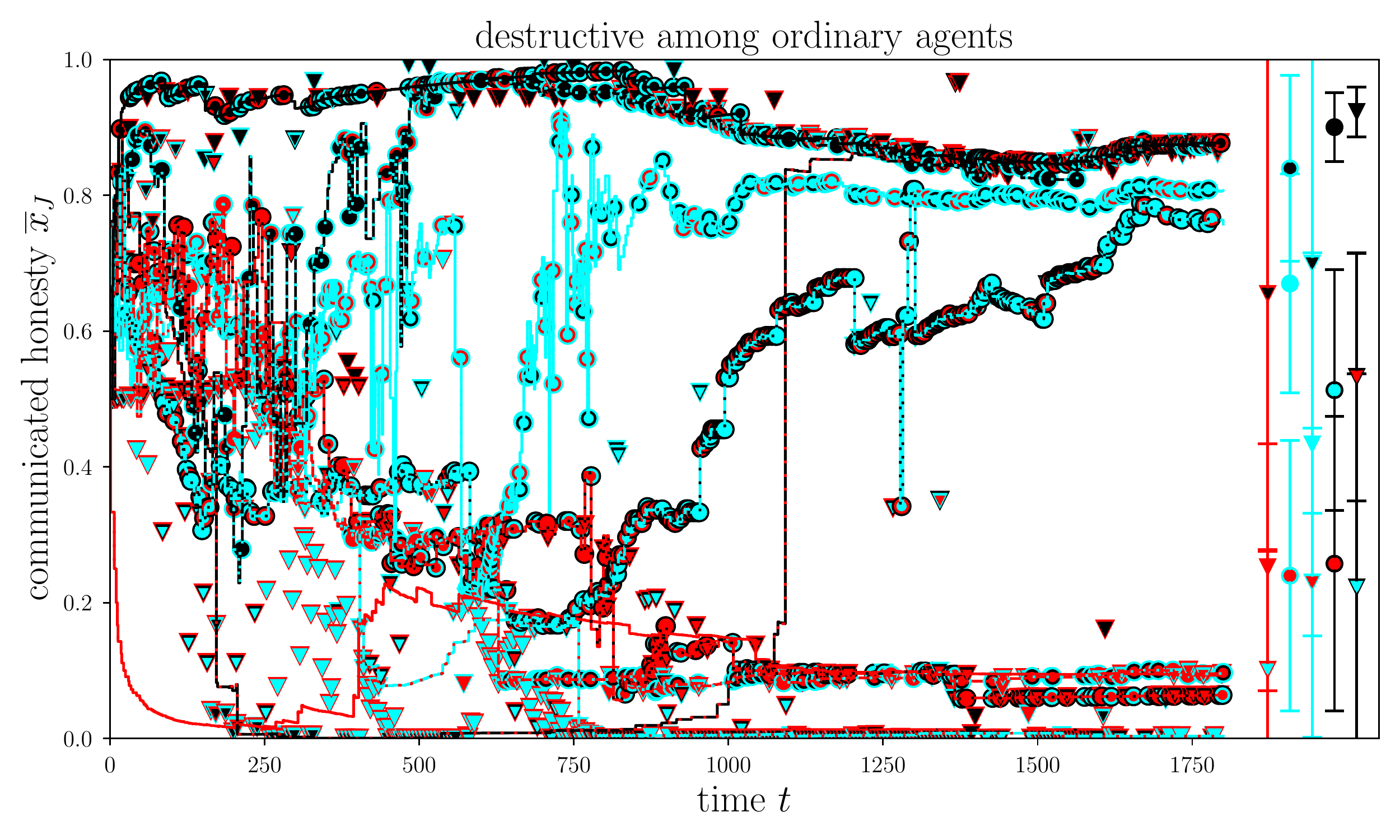}

\caption{As Fig.\ \ref{fig:Basic-communication-strategies-2}, just for agent
red being clever (first row), manipulative (second row), dominant
(third row), and destructive (fourth row). The left column shows simulations
with the random sequences No.\ 1, and the right with No.\ 2. \label{fig:Special-communication-strategies-1}}
\end{figure*}

\begin{figure*}[!t]
\begin{centering}
\includegraphics[width=0.3\textwidth]{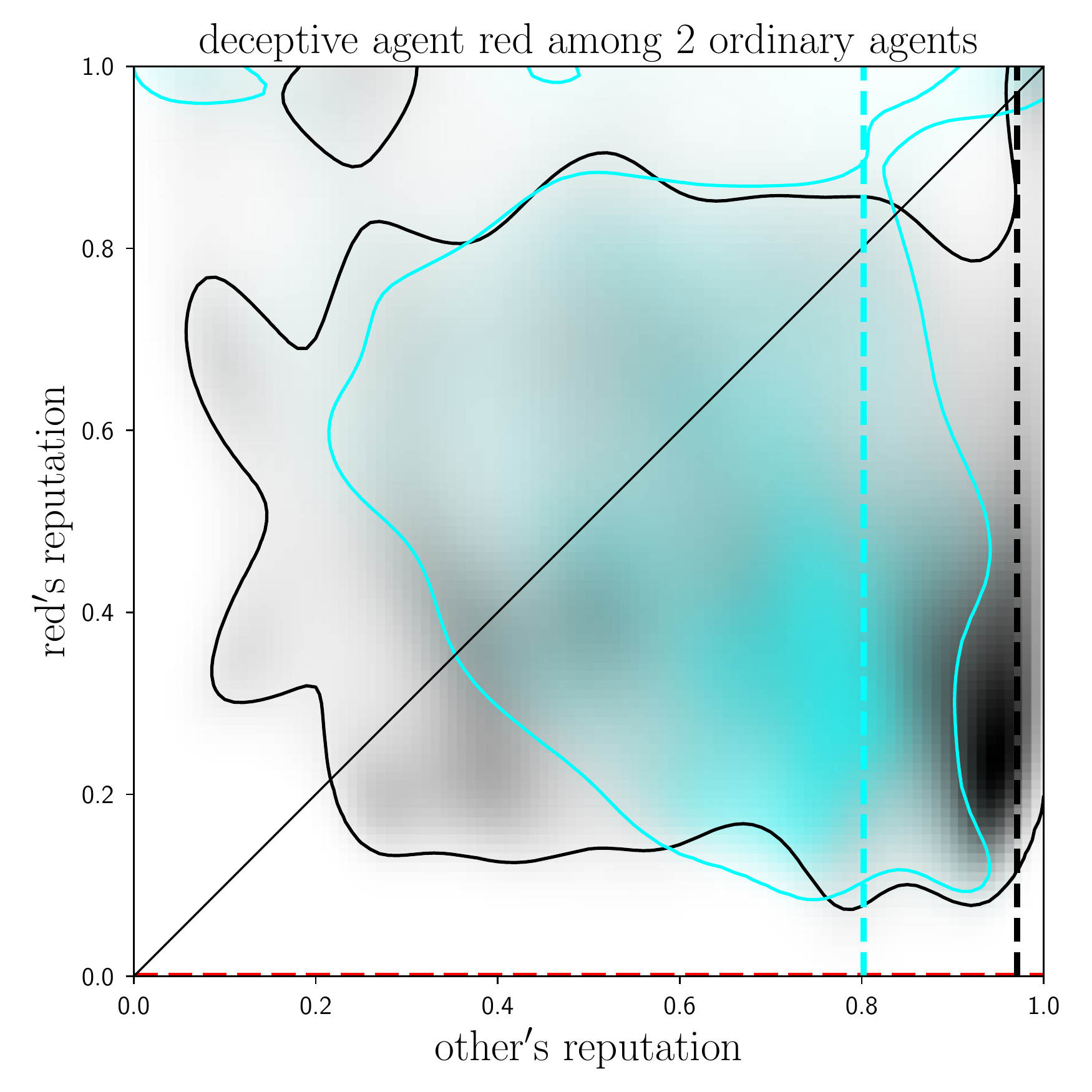}\includegraphics[width=0.3\textwidth]{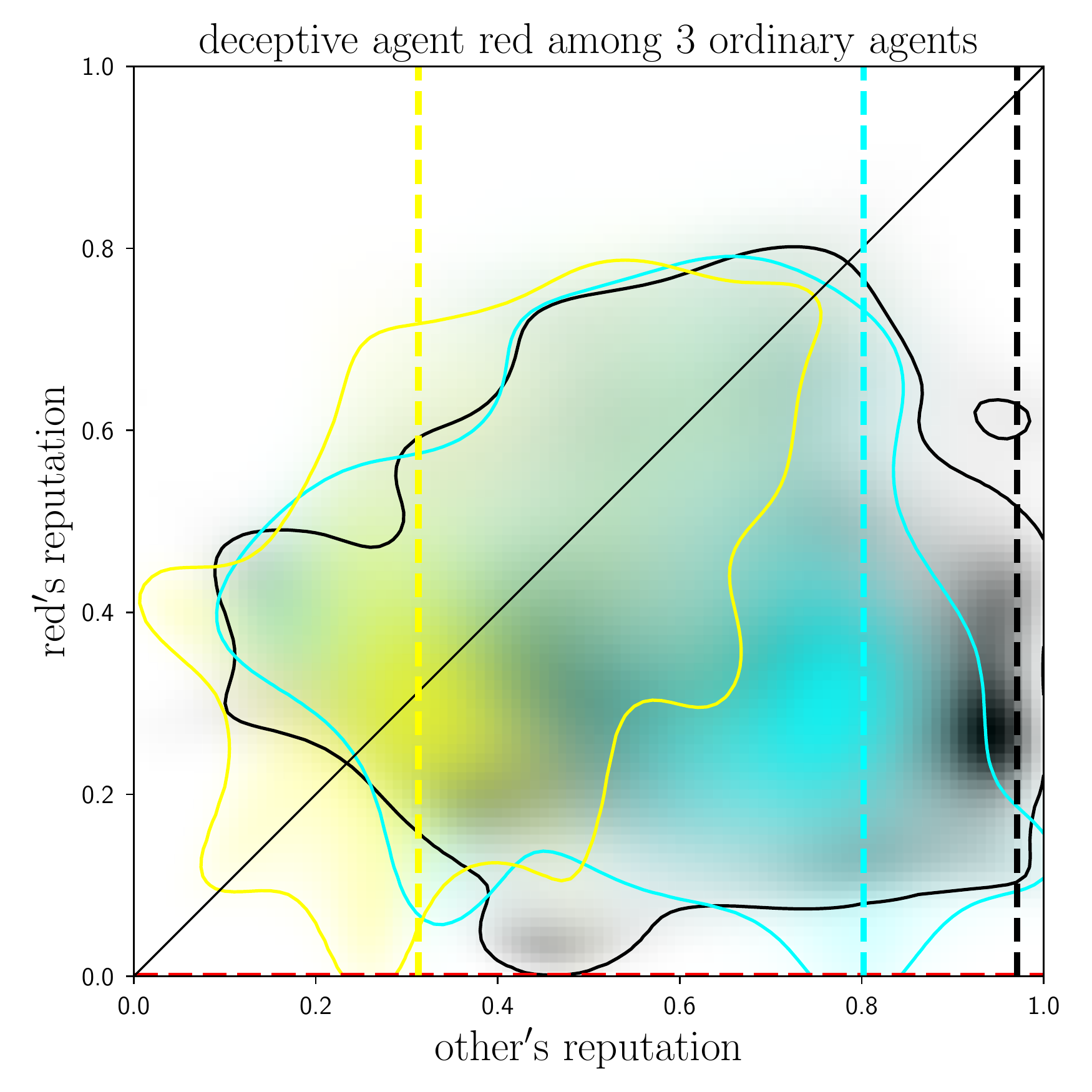}\includegraphics[width=0.3\textwidth]{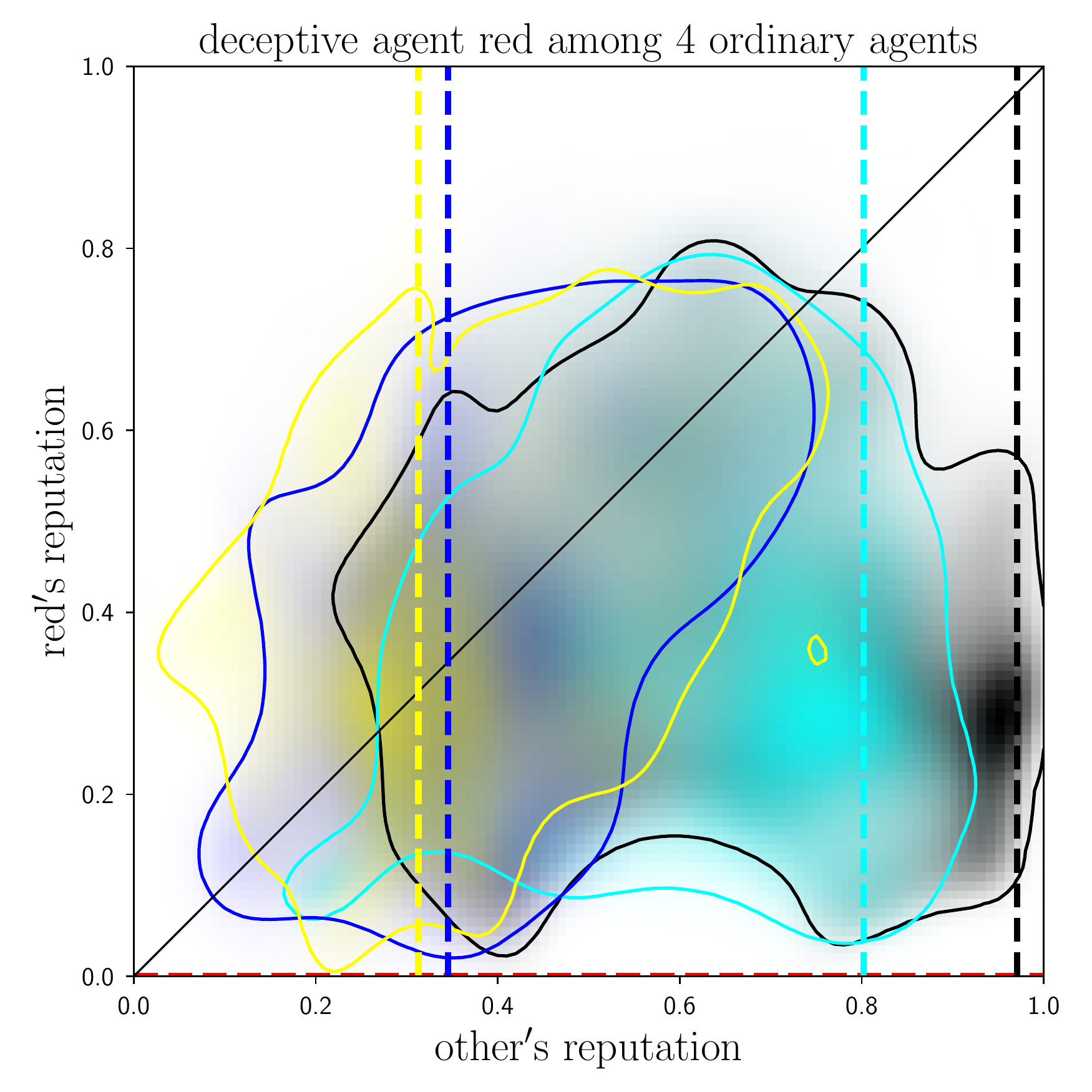}
\par\end{centering}
\begin{centering}
\includegraphics[width=0.3\textwidth]{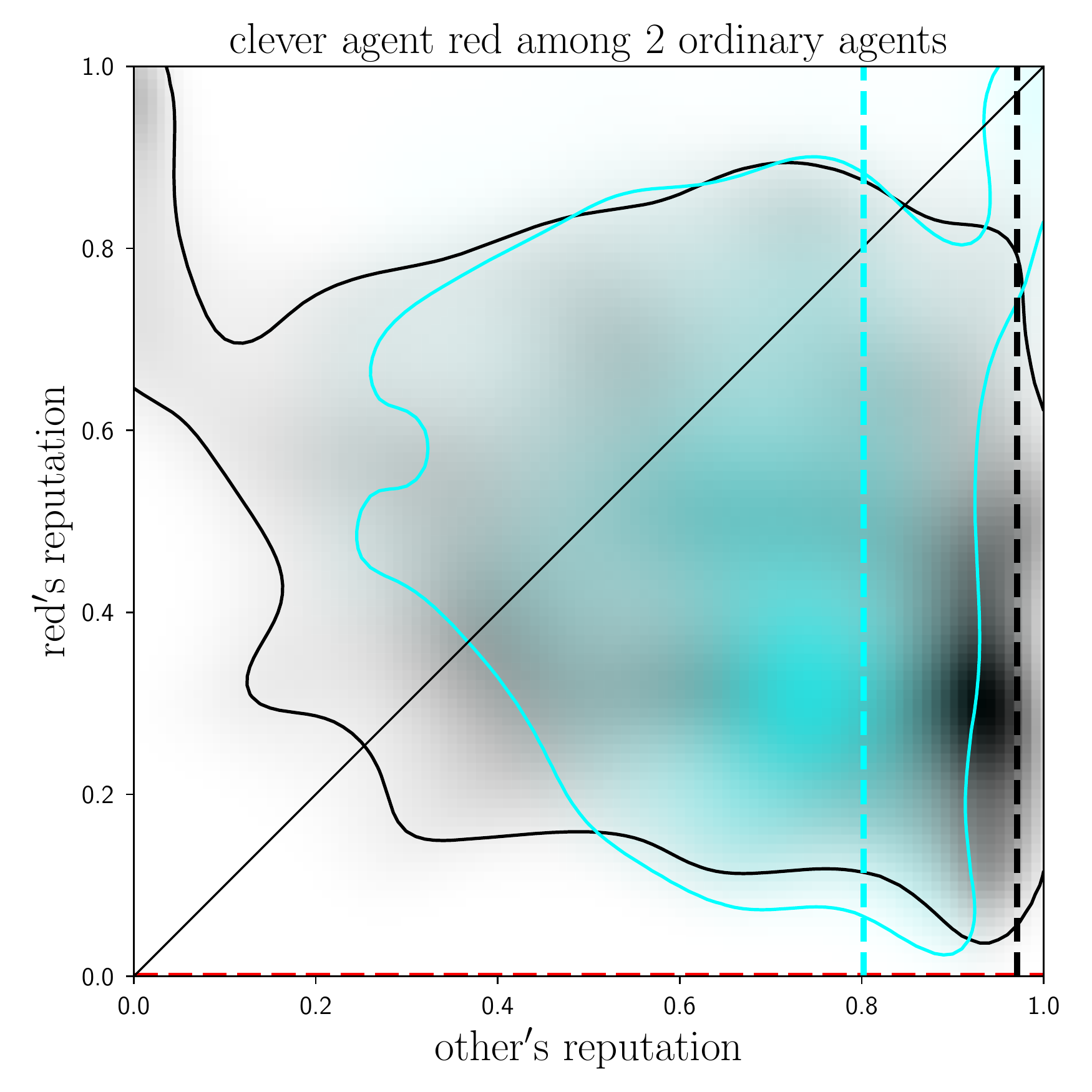}\includegraphics[width=0.3\textwidth]{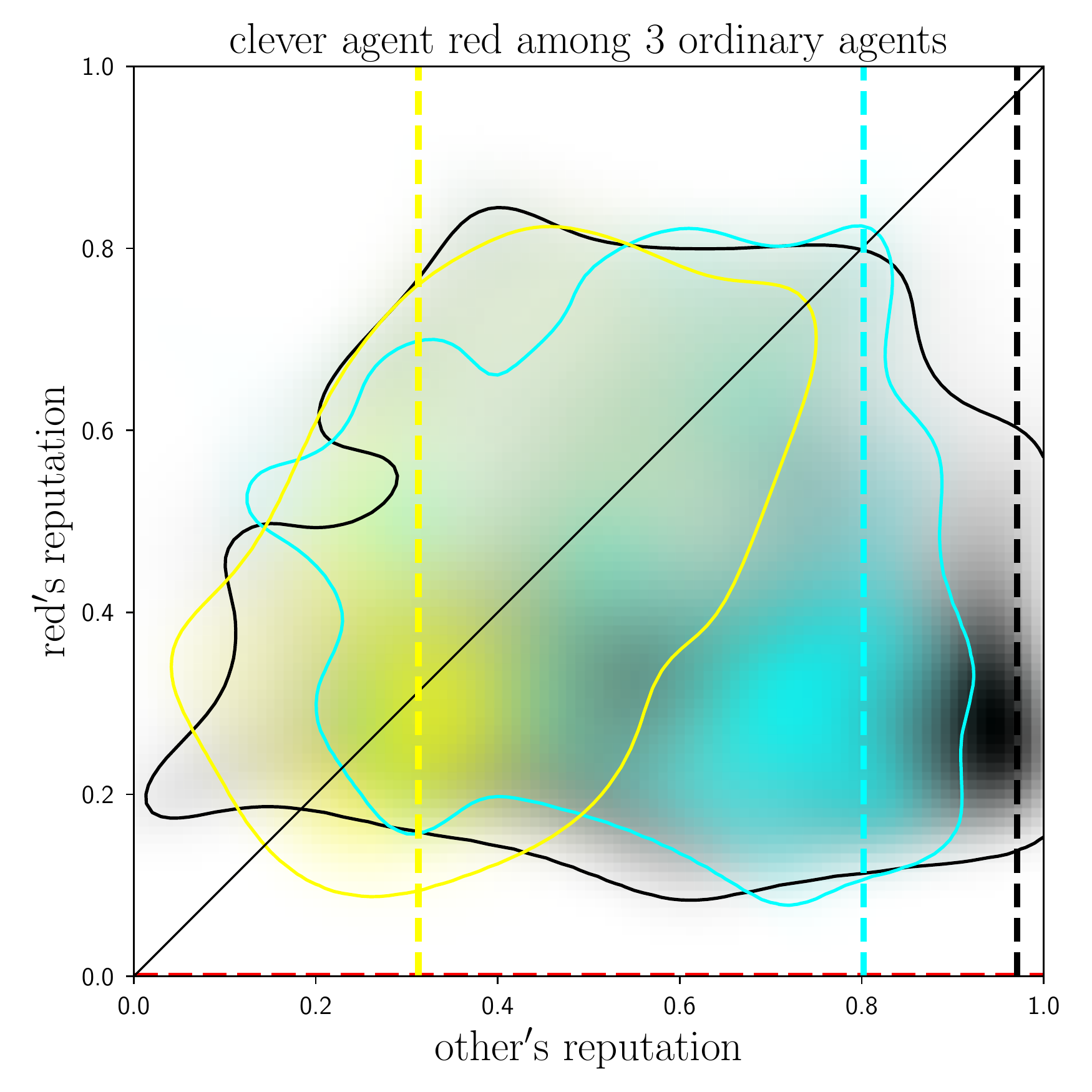}\includegraphics[width=0.3\textwidth]{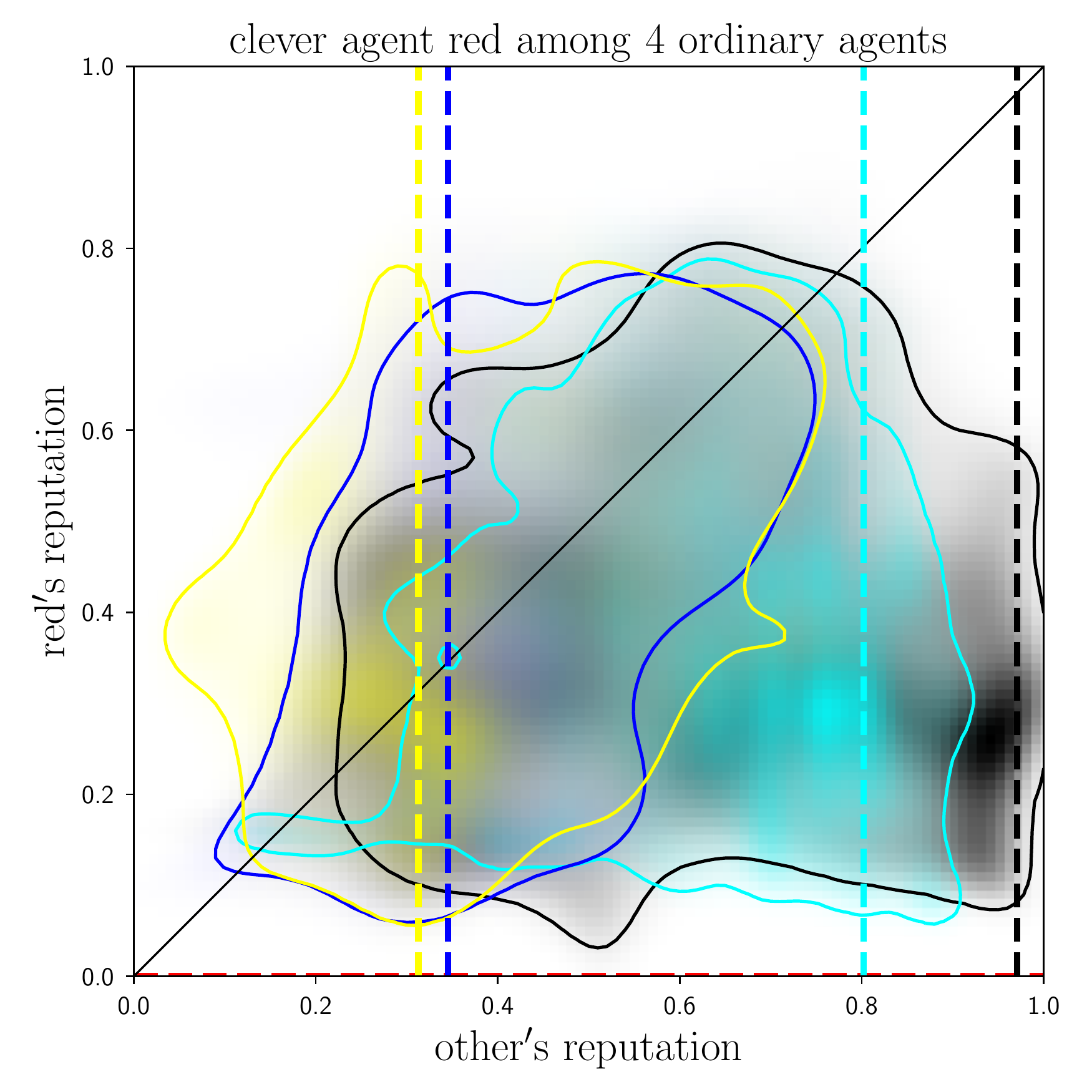}
\par\end{centering}
\begin{centering}
\includegraphics[width=0.3\textwidth]{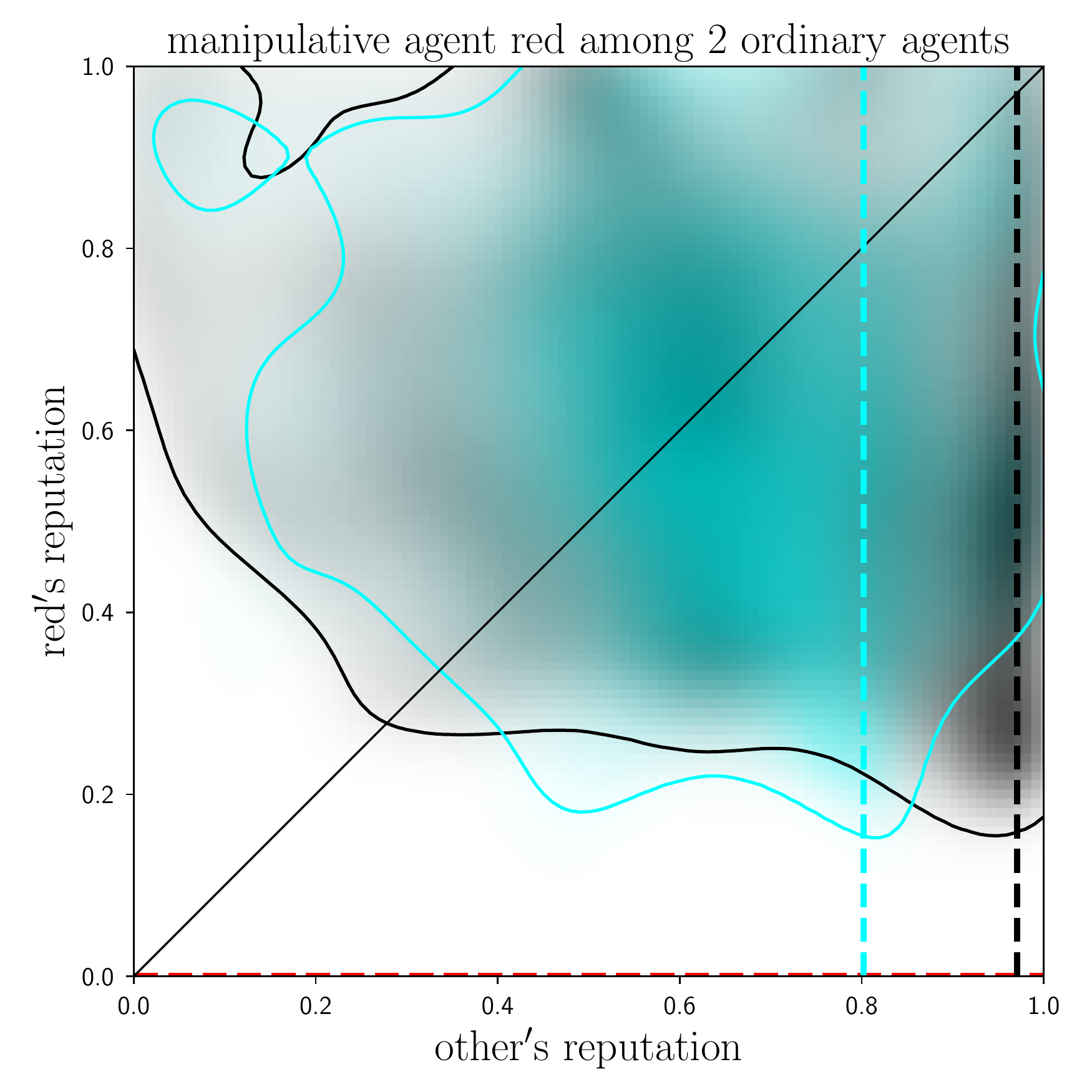}\includegraphics[width=0.3\textwidth]{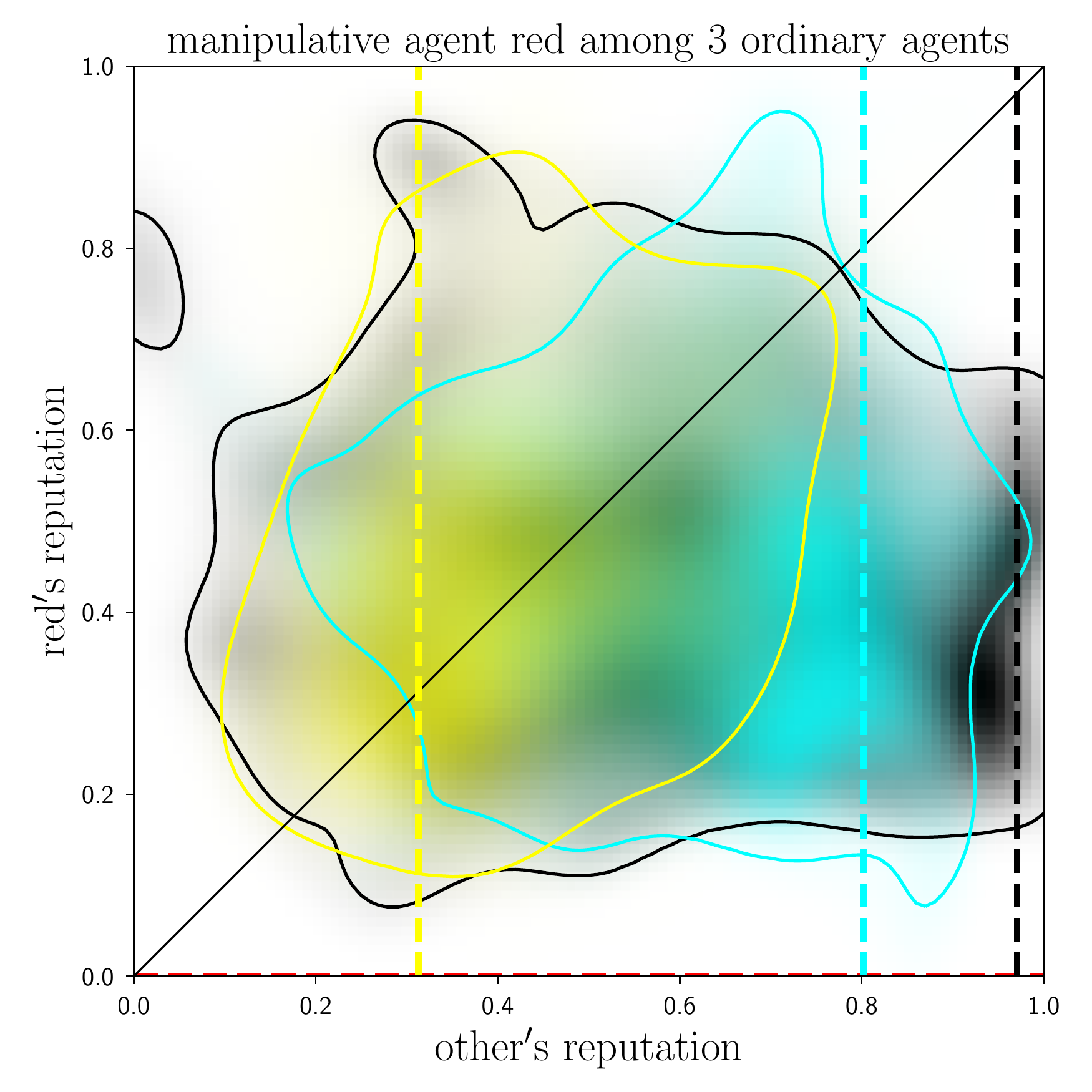}\includegraphics[width=0.3\textwidth]{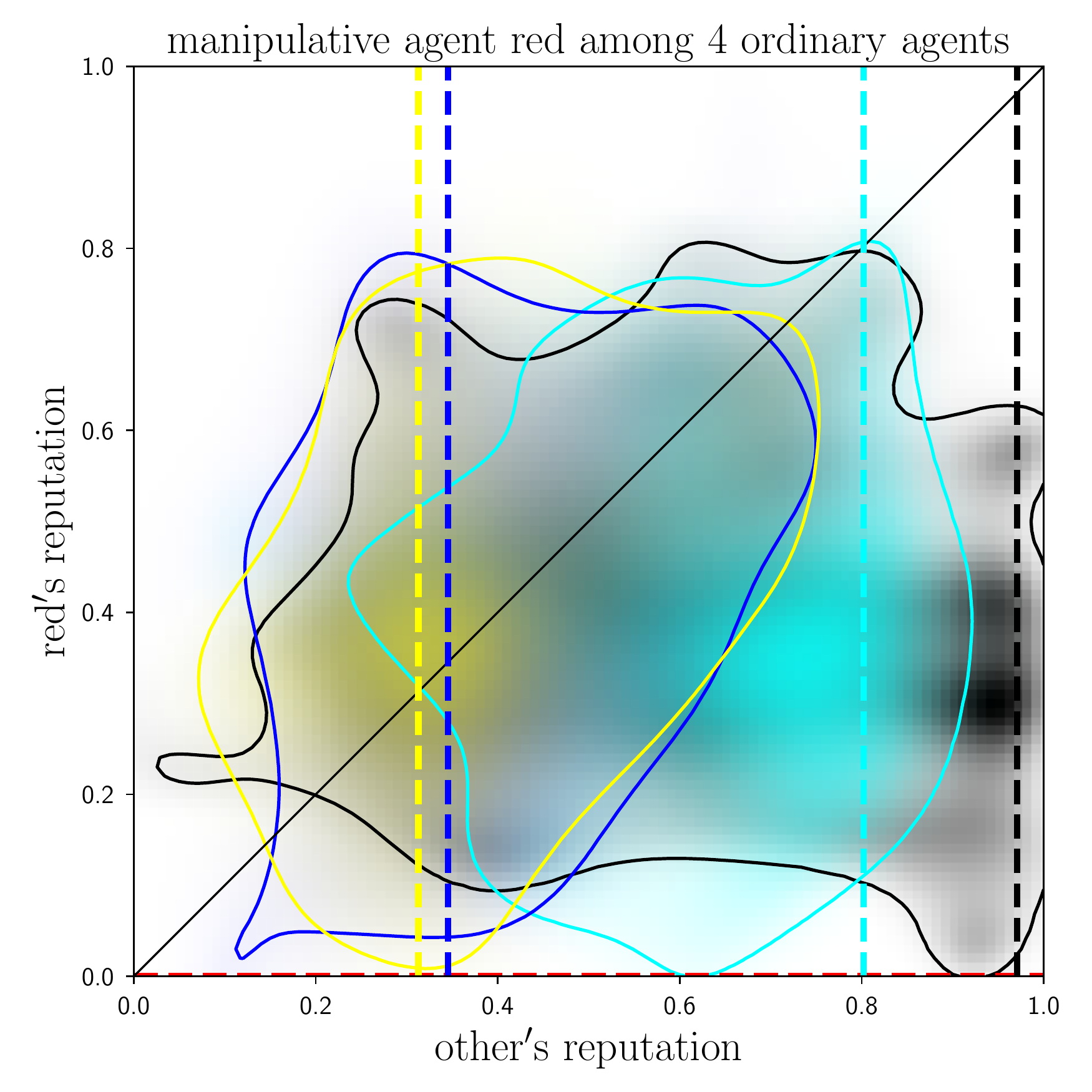}
\par\end{centering}
\begin{centering}
\includegraphics[width=0.3\textwidth]{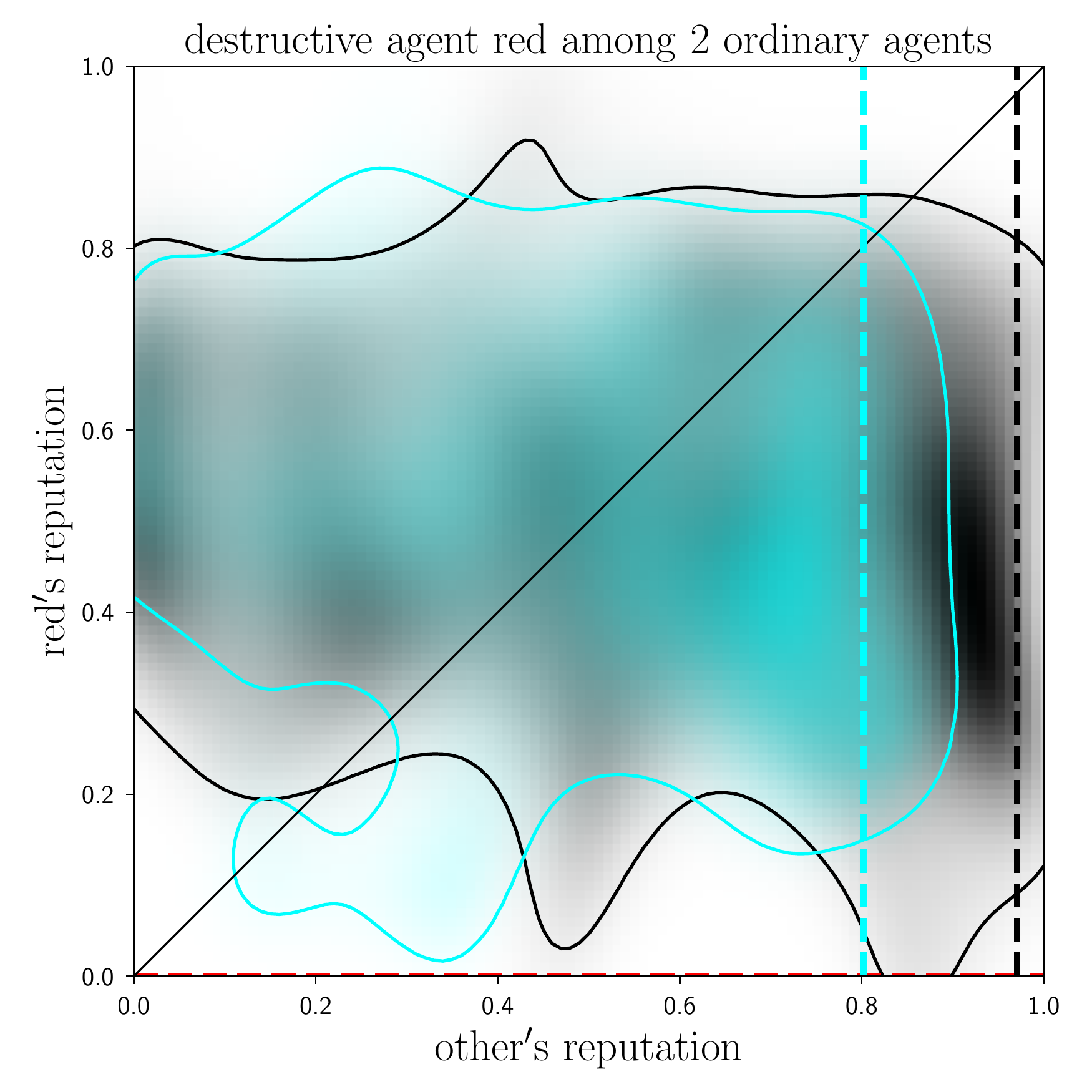}\includegraphics[width=0.3\textwidth]{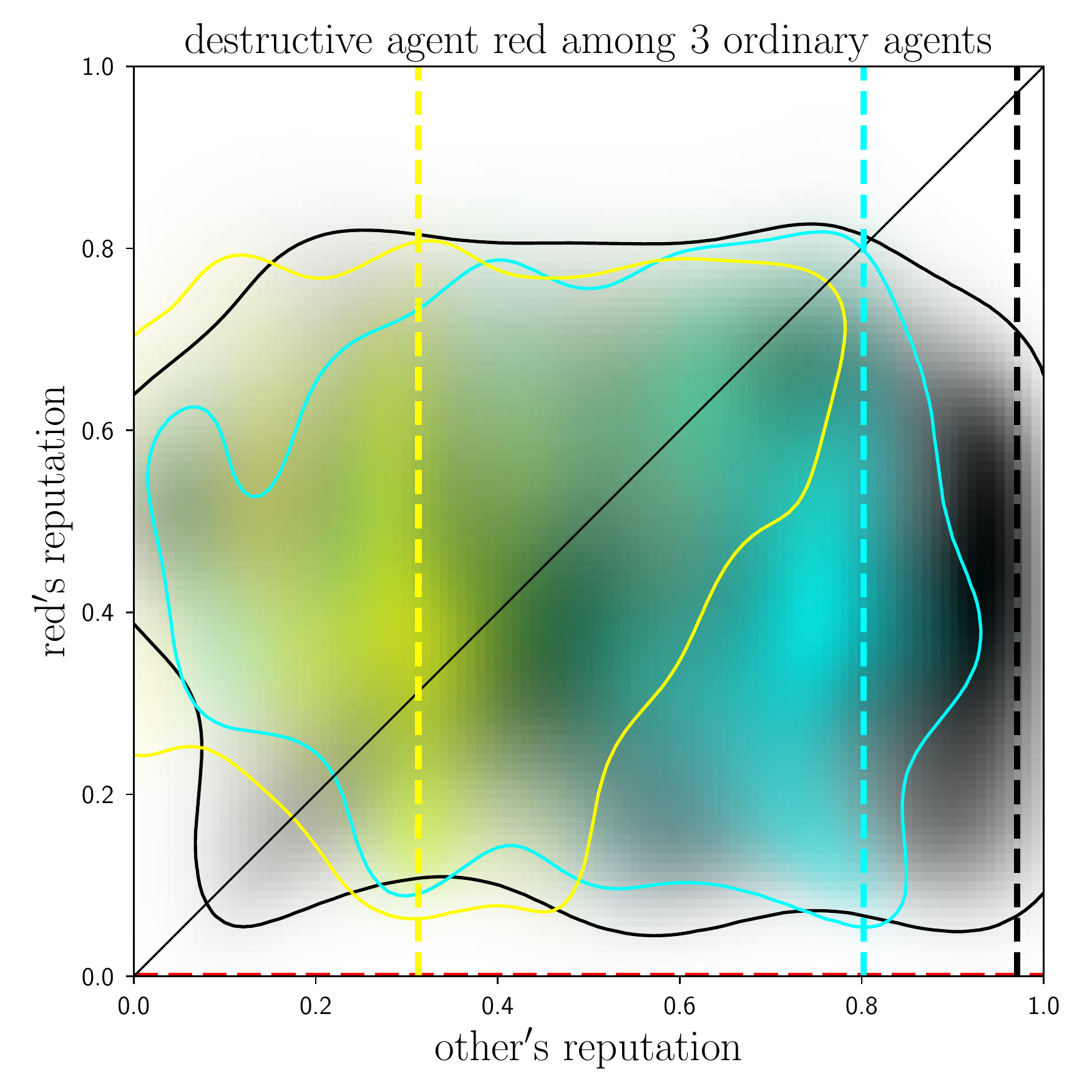}\includegraphics[width=0.3\textwidth]{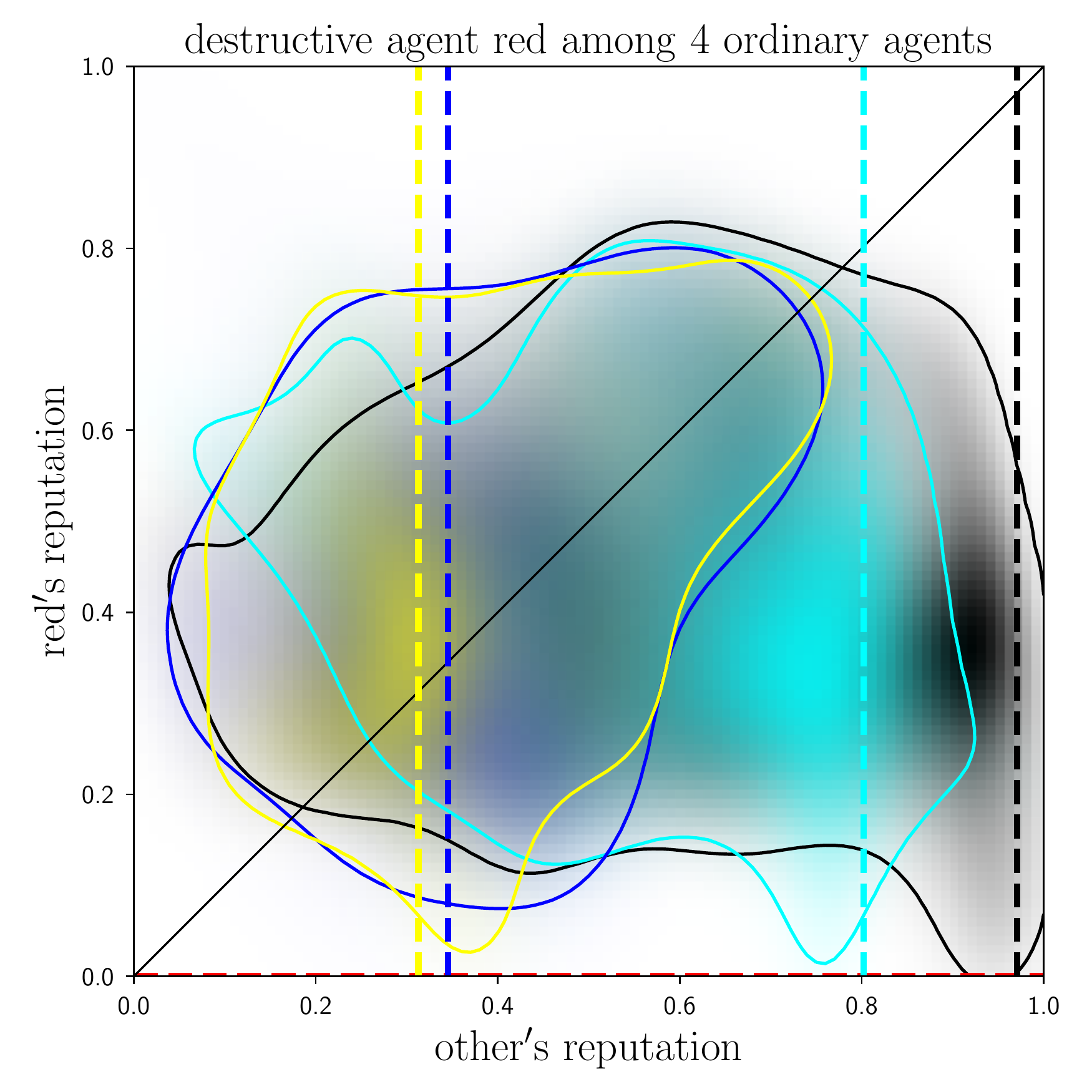}
\par\end{centering}
\caption{Like Fig.\ \ref{fig:Statistics-scatter-3A}, just for agent red being
deceptive, clever, manipulative, and destructive from top to bottom,
respectively. \label{fig:Statistics-scatter-3A-1-2}}
\end{figure*}

Fig.\ \ref{fig:Basic-communication-strategies-2} shows the communication
patterns for the basic communication strategies for random sequences
No.\ 1 and Fig.\ \ref{fig:Special-communication-strategies-1} for
special strategies for random sequences No.\ 1 and No.\ 2. Fig.\ \ref{fig:Statistics-scatter-3A-1-2}
show two dimensional reputation distributions for different agent
types for simulation runs with three to five agents, respectively.

\begin{figure*}[t]
\begin{centering}
\includegraphics[width=0.3\textwidth]{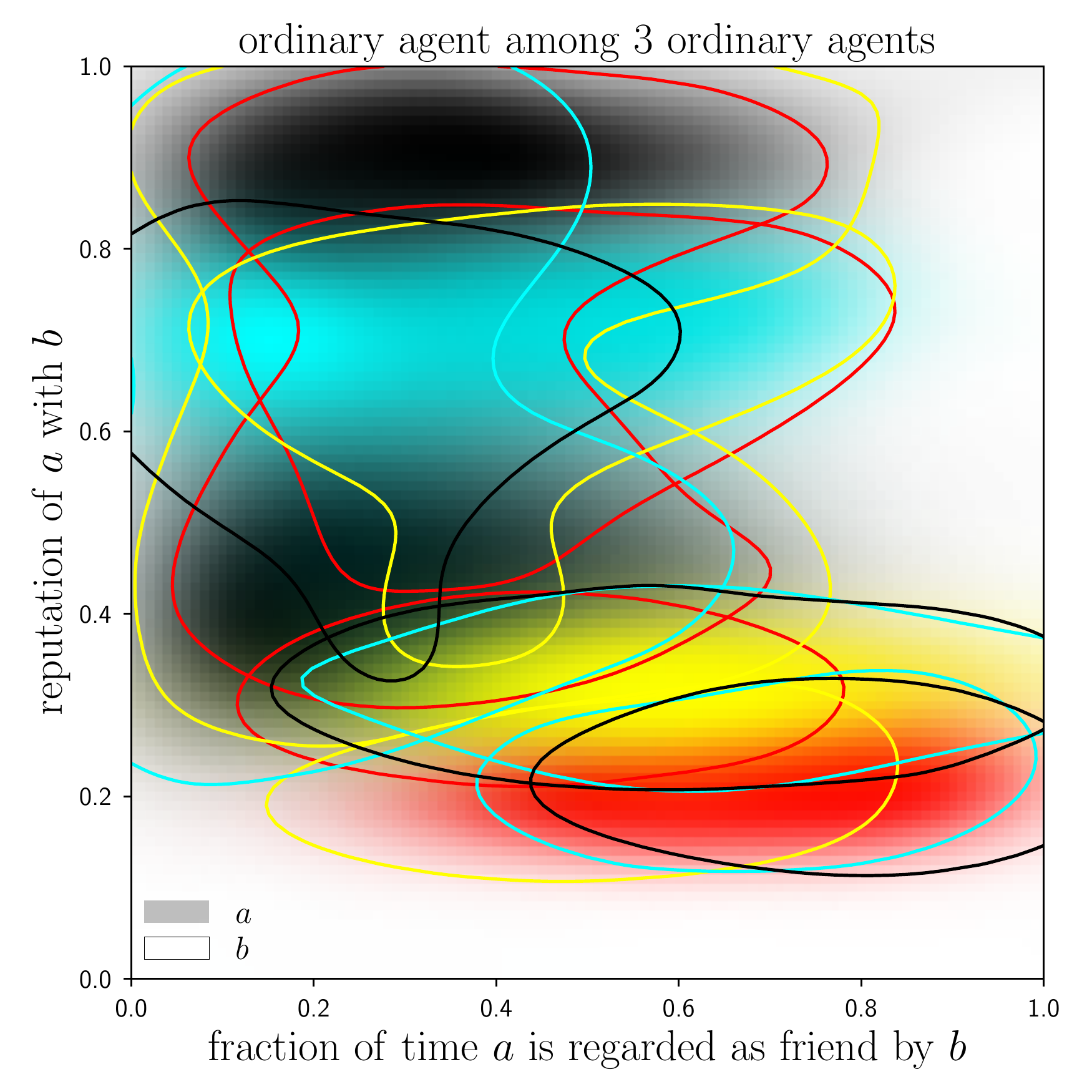}\includegraphics[width=0.3\textwidth]{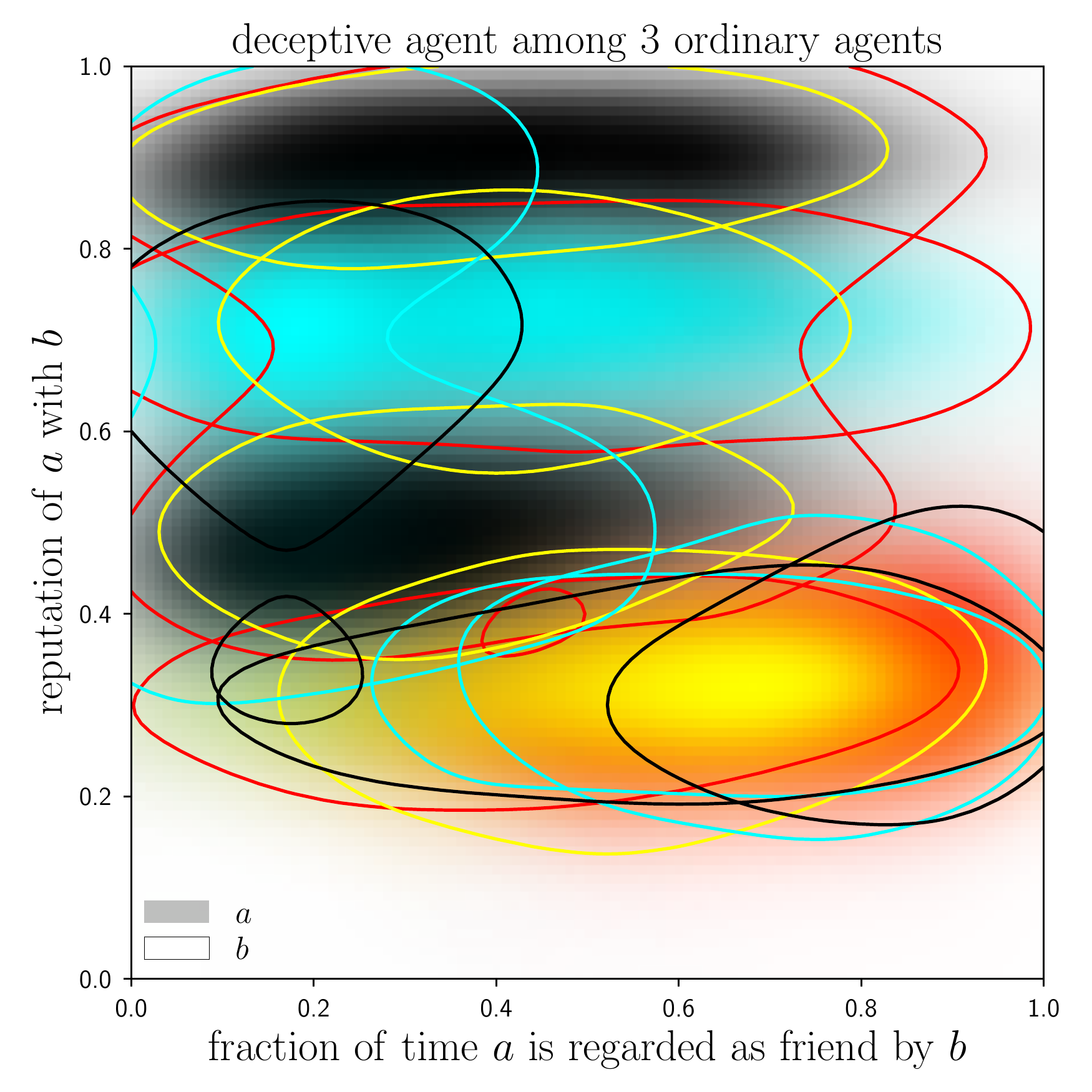}\includegraphics[width=0.3\textwidth]{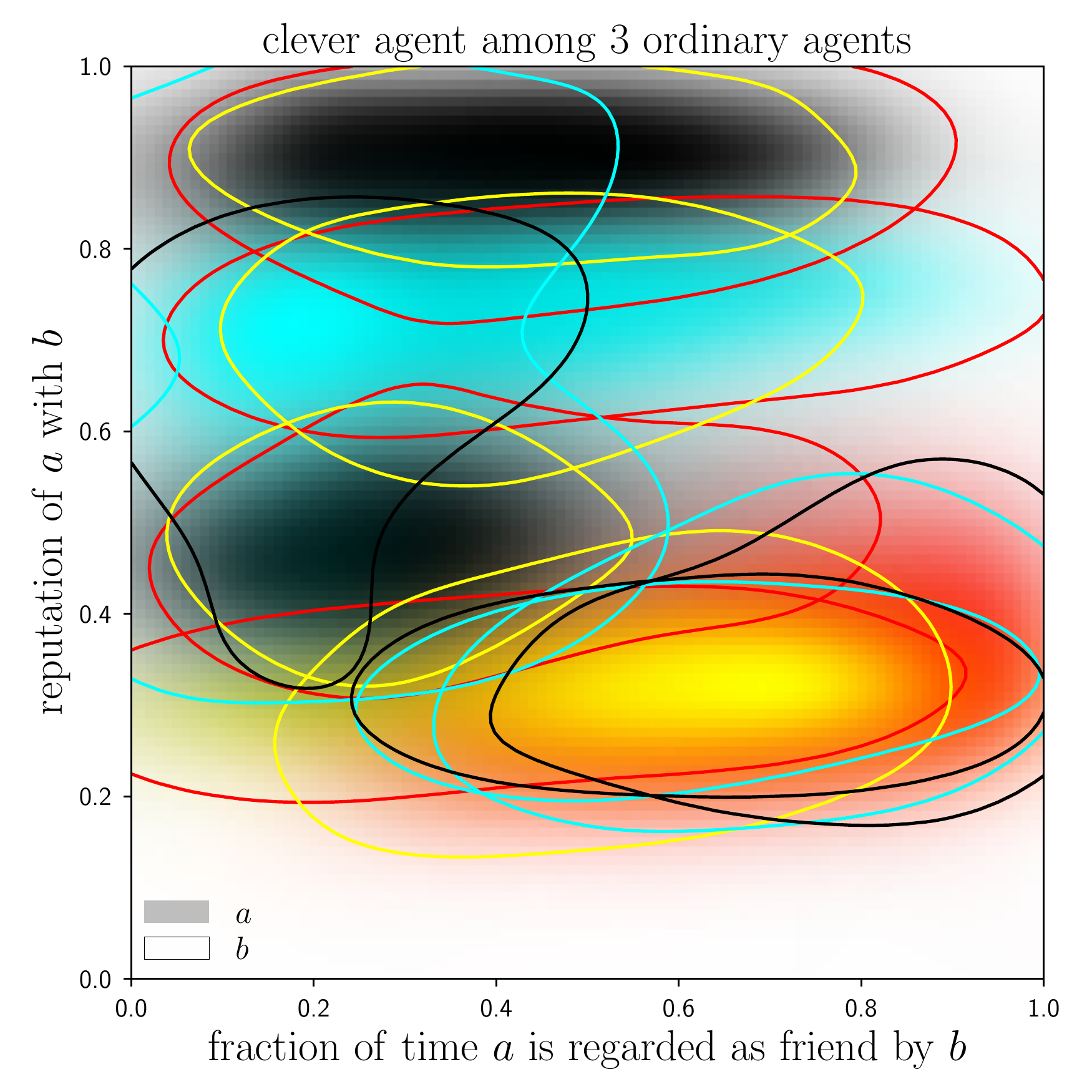}
\par\end{centering}
\begin{centering}
\includegraphics[width=0.3\textwidth]{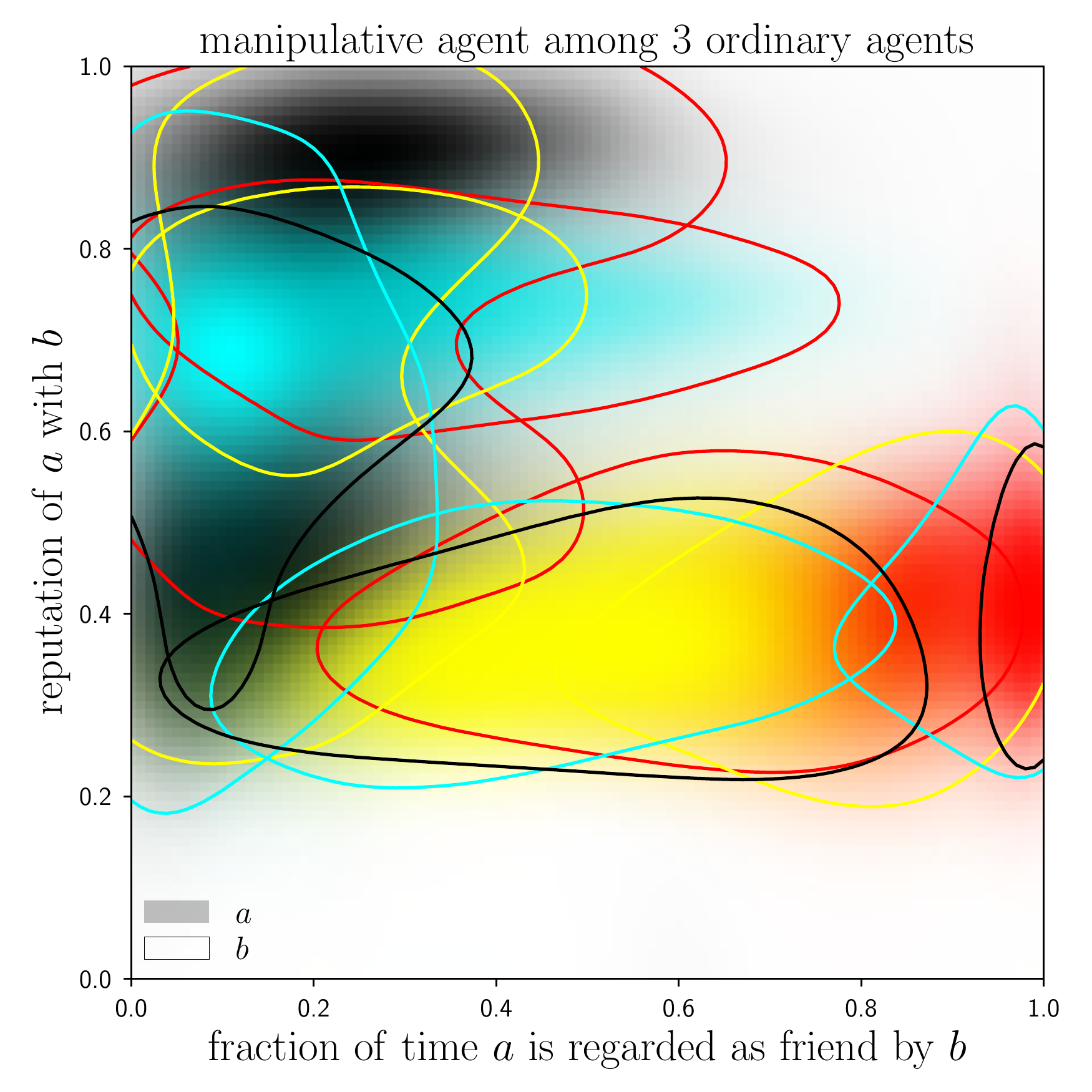}\includegraphics[width=0.3\textwidth]{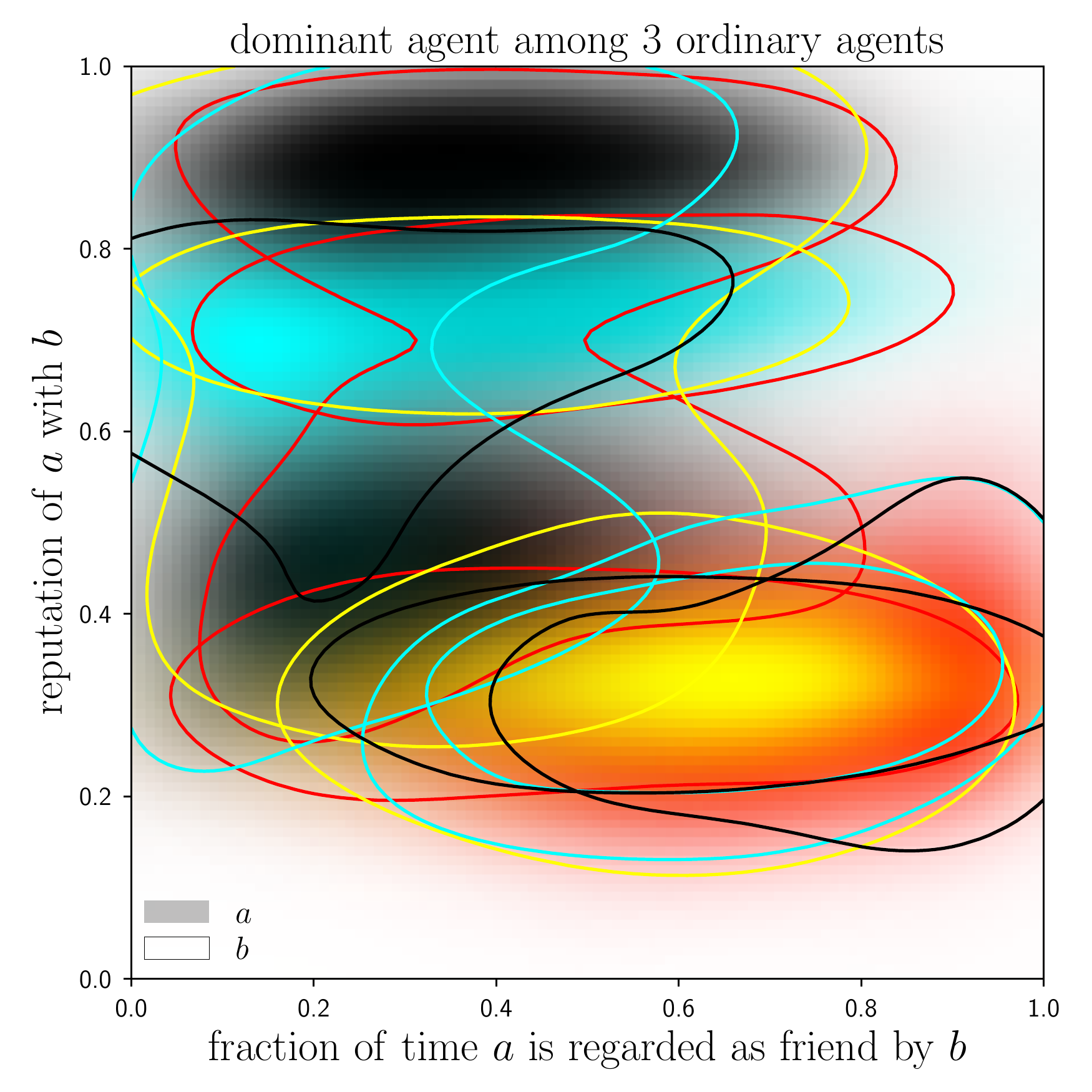}\includegraphics[width=0.3\textwidth]{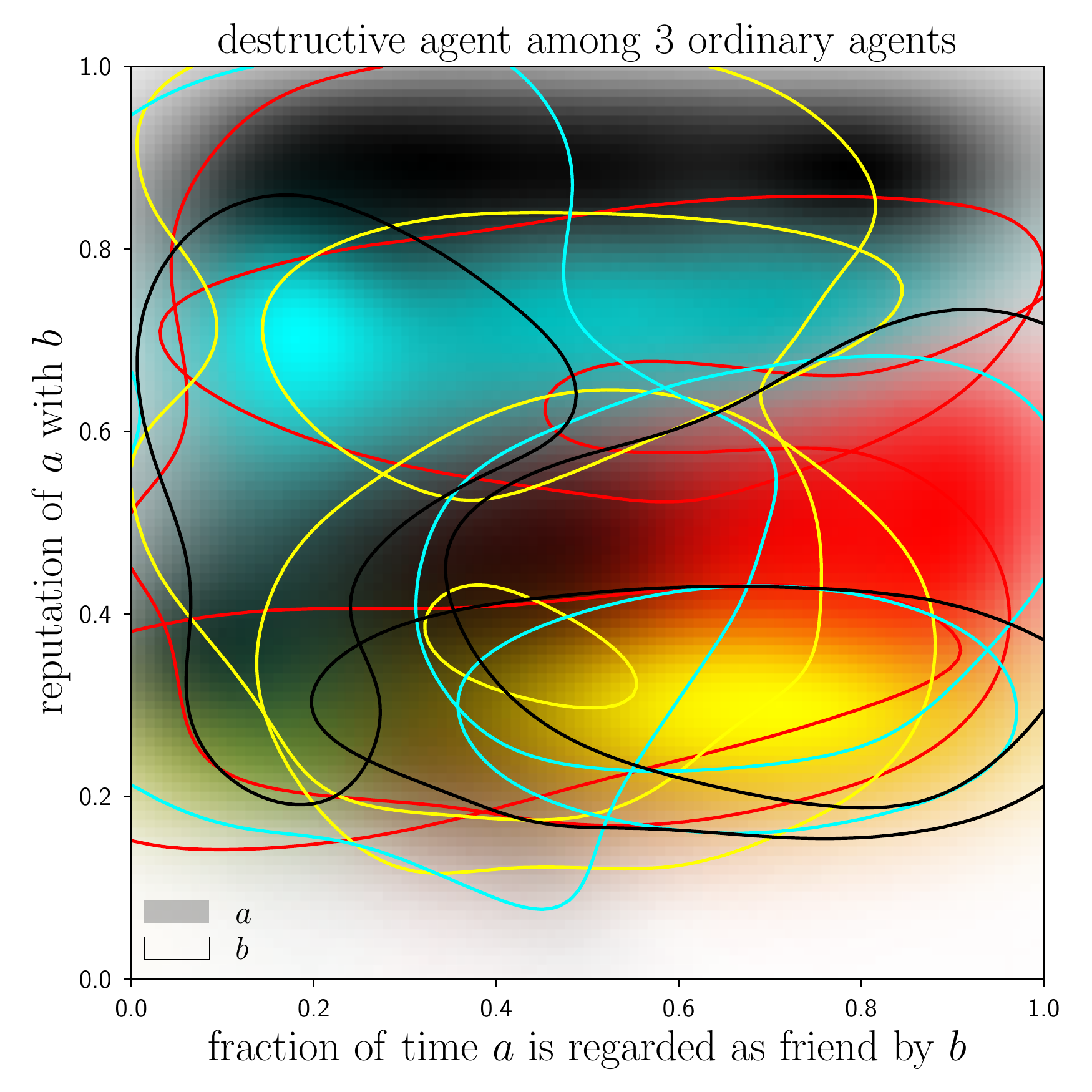}
\par\end{centering}
\begin{centering}
\includegraphics[width=0.3\textwidth]{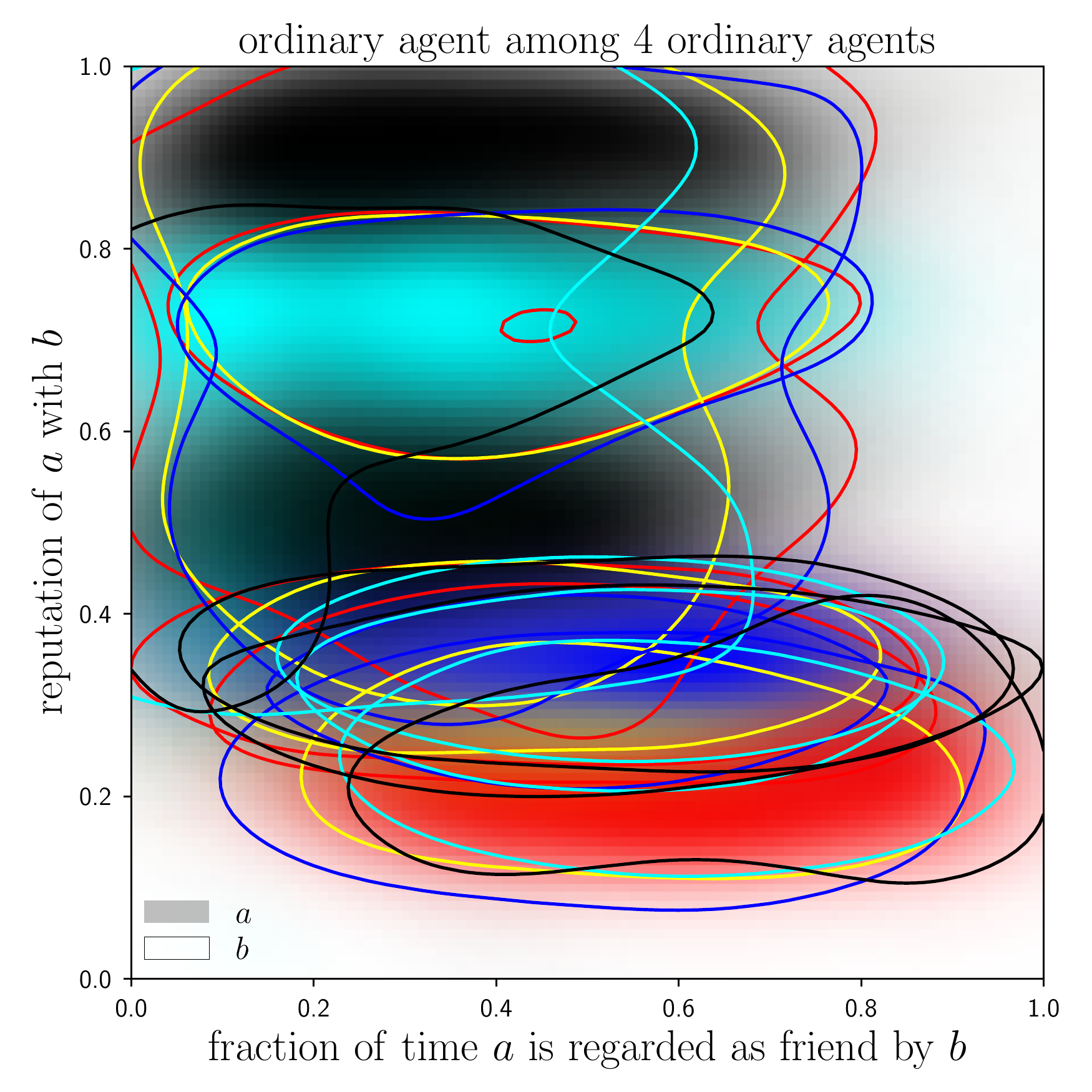}\includegraphics[width=0.3\textwidth]{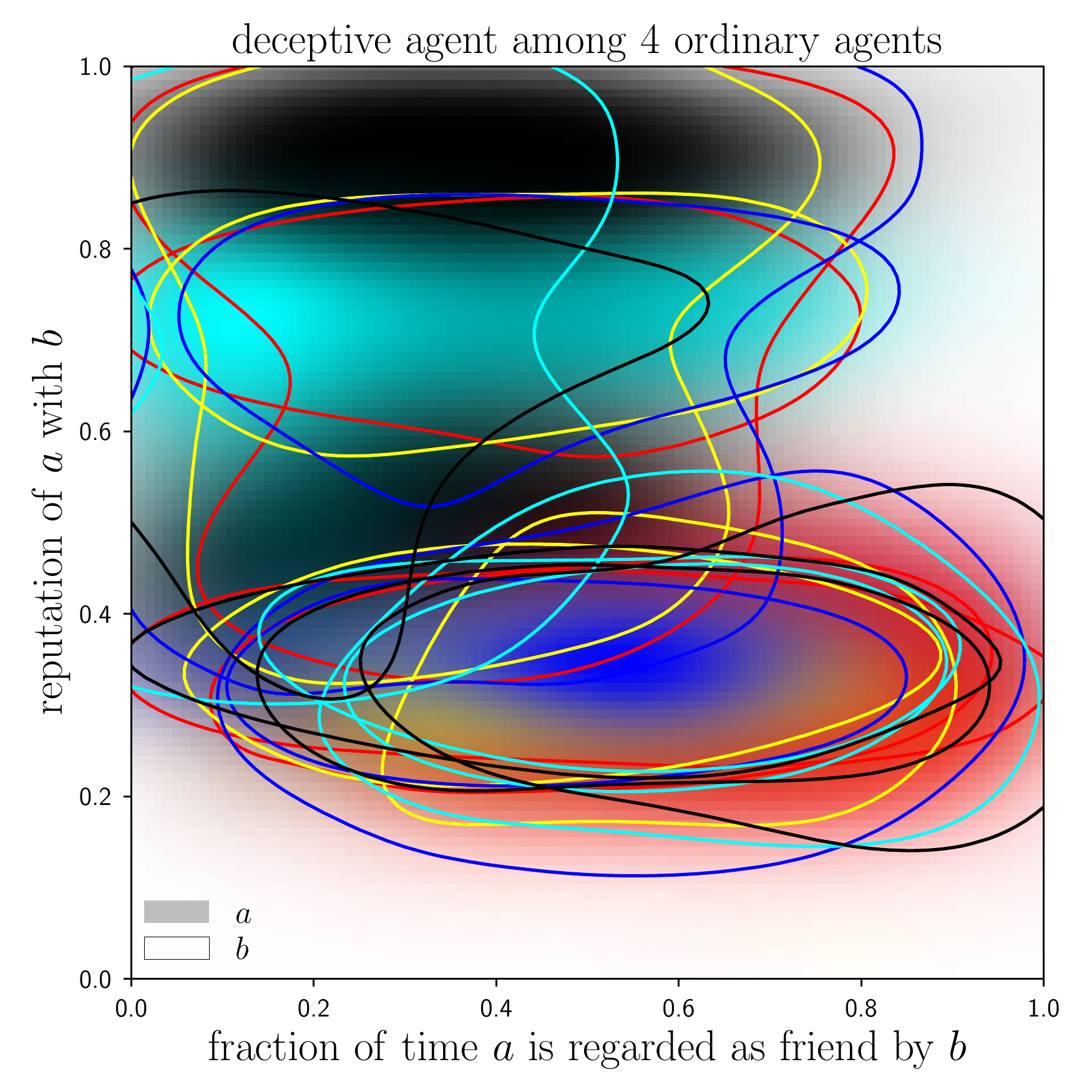}\includegraphics[width=0.3\textwidth]{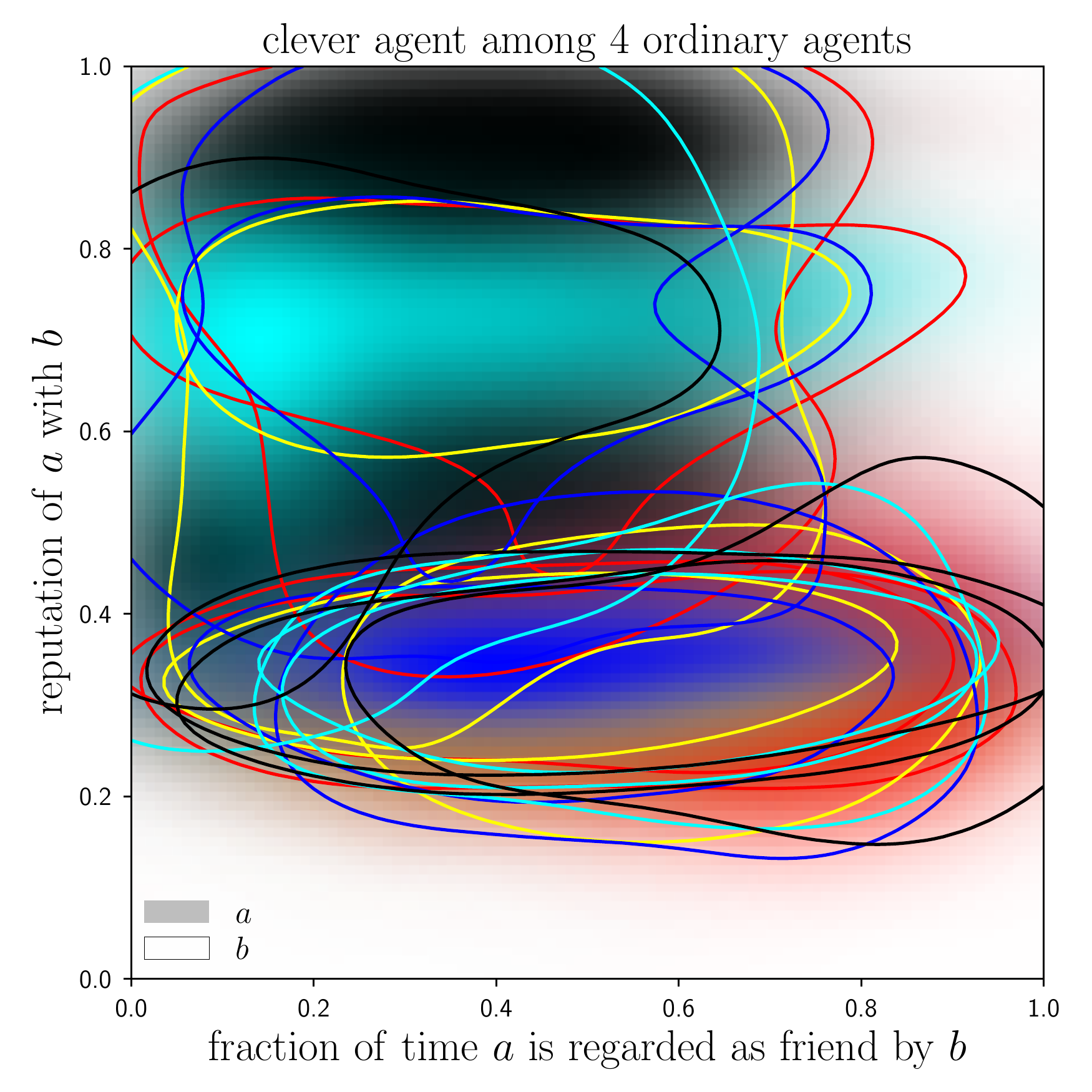}
\par\end{centering}
\begin{centering}
\includegraphics[width=0.3\textwidth]{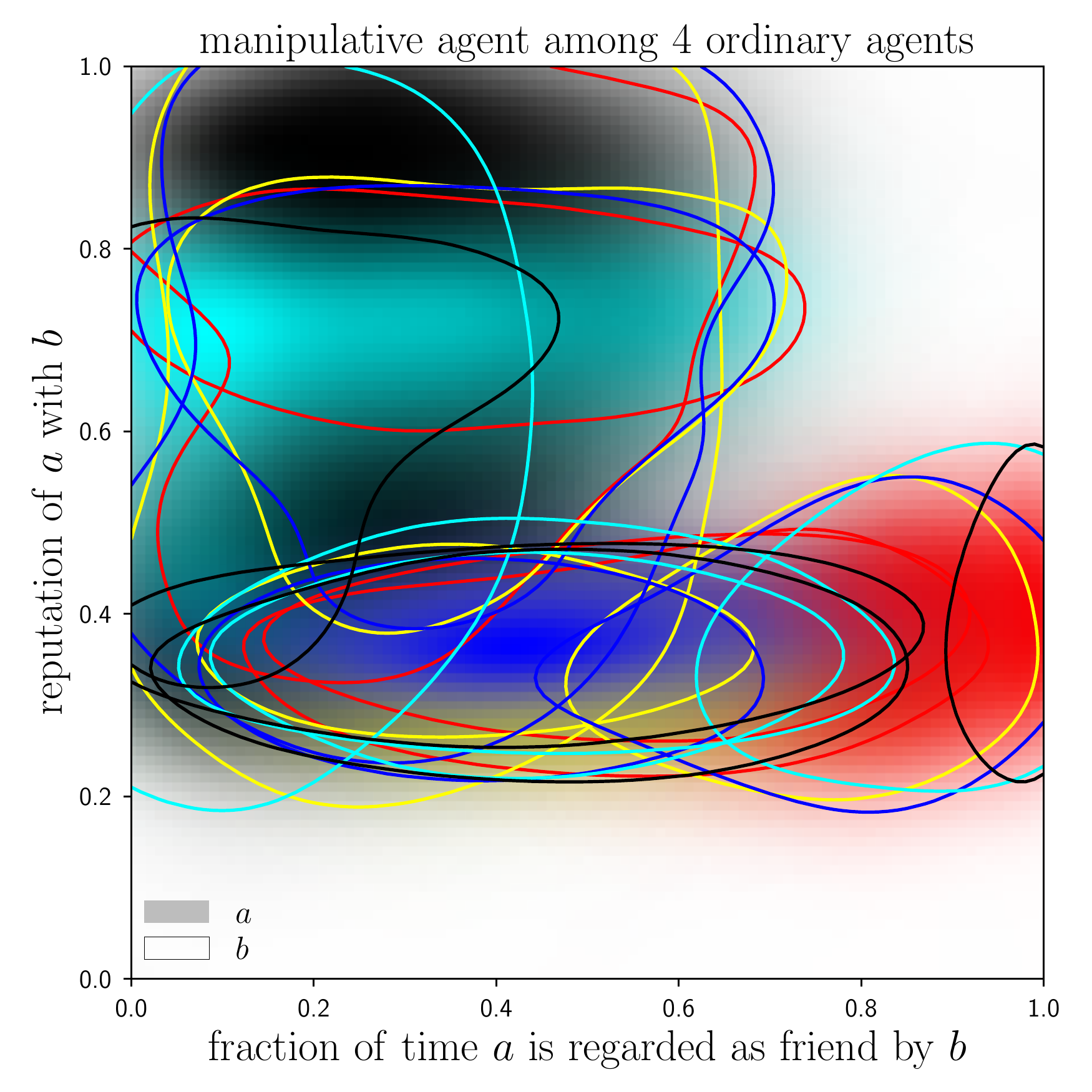}\includegraphics[width=0.3\textwidth]{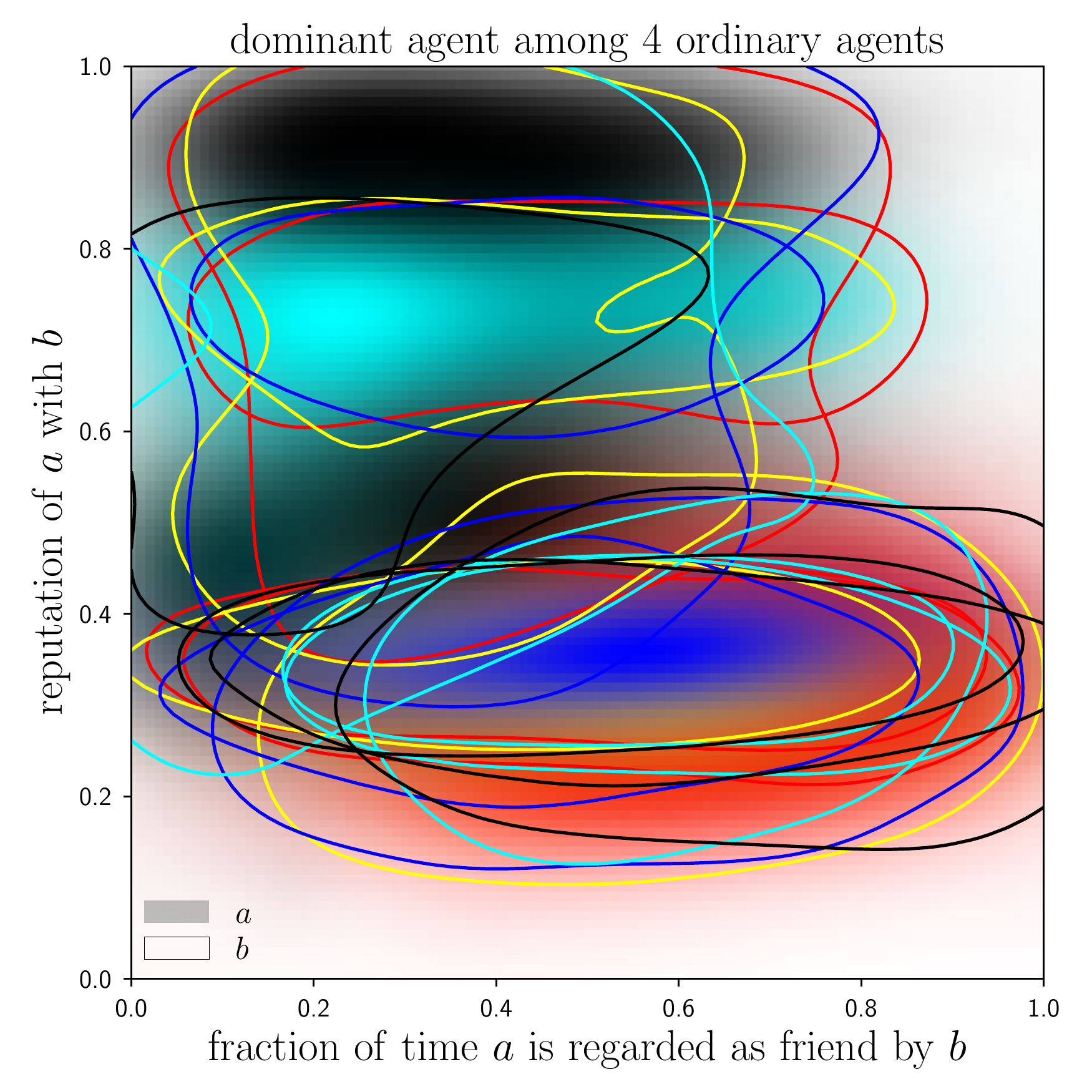}\includegraphics[width=0.3\textwidth]{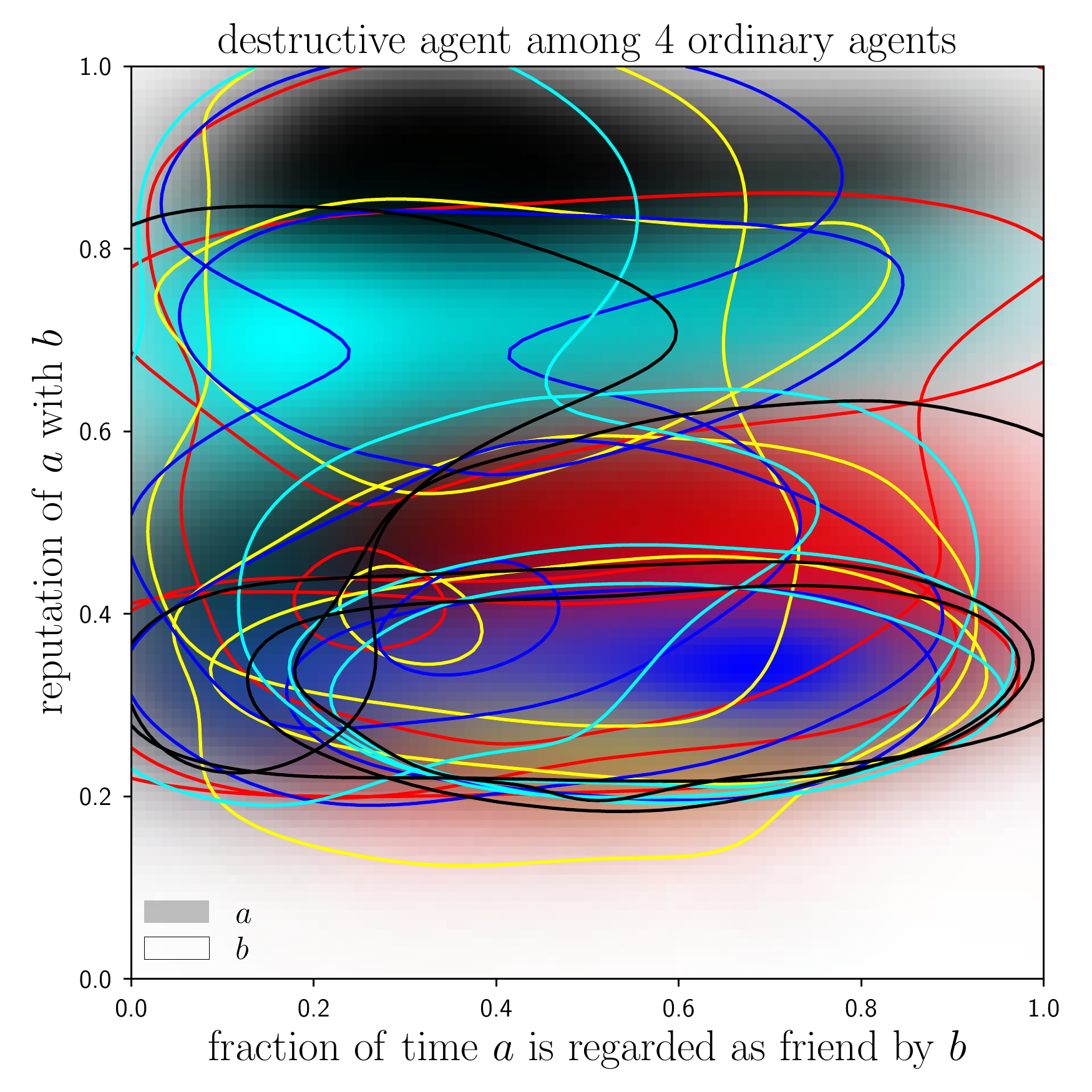}
\par\end{centering}
\caption{Like Fig.\ \ref{fig:Distribution-of-reputation}, just for simulations
with four (upper rows) and five (lower rows) agents. \label{fig:Distribution-of-reputation-1}}
\end{figure*}

\begin{figure*}[!t]
\includegraphics[width=0.3\textwidth]{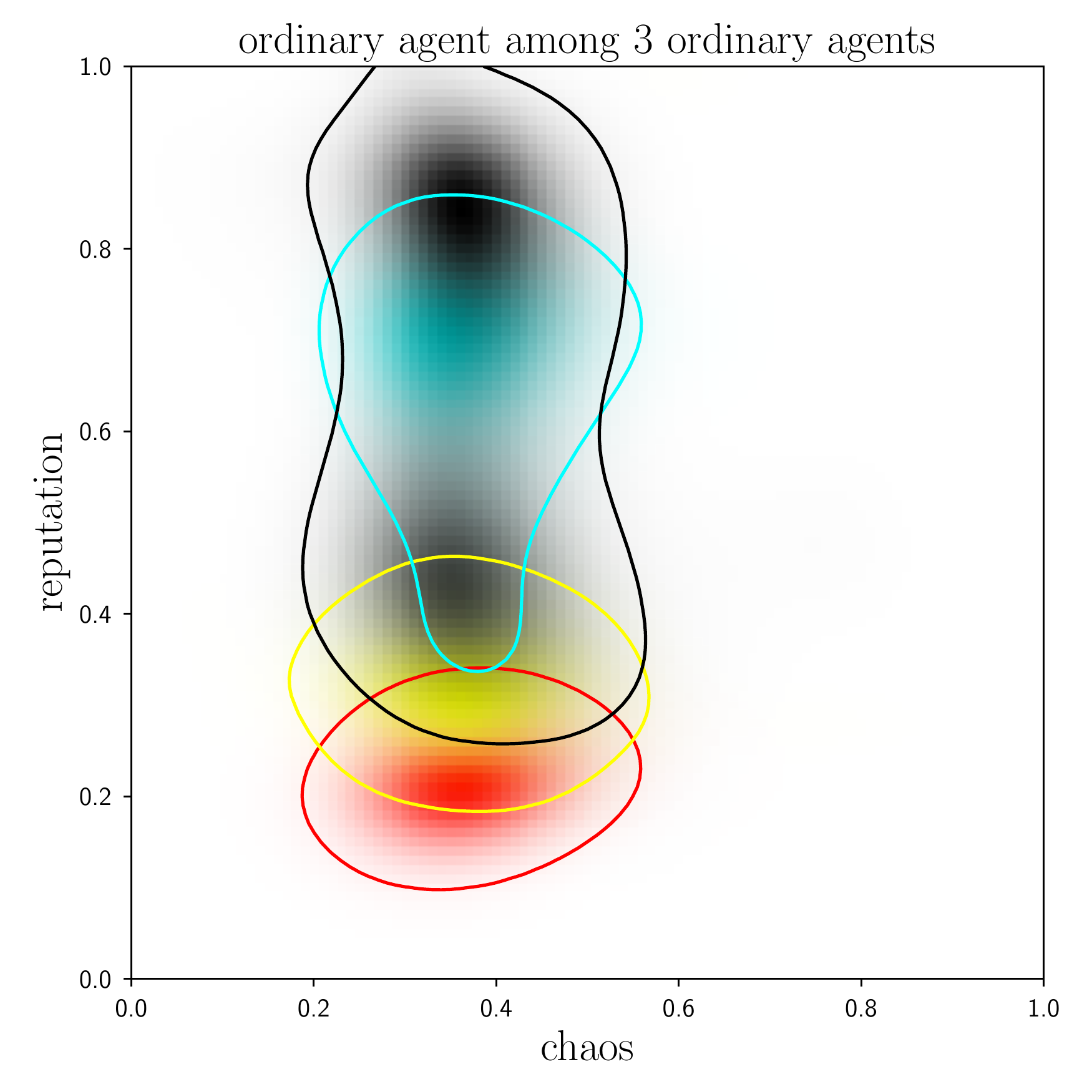}\includegraphics[width=0.3\textwidth]{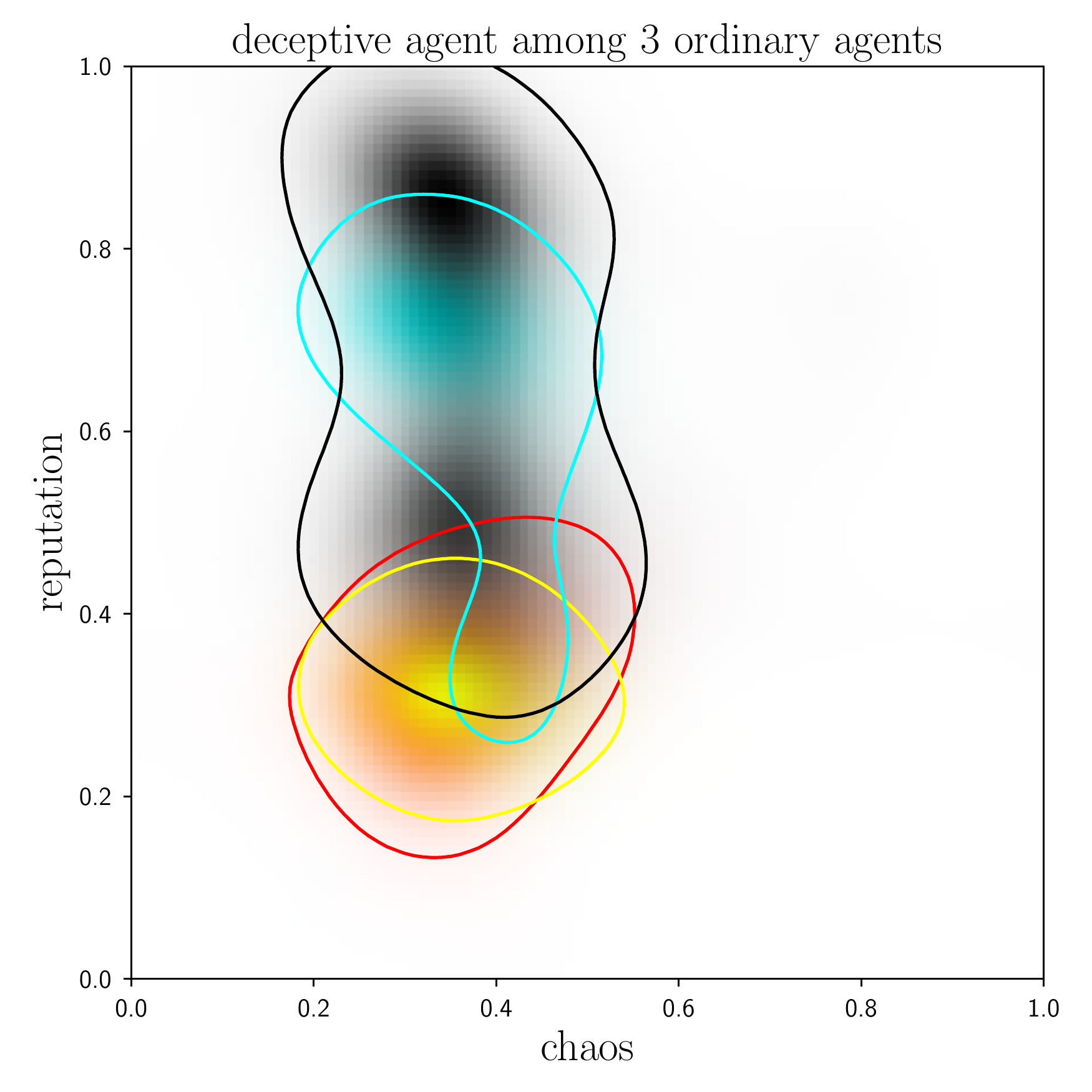}\includegraphics[width=0.3\textwidth]{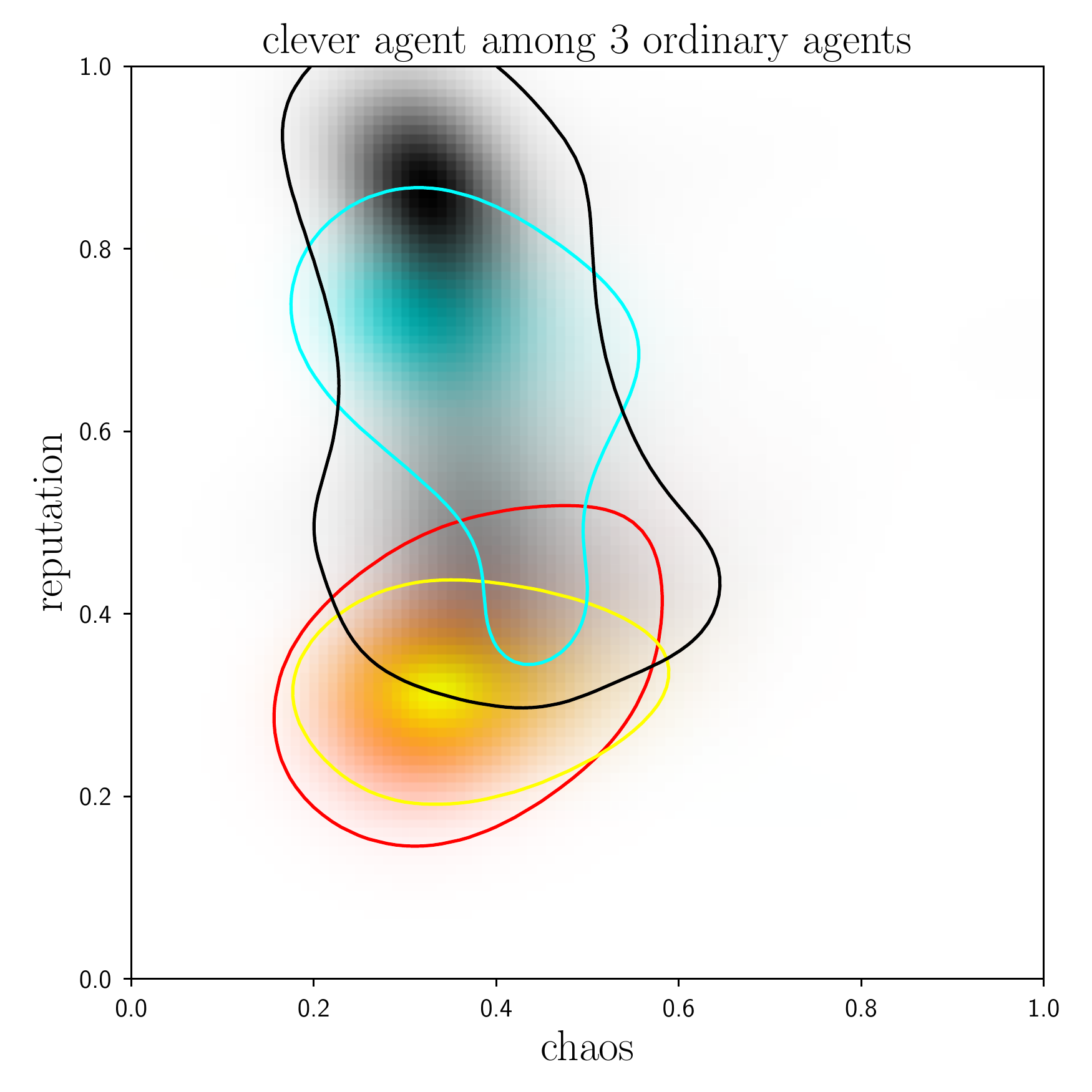}
\begin{centering}
\includegraphics[width=0.3\textwidth]{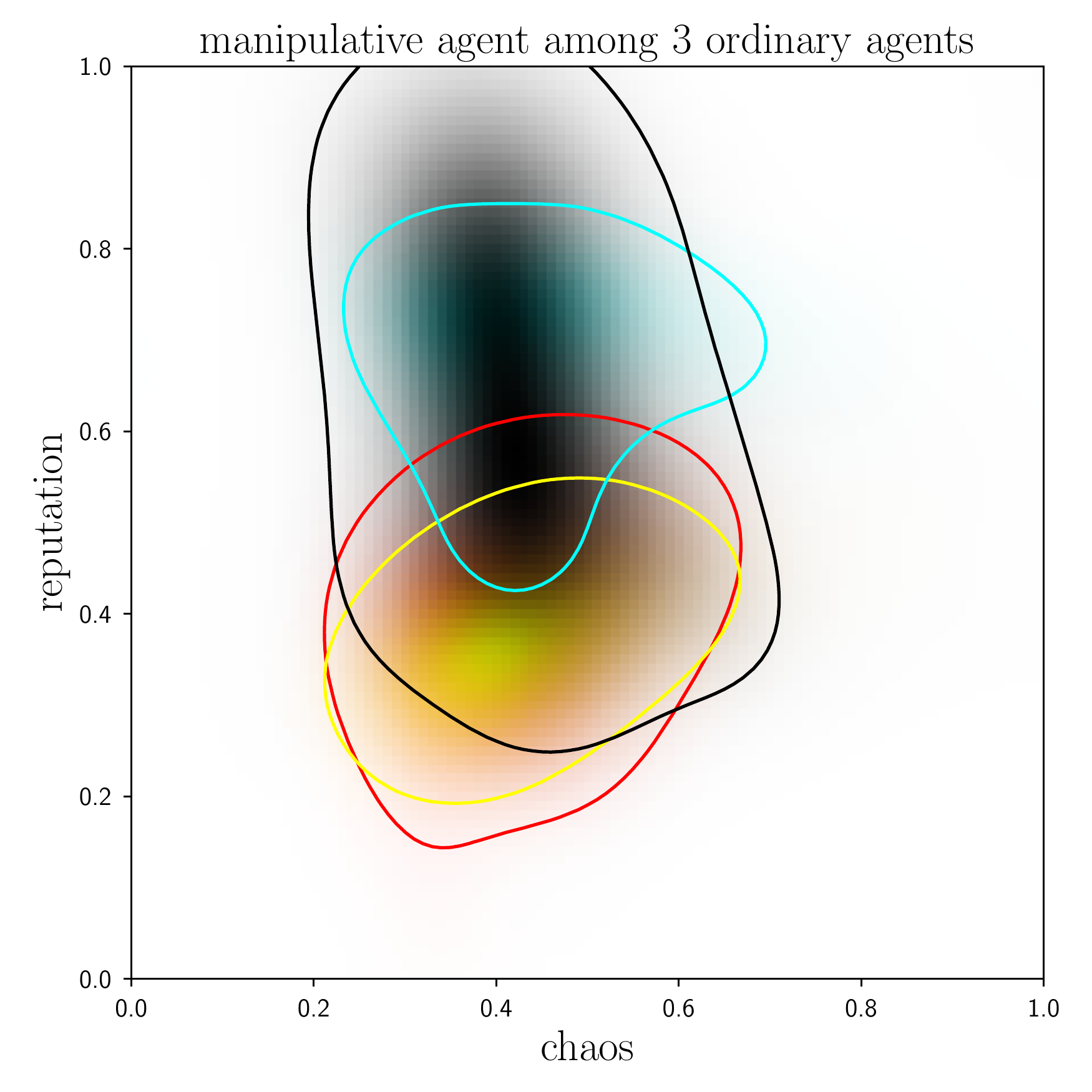}\includegraphics[width=0.3\textwidth]{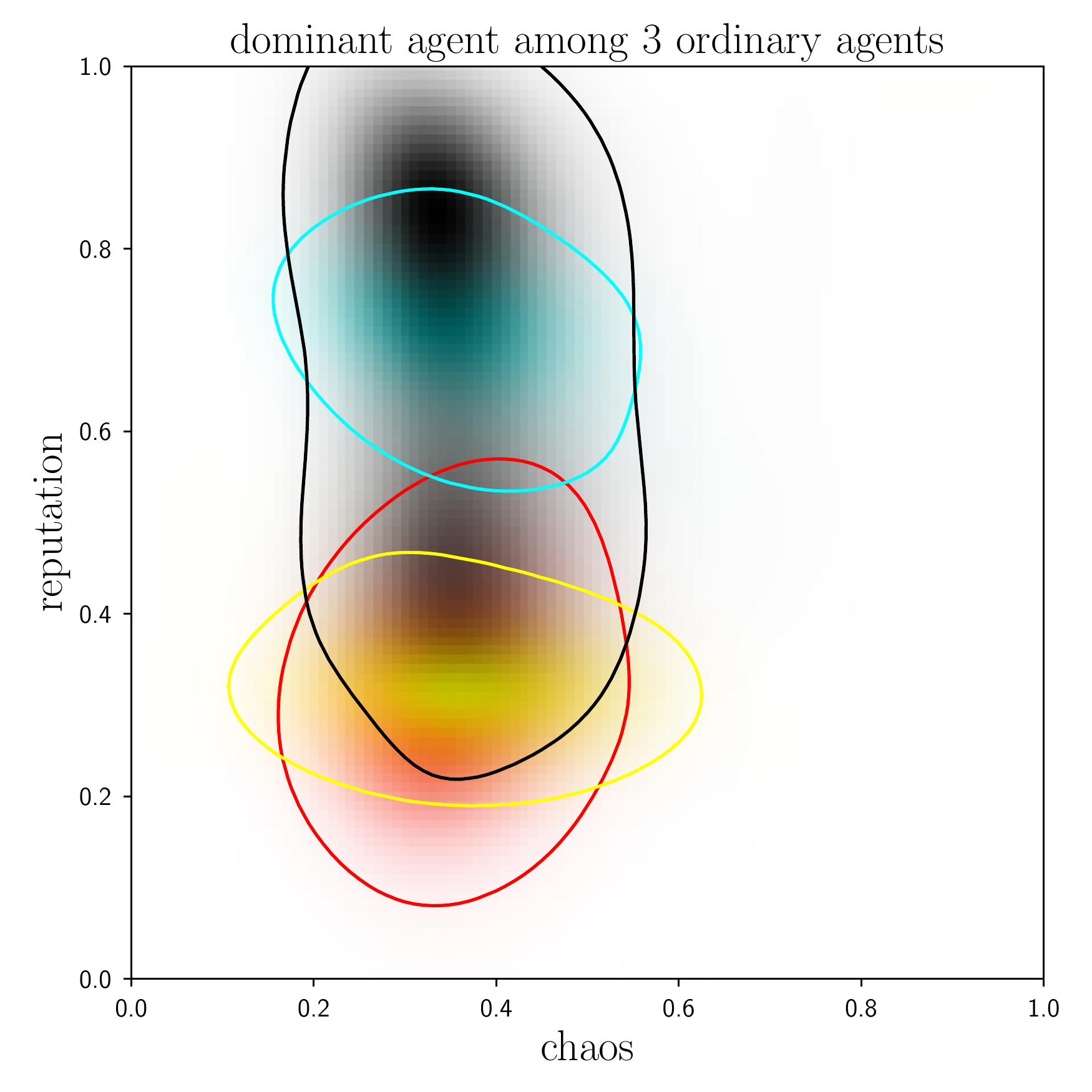}\includegraphics[width=0.3\textwidth]{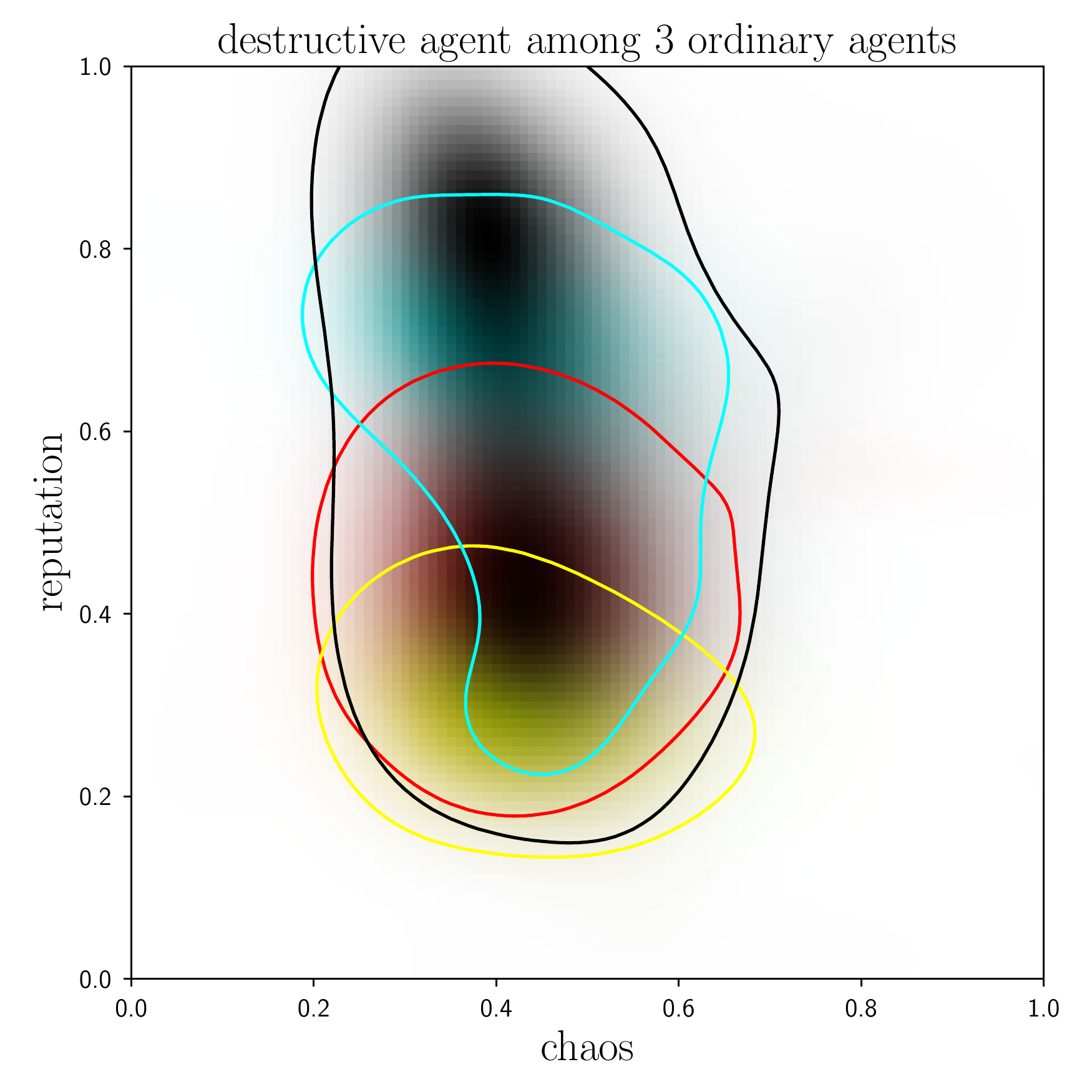}
\par\end{centering}
\begin{centering}
\includegraphics[width=0.3\textwidth]{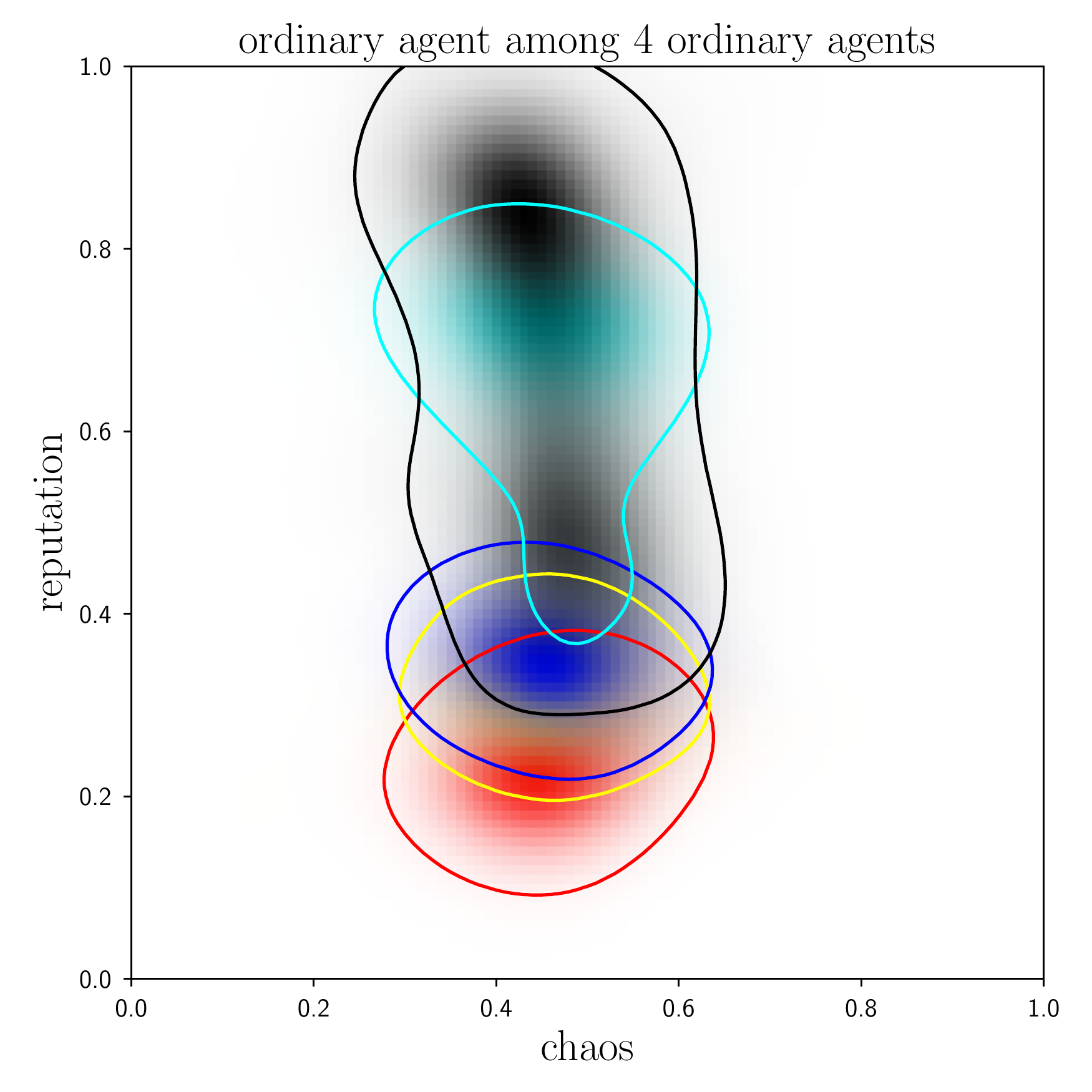}\includegraphics[width=0.3\textwidth]{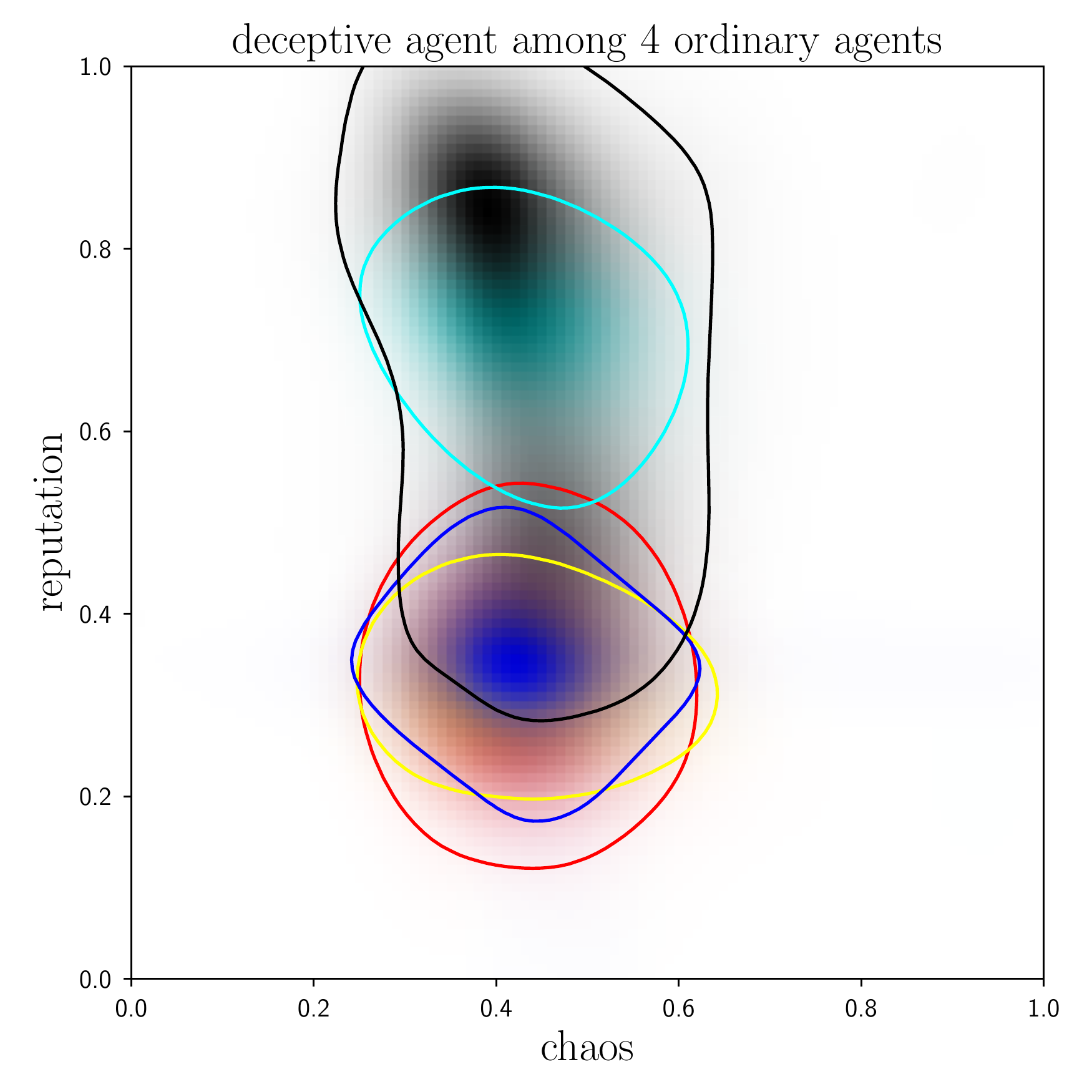}\includegraphics[width=0.3\textwidth]{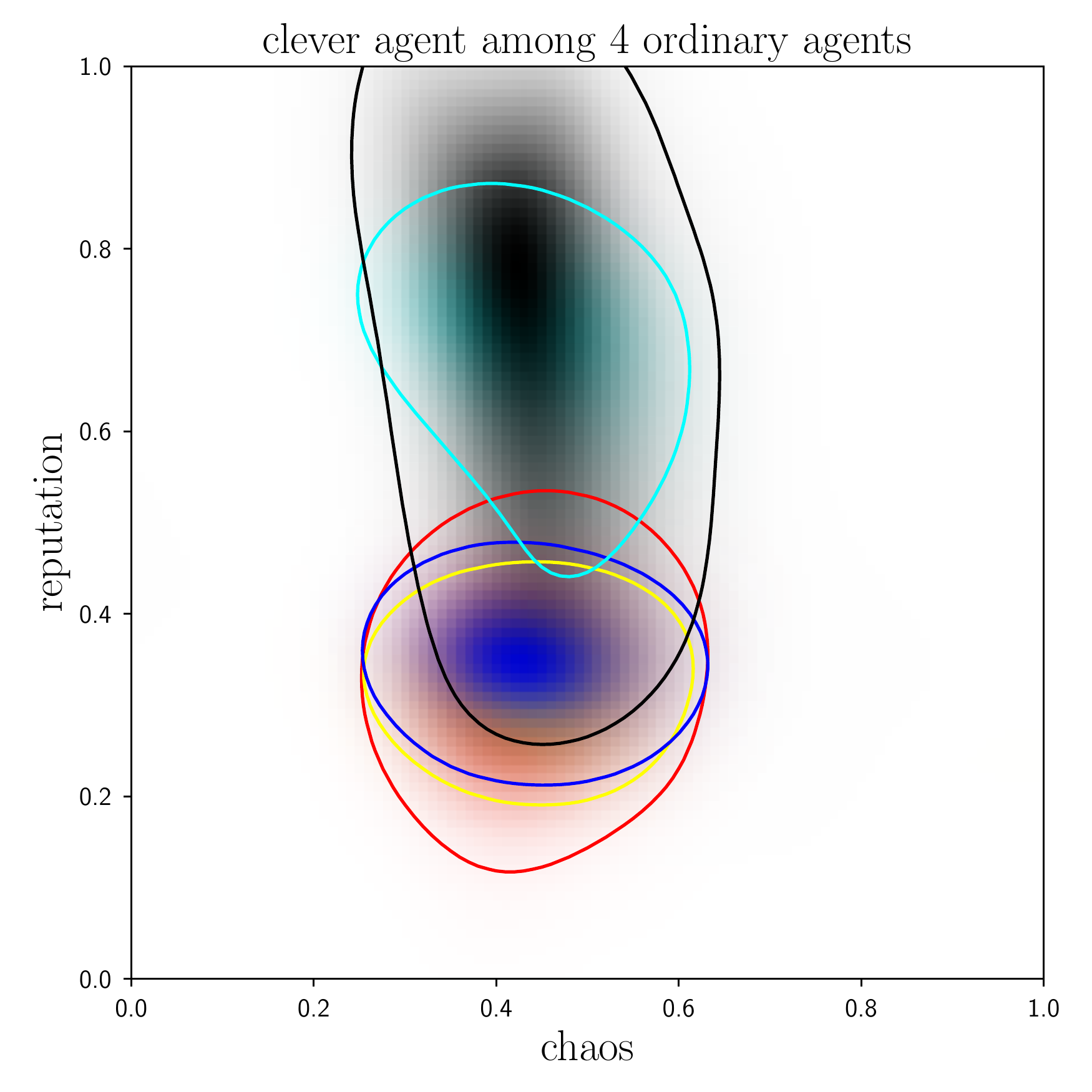}
\par\end{centering}
\begin{centering}
\includegraphics[width=0.3\textwidth]{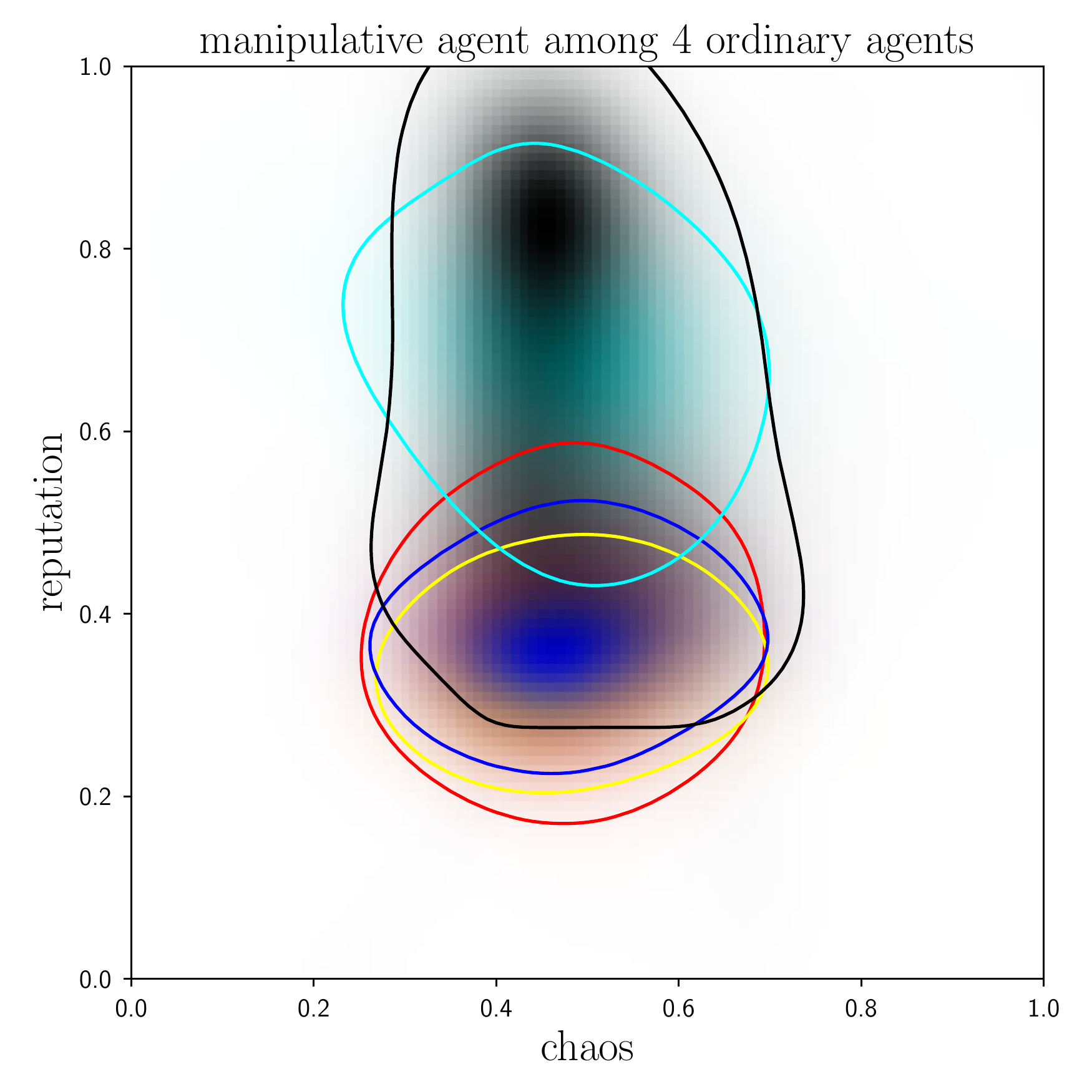}\includegraphics[width=0.3\textwidth]{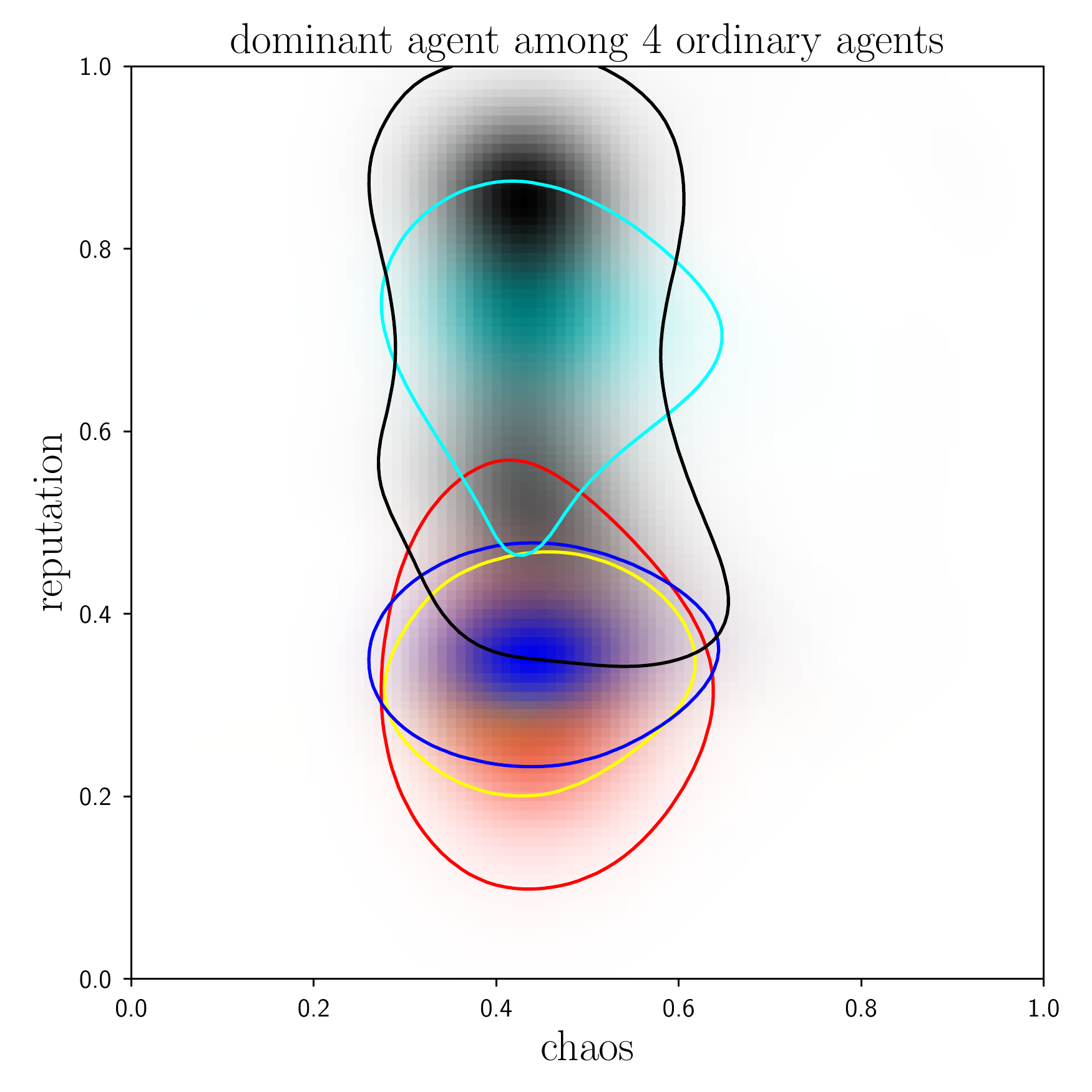}\includegraphics[width=0.3\textwidth]{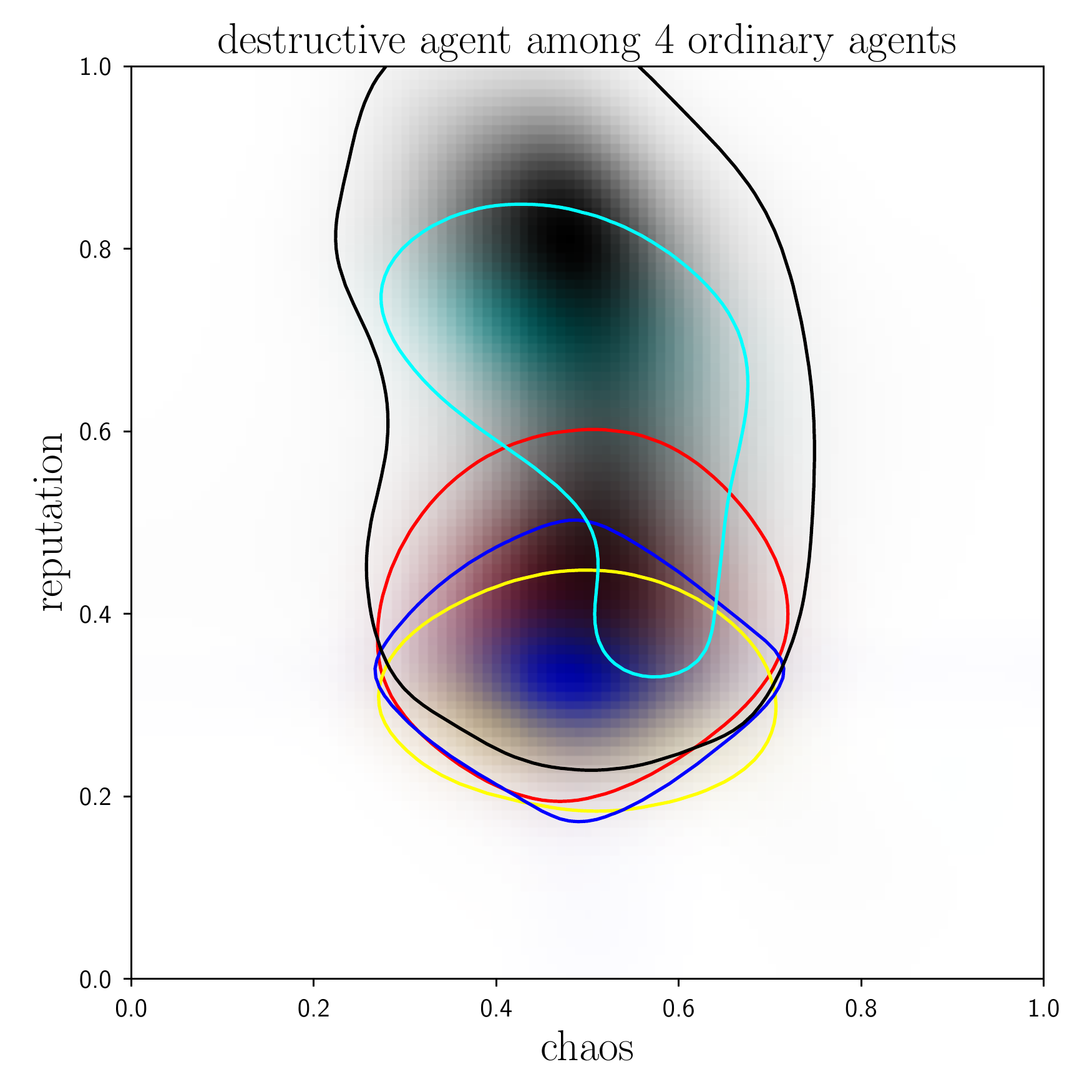}
\par\end{centering}
\caption{Like Fig.\ \ref{fig:Statistics-chaos-3A}, just for simulations with
four (upper rows) and five (lower rows) agents. \label{fig:Statistics-chaos-4A}}
\end{figure*}

Fig.\ \ref{fig:Distribution-of-reputation-1} shows the run averaged
relation of reputation and friendship between agents in the four and
five agent simulations, respectively. Fig.\ \ref{fig:Statistics-chaos-4A}
displays the relation between the run averaged reputation of an agent
and the level of social chaos.

\end{document}